\documentclass[draft,final]{vutinfth} 

\usepackage{lmodern}        
\usepackage[T1]{fontenc}    
\usepackage[utf8]{inputenc} 

\usepackage{amsmath}    
\usepackage{amssymb}    
\usepackage{mathtools}  
\usepackage{microtype}  
\usepackage{multirow}   
\usepackage{booktabs}   
\usepackage{subcaption} 
\usepackage[ruled,linesnumbered,algochapter]{algorithm2e} 
\usepackage[usenames,dvipsnames,table]{xcolor} 
\usepackage{nag}       
\usepackage{todonotes} 
\usepackage{hyperref}  
\usepackage[acronym,toc]{glossaries} 
\usepackage{url}
\usepackage{listings}
\usepackage[sort&compress,numbers]{natbib}
\usepackage{graphicx}
\usepackage{verbatim}
\usepackage{adjustbox} 
\usepackage{makecell}
\usepackage{cancel} 
\usepackage{mdwlist} 
\usepackage{xurl} 

\newcommand{\authorname}{Jovan Jeromela} 
\newcommand{\thesistitle}{Context-Based Tweet Engagement Prediction} 

\hypersetup{
    pdfpagelayout   = TwoPageRight,           
    linkbordercolor = {Melon},                
    pdfauthor       = {\authorname},          
    pdftitle        = {\thesistitle},         
    pdfsubject      = {Subject},              
    pdfkeywords     = {a, list, of, keywords} 
}

\setpnumwidth{2.5em}        
\setsecnumdepth{subsection} 

\nonzeroparskip             
\makeindex      
\makeglossaries 

\setauthor{}{\authorname}{BSc}{male}
\setadvisor{Associate Prof. Dipl.-Ing. Dr.techn.}{Peter Knees}{}{male}

\setfirstreviewer{Pretitle}{Forename Surname}{Posttitle}{male}
\setsecondreviewer{Pretitle}{Forename Surname}{Posttitle}{male}

\setsecondadvisor{Pretitle}{Forename Surname}{Posttitle}{male} 

\setregnumber{01528091}
\setdate{25}{05}{2023} 
\settitle{\thesistitle}{Context-Based Tweet Engagement Prediction} 

\setthesis{master}
\setmasterdegree{dipl.} 
\setcurriculum{Data Science}{Data Science} 

\setfirstreviewerdata{Affiliation, Country}
\setsecondreviewerdata{Affiliation, Country}

\begin{document}

\frontmatter 

\addtitlepage{naustrian} 
\addtitlepage{english} 
\addstatementpage


\begin{acknowledgements*}
Foremost, I want to thank my supervisor Associate Prof. Dipl.-Ing. Dr.techn. Peter Knees for his valuable guidance during the entire process of writing this thesis, for his quick responses to my questions, and for his understanding regarding obstacles that prolonged the completion of this work.

I also want to thank my entire family -- especially my parents and grandparents -- from the bottom of my heart for the encouragement and support, both financial and emotional, without which my graduation would not have been possible.
\end{acknowledgements*}

\begin{kurzfassung}
Twitter ist aktuell eine der größten Social-Media-Plattformen. Seine Benutzerinnen und Benutzer können kurze Beiträge -- sogenannte Tweets -- teilen, lesen und mit ihnen interagieren. Für die ACM Recommender Systems Conference 2020 veröffentlichte Twitter einen rund 70 GB großen Datensatz für die jährliche RecSys Challenge. Im Jahr 2020 lud die RecSys Challenge teilnehmende Teams ein, Modelle zu erstellen, die die Wahrscheinlichkeit von Interaktionen für bestimmte Benutzer-Tweet-Kombinationen vorhersagen würden. Die eingereichten Modelle zur Vorhersage von Like-, Reply-, Retweet- und Quote-Interaktionen wurden anhand von zwei Metriken bewertet: die Fläche unter der Precision-Recall-Kurve (PRAUC) und die relative Kreuzentropie (RCE). 

In dieser Diplomarbeit haben wir den Datensatz und das Evaluationsverfahren der RecSys 2020 Challenge verwendet, um zu untersuchen, wie gut der Kontext allein zur Vorhersage der Wahrscheinlichkeit einer Tweet-Interaktion verwendet werden kann. Dabei haben wir die Spark-Engine auf dem Little Big Data Cluster der TU Wien eingesetzt, um skalierbare Pipelines für Datenvorverarbeitung, Feature Engineering, Feature Selection und maschinelles Lerner zu erstellen. Manuell haben wir knapp 200 zusätzliche Features gestaltet, um den Tweet-Kontext zu beschreiben. 

Die Ergebnisse zeigen, dass Features, die die bisherige Interaktionshistorie der Benutzerinnen und Benutzer und die Popularität von Hashtags und Links im Tweet beschreiben, am informativsten waren. Wir haben außerdem festgestellt, dass Faktoren wie der Vorhersagealgorithmus, die Größe des Trainingsdatensatzes, die Stichprobenmethode des Trainingsdatensatzes und die Feature Selection die Ergebnisse signifikant beeinflussen. Der Vergleich der besten Ergebnisse unserer rein kontextbasierten Vorhersagemodelle mit rein inhaltsbasierten Modellen und mit von den Challenge-Gewinnern erstellten Modellen zeigt, dass die kontextbasierten Modelle schlechtere RCE-Werte erzielt haben. Diese Arbeit schließt ab, indem sie diese Diskrepanz erwähnt und potenzielle Verbesserungen für unsere Implementierung, die in einem öffentlichen Git-Repository geteilt wird, vorschlägt. 
\end{kurzfassung}

\begin{abstract}
Twitter is currently one of the biggest social media platforms. Its users may share, read, and engage with short posts called tweets.  For the ACM Recommender Systems Conference 2020, Twitter published a dataset around 70 GB in size for the annual RecSys Challenge. In 2020, the RecSys Challenge invited participating teams to create models that would predict engagement likelihoods for given user-tweet combinations. The submitted models predicting like, reply, retweet, and quote engagements were evaluated based on two metrics: area under the precision-recall curve (PRAUC) and relative cross-entropy (RCE). 

In this diploma thesis, we used the RecSys 2020 Challenge dataset and evaluation procedure to investigate how well context alone may be used to predict tweet engagement likelihood. In doing so, we employed the Spark engine on TU Wien’s Little Big Data Cluster to create scalable data preprocessing, feature engineering, feature selection, and machine learning pipelines. We manually created just under 200 additional features to describe tweet context. 

The results indicate that features describing users’ prior engagement history and the popularity of hashtags and links in the tweet were the most informative. We also found that factors such as the prediction algorithm, training dataset size, training dataset sampling method, and feature selection significantly affect the results.  After comparing the best results of our context-only prediction models with content-only models and with models developed by the Challenge winners, we identified that the context-based models underperformed in terms of the RCE score. This work thus concludes by situating this discrepancy and proposing potential improvements to our implementation, which is shared in a public git repository. 

\end{abstract}

\selectlanguage{english}

\tableofcontents 

\mainmatter

\chapter{Introduction}
\label{chIntroduction}

This diploma thesis investigates tweet engagement prediction, i.e. forecasting the probability that users of a social media platform Twitter would interact with the posts they see. We propose, implement, and evaluate classification models that are based solely on the tweet's context, disregarding the text of the tweet entirely. These models revolve around data published for the RecSys 2020 Challenge\footnote{\label{note1}\url{https://recsys-twitter.com/previous_challenge} (last access: 2021-05-22)}, presented in \cite{tweetPap}. This dataset was provided by Twitter and it contains information about a set of users and posts on this social network. The goal of the surrounding recommendation systems challenge was to predict engagement probabilities for provided combinations of individual users and tweets for four distinct engagement types: likes, replies, retweets, and retweets with comments (also called quotes).

\section{Motivation}
\label{secMotivation}

With more than 350 million active users\footnote{\url{https://www.statista.com/statistics/272014/global-social-networks-ranked-by-number-of-users/} (last access: 2021-05-22)}, Twitter is one of the most widely used social networks. As such, Twitter has inspired copious amounts of research focusing on the relation between activity on the website and influenza outbreaks (e.g. \cite{twitterInfluenza}), elections (e.g. \cite{twitterElections}), suicidality (e.g. \cite{twitterSuicidality}), natural disasters (e.g. \cite{twitterDisasters}), marketing (e.g. \cite{twitterMarketing}), stock market developments (e.g. \cite{twitterStockMarket}), etc. Such a wide spectrum of contexts and topics discussed on Twitter means that providing relevant content to users is a task that is as important as it is complex, as elaborated in \cite{tweetPap, recsys2020overview}. While users can opt to see tweets of users they chose to follow in a counter-chronological order, Twitter's default Home timeline displays first those tweets which were deemed most relevant for the target user. By providing a dataset and the task to the 2020 ACM Recommender Systems Challenge, Twitter got the opportunity to motivate novel approaches from both the industry and academia, that it can then use to improve this tweet recommendation mechanism. 

For participants of the challenge, Twitter's dataset and task were exceptional for a number of reasons. First, the dataset itself is significantly larger than those provided in previous RecSys challenges (cf. e.g. \cite{RecSys2014overview, RecSys2018overview, RecSys2019overview}), prompting the need to use appropriate system architecture (distributed file systems, multiple GPUs,...), adequate frameworks (e.g. Apache Spark, Amazon Web Services, etc.), fitting sampling techniques, and computationally efficient algorithms. Second, the dataset contains some of the features which are not normally accessible through Twitter's API, leading to the possibility of gaining novel insights into the communication and engagement patterns of the social network's users. Last, while it is possible to use a different model for each of the four engagement types (replies, likes, retweets, and quotes), each model is evaluated using two evaluation metrics (PRAUC and RCE, cf. section \ref{secMetrics}) propelling the need to develop suitable hyper-parameter tuning and optimisation techniques. 

Working on this task after the challenge deadline had passed and the winning solutions had been published, allowed us to examine and compare best-performing approaches. Insights stemming from this examination, combined with original ideas, were then applied to a subtly, yet decidedly different prediction task which focuses exclusively on the contextual features. In this manner, we were able to investigate the extent to which the importance of individual explanatory variables and the appropriateness of proposed algorithms remains unaffected by the lack of tweet text. Simultaneously, the importance of tweet context per se was evaluated and compared to that of tweet content. This was achieved by contrasting the performance of context-only models presented in this thesis to that of content-only models presented in \cite{davidsThesis}. Moreover, as the implementation of the proposed models was done primarily on ``Little Big Data'', a cluster computer of TU Wien (cf. section \ref{secLBD}) using a regular student account antecedently created for the purposes of a university course, this thesis can serve as a testament to computation capabilities available to data science students at the university at the time.

\section{Problem Statement}
\label{secProblemStatement}
The dataset of the RecSys 2020 Challenge constituted the foundation of the problem investigated in the thesis. This dataset, described in detail by the challenge organisers in \cite{tweetPap} and further investigated by Gradinariu in \cite{davidsThesis}, comprises information about Twitter users, tweets, and the so-called engagement features -- containing a feature indicating whether the author of the tweet follows the potential engager as well as timestamps of individual engagement actions if these actions had indeed occurred. The data is split by the organisers into three distinct and disjoint subsets: training (or train, for short), validation (or val), and test dataset. The training dataset contains approximately 160 million tweets sampled within a week in early 2020 while testing and validation sets contain a further 40 million in total, which were sampled during the following week.

Each instance (row) of the training dataset contains features listed in table \ref{tabRows}. While test and validation datasets contain all tweet and user features from table \ref{tabRows}, the only engagement feature in these two datasets that had not been hidden prior to the end of the challenge was whether the author's account followed the account that had potentially engaged with the tweet. In essence, the remaining four engagement features served as the target variables for the classification task at hand.

\begin{table}[ht]
\centering
\begin{adjustbox}{width=1\textwidth}
\begin{tabular}{|l|l|l|}
\hline
\textbf{}                                                         & \textbf{Feature Name}                                                                                                                                           & \textbf{Feature Description}
\\ \hline
\textbf{Tweet}                                                    & \begin{tabular}[c]{@{}l@{}}Text tokens\\ Hashtags\\ Tweet id\\ Present media\\ Present links\\ Present domains\\ Tweet type\\ Language\\ Timestamp\end{tabular} & \begin{tabular}[c]{@{}l@{}}Ordered list of Bert ids corresponding to Bert tokenization of Tweet text\\ Tab separated list of hastags (identifiers) present in the tweet\\ Tweet identifier\\ Tab separated list of media types. Type can be in (Photo, Video, Gif)\\ Tab separeted list of links (identifiers) included in the Tweet\\ Tab separated list of domains included in the Tweet\\ Tweet type, can be either Retweet, Quote, Reply, or Toplevel\\ Identifier corresponding to the inferred language of the Tweet\\ Unix timestamp, in sec of the creation time of the Tweet\end{tabular} 
\\ \hline
\textbf{\begin{tabular}[c]{@{}l@{}}User\\ (Author)\end{tabular}}  & \begin{tabular}[c]{@{}l@{}}User id\\ Follower count\\ Following count\\ Is verified?\\ Account creation time\end{tabular}                                       & \begin{tabular}[c]{@{}l@{}}User identifier\\ Number of followers of the user\\ Number of accounts the user is following\\ Is the account verified?\\ Unix timestamp, in seconds, of the creation time of the account\end{tabular}
\\ \hline
\textbf{\begin{tabular}[c]{@{}l@{}}User\\ (Engager)\end{tabular}} & \begin{tabular}[c]{@{}l@{}}User id\\ Follower count\\ Following count\\ Is verified?\\ Account creation time\end{tabular}                                       & \begin{tabular}[c]{@{}l@{}}User identifier\\ Number of followers of the user\\ Number of accounts the user is following\\ Is the account verified?\\ Unix timestamp, in seconds, of the creation time of the account\end{tabular}                                       \\ \hline
\textbf{Engagement}                                               & \begin{tabular}[c]{@{}l@{}}Engagee follows engager?\\ Reply timestamp\\ Retweet timestamp\\ Retweet+comment\\     timestamp\\ Like timestamp\end{tabular}       & \begin{tabular}[c]{@{}l@{}}Does the account of the author follow the account that has engaged?\\ If there is at least one, unix timestamp, in s, of one of the replies\\ If there is one, unix timestamp, in s, of the retweet of th\\ If there is at least one, unix timestamp, in s, of one of the \\     retweet with comment of the tweet by the engaging user\\ If there is one, Unix timestamp, in s, of the like
\end{tabular}
\\ \hline
\end{tabular}
\end{adjustbox}
\caption[ignore this error]{Details about the dataset, as seen on RecSys 2020 challenge homepage}
\label{tabRows}
\end{table}

Formally, the task of the challenge was to create predictions in the format $<$Tweet\_ID$>$, $<$User\_ID$>$\footnote{The user ID refers to the engager, i.e. the account that has seen the tweet. While the same tweet can be seen by multiple users, each of those users can see it at most once (from the perspective of the data sampled in this dataset). Similarly, it was unnecessary to also submit the author's user ID, since each tweet is uniquely linked to a single author.}, $<$Prediction$>$. The values for the predictions were to take to a value between zero and one representing the esitmated likelihood that the given user had engaged in the respective manner (like, reply,  retweet,  retweet  with  comment) with the given tweet.

\section{Aim of the Work}
\label{secAim}

The main aim of the thesis was to develop meaningful (w.r.t. the selected evaluation metrics, cf. section \ref{secMetrics}) models based on all features with the sole exception of text tokens (i.e. tweet text). We dubbed this subset of features the \emph{context} of the tweet, in contrast to the \emph{content} of the tweet. Conversely, the methods that rely solely on the tweet's content were investigated in \cite{davidsThesis}. 

One might argue that some of these features (such as the hashtags or the language) could be seen as the tweet's content rather than context. However, a good counterargument would be that those aspects of the tweet are emblematic of a larger conversation the tweet is a part of (i.e. the context) rather than just of the tweet's content. Similarly, the number of media elements, the hashtags, and the links are descriptive of the type of conversation that had encompassed the tweet. Lastly, one could coherently claim that certain aspects of a dataset ought to be seen as parts of both the tweet's content and context.

In addition, this thesis investigated the efficiency and scalability of all of the developed approaches, as further elaborated in chapter \ref{chImplementation}.

\section{Methodological Approach}
\label{methodologicalApproach}

The main motivation of the diploma thesis is to investigate the role and significance of context for tweet engagement prediction. Thus the central point of the work was engineering additional features from the features provided by the challenge organisers. From 17 explanatory features (excluding the three ID fields and four target engagements), we ended up with over 300 explanatory features, which were then grouped into two feature sets. The two feature sets differ in whether they include the ``oracle'' features, i.e. engineered features that contain information about the events that took place after the user had seen the target tweet. The performance of models based on each of these two feature sets was compared.

These new features were created for the full test, val, test, and combined test+val datasets as well as for $1\%$, $2\%$, $5\%$, and $10\%$ subsets of each of these datasets. The subsets were sampled in five distinct manners: completely randomly; preserving the ratios of individual counts of engaging users to the number of total instances in the dataset (i.e. the size of the dataset); preserving the ratio of the engaged-with users to total instances; preserving the ratios of both the engaging users and engaged-with users to total instances; as well as preserving the ratio of individual tweets to total instances in the dataset. This resulted in a total of $4 \text{ subset sizes} \cdot 5 \text{ sampling methods} \cdot 4 \text{ dataset sources} + 4 \text{ unsampled full datasets}  = 84$ sets of data.

Initially, we implemented a baseline prediction baseline. It was based solely on the basic 17 features pre-provided by the event organisers. The classification was done by applying logistic regression and decision trees. Thereafter, we ran the main evaluation that relied on the engineered features. Firstly, we tested only the non-oracle features by applying random forests, decision trees, gradient boosting, and na\"ive Bayes. We also tested whether the $\chi^2$ (Chi-Squared) feature selection was effective. The second part of the main round was calculated only locally due to the closure of the Little Big Data cluster (cf. section \ref{secLBD}). Therefore, in this second part of the main evaluation, we could only use $1\%$ and $2\%$ subsets. Still, we tested both feature sets (including and excluding oracle features). We classified using support vector classifiers, logistic regression, and multi-layer perception in addition to random forests, decision trees, gradient boosting, and na\"ive Bayes. In the main round, all hyper-parameters classifiers were tuned using cross-validation, except for the multi-layer perception, due to computational constraints. 

Finally, we used statistical testing (the non-parametric Friedman for paired data with multiple groups followed by posthoc non-parametric pairwise tests) to understand what factors (sampling size, sampling technique, subset used for training, feature selection, the importance of oracle features, and the used algorithm) influenced the end results. In addition, we also looked into the most significant individual models to understand what features played the most significant roles in classification. 

\section{Structure of the Work}
\label{secStructure}

Chapter \ref{chSotA} presents the theoretical background of our work by introducing recommender systems in general and within the context of tweet engagement prediction. It also describes approaches and findings from the teams participating in RecSys 2020 Challenge that published papers on their findings. Chapter \ref{chMethodology} presents guiding principles and design decisions concerning the set-up of the experiments. Details concerning the implementation of models can be found in chapter \ref{chImplementation}. The same chapter also contains details about the environment the models were trained and evaluated in. Evaluation metrics and most significant experiment results are provided in chapter \ref{chEvaluation}, with the remaining results provided in appendix \ref{appAdditionalResults}. Lastly, chapter \ref{chCriticalReflection} discusses the importance of findings and proposes potential further research directions.
\chapter{Related Work}
\label{chSotA}

This chapter contextualises the approach taken in this thesis. Section \ref{secML} introduces machine learning, recommender systems, and classification. In doing so, the section relates the concepts from these research areas to the RecSys 2020 Challenges. The section also notes that the Challenge is, in fact, a classification task, as it asked the participants to provide a set of class assignments (whether the particular user engaged with the given tweet or not) rather than ranked lists of items. 

Section \ref{secTweetEngagementPrediction} deals with the challenge of recommending tweets in the wild. It should be noted that between the completion of the thesis and the time of the challenge, Twitter has changed its ownership structure, as elaborated in \cite{doesMuskTwitterTakeoverMatter}. Before the takeover, Twitter and its API were extensively used in research \cite{twitterAndResearch}. Section \ref{secTweetEngagementPrediction} examines such research focusing on the tweet recommendation. Contrastingly, section \ref{secTwittersCurrentRS} shortly introduces the actual Twitter recommendation algorithm, which was partially made public in early 2023, after the takeover. 

Lastly, section \ref{secRecSys2020Conference} elaborates on the results of the Challenge presented during the ACM Recommender Systems Conference 2020. The section summarises all of the approaches which the participating teams published. This summary includes data partitioning, feature engineering, the teams' guiding principles and classification models, as well as the used computing architecture and the achieved results.

\section{Machine Learning}
\label{secML}

Machine learning is a study of using data-driven techniques to process information, make predictions, learn hidden patterns, and aid decision-making \cite{mlbBayesianReasoningAndML2012,mlbLearningPyspark}. Stated even more succinctly, machine learning aims to extract knowledge from data \cite{mlbIntroductionToMLWithPython2016} without explicit prior programming \cite{mlbMLWithPyspark2019chapter2introToML}. Historically, machine learning aimed to mimic human reasoning and biological systems, but nowadays, it also includes a number of mathematical models aimed at solving very specific predictive tasks \cite{mlbBayesianReasoningAndML2012,mlbLearningPyspark}. In recent years, machine learning has become ubiquitous in everyday life and discourse \cite{mlbIntroductionToMLWithPython2016,mlbMLWithPyspark2019chapter2introToML}. Its applications thus include time series analysis (e.g. for predicting stock prices \cite{mlbIntroductionToMLWithPython2016}), generating content from user prompts (such as response messages or images \cite{generativeAI}), outcome prediction (e.g. forecasting yield levels of specific crops \cite{rapeseedYieldPrediction}), retrieving information (including for web search engines such as Google \cite{mlbIntroductionToMLWithPython2016}), controlling intelligent agents (such as virtual assistants \cite{jovanDC} or autonomous vehicles \cite{jasReccover}), and creating ranked lists (including Tweets to be displayed on a user's feed \cite{twitterAndResearch}, or, for instance, creating personalised music playlists \cite{Schedl2022musicRecommenders}). Sing \cite{mlbMLWithPyspark2019chapter2introToML} recognises four major categories of machine learning:

\begin{enumerate}
    \item \textit{Supervised machine learning}, which includes \textit{training} models based on training data that has correct answers or correct outcomes (which are referred to as labels). The models learn to predict these labels, or \textit{target variables}, given a set of other features, called \textit{explanatory variables}. 
    \item \textit{Unsupervised machine learning}, in which case the data contains no labels or target variables. Instead, the algorithm's goal is to find useful hidden patterns or signals within data.
    \item \textit{Semi-supervised machine learning} is (self-evidently) situated between the first two approaches and typically revolves around data which is only partially labelled. Usually, part of the data is labelled manually (i.e. by human labellers) and machine learning is used to label the rest. Then, the whole dataset, including both the manually and algorithmically labelled parts, is used as the basis for further machine-learning tasks.
    \item \textit{Reinforcement machine learning}, which typically does not include learning based on historical (past) data like the first three approaches, but instead learns based on rewards given for its actions within a given environment. The goal of machine learning is typically for an agent to learn strategies that tell it which actions to perform in a certain state to maximise its total reward gain.
\end{enumerate}

Within supervised learning, there are two major groups of tasks based on the type of target variable to be predicted \cite{mlbMLWithPyspark2019chapter2introToML,mlbTheElementsOfStatisticalLearning2009Chapter1intro}. If the outcome variable is of categorical type (i.e. has a value from a discrete and final set of values, cf. subsection \ref{secScales} for a more detailed overview of types of variables), then it is known as a \textit{classification} task. Examples include programs that determine if an e-mail is SPAM or not, as well as computer vision programs that can recognise hand-written digits. In contrast, if the target variable is numerical/quantitative, then the task is called \textit{regression}. An example of a regression task is predicting a person's salary, given their age and profession. The RecSys 2020 Challenge is based on supervised machine learning. As later chapters explore further, it is, in fact, an offline classification task. However, since recommending tweets to be seen by a user in their timeline is a typical recommender system task, the next subsection explores this area further.

\subsection{Recommender Systems}
\label{secRecSys}

This section describes recommender systems in general and is thus primarily based on the first chapter of the Recommender Systems Handbook (2022) by Ricci et al. The book's authors \cite{recSysBook2022chapter1}, in reference to \cite{Burke2007, Resnick1994, resnick1997recommender}, define recommender systems to be both techniques as well as software tools that suggest \textit{items} to \textit{users}. These items may be products to buy \cite{lee2021entry}, films to watch \cite{netflixPrize}, music to listen to \cite{Schedl2022musicRecommenders}, news to read \cite{Resnick1994}, tweets to engage with \cite{recsys2020overview}, etc. The authors \cite{recSysBook2022chapter1} note that the main utility of the recommender system for the user is to provide the most relevant subsection of items from a much bigger set. Without this service, the number of items to choose from might be overwhelming for the user \cite{recSysBook2022chapter1}. Ricci et al. \cite{recSysBook2022chapter1} also note that, in its simplest form, the output of a recommender system is a ranked list of items. Applying this definition to Twitter, the items would be the tweets, the users would be the registered Twitter users for whom the website maintains user models, and the suggestion would be the ``timeline'' (i.e. a personalised scrollable ranked list) with the tweets. 

The simplest recommender systems provide rankings in a non-personalised fashion (e.g. they simply list all available songs to everyone in order of the number of plays within the past day) \cite{recSysBook2022chapter1}. However, in the case of more sophisticated systems that are commonplace in e-commerce and social media websites nowadays, recommendations are personalised, thus offering different items for specific users or user groups and adapting rankings accordingly \cite{recSysBook2022chapter1}. For its registered users, Twitter too offers personalised recommendations in its primary feed,  while a non-personalised reverse-chronologic list of tweets authored by the users the target user is following is also offered \cite{recsys2020overview}. This is further discussed in subsection \ref{secTwittersCurrentRS}. 

Ricci et al. \cite{recSysBook2022chapter1}, in a preview of the book's following chapters \cite{Abdollahpouri2022,Jannach2022}, note that there are three important stakeholders of recommender systems: consumers/users, providers/suppliers, and system owners. We can observe that Twitter is an interesting example to consider from this perspective. This is because, while the system owner clearly is Twitter as a company, the social media's registered users are both consumers/users and providers/suppliers of the recommended items (tweets). In other words, the users are both those writing tweets to be recommended to others while also using the platform to read other users' tweets in their own timelines.

The Recommender System Handbook \cite{recSysBook2022chapter1, Jannach2022} also identifies the benefits of employing a recommender system for the system owners. Both if we were to accept Twitter's proclaimed purpose of wanting to enable global public conversation \cite{twitterRules} or if we were to assume the more conspicuous desire to make Twitter more attention-grabbing \cite{socialMediaAddiction}, three of these benefits appear particularly enticing. The first is increasing user satisfaction, which means that the user sees tweets that she finds more enticing and more relevant to her interests. The second is increasing user fidelity which allows the user to feel like a valued guest of the website thanks to recommended tweets that correspond to her previous reaction patterns and make her more likely to visit the social media platform again. Lastly, by recommending tweets and noting to which tweets the user reacted, the system owner can better understand what the user wants. This knowledge can then be used for the owner's other goals, like serving the users ads she is more likely to engage with, leading to more revenue for the company. 

The recommender system's output is based on a diverse set of data types and sources. Conceptually, it can stem from information on three kinds of objects: items, users, and interactions \cite{recSysBook2022chapter1}. In the context of the RecSys 2020 Challenge, the item descriptions would be the features in fields ``Tweet'' and ``User (Author)'' of table \ref{tabRows}. The latter would also be classified as information about the item and not the user since it does not describe the user for whom the ranked list (or, here, engagement prediction) is to be made. Thus, only features from ``User (Engager)'' would be user information. As is further elaborated in section \ref{secTwittersCurrentRS}, Twitter also collects and utilises other information about its users in its recommendations, but there were likely omitted to preserve user privacy, which was one of the proclaimed primary goals of assembling the dataset for the challenge \cite{recsys2020overview}. Similarly, the only interaction data the challenge provides is the target of the prediction itself, i.e. features in the field ``Engagement'' (apart from the features ``Engagee follows engager?'', which is a piece of information about the user). 

Information about interactions is further explored in \cite{recSysBook2022chapter1, Jannach2022SessionBasedRS} within the context of recommender systems. The authors first note that the information about the interaction between users and items can be explicit (that is, user-provided numerical, ordinal, or binary item ratings, cf. subsection \ref{secScales}) and implicit. Implicit feedback is comprised of all actions the user performs on the item: in the case of Twitter, that would, of course, include the user liking the tweet, replying to the tweet, or retweeting it, but it would also include the user expending tweet details, opening the author's profile, etc. The authors note that implicit information about the interaction is much more abundant than explicit interaction, which is also true in the case of Twitter\footnote{Based on the thesis author's personal Twitter use, the platform only offers two forms of explicit interaction feedback. The first is a rare option to say whether a promoted tweet written by a user the target user does not follow is relevant. The second form of explicit feedback available to the user is reporting that a displayed tweet is irrelevant or in breach of Twitter rules}). Yet, the authors of the handbook \cite{recSysBook2022chapter1, Jannach2022SessionBasedRS} also underline that a downside of relying on implicit feedback is that it, as a ``unitary rating'', cannot be used to indicate a specific level of user preference.

Lastly, Ricci et al. \cite{recSysBook2022chapter1} also elaborate on some of the specific techniques recommender systems employ to perform their base functionality -- predicting what items are best to suggest for the target user. The authors elaborate on content-based prediction, collaborative filtering, knowledge-based recommender systems, and others. However, these algorithms could not be directly used by the author of this thesis or by the challenge's winning teams, as is elaborated in section \ref{secRecSys2020Conference}. This is because the challenge does not ask the participants to create a full recommender system; the expected output of the prediction is not a ranked list of suggested tweets. Instead, the expected output is classifications corresponding to the likelihood of the user interacting with the given tweets. This information indeed signals the utility of an item for the user and can thus be a component of a recommender system. Nevertheless, the RecSys2020 Challenge is to create a classifier, not a recommender system. Therefore, the next section looks into classification.

\subsection{Classification}
\label{secClassification}

Following the formal definition from \cite[chapter 4]{mlbPatternRecognitionAndML2006}, the goal of classification tasks is to assign the target variable one of $n$ discrete classes $C_k$, where $k \in \{1,2,3\dotsc n\}$ given a vector of explanatory variables $x$. The goal of classifiers (algorithms that solve classification tasks) is to define decision boundaries that divide the input space into decision regions. If these decision boundaries can be defined by linear functions, then the classifiers are called linear classifiers and the instances linear models. Bishop and Nasrabadi \cite{mlbPatternRecognitionAndML2006} recognise three main approaches to designing linear classifiers, depending on whether they are based on:

\begin{enumerate}
    \item \textit{A discriminant function} which assigns a target value for each specific input vector \textit{x};
    \item \textit{A conditional probability distribution} $p(C_k | x)$, which is modelled on training data in an inference stage and then applied to new input vectors to make predictions; or
    \item \textit{A generative approach} that models class-conditional densities $p(x|C_k)$ and the prior probabilities $p(C_k)$ for the classes and then computes the require posterior probabilities for new vectors by applying the Bayes' theorem $p(C_k|x) = \frac{p(x|C_k)p(C_k)}{p(x)}$.
\end{enumerate}

Logistic regression, which will be examined in more detail in section \ref{secLR} and which was one of the algorithms used for this challenge, belongs to this last group of linear classifiers \cite[chapter 4]{mlbPatternRecognitionAndML2006}. Each classifier has a set of parametres which affect how it fits the training data. In the case of logistic regression, one of these parametres is the regularisation coefficient $\lambda$ that regulates the model's flexibility to capture trends within data \cite[chapter 3]{mlbPatternRecognitionAndML2006}. In general, the more features of the input vector are considered in the fitted model (i.e. the more complex the model is), the better its performance will be on the training set. However, if the linear (or other) classifier is so complex that it \textit{overfits} the training data, its ability to generalise decreases, that is, the model's \textit{variance} increases\footnote{As is further elaborated in \cite[chapter 3]{mlbPatternRecognitionAndML2006}, the bias--variance is originally formulated for the least-squares regression, where the target variable is a number rather than a class. But the trade-off also exists in the case of other classifiers. The chapter showcases how this concept can be generalised to all binary classification tasks. Binary classification tasks are those where there are only two possible classes for the target class to be assigned into. The generalisation from the chapter is based on the fact that the binary classifiers' outputs can be seen as the likelihood that the instance belongs to one of the two classes. Thus their output is a number, so the definition from the least-squares regression can be extended. Following this line of reasoning, generalisations of the bias and variance for classifiers with multiple target classes are also possible.}. This would result in the variance of the model increasing for the training set. Yet if the model's flexibility is contained so much that it becomes too rigid to capture the complexity of the data, then its \textit{bias} increases. Finding a perfect trade-off between model flexibility to capture the data's complexity and its robustness that allows it to generalise its predictions to unseen data is known as a bias-variance trade-off.

The type of classification task that the RecSys 2020 Challenge presents is a binary classification task, as each of the four engagement types has only two classes \cite{recsys2020overview}. Apart from this logistic regression, we have also tested Na\"ive Bayes, a simple classifier employing Bayesian statistics. In addition, we also tried employing decision trees, which split the decision space using a tree graph where each node checks the value of one of the input variables. Random Forests and Gradient Boosting Trees are two algorithms built on decision trees that may be less prone to over-fitting. Lastly, we have also tried Support Vector Machines Classifier. Section \ref{secAlgorithms} describes how each of the employed machine learning algorithms fits a model to the training data and then creates predictions.

\section{Tweet Engagement Prediction}
\label{secTweetEngagementPrediction}

Many prior studies have been published regarding tweet engagement predictions. For example, \cite{wadhwa2017maximizing} found that factors such as time of the day, day of the week, and whether a tweet includes a hashtag or an image can all affect the engagement rate. Similarly, \cite{han2019analysis} analysed the effects of pictures, URLs, and mentions on engagement rates for tweets of corporations in different fields. A multi-year, multi-track competition revolving around Twitter, an overview of which is available at \cite{bellot2014overview}, offers many further insights. There are also numerous papers such as \cite{shirdastian2019using, hill2013does, bakshy2011identifying} that look into the relationship between tweet engagement rates and brand perception. The relation between communication on Twitter and national elections was studied in \cite{younus2014election, gayo2012wanted, gaurav2013leveraging}.

The organisers of the RecSys 2020 Competition in \cite{tweetPap} introduced several state-of-the-art approaches for predicting tweet engagement. Many of these approaches rely fully or partially (e.g. \cite{tp3}  also uses a linear model and \cite{tp4} incorporates factorisation machines) on neural networks. However, by excluding tweet text, the need for neural NLP solutions was alleviated. Therefore neural networks and the solutions presented in \cite{tweetPap} were, for the most part, inapplicable for the proposed problem definition.

On the other hand, a significantly more relevant starting point and a vast source of papers detailing possible feature engineering approaches and prediction models is the RecSys 2014 Challenge\footnote{\url{http://2014.recsyschallenge.com/} (last access: 2020-05-20)} -- which involved predicting tweet engagement for tweets containing film reviews generated via a movie-reviewing website IMDb. An analysis of the challenge can be found in \cite{loiacono2014analysis}. There are multiple discrepancies between this and the RecSys 2020 Challenge. The two most important disparities relative to the RecSys 2014 Challenge's dataset are its vastly smaller size (170 thousand instead of 160 million tweets for the training dataset) and significantly less diverse tweets (tweets with just movie reviews in contrast to randomly sampled tweets on any topic). However, many lessons and methods from this challenge remained germane. For example, \cite{Wasilewski} used logistic regression and item clustering, while \cite{guillou2014user} described an approach using LambdaMART and random forests. Several solutions, such as \cite{palovics2014recsys, saha2014popular, abdollahi2014two}, incorporated matrix factorisation. Many approaches, such as \cite{magalhaes2014recommender, Wasilewski, abdollahi2014two, guillou2014user, loiacono2014analysis, palovics2014recsys}, noted that an important step of the challenge was predicting whether a tweet would get any engagements or not (a binary classification problem). Lastly, \cite{loiacono2014analysis, ZamaniSM15, singh2014ranking} described a way of feature extraction and selection underlying the importance of feature engineering. 

However, like in the case of the papers published following the RecSys 2020 Challenge, conclusions from neither the papers from the RecSys 2014 Challenge nor the previously named papers on tweet engagement, in general,  are directly applicable to the problem at hand, given the limitation of the scope to tweet context. On the one hand, excluding text simplified the task by limiting the necessity of computationally expensive natural text processing. On the other hand, it made extracting additional contextual features paramount.

\section{Twitter's Current Recommender Systems}
\label{secTwittersCurrentRS}

In March 2023, Twitter released a part of its recommendation system algorithm on GitHub \cite{twitterRecommenderRepository, twitterRecommenderRepositoryMLcomponents} and a high-level explanation of the system principles on its blog \cite{twitterRecommenderBlog}. This section summarises information provided in these sources and highlights elements which overlap with the work done in the technical part of this dissertation. 

The blog post revealed that roughly half of the average Twitter user's timeline stems from authors the user follows (called the in-network sources) and the other half from everyone else (the out-of-network sources). As is stated in \cite{twitterRecommenderBlog}, the main component in ranking tweets for the in-network is the RealGraph framework, originally published by Kamath et al. in \cite{realgraph}. As elaborated in \cite{realgraph}, the framework predicts the probability of any engagement between the two users and consists of three parts: generating the graph of users, scoring the graph features, and using a logistic regression algorithm to predict engagement probabilities. Regarding the first step and the creation of a directed graph between users: an edge from user A to user B is created if A follows B, if A has B in their address book (provided that both A and B allowed Twitter access to their respective address books), or if A had interacted with B in the past. The edges of the graph are updated periodically but not in real time. Quantities such as the number of days since A followed B, the number of days when A interacted with B, the mean and variance of interactions between A and B, and the number of days since the first and last interaction for all of the interaction types. The ``exponentially-decayed interaction count'' is saved as a graph edge feature indicating interactions between A and B specifically and as a user feature indicating the sum of all outgoing interactions for user A. Further user features are also utilised: ``[t]hese include number of tweets in the last week, language, country, number of followers, number of people they follow, and PageRank on the follow graph'' \cite{realgraph}. Much like the RecSys 2020 Challenge, the goal of the training task is binary classification, predicting the existence of interactions of any interaction\footnote{As we established, for RecSys 2020 challenge, predicting the four interaction types (like, reply, retweet, quote) was done separately. Predicting any engagement interaction corresponds to predicting the ``react'' target (which is defined as logical or of the four original engagement types). This fifth type was not part of the challenge but was part of the thesis experiment, and it is that engagement type that corresponds to the training goal of the Realgraph.} based on a combination of features. The result is that different features are assigned various weights by RealGraph, with the weights corresponding to the likelihood that user A would interact with user B. The algorithm was implemented in Pig on Hadoop distributed file system, enabling predilections on the order of magnitude of hundreds of millions of users \cite{realgraph,pig,lsmltwitter}. To evaluate the Realgraph, Twitter uses stochastic gradient descent logistic regression with L2-regularisation. As the evaluation metric, they use the area under the curve (AUC), a metric similar to that used in the challenge (see section \ref{secPRAUC} for more details). As is stated in \cite{twitterRecommenderBlog}, the higher the score from the RealGraph, the higher the likelihood of engagement between two users is considered to be, and the more tweets from that author would be included in the In-Network of the target user. The same logistic regression classification approach was still in use in 2023 as well \cite{twitterRecommenderBlog}.

Regarding the out-of-network sources, the Twitter team in their blog entry \cite{twitterRecommenderBlog} notes that short-listing tweets based on those the user does not follow, does not have in their address book, and had not previously interacted with is a more challenging problem. They thus employ two sources for this: social graphs and embedding spaces. The former estimates potentially relevant matches by considering the engagements of users similar to the target user who thus build their social graph. The blog entry does not provide more details on how the similarity of users for this use case was measured. In contrast, the only out-of-network branch in \cite{twitterRecommenderRepository} is the CR-Mixer, ``a candidate generation service proposed as part of the Personalization Strategy vision for Twitter''. Its description further states that it uses data from the RealGraph and the UserProfileService, indicating that similarity may be calculated using the weighted features from the RealGraph and the information the users entered in their Twitter profiles. In any case, \cite{twitterRecommenderBlog} further states that the social graph also considers whom the people from the in-network interact with and who else interacted with the tweets similar to those that the target user had interacted with (plus what else did those users interact with). On the other hand, the embedding spaces approach aims to learn and represent the target user's interests and to find tweets and users that relate to those. This is achieved primarily using SimClusters, a community discovery algorithm based on neighbourhood-aware Metropolis-Hastings sampling \cite{twitterRecommenderBlog, simClusters, namha}. The authors of \cite{twitterRecommenderBlog} note that there are around $145 000$ communities in 2023 ranging in size from thousands to millions. Communities are updated periodically and, for each of the communities, individual influential users were identified using Sparse Binary Factorisation (SBF), a publicly available Twitter-adapted matrix factorisation algorithm \cite{twitterRecommenderBlog, sbfTwitter, matrixFactorisation}. Each user may belong to multiple communities, and tweets that were engaged with by the members of those communities would be considered for the ranking.

The ranking of tweets to be shown to the target user begins with around $1 500$ candidate tweets from the in-network and out-of-network sources \cite{twitterRecommenderBlog}. The blog post \cite{twitterRecommenderBlog} only specifies that the ranking is based on a neural network with 48 million parametres that predicts the likelihood of user engagement whose input is ``thousands of features'' and that the source of the tweet (in-network or out-of-network) is not considered at this stage. As can be seen in \cite{twitterRecommenderRepository, twitterRecommenderRepositoryMLcomponents}, the ranking algorithm is based on Earlybird \cite{earlybird} and MaskNet \cite{masknet}.

Lastly, as described in \cite{twitterRecommenderBlog, twitterRecommenderRepository}, the Home mixer performs final filtering and assembles the filtered candidate tweets into the target user's main timeline. This includes visibility filtering, such as removing the tweets from the users the target user blocked \cite{twitterRecommenderBlog}. Consecutive tweets from the same authors are also removed. Moreover, tweets are removed to balance in-network and out-of-network sources. Furthermore, tweets are excluded ``if  the viewer has provided negative feedback around it'' \cite{twitterRecommenderBlog} or if there isn't at least a second-degree connection between the author and the target user. Effectively, this means tweets written by authors who would have more than two degrees of connection in the RealGraph are removed. Some responses to candidate tweets  are also bundled together with those tweets for the last step. A further step assuring that the tweet to be shown had not been edited in the meantime (and replacing them with the updated version if it had) is also performed. Ultimately, the filtered tweets to be shown are mixed with other content, such as ads, onboarding prompts, and recommendations of other Twitter users for the target user to follow. 

\section{ACM Recommender Systems Conference 2020}
\label{secRecSys2020Conference}
In September 2020, eight papers presenting solution approached for the RecSys 2020 Challenge were published at the ACM Recommender Systems Conference 2020\footnote{\url{http://www.recsyschallenge.com/2020/} (last access 2021-03-12)}. Among them are publications written by all teams whose solutions had taken the top five places on the final leaderboard. This section compares some of the findings from these publications.

\subsection{Data Partitioning}
\label{secRecSys2020DataPartitioning}

Sampling data to simulate test and val subsets was an important aspect of the Challenge. Deciding how to do this was a non-trivial task given that the original training dataset was sampled from tweets from within one week, whereas the original test and validation datasets were sampled from the following week. Moreover, validation and test datasets had the engagement features removed (as they were to serve as target classes) until the end of the challenge. Lastly, the test dataset (roughly as big as the val dataset) was released just two weeks before the challenge deadline. 

In \cite{[CP2], [CP6], [CP7]}, around 10\%, 12\%, and 50\% of tweet ids, respectively, were randomly selected from the train set to simulate validation set, ensuring that there is no overlap of tweets between validation and train subsets (as is the case with the original train and val datasets). Shifting timestamps by seven days to simulate tweets from the following week was then done in \cite{[CP6]}. Additionally, the authors of \cite{[CP6]} ensured that the ratio of cold start users remained consistent in their validation subset with one of the actual val datasets. 10\% of the training set was also used in \cite{[CP3]} but here the focus was on ensuring that tweets come from all seven days of the first week. To get a data sample with an equal number of positive and negative tweets for each engagement type, the team from \cite{[CP4]} performed negative under-sampling before sub-sampling even further to make the training process less computationally intensive. 

Taking just the last day for the validation subset was criticised by \cite{[CP6]} as it would eliminate the possibility of effectively extracting the day of the week as an explanatory variable. However, taking the last two days for validation was one of the two splitting methods used by the winning team in \cite{[CP7]}. 

Moreover, authors of \cite{[CP7]} note that due to data leakage which occurred during the creation of the dataset, the engagements in the last four hours were irregular and inconsistent with the remainder of the dataset. They therefore warn that it was misleading to simply take the last four hours for the validation subset because of the misleading ground truths (a lower ratio of positive instances than what would be expected). However, the second-place-winning team stated in \cite{[CP9]} that they used exactly the last four hours as the validation subset. To create a training subset, this team used a 24-hour sliding window so that the training is based on predicting engagements from one window at a time, with the remainder of the train dataset (minus the very last four hours) being used for feature engineering and context extraction. Similarly, the last 16 hours of the train dataset were taken to be the validation subset for \cite{[CP8]}. The authors defended this partition choice by stressing that it prevents ``data leakage through time dependence''. Moreover, \cite{[CP8]} wanted to preserve the cold-start distribution of the training and validation datasets by, on the one hand, selecting 27\% of the users from the validation subset and deleting their appearances from the train subset and, on the other hand, by removing certain previously extracted features for 7\% of the users from the train subset.

While \cite{[CP3]} argued that train and validation datasets come from the same distribution, this was disputed by \cite{[CP4], [CP7]}. Namely, the authors of \cite{[CP4], [CP7]} stated that the adversarial validation models they had built based on \cite{pan2020adversarial, avm1, avm2} had an AOC of almost $1$ when predicting whether a tweet is taken from the training or validation dataset, suggesting a significantly different distribution of the two. To account for this, \cite{[CP4]} selected those features for their models that do not deviate significantly between the two datasets. Moreover, \cite{[CP7]} noted that the approaches on their validation subsets (based on 50\% randomly selected tweet ids and the last two days of the first week) were not perfectly correlated with the results from the actual validation dataset, so they tuned hyperparametres based on their public leaderboard scores.

\subsection{Feature Engineering}
\label{secRecSys2020FeatureEngineering}

Most of the published solutions contain descriptions of feature engineering processes. In \cite{[CP2]}, the authors grouped the inferred features into three groups:
\begin{enumerate}
    \item Tweet features: number of media elements, number of each media element, hour, tweet type, number of links, hashtags, local hashtag popularity calculated within a two-hour window, global hashtag popularity calculated over the whole dataset, language, text length, whether there are question marks in the tweet, whether the tweet contains words ``RT'' and ``Retweet'' as elaborated in \cite{SuhRetweeted}, number of appearances in the dataset, 50-dimensional text embedding, engagement probabilities calculated on a model based solely on tweet language;
    \item Engaging user features: logged numbers of followers and followings, following ratio, logged number of interactions of this user with other users in the dataset, whether the user is verified, number of weeks since the user account was created, the ratio of positive to all interactions for the engaging user and the engaged user for the given interaction instance, the ratio of positive to all interactions for engaging the user and all engaged users;
    \item Engaged user features: the number of times the user appears in the dataset, whether the engaging user follows the engaged, whether the user has ever interacted with a tweet in the language of the current tweet.
\end{enumerate}

In \cite{[CP5]}, the authors extracted features that could be grouped in the same manner. For each group of features, they then applied feature hashing to map these features into a lower dimensional space.

The solution presented by \cite{[CP4]} used frequency encoding for categorical features as well as groups of categorical features in order to represent relationships between them. Moreover, they created graph features to represent relationships between users. Specifically, first and second-degree connections were selected, as was the user influence, which was created by employing the PageRank algorithm from \cite{page1999pagerank} over the engagements. They also extracted user similarities based on user preferences for individual topics extracted from the graph built based on the like engagements using walk with restarts from \cite{tong2008random}. The authors also created the following features: text features using BERT algorithm \cite{bert}, number of sentences, number of mentions, account age, the ratio between followers and following users, relative active time of users, the main language of users, a combination of user IDs, hashtags, media, and other categorical features. One of the key findings this team presented and that they considered crucial for their high ranking (three) was the importance of incorporating intermediary predictions for the other three engagement types in making the final prediction. 

The authors of \cite{[CP6]}, who ranked fourth, also stated that feature engineering had been essential for achieving competitive results. They claimed to have extracted 300 additional features, half of which had been found to be informative enough to be preserved in the final model. Firstly, the authors extracted text features such as unique word frequency, tweet topics (by manually assigning individual topics to lists of words and then counting the number of words for each topic), and text embeddings  produced by the DistilBERT algorithm \cite{distilbert}. Secondly, they extracted features that model user behaviour. Each of those features had both a timestamp-aware (modelling a sequential engagement pattern based on all engagements that occurred before the given user-tweet instance) and a cumulative version (insensitive to sequential patterns and computed using the entire user history). These features are then further divided into three subgroups: number of active and passive engagements, number of engagements with language/hashtag, and user similarity. This last division was computed based on an undirected graph whose nodes were the users, the edges indicated that there had been an engagement between the two users, and the weights were to represent the counts of occurrences of each engagement type. From here, they created four features for each engagement type: based on direct connections and based on connection via a path of length equal to two, as well as based on binary representations (indicating whether there had been at least one engagement) and based on weights (counts of such engagements). A feature the authors especially highlighted was the number of previous negative engagements, which had been selected for by all four final ensemble models. 

The authors of \cite{[CP8]} regarded tweet text as the primary source of information, so they attempted to extract topics discussed in tweets by reducing the dimensionality of BERT embeddings using principle component analysis and then clustering the result using k-means clustering. Each of the 150 clusters was then assigned a topic, and for each user five most relevant topics were found. The authors also created embeddings for user behaviour. To account for the fact that many users have only a handful of appearances, they grouped users with fewer than 71 appearances in 400 buckets of at least 450 appearances. The model then learnt user embeddings for each of the often appearing users individually and for users in each of the buckets collectively. Lastly, the authors also extracted simpler features presented in table 1 of \cite{[CP8]}. Among these features are logarithms of the follower differences between the engager and the author, quantiles of account creation date, logarithms of the number of followers and accounts following users, the language of the tweet, as well as the day of the week and the hour when the tweet was posted.  

The second-placed team stated in \cite{[CP9]} that they had extracted 467 new features that could be divided into four groups enumerated below.

\begin{enumerate}
    \item Engaging user: represented engaging user's historical and current account information, including time-based features which had been found to be particularly informative. The features included trends regarding the follower and following users count, account age, verification status, and the engaged-with and posted tweets.
    \item Engaged-With user:  analogous to the engaging user.
    \item Tweet: features similar to those in \cite{[CP8], [CP2]}. The authors noted that using historical data for tweets was impossible because the overlap between tweets in training and datasets was an empty set.
    \item Interactions: collaborative filtering for user-tweet and engaged with user-engaging user pairs. For the former, the focus was on the similarity between the tweets that the user had previously engaged with or had created and the target tweet. In contrast, for the latter focused on the joint engagement history, summarising prior engagements between the two users. 
\end{enumerate}

The winning team provided the lists of the 15 most important features based on XGBoost (cf. subsection \ref{secRecSys2020Models}) for each engagement type in appendix C of \cite{[CP7]}. Judging by these lists, one of the most fruitful design choices was to create target encodings, i.e. ``statistics from a target variable segmented by the unique values of one or more categorical features'' \cite{[CP7]}, where the target variable was one of the four engagements and the chosen statistics was mean. They also created a number of text-based features such as the counts of words, characters, the symbol ``@'', the most popular and least popular words in tweets, and the first username mentioned in the tweet. A group of new features was the count of engagements and own tweet posts in the preceding $5$, $60$, $240$, $480$, and $1440$ minutes. Lastly, they too created graph features for direct connections and second-degree connections. Unlike \cite{[CP6]}, the edges were not based on engagements but on follows. By creating this graph, they managed to create a more complete overview of the social relation between the users (the original dataset provided explicit information on whether the author had followed the account of the engaging user). Both the first and second-degree follower counts in both directions often appeared among the top 15 most important features. Moreover, simply counting the number of characters and words in the tweet text also seemed to be useful and was often among the top 15. All other most informative features were different combinations of categorical values combined via target encoding. One observation the authors noted was that number of interactions per user was quite low, which is why they could not model additional features that might be interesting.

Intriguingly, the team from \cite{[CP7]} did not use BERT or DistilBERT embeddings as features, given that the tweet text itself was not significant enough to justify the great computational overhead. This contrasts the works of \cite{[CP4], [CP6], [CP8], [CP9]}, where text embeddings were the cornerstone of their models. Nevertheless, it was actually \cite{[CP7]} that achieved the best results for all evaluation metrics with the sole exception of PRAUC for reply engagement, where \cite{[CP9]} was better, as further discussed in section \ref{secRecSys2020Results}.

\subsection{Guiding Principles and Selected Models}
\label{secRecSys2020Models}

Authors of \cite{[CP1]} observed that, on the one hand, recent academic papers in recommender systems often focus on deep learning, presenting novel approaches that consistently beat the baseline. On the other hand, the authors also note that deep learning had not reached this level of success in recent recommender systems competitions, such as the annually organised RecSys challenge.

The 2020's challenge results partially refuted this hypothesis. Models from \cite{[CP9], [CP6], [CP8], [CP2]} (winning ranks two, four, six, and nine) included neural networks with at least two layers on top of the text embeddings and other features in their final model architectures. The winning team stated in \cite{[CP7]} that they had also experimented with a feed-forward neural network, but that they had had to exclude it due to the approaching challenge deadline even though the approach had shown promising early results. Conversely, teams from \cite{[CP4], [CP3], [CP5]} (winning ranks three, four, and twenty-eight) did not use neural networks in their final models (excluding potentially the use of BERT-based embedding, which was originally created using neural networks). In the following paragraph, we summarise the top five best-ranking teams' main ideas and best-performing models.

In \cite{[CP7]}, the team which won seven out of eight categories (being outmatched only in the case of the area under the precision-recall curve for predicting replies by \cite{[CP9]}) presented their approach. The team emphasised the importance of developing GPU-powered workflow, stating that the consequent dramatic decrease in computational time had allowed them to tune their models based not only on train dataset splits but on the publicly available leaderboard scores as well. This optimisation also allowed them to tune their hyperparametres using a grid-search algorithm. The tuning of parametres was based on RCE score only since even the trivial model that simply predicted no engagements achieved very high PRAUC scores. The authors also underlined the importance of feature engineering, stating that focusing on inferring features had been more beneficial than tweaking model architecture. The models this team experimented with are based on gradient boosting trees, FTRL-proximal regularisation (an online algorithm presented in \cite{do2009proximal}, which the authors chose for its speed and good performance for high-cardinality variables), alternating least squares matrix factorisation (ALS, based on \cite{ALS}), DistilBERT (again based on \cite{bert, distilbert}), feed-forward neural networks, and XGBoost (from \cite{chen2016xgboost}). For the tree-based approaches, the team experimented with LightGBM from \cite{ke2017lightgbm} and CatBoost from \cite{CatBoost}, but abandoned them because the results of these models were similar to those produced by XGBoost models that had the additional advantage of being runnable on GPU. Their ALS approach was based on a dot-product between user and item embeddings. The authors concluded that including more positive examples in the training subset increased the PRAUC score of the ALS model. A problem the team encountered was that the embeddings the ALS learnt from the train dataset were rendered unusable for test instances by the fact that there were no overlapping tweets between the two datasets. They tried to overcome this by modelling hashtags and links, rather than tweets, as items. The ALS features created this way did increase the PRAUC score by around 1.3 for the like target of the validation dataset, but this was deemed an insufficient improvement for the team to include the ALS features in the final model. Yet they hypothesised that matrix factorisation could be more useful in denser datasets. The authors used the pre-trained DistilBERT transformer to produce 768 features representing tweet text. This embedding was then reduced to 32 dimensions which were included in gradient boosting models, but the consequent prediction improvements were minor. Using the embeddings alone to predict engagements did not lead to high scores. This led the authors of the winning solutions to conclude that the tweet text is not informative enough to justify the additional computational expense. Thus, they focused more on contextual features, similar to what was done in this thesis. The authors stated that feed-forward neural networks could be trained quickly on the GPU and that they had obtained promising preliminary results by employing such neural models. Nevertheless, this team could not include neural networks in their final prediction pipelines due to the approaching challenge deadline. When it comes to finding the optimal encoding of categorical features, Schiffere et al. \cite{[CP7]} concluded that target encoding, described in section \ref{secRecSys2020FeatureEngineering}, was undoubtedly the best choice enabling better predictions than ordinal, frequency/count, hash, and multi-feature target encodings. Ultimately, the best-performing model was an ensemble of three individual XGBoost models trained on different subsets of features and different samples of the train dataset. XGBoost was also used for calculating the relative importance of the extracted features.

The team, which ranked second, stated in \cite{[CP9]} that their motivation was to create a pipeline comprising collaborative filtering and a deep learning architecture that capitalised on recent advances in the field of neural networks. In contrast to the winning team, here a special emphasis was put on tweet content. Their final model had three groups of input features. The first set of extracted features, discussed in part in section \ref{secRecSys2020FeatureEngineering}, described various aspects of both the engaging and the engaged-with user and the tweet itself. To make the format of the features appropriate for the neural network training, the authors of \cite{[CP9]} applied one-hot encoding for all categorical features and transformed all numerical features using the Yeo-Johnson power transformation from \cite{yeojohnson2000}. The second set of input features were embeddings created by a language model. These content-based features were indicative of the theme and the topic of the tweet. In order to create a more nuanced understanding of the tweet's semantic sentiment and structure, the authors applied pretrained Hugging Face masks\footnote{The libraries were downloaded from \url{https://huggingface.co/bert-base-multilingual-cased} and \url{https://huggingface.co/xlm-roberta-large} (access date: 2021-04-12).} for multilingual version of BERT (based on \cite{bert}) and XLR-R (from \cite{XLRR}) models. This allowed the team to fine-tune language models to tweet-specific writing style containing many abbreviations and slang terms. The third set of features comprised embeddings that described the engaging user's engagement history.  Lastly, the transformer from \cite{vaswani2017attention} that utilises the so-called attention mechanisms was combined with the three groups of features to provide the final input for the model. The resulting features with attention were then fed to multiple feed-forward neural networks with one fully connected layer and ReLU activation fictions. The dimensions of these one-layer neural networks were 5000, 3000, 2000, 1000, 500, and finally four -- corresponding to the four engagement types. Since all of the input features were differentiable, the pipeline could be trained end-to-end.

Team Wantedly that authored \cite{[CP4]} ranked third. Like \cite{[CP7]}, they relied mostly on gradient boosting models and target encoding. Unlike the winning team, however, they used LightGBM from \cite{ke2017lightgbm} rather than XGBoost as it was deemed a better fit for the computational architecture they had access to. It is also significant to say that they had used adversarial validation to discover which features did not deviate significantly between test and train datasets and then build their models based on information inferred from those features. A cornerstone of their model architecture was having first-stage models that provided preliminary predictions for each of the four engagement types based on target-independent features. The first stage comprised three models for each target, and the outputs were averaged predictions. These preliminary engagement predictions together with target-dependent features were then fed to the second stage model that provided final engagement predictions. The second stage had five models for each of the targets and the final prediction was defined to be the mean of the five predictions. This two-stage architecture was motivated by the engagement types' high co-occurrence, meaning that predicting other engagement types before making the final prediction for each of the four was very beneficial. It is very interesting that most of the teams with published papers -- including the two that achieved better final rankings -- did not include this kind of cross-modelling.

Sumit Sidana, the sole author of \cite{[CP3]} achieved rank four. His final models were based primarily on factorisation machines (FM) and field-aware factorisation machines (FFM) from \cite{factorisationMachines, fieldAwareFactorisationMachines}. The latter model performed better for all engagements but for replies, for which the standard FM performed better. It is also worth noting that Sidana's prediction for quotes failed to beat the baseline, as the achieved RCE was negative (although only slightly below zero). Similarly as in the case of \cite{[CP7]}, the author of \cite{[CP3]} chose to focus on context-based approaches due to poor scores achieved after initial experiments with BERT-based models that represented tweets as 760-dimensional embedding vectors. In this approach, Sidana then modelled all engaging users by averaging the embedding vectors of all tweets they had positive engagements with. The prediction was then based on the dot-product of the engaging users' averaged vectors and the target tweets' embedding vectors. In contrast, the models used for final predictions were based on collaborative filtering with a focus on the tuning of parametres such as the learning rate, the regularisation factor, and the number of iterations. In final models for all four engagement types, stochastic gradient descent proved to be a better-performing optimisation method than alternating least squares or Markov chain Monte Carlo.

The authors of \cite{[CP6]} shared the rank four with \cite{[CP3]}, whereas their overall model architecture based on gradient boosting for decision trees and neural network was somewhat reminiscent of the final architecture from \cite{[CP7]}. Like \cite{[CP7]}, the team from \cite{[CP6]} also used XGBoost and observed that the PRAUC metric was not a good choice for parametre tuning. However, instead of focusing solely on RCE, this team created a new hybrid evaluation metric based on both original target metrics. In addition to XGBoost, the team from \cite{[CP6]} also used LightGBM, as in the case of \cite{[CP4]}. A fine-tuned DistilBERT model (presented in \cite{distilbert}) capable of representing the semantics of domain-specific elements (such as hashtags and mentions) was applied, similarly as in \cite{[CP9]}. Specifically, the embeddings from the pre-trained DistilBERT model were ran through two pairs of dense and drop-out layers. Other features (50 for the quote engagement and 74 for the rest) were also included as the input to the first dense layer. Akin to \cite{[CP4]}, Felicioni et al. \cite{[CP6]} also realised the importance of having two-stage models which enable preliminary predictions for all four engagement types to be used for the final prediction for each engagement. Their final model architectures differed for like class on the one hand and the remaining three engagements on the other. The model for predicting likes takes extracted features and the four preliminary engagement predictions from the neural network as the input. The final models for replies, retweets, and retweets with comments also take preliminary predictions from the gradient boosting trees model, rather than just those stemming from the neural network. 

\subsection{Architecture}
\label{secRecSys2020Architecture}

Table \ref{tabArchitecture} contains the system architectures the teams whose solutions were published had worked on.

\begin{table}[h]
    \centering
    \begin{tabular}{lll}
\toprule
Reference & Rank & \makecell{Reported architecture} \\
\midrule
\cite{[CP7]} &    1 &    \makecell{A single machine with four NVIDIA Tesla V100 GPUs} \\
\cite{[CP9]} &    2 &    \makecell{Ubuntu servers with IBM POWER9 CPUs @ 3.8GHz,\\ 600GB of RAM, and NVIDIA Tesla V100 GPUs} \\
\cite{[CP4]} &    3 &    \makecell{Machines with 64 vCPUs and 600GB of memory} \\
\cite{[CP3]} &    4 &     \makecell{AWS instance on Ubuntu 18.04 having 96 cores and 192GB \\of RAM} \\
\cite{[CP6]} &    4 &    \makecell{AWS (no further details provided)} \\
\cite{[CP8]} &    6 &    \makecell{AWS c5d.4xlarge EC2 instance with 16 vCPU cores and no \\ GPU} \\
\cite{[CP2]} &    9 &     \makecell{GeForce RTX 2080 Ti GPU} \\
\cite{[CP5]} &   28 &    \makecell{Spark cluster with 28-core 2.00Ghz and 200GB of RAM} \\
\bottomrule
    \end{tabular}
    \caption{Reported machine architectures}
    \label{tabArchitecture}
\end{table}

In contrast to all other teams, the authors of \cite{[CP6]} did not report any details on their system architecture. However, from the code they published, it could be seen that their model was likely ran on an Amazon Web Services (AWS) instance, given the name of the top-level folder. Details on the AWS instance are, however not available, so we cannot say whether they relied on multiple CPUs or on GPUs.

The winning team was a group of engineers at NVIDIA and in their paper \cite{[CP7]} they stated that their usage of NVIDIA's GPUs had allowed them to achieve a significant speed-up enabling them to tune the parametres not only on subsets of the train dataset but also based on the results from the challenge leaderboard. Specifically, authors note that their GPU-oriented code ran end-to-end with a speed-up of factor 28 compared to their CPU-optimised code, while the training part was sped up by a factor of 120. The team from \cite{[CP9]}, who ranked second, stated that they too had used NVIDIA Tesla V100 GPUs. This is contrasted by the team which achieved rank six and which stated in \cite{[CP8]} that they chose an Amazon Web Services (AWS) plan including 16 vCPUs and no GPU given the premise that the bottleneck was loading the data to feed the model, rather than forwarding of back-propagation of steps. 

\subsection{Results}
\label{secRecSys2020Results}

Tables \ref{tabLeaderBoard1}, \ref{tabLeaderBoard2}, and \ref{tabLeaderBoard3} contain the final ranking and scores of the teams which published papers at the ACM Recommender Systems Conference 2020 and which were discussed in this section. As we can see, multiple teams could share the same rank if their scores are equal. Somewhat unconventionally, in the case of a tie the subsequent ranks remain successive without a proportional skip in the ranking values. For instance, after two teams with a score of 20 got the rank four, the team wsm\_LLLLL\footnote{Full final leaderboard can be seen at \url{https://recsys-twitter.com/previous_challenge} (access date: 2021-04-06).} (which had not published a paper on their approach and is thus not included in the tables below) with the score of 21 got rank five, rather than six. 

\begin{table}[h]
\centering
\begin{tabular}{llllll}
\toprule
Team/User Name &     Reference & Rank & Score & Mean PRAUC & Mean RCE \\
\midrule
NVIDIA RAPIDS.AI &  \cite{[CP7]} &    1 &     9 &         \textbf{0.455} &     \textbf{33.1375} \\
learner          &  \cite{[CP9]} &    2 &    14 &      0.403275 &     23.2325 \\
Team Wantedly    &  \cite{[CP4]} &    3 &    18 &        0.3906 &       22.53 \\
learner\_recsys   &  \cite{[CP3]} &    4 &    20 &        0.4487 &       12.79 \\
BanaNeverAlone   &  \cite{[CP6]} &    4 &    20 &        0.3915 &      18.865 \\
Los Trinadores   &  \cite{[CP8]} &    6 &    22 &      0.528875 &       9.245 \\
Not\_Last\_Place   &  \cite{[CP2]} &    9 &    28 &      0.366975 &      17.085 \\
saaay71          &  \cite{[CP5]} &   28 &    63 &       0.20685 &    -14.8425 \\
\bottomrule
\end{tabular}
\caption{Leaderboard overview for the teams that published papers}
\label{tabLeaderBoard1}
\end{table}

\begin{table}
    \centering
\begin{tabular}{lllll}
\toprule
Team/User Name & PRAUC Like & RCE Like & PRAUC Reply & RCE Reply \\
\midrule
NVIDIA RAPIDS.AI &     \textbf{0.9108} &    \textbf{53.01} &      0.2185 &     \textbf{24.23} \\
learner          &     0.7761 &    25.42 &      \textbf{0.2218} &     21.94 \\
Team Wantedly    &     0.7716 &    24.76 &      0.1918 &     20.44 \\
learner\_recsys   &     0.7221 &    18.28 &      0.1161 &     10.42 \\
BanaNeverAlone   &     0.7531 &     21.2 &       0.185 &     18.71 \\
Los Trinadores   &      0.731 &    20.09 &      0.5135 &     -0.04 \\
Not\_Last\_Place   &     0.7511 &    17.64 &      0.1761 &      18.6 \\
saaay71          &     0.5879 &     7.21 &      0.0593 &     -6.27 \\
\bottomrule
\end{tabular}
\caption{Scores for like and reply; notice the only case where \cite{[CP7]} did not achieve the best score}
\label{tabLeaderBoard2}
\end{table}

\begin{table}
\centering
\resizebox{\textwidth}{!}{%
\begin{tabular}{lllll}
\toprule
Team/User Name & PRAUC Retweet & RCE Retweet & PRAUC Quote & RCE Quote \\
\midrule
NVIDIA RAPIDS.AI &        \textbf{0.6111} &       \textbf{37.91} &        \textbf{0.0796} &        \textbf{17.4} \\
learner          &        0.5395 &       30.96 &        0.0757 &       14.61 \\
Team Wantedly    &        0.5266 &       30.06 &        0.0724 &       14.86 \\
learner\_recsys   &        0.4529 &        22.5 &        0.5037 &       -0.04 \\
BanaNeverAlone   &        0.5042 &       27.21 &        0.1237 &        8.34 \\
Los Trinadores   &        0.3673 &       16.93 &        0.5037 &          -0 \\
Not\_Last\_Place   &        0.4891 &        21.8 &        0.0516 &        10.3 \\
saaay71          &        0.1703 &        2.67 &        0.0099 &      -62.98 \\
\bottomrule
\end{tabular}
}
\caption{ Scores for retweet and retweet with comment (recently renamed to quote)}
\label{tabLeaderBoard3}
\end{table}

\chapter{Proposed Method}
\label{chMethodology}

In this chapter, we detail the proposed method of assessing the role of contextual features in tweet engagement prediction. In section \ref{secSpark}, we introduce PySpark, the Python version of the Apache Spark high-performance computing framework and engine that was used for the technical part of this work. 

Section \ref{secDataSet} then provides an overlook of key aspects of the RecSys2020 Challenge dataset, which served as the starting point of the diploma thesis. The big size of the dataset meant that data sub-sampling was required to create and assess initial model designs within a reasonable amount of time. Given that the tweet's context depends on aspects like the relative frequency of appearances of individual users which would not be preserved in random sub-sampling, splitting data was not a trivial task. The methods used for this purpose are described in detail in section \ref{secSampling}. 

After presenting details about the dataset and a number of attempted sampling methods, section \ref{secFeatureEngineering} presents the central aspect of the proposed approach -- feature engineering. As underlined in many publications by the best-performing teams at the RecSys 2020 Challenge, defining new features to represent user relations and behavioural patterns was key to achieving good prediction scores. This task had been made even more important given this thesis' exclusion of tweet text. The manner in which the most informative among these numerous features were then selected is described in section \ref{secFeatureSelection}.

Section \ref{secAlgorithms} follows with a description of the prediction algorithms we used for solving the task at hand. Lastly, we present the method we employed to tune the hyperparametres and fit the prediction models in section \ref{secMachineLearningPipelineDesign}. This section also describes how the predictions of the fitted models were evaluated.

\section{Spark}
\label{secSpark}

Spark is an open-source processing engine for machine learning and data science managed by Apache\footnote{The project website can be seen at \url{https://spark.apache.org/} (last access: 2023-05-21)}. The engine was originally developed in the doctoral dissertation by Matei Zaharia \cite{SparkPhd} and has since become one of the largest and most active open source communities \cite{mlbLearningPyspark, bookSparkTheDefinitiveGuide}. The leading motivation for the development of the engine was that prior large-scale data-intensive processing implantations, the most popular of which was MapReduce \cite{mapReduce}, were built for acyclic operations that are unsuitable for applications that reuse the same set of data across multiple parallel operations \cite{spark}. These applications include machine learning algorithms and interactive machine learning algorithms \cite{spark}, such as the ones implemented for this thesis as well. By using a new lazily evaluated abstraction called \textit{resilient distributed datasets (RDDs)} that allow data sharing across multiple iterative computations, Spark may outperform Hadoop by a factor of ten in machine learning and other iterative algorithms \cite{SparkPhd,spark}. When data used is fully stored in memory, this processing speed may be up to 100 times faster \cite{mlbLearningPyspark}. While doing so, Spark retained the extendability, scalability, and fault tolerance of MapReduce \cite{spark, SparkPhd}.

The backbone of Spark operations, the RDDs, are a collection of Java Virtual Machine objects that are parallelised, i.e. divided based on some key and distributed to executor noted \cite{mlbLearningPyspark}. The key to the efficiency and fault-tolerance of RDDs is that their interface is based on coarse-grained transformations (e.g. map, filter, join) that operate on the lineage of data transformations rather than the data itself \cite{SparkPhd}. Thus, if a partition of data RDD is lost, only the lost partition needs to be recovered, rather than the whole of the dataset \cite{SparkPhd}. As mentioned before, RDDs and dataframes (which we introduce in the next paragraph) are lazily evaluated \cite{bookSparkTheDefinitiveGuide}. This implies that Spark does not perform data transformations in the order in which it was written in the code, but instead ``waits'' until the last possible moment, e.g. when the results of a transformation need to be \textit{collected} into a file, or aggregated into a statistic to be displayed at the screen \cite{bookSparkTheDefinitiveGuide}. This allows Spark to build and optimise plans of transformations to be performed end-to-end before actually computing on those plans \cite{bookSparkTheDefinitiveGuide}. RDDs are \textit{schema-less} or \textit{schema-agnostic} data structures,  meaning that instances contained in it can be of different (inconsistent) data types \cite{mlbLearningPyspark, SparkPhd}. For instance, in the case of PySpark, a list containing tuples, dictionaries, and lists could be parallelised into an RDD \cite[chapter 2]{mlbLearningPyspark}.

In contrast, dataframes are distributed datasets organised into named and typed columns that offer advantages when processing structured data in Spark \cite[chapter 3]{mlbLearningPyspark}. Dataframes are similar to the data type of the same name in the Python pandas package or in R, and in the case of PySpark and Spark for R, the conversion between datatypes is straightforward \cite{mlbLearningPyspark, bookSparkTheDefinitiveGuide}. The main difference between dataframes in pandas or R and dataframes in Spark is that the latter does not have to be stored and processed on a single machine but can instead be \textit{partitioned} to thousands of computers \cite{bookSparkTheDefinitiveGuide}. While it is possible to manipulate this partitioning of dataframe and how it would physically be stored on individual executors on the cluster, a more common approach is for the developer to specify high-level transformations of dataframes, and then Spark automatically optimises the distribution \cite[p. 18]{bookSparkTheDefinitiveGuide}. The dataframe data type was introduced in Apache Spark 1.0\footnote{Before Spark 1.3, dataframes used to be called SchemaRDD \cite[p. 56]{mlbLearningPyspark}.} and has since become the dominant structured API in Spark \cite{mlbLearningPyspark, bookSparkTheDefinitiveGuide}. The main advantages of dataframes over RDDs can be divided in two groups: ease of use and better performance \cite{mlbLearningPyspark}. Regarding the former, the structured nature of data allowed for more ready-to-use functions to be developed for dataframes than to RDDs and the main Spark machine learning library \textit{MLLib} \cite{mllibThePaper} transitioned from RDD to dataframe-based API in Spark 2 \cite{mlbLearningPyspark, pysparkMLLIBdocumentation}. On the other hand, the set structure of the data allowed Apache Spark to optimise the execution queries even further \cite[p. 33]{mlbLearningPyspark}. This improvement is especially noticeable in the case of PySpark, i.e. Spark for Python, since it allowed for a smaller communication overhead with the Java Virtual Machine. Consequently, in the implementation of this thesis, we relied exclusively on dataframes to store data in our Spark jobs. Later in this thesis, we specify further how exactly this was done, and in which manner we used the functions from MLLib, Spark's machine learning library \cite{mllibThePaper}.

\section{Dataset}
\label{secDataSet}

To better understand the proposed method in this thesis, it is necessary to provide insight into dataset size, the ratios of unique users and tweets to all instances, tweet timestamps, and the target engagements. These dataset elements shaped our approaches to dataset sampling, feature engineering, and the interpretation of results.

As previously mentioned, one of the defining aspects of the RecSys 2020 Challenge dataset \cite{tweetPap,recsys2020overview} is its size. This dataset was split by the organisers into three parts: train, val, and test. Tables \ref{tabDataset1} and \ref{tabDataset2} show the number of instances in each dataset. The tables also contain the ratios of distinct engaging users (tweet viewers), engaged-with users (tweet authors) and tweets to the total number of instances in the dataset.

\begin{longtable}{|l|l|l|l|}
\caption{Number of instances and unique tweets as well as the ratios between the two for each dataset}
\label{tabDataset1}\\ 
\hline
\textbf{Dataset} & \textbf{no. of instances} & \textbf{no. of tweets} & \textbf{\#inst./\#tw.}  \endfirsthead 
\hline
\textbf{train}   & $141,748,660$         & $69,680,733$       & $2.03$                \\ 
\hline
\textbf{val}     & $10,271,157$          & $7,042,109$        & $1.46$                \\ 
\hline
\textbf{test}    & $10,350,033$          & $7,096,161$        & $1.46$                \\
\hline
\end{longtable}

\begin{longtable}{|l|l|l|l|l|}
\caption{Number of unique authors and viewers (engaged-with and engaging uswers) and the ratio of these to the total number of instances in each dataset}
\label{tabDataset2}\\ 
\hline
\textbf{Dataset} & \textbf{no. of authors} & \textbf{\#inst./\#auth.} & \textbf{no. of viewers} & \textbf{\#inst./\#vie.}  \endfirsthead 
\hline
\textbf{train}   & $14,471,617$        & $9.79$                 & $24,929,665$        & $5.69$                 \\ 
\hline
\textbf{val}     & $2,772,099$         & $3.71$                 & $5,597,372$         & $1.83$                 \\ 
\hline
\textbf{test}             & $2,794,081$         & $3.70$                 & $5,633,471$         & $1.84$                 \\
\hline
\end{longtable}

The train datasets contain sampled tweets written between 00:00 UTC on Thursday, 2020-02-06, to 23:59 UTC on Wednesday, 2020-02-12. The val and test datasets were collected the following week, from Thursday, 2020-02-13 to Wednesday, 2020-02-19. This information can be seen in the column \texttt{Time\-stamp}. The column does not contain information about the time the tweet was seen, nor what the timezone of the viewer or of the authors is. 

\begin{figure}[htp]
\centering
\includegraphics[width=.45\textwidth]{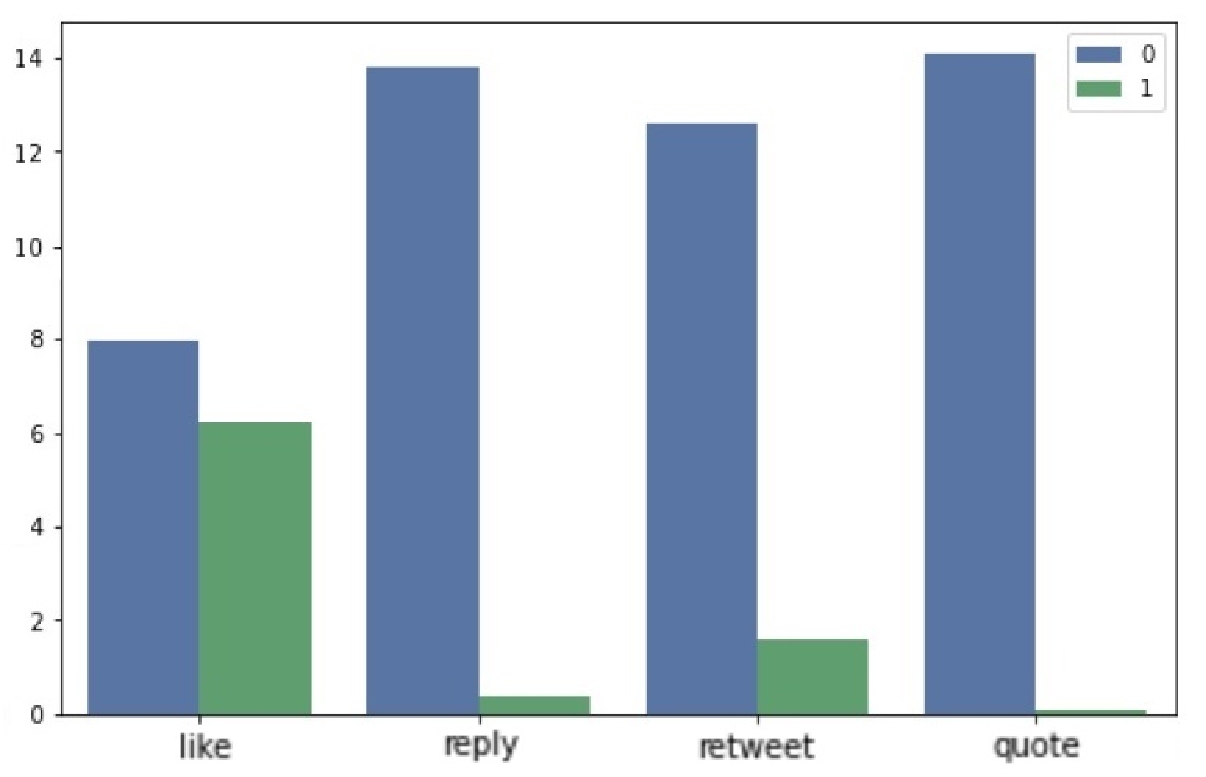} 
\caption{Total counts of positive ($1$) and negative ($0$) engagements in the train dataset for each engagement type; the y-axis' order of magnitude is ten million ($10^7$).}
\label{figDisbalance}
\end{figure}

Lastly, the dataset engagement targets -- like, reply, tweet, and quote -- were very imbalanced, particularly for engagement types other than like. Figure \ref{figDisbalance} illustrates the proportions of positive ($1$) and negative ($0$) interactions in the train dataset. This visualisation is inspired by a similar graph in Gradinariu's thesis \cite{davidsThesis}. Gradinariu also provided further insights into the dataset, focusing primarily on the tweet text, which we here omit due to our emphasis on context-based prediction.

Pre-provided feature attributes can be seen in table \ref{tabRows} in chapter \ref{chIntroduction}. The following subsection contains more details on the attribute types contained in the dataset.

\subsection{Attribute Types}
\label{secScales}

Understanding and correctly accounting for different attribute types, or attribute interval scales, can significantly improve the performance of algorithms \cite[p. 155]{mlbBayesianReasoningAndML2012}. Moreover, applying data analysis methods inappropriate for the data at hand had been recognised as a common mistake \cite{introToBistatistics}. Three central data encodings are categorical, ordinal, and numerical \cite[p. 155]{mlbBayesianReasoningAndML2012}. Numerical data is also referred to as quantitative data \cite[p. 456]{mlbTheElementsOfStatisticalLearning2009Chapter14unsupervisedLearning} and can be divided even further \cite{introToBistatistics}. This subsection defines each of these attribute scales and assigns each of the pre-provided features of the dataset into its type.

Categorical data is also called nominal data, and its defining characteristic is that the attribute values are distinct values with no intrinsic ordering \cite[p. 155]{mlbBayesianReasoningAndML2012}. The feature \emph{tweet\_type}, which can take the values of ``Retweet'', ``Quote'', ``Reply'', or ``Toplevel'', is of this type. For machine learning algorithms that require numerical input, categorical data could be one-hot-encoded, which is the process of turning one categorical attribute into multiple binary attributes \cite[p. 155]{mlbBayesianReasoningAndML2012}. More specifically, for each category of the attribute, there would be a binary feature which is set to True (or one) when the attribute for the given instance belongs to that category, and all other binary features are set to False (or zero). As is noted in  \cite[p. 155]{mlbBayesianReasoningAndML2012}, the benefit of this approach over the simple mapping of textual categories into numerical representations is that it creates no false ordered relations, while the downside is that it creates dependencies in the resulting attributes (when one of the resulting one-hot-encoded columns is one, others must be zero). Another simpler manner of transforming categorical data originally encoded using text into numerical representation is to assign each category a numerical value. This approach also avoids the additional downside of the one-hot-encoding, which is that, if there are many categories, there would have to be equally as many new attributes. Since this is the case with the other categorical attribute in the dataset, namely \emph{language}, we opted for this approach in our implementation. More details can be seen in Subsection \ref{secFinalFeatureEngineeringAndCategorisation}.

Ordinal data, in contrast to categorical data, has an inherent ordering and is often represented as contiguous integers \cite[p. 456]{mlbTheElementsOfStatisticalLearning2009Chapter14unsupervisedLearning}. If ordinal data is originally presented in a textual form, any mapping into numerical representation should preserve the ordering. For instance, a three-part satisfiability ranking could be mapped as: ``satisfied'' $\rightarrow$ $1$, ``neutral'' $\rightarrow$ $0$, ``unsatisfied'' $\rightarrow$ $-1$, as suggested in \cite[p. 155]{mlbBayesianReasoningAndML2012}. While none of the features in the dataset is ordinal, we chose to present the $timestamp$ attribute, indicating when the tweet was written, as two features: one being the day of the week and the other the hour of the day (both train, as well as test and val, were sampled over exactly one week). We then mapped the first day of the week in the dataset (Thursday) as $1$, the second present day (Friday) as $2$, etc. The hours in the day have an obvious representation, which is to represent them as integers between $0$ and $23$, which correspond to full hours during which they were written, after omitting the minutes\footnote{As Twitter is a global service, ideally the hours should be in the time zone of the engaging author. Unfortunately, we do not have this information in the dataset, although Twitter presumably can infer its users' location and time zone based on their IP or device settings. Further work may try to estimate the time zone based on the tweet's language, but this approach, too, would be imperfect, especially for global languages such as English and French.}.

Numerical or quantitative attributes are represented by real numbers, such as the temperature or a salary \cite[p. 155]{mlbBayesianReasoningAndML2012}. Yet, a further division of numerical values is possible: into interval scale and ratio scale \cite{introToBistatistics}. The interval scale can be seen as ``a step more sophisticated than the ordinal scale'' \cite{introToBistatistics} as the magnitude between any two values is always defined and consistent. For instance, the difference between two scales on a thermometer (interval scale) is always consistently defined, e.g. $1^{\circ}C$ and $2^{\circ}C$ is the same as between $4^{\circ}C$ and $5^{\circ}C$ - it is $1^{\circ}C$ in both cases. On the other hand, it is unclear whether the same can be said for a difference between two grades an Austrian student may be awarded. That is, the grades 1 (\textit{sehr gut}) and 2 (\textit{gut}) are potentially not as different as the grades 4 (\textit{genügend}) and 5 (\textit{nicht genügend}), which is typical for the ordinal scale. Another difference between the interval scale and the ratio scale is that the ratio scale contains an absolute zero, whereas, for the interval scale, the zero on the scale is arbitrarily chosen \cite{introToBistatistics}. Thus, on the one hand, temperature expressed in degrees Celsius is in interval scale, because the zero is arbitrarily set to the freezing temperature of water at sea level altitude, and negative values are possible. Temperature measured in Kelvin, on the other hand, is an example of the ratio scale, as the zero corresponds to absolute zero, the lowest temperature physically possible \cite{introToBistatistics}. For the dataset at hand, attributes that represent followers and following counts are ratio features. As elaborated in subsection \ref{secEncoding}, present media, links, and domains were also transformed in simple counts of present media elements of the type, making these into examples of ratio features as well.

\section{Data Sampling Methods}
\label{secSampling}

One of our motivations was to investigate how the change in dataset size affects the correctness of classifications. Dataset sampling in the context of this Challenge is not a trivial task, however. This is because the ratio of unique authors, unique viewers, unique users (i.e. authors and viewers), and tweets to the number of data instances is not maintained when a randomly selected sample is taken. Therefore, we have decided to have five distinct sampling methods or \textit{sampling techniques}:

\begin{enumerate}
    \item \texttt{random}: Completely random sampling (without replacement);
    \item \texttt{EU}: Random sampling that maintains the approximate ratio of the viewers (Engaging Users) to the number of instances;
    \item \texttt{EWU}: Random sampling that maintains the approximate ratio of the authors (Engaged-With Users) to the number of instances;
    \item \texttt{inter\_EWU+EU}: Random sampling that maintains the approximate ratio of both viewers and authors to the number of instances (the abbreviation alludes to the fact that we used set intersection to sample instances, as described in subsection \ref{secFeaturePrepImplementation});
    \item \texttt{tweet}: Random sampling that maintains the approximate ratio of unique tweets to the number of instances.
\end{enumerate}

In addition to working with full datasets, we created subsets of four distinct sizes. As elaborated in chapter \ref{chImplementation}, this allowed the code to be run on a local computer as well. These \textit{sample sizes} are as follows:

\begin{enumerate}
    \item \texttt{sample\_1pct}: Approximately 1\% of the dataset;
    \item \texttt{sample\_2pct}: Approximately 2\% of the dataset;
    \item \texttt{sample\_5pct}: Approximately 5\% of the dataset;
    \item \texttt{sample\_10pct}: Approximately 10\% of the dataset.
\end{enumerate}

Lastly, as described in section \ref{secDataSet}, the training dataset differs significantly in values of its explanatory features from those in val and test datasets. Therefore, we sampled our subsets from four distinct \textit{sources}:

\begin{enumerate}
    \item \texttt{train}: From the original train dataset;
    \item \texttt{val}: From the original val dataset;
    \item \texttt{test}: From the original test dataset;
    \item \texttt{val+test}: From the merged val and test datasets.
\end{enumerate}

Thus, we created a total of $5 \cdot 4 \cdot 4 = 80$ subsets in addition to the 3 original datasets and the combined \texttt{val+test} dataset. The naming convention for these subsets was \texttt{identifyingPrefix\_source\_technique\_sampleSize}. The complete list of all $84$ datasets and subsets (with the identifying prefix \emph{Final\_}, that indicates the finalised feature engineering together with actual sizes of the datasets and subsets) can be seen in section \ref{appSubsets} in the appendix. The implementation of this data preprocessing and data sampling can be seen in notebook \texttt{PS\-\_Data\-Prep\-/Py\-Spark\--Data\-Prep\--00\--Data\-Sampling.ipynb} that can be found in the publically shared git repository linked in subsection \ref{secCode}.

\section{Feature Engineering}
\label{secFeatureEngineering}

One of the central points of this thesis was inferring new features\footnote{Following the example in, e.g. \cite{mlbIntroductionToMLWithPython2016}, we interchangeably refer to this process of extracting new features as \textit{feature engineering} or \textit{feature extraction}. It should be noted that some works, such as \cite{mlbLearningPyspark}, differentiate between the two and consider \textit{feature engineering} to be a broader term that includes both feature extraction and feature selection.} that would enable more effective modelling of the tweet's context. As elaborated in subsection \ref{secRecSys2020FeatureEngineering}, most of the authors of the best-performing Challenge solutions recognised extracting new features as a decisive step towards creating models that achieved high prediction scores.

First, we had to convert each of the four target engagements from the original form consisting of timestamps indicating when the interaction had happened into a simpler Boolean form, more suitable for the classification task at hand. Subsequently, we added a new target label \emph{react}, which is set to be true when at least one of the ``actual'' engagements (like, reply, retweet, quote) took place. Throughout this section, we provide an interrupted (with text in-between some items) enumerated list with feature names and short descriptions for a better overview. Thus, the first five extracted features are:

\begin{enumerate}
    \item \emph{like}: a feature whose value is set to $0$ if \emph{like\_timestamp} is \emph{None} (an empty filed), and $1$ if it is a timestamp.
    \item \emph{reply}: analogues of \emph{like} for \emph{reply\_timestamp}.
    \item \emph{retweet}: analogues of \emph{like} for \emph{retweet\_timestamp}.
    \item \emph{quote}: analogues of \emph{like} for \emph{retweet\_with\_comment\_timestamp}.
    \item \emph{react}: true ($=1$) if any of the four engagements (\emph{like}, \emph{reply}, \emph{retweet}, \emph{quote}) occurred (are equal to $1$) and false ($=0$) otherwise.
\suspend{enumerate}

The implementation of these preprocessing steps can be seen in notebook \texttt{PS\-\_Feature\-Eng/Py\-Spark\--Feature\-Eng\--00\--Feature\-Eng.ipynb}. The remaining extracted features can be grouped in several categories corresponding to the subsections that follow. 

\subsection{Encoding}
\label{secEncoding}

As elaborated in section \ref{secScales}, most machine learning algorithms that we used, especially in how they were implemented in PySpark, could only accept numerical explanatory variables. Therefore, the categorical features needed to be either one-hot-encoded or encoded as a number. Moreover, scaling or binning of timestamps was required.

Present media, or \textit{media\_count}, was a feature comprising lists of media types in the tweet. Media type could be either a ``Photo'', a ``Video'', or a ``Gif'', and each tweet could contain none, one, or multiple of each of the media types. We had no information on what the media files had actually depicted. Therefore, we extracted four new features, three of which were simple counts for each of the three media types, while the last feature was the sum of all media files.

\resume{enumerate}
    \item \emph{photos\_count}: number of photos in the tweet.
    \item \emph{videos\_count}: number of videos in the tweet.
    \item \emph{gif\_count}: number of gifs in the tweet.
    \item \emph{media\_count}: sum of the three features above (this replaces the column of the same name in the original datasets, e.g. a cell with value $[$Photo Photo Video$]$ is replaced with $3$).
\suspend{enumerate}

We also added simple counts of the number of hashtags and links present in the target tweet.

\resume{enumerate}
    \item \emph{hashtags\_count}: number of hashtags in the target tweet.
    \item \emph{links\_count}: number of links in the target tweet.
    \item \emph{domains\_count}: number of domains in the target tweet. 
\suspend{enumerate}

We created two tweet-timestamp-based features.

\resume{enumerate}
    \item \emph{tweet\_weekday}: day of the week of the publication of the tweet (cf. subsection \ref{secScales} on the exact numbering).
    \item \emph{tweet\_hour}: the hour of the day of the publication of the tweet (UTC).
\suspend{enumerate}

Moreover, we added five features based on timestamps concerning the user. Four of these relate to the individual users (two for each the author and the viewer, as well as one for the relative age of the two). Note that we decided to model the relative age as a difference rather than a ratio. This design choice was made as the age from the creation of the account is an ordinal column making the difference a meaningful measure that is semantically more immediately understandable. 

\resume{enumerate}
    \item \emph{engaged\_creation\_year}: the year in which the viewer of the target tweet has created their Twitter account.
    \item \emph{engaging\_creation\_year}: the year in which the author of the target tweet has created their Twitter account.
    \item \emph{engaged\_age}: the age of the viewer of the target tweet represented as the number of months that passed between March 2006 and the month of the creation of the account.
    \item \emph{engaging\_age}: as \textit{engaging\_age} but for the author of the target tweet.
    \item \emph{creation\_age\_difference}: the arithmetic difference between \textit{engaged\_age} and \textit{engaging\_age}.
\suspend{enumerate}

The implementation of these feature encodings can be seen in notebook \texttt{PS\_FeatureEng
/PySpark-FeatureEng-01-Encoding.ipynb}.

\subsection{Graph-based Features}
\label{secGraphBasedFeatures}

The original dataset contained a feature indicating whether the account of the author\footnote{also referred to as engagee, engaged user, and engaged-with user} followed the account that had viewed their tweet\footnote{i.e. the viewer, the engager or the engaging user}. The opposite direction is not provided since, by the design of the dataset, the viewer always follows the author. This has been done in order not to reveal non-public information regarding which tweets a user had seen and chosen not to engage with. So, as described in \cite{tweetPap}, the authors created the pseudo-negative dataset by taking a subset of all tweets written by users the engaging users follow with which they did not interact (either because the engager actually has seen the tweet and chose not to interact with it or because the engager has not seen the tweet at all). 

Based on these provided first-degree connections, we added second-degree connections by, essentially, joining the list of following relations with itself (with the joining key being the followed users on one side and following users on the other) and then adding the new list's followed users as the old list's following users second-degree connections. We see the follows-relation as more static than other types of user-to-user connection. Therefore, we chose to use here use all information available to get a more complete picture of the follow-relation. Specifically, we used:

\begin{itemize}
    \item The whole train dataset to get the relations for the instances in the training dataset;
    \item The whole train and the whole val dataset for the val dataset;
    \item The whole train, val, and test datasets for the test and val+test datasets.
\end{itemize}

We do not consider this to mean that these features are partially ``looking into the future''. It is the case that, for instance, the follows-information that we get from a tweet from Saturday might  be ``transferred'' into a feature for a targeted tweet for Friday. Nevertheless, we take this as-is (we do not consider the graph-based features in this subsection to be in the ``oracle'' category, as we do for the features below). One reason for this decision is the aforementioned relative static nature of the relationship -- following and unfollowing users does not happen as frequently as engaging with a hashtag or liking a tweet. Another reason is that all of the graph-based features we introduce in this subsection are information Twitter would likely save for each user rather than for each tweet. So, in essence, here we are approaching the engineering of this set of features as rebuilding that user profile information. 

In case of inconsistencies (where for one tweet, A follows B, but in another, A does not follow B), we take the positive case (i.e. we set that A follows B). An alternative to this approach would be to take the later follows relation (rather than taking the positive relation if it exists in at least one instance). There are two reasons we chose not to do this. The first one is technical, as we wanted to have the memory requirements be as low as possible (this was already one of the most memory-intensive parts of feature engineering). The second was that we assume that the follows-graph we create will be negatively skewed anyways and that we would have fewer positive relations than is actually the case due to the limited amount of data we have. So this design choice might have a likely positive side-effect of decreasing this anticipated skewness. We created the following two follows-relations:

\resume{enumerate}
    \item \emph{graph\_engagee\_follows\_engager\_2d}: did the author of a tweet follow a third user who followed the viewer?
    \item \emph{graph\_engager\_follows\_engagee\_2d}: did the user who has seen the author's tweet follow a third user who followed the author?
\suspend{enumerate}

We also created two features that relate to the ratios of the number of followers. 

\resume{enumerate}
    \item \emph{ratio\_engaged\_to\_engaging\_follower\_counts}: the ratio of numbers of followers between the author and the engager (based on \textit{engaged\_with\_user\_follower\_count} and \textit{engaging\_user\_follower\_count}, which are features that were given in the original datasets).
    \item \emph{ratio\_engaged\_to\_engaging\_following\_counts}: the ratio of the number of followers (of the first degree) of the target tweet's author compared to the tweet's viewer (based on given features \textit{engaged\_with\_user\_following\_count} and \textit{engaging\_user\_following\_count}).
\suspend{enumerate}

For the five engagement types (four pre-provided engagements plus \emph{reacts}), we created two features indicating whether the engager had ever (not just for the tweet in question) interacted with the engagee and vice versa. We also created second-degree connections in a manner analogous to what was described for the followers\footnote{These graph features are very similar to those in the RealGraph that Twitter actually applies, as reported in \cite{twitterRecommenderBlog,realgraph} and discussed in subsection \ref{secTwittersCurrentRS}. At the time of feature engineering, however, we were unaware of this fact.}. 

As discussed for the following relation, these graph relations too are seen as static and user-based rather than tweet-based. For this reason, we did not concern ourselves with the potential time inconsistencies within the training set and used the whole train dataset for feature engineering for itself. For the remaining datasets, we used the whole train dataset as well, but we did not use insights from the val and test datasets because this would not have been possible during the challenge itself. Namely, the engagement ground truth was released only after the challenge had finished.

Again, in case of inconsistencies, we took the relation to be positive. This time this is the only meaningful choice, as the features model whether there has ever been an engagement relation for the user. In this manner, we created a total of 20 binary engagement features:

\resume{enumerate}
    \item \emph{graph\_engaged\_flag\_\textbf{engagement}\_engaging\_1d}: a set of features with \emph{engagement} $\in\ \{$\emph{liked, replied, retweeted, quoted, reacted}$\}$ indicating whether the author had engaged in that manner with the person who had seen the tweet in the past.
    \item \emph{graph\_engaging\_flag\_\textbf{engagement}\_engaged\_1d}: a set of features with \emph{engagement} $\in\ \{$\emph{liked, replied, retweeted, quoted, reacted}$\}$ indicating whether the person who saw the tweet had engaged in that manner with the author in the past.
    \item \emph{graph\_engaged\_flag\_\textbf{engagement}\_engaging\_2d}: a set of features indicating whether the author of the tweet had \emph{engagement} $\in\ \{$\emph{liked, replied, retweeted, quoted, reacted}$\}$  a tweet of someone who in turn engaged in the same manner to the tweet of the person who saw the author's target tweet.
    \item \emph{graph\_engaging\_flag\_\textbf{engagement}\_engaged\_2d}: a set of features indicating whether the user who saw the target tweet had \emph{engagement} $\in\ \{$\emph{liked, replied, retweeted, quoted, reacted}$\}$ a tweet of someone who in turn had engaged in the same manner to the tweet of this author.
\suspend{enumerate}

We also created a version of the engagement features above that actually counts the engagements (rather than building a binary flag). This offers a context-based proxy for the shared interest between the two users; the more engagements there are in their shared engagements graph, the more similar their interests. Thus we got 20 new engagement features:

\resume{enumerate}
    \item \emph{graph\_engaged\_count\_\textbf{engagement}\_engaging\_1d}: a set of features with \emph{engagement} $\in\ \{$\emph{liked, replied, retweeted, quoted, reacted}$\}$ indicating how many times the author had engaged in that manner with the person who had seen the tweet in the past.
    \item \emph{graph\_engaging\_count\_\textbf{engagement}\_engaged\_1d}: a set of features with \emph{engagement} $\in\ \{$\emph{liked, replied, retweeted, quoted, reacted}$\}$ indicating how many times the person who saw the tweet had engaged in that manner with the author in the past.
    \item \emph{graph\_engaged\_count\_\textbf{engagement}\_engaging\_2d}: a set of features indicating how many times the author of the tweet had \emph{engagement} $\in\ \{$\emph{liked, replied, retweeted, quoted, reacted}$\}$  a tweet of someone who in turn engaged in the same manner to the tweet of the person who saw the author's target tweet.
    \item \emph{graph\_engaging\_count\_\textbf{engagement}\_engaged\_2d}: a set of features indicating how many tymes the user who saw the target tweet had \emph{engagement} $\in\ \{$\emph{liked, replied, retweeted, quoted, reacted}$\}$ a tweet of someone who in turn had engaged in the same manner to the tweet of this author.
\suspend{enumerate}

The implementation of the creation of these features can be seen in notebook \texttt{PS\_FeatureEng
/PySpark-FeatureEng-02-GraphBased.ipynb}. Yet, even after multiple attempts of graph creation optimisations (such as saving partial results into files on the disk and slimming down graphs to single columns before merging), the creation of engagement features for the second degrees failed on the LBD cluster for the full dataset due to insufficient memory. Therefore, as a proxy, we created these features for the full datasets based on the 10\% counterparts. This version of the implementation can be seen in the notebook \texttt{PS\_FeatureEng/PySpark-FeatureEng-02-GraphBased-Proxy4FullDs.ipynb}

\subsection{Current Trends}
\label{secShortTimePopularityEstimates}

This group of features looks at events that happened in the last 30 minutes, 60 minutes, and 120 minutes, as well as 12 hours, 24 hours, and 48 hours. The motivation for this set of features is Twitter's motto: ``It's what's happening'' \cite{recsys2020overview}. These features are thus our proxies for what is currently popular, or ``trending'' in the slang of social media. Since we are focusing on context, we investigated both non-user-specific features (hashtags, domains, and links) as well as user-specific features (measuring the popularity of authors and engagement activity levels of viewers).

Since val and test datasets were sampled from the week that followed the week in which the train dataset was sampled, we first needed to append  the last two days of the train dataset to val, test, and val+test datasets (cf. section \ref{secSampling}). We did this for all subsets as well by appending the same sampling methods together. For instance, the last two days of \textit{train\_EWU\_sample\_10pct} were appended to \textit{val\_EWU\_sample\_10pct} , etc. 

The user-specific features we created in this subsection were again inspired by \cite{[CP9]}. Specifically, we created new features representing counts of tweet views. We slightly changed the time windows: they looked into 5 minutes, 1h, 4h, 8h, and 24h, whereas we had 30 minutes, 1h, 2h, 12h, 24h, and 48h. We chose not to look into 5 minutes, as the feature would be mostly empty for smaller dataset samples. The remaining differences were made as we believed they would allow us for an easier generalisation of patterns: we have three short time windows, and three longer time windows, each with three values that differ by a factor of two. And as stated above, we also looked into the whole time dataset. Since for each instance we save both the number of tweets the engaging user has seen and the number of views the engaged-with user has gotten, we thus get $6 \cdot 2 =  12$ new features. 

\resume{enumerate}
    \item \emph{engaging\_saw\_tweets\_count\_\textbf{no\_hours}}: a set of features with \emph{no\_hours} $\in$ $\{0.5h,$ $1h,$ $2h,$ $12h,$ $24h,$ $48h\}$ representing how many tweets the engaging user has seen within the given time period.
    \item \emph{engageds\_tweets\_views\_count\_\textbf{no\_hours}}: a set of features with \emph{no\_hours} $\in$ $\{0.5h,$ $1h,$ $2h,$ $12h,$ $24h,$ $48h\}$ representing how many times tweets of the engaged-with user have been seen within the given time period.
\suspend{enumerate}

The dataset contained features with extracted hashtags, links, and domains present in the tweet (if there were any). Unlike the tweet text, however, these three features could not be decoded to retrieve their meaning, as the fields contained hashes of the text rather than BERT tokens or clear text, as elaborated in \cite{tweetPap}. While the input of hash functions is unobtainable from their output, we still know that the same input always leads to an identical output. For each target tweet, we looked at all its hashtags, links, and domains present in that tweet and counted how many other tweets in the past  0.5, 1, 2, 12, 24, and 48 hours there have been with at least one of these tweet elements and summed the appearances in the corresponding three sums. Thus calculated short-term hashtag, domain, and link popularity, similarly as was done in \cite{[CP2]}. These are the $6 \cdot 3 = 18$ non-user-specific short-time popularity estimates. 

\resume{enumerate}
    \item \emph{hashtags\_frequency\_\textbf{no\_hours}}: sum of the counts of appearances for each of the hashtags in the tweet over the last \emph{no\_hours} $\in\ \{0.5h, 1h, 2h, 12h, 24h, 48h\}$ in the dataset.
    \item \emph{links\_frequency\_\textbf{no\_hours}}:  sum of the counts of appearances for each of the links in the tweet over the last \emph{no\_hours} $\in\ \{0.5h, 1h, 2h, 12h, 24h, 48h\}$ in the dataset.
    \item \emph{domains\_frequency\_\textbf{no\_hours}}:  sum of the counts of appearances for each of the domains in the tweet over the last \emph{no\_hours} $\in\ \{0.5h, 1h, 2h, 12h, 24h, 48h\}$ in the dataset.
\suspend{enumerate}

We also collected how many other tweets with at least one of the hashtags, links, or domains from the target tweet were seen (in the dataset) by the viewer of the target tweet. We do this again for the same six short time windows. This approach may be considered a rudimentary proxy for whether the viewer is genuinely interested in the area. Thus we created $6 \cdot 3 = 18$ user-specific short-term popularity features.

\resume{enumerate}
    \item \emph{user\_hashtags\_frequency\_\textbf{no\_hours}}: sum of the counts of appearances for each of the hashtags in the tweets in the dataset that the engaging user has seen over the last \emph{no\_hours} $\in\ \{0.5h, 1h, 2h, 12h, 24h, 48h\}$.
    \item \emph{user\_links\_frequency\_\textbf{no\_hours}}:  sum of the counts of appearances for each of the links in the tweets in the dataset that the engaging user has seen over the last \emph{no\_hours} $\in\ \{0.5h, 1h, 2h, 12h, 24h, 48h\}$.
    \item \emph{user\_domains\_frequency\_\textbf{no\_hours}}:  sum of the counts of appearances for each of the domains in the tweets in the dataset that the engaging user has seen over the last \emph{no\_hours} $\in\ \{0.5h, 1h, 2h, 12h, 24h, 48h\}$.
\suspend{enumerate}

The time-window-based features from this section are implemented in the notebook with the title \texttt{PySpark-FeatureEng-03-TimeFeatures.ipynb}. It should be noted that extracting these features was very time-intensive, as is discussed in section \ref{secComputationalPerformance}.

\subsection{Oracle Features}
\label{secOracleFeatures}

We also created a version of features from subsection \ref{secShortTimePopularityEstimates} that does not only look at a short time window before the target tweet but the whole dataset on which the features are engineered. We dubbed these features ``oracle'' because they provide an estimate of popularity that also looks into the future and is a dynamic one (unlike the graph features in subsection \ref{secGraphBasedFeatures}). As we will see later, we chose to exclude these $6$ features from a set of experiments, as this is a piece of information that would not be available for online predictions. We then compared these results with the results we got for all features, including these oracle features, to get an insight into whether "looking into the future" helps make the predictions better. We thus have the following eight features:

\resume{enumerate}
    \item \emph{hashtags\_frequency}: sum of the counts of appearances for each hashtag in the tweet.
    \item \emph{links\_frequency}: sum of the counts of appearances for each link in the tweet.
    \item \emph{domains\_frequency}: sum of the counts of appearances for each domain in the tweet.
    \item \emph{user\_hashtags\_frequency}: sum of the counts of appearances for each hashtag in the tweets in the dataset that the engaging user has seen.
    \item \emph{user\_links\_frequency}: sum of the counts of appearances for each of the links in the tweets in the dataset that the engaging user has seen.
    \item \emph{user\_domains\_frequency}: sum of the counts of appearances for each of the domains in the tweets in the dataset that the engaging user has seen.
    \item  \emph{engaging\_saw\_tweets\_count}: how many tweets the engaging user has seen in total.
    \item \emph{engageds\_tweets\_views\_count}: how times in total have tweets of the engaged-with user been seen.
\suspend{enumerate}

These features, like those in subsection \ref{secShortTimePopularityEstimates} are also implemented in the notebook \texttt{PySpark-FeatureEng-03-TimeFeatures.ipynb}.

\subsection{Tweet Elements Engagement History}
\label{secTweetElementsEngagementHistory}

We also wanted to see the engagement history for all tweet elements (hashtags, domains, and links). And we want to do this both in general and for each user. In subsection \ref{secShortTimePopularityEstimates}, we just counted the hashtags, domains, and links a user has seen in the previous hours from the same dataset). Yet, in this subsection, we want to see what ratio of the specific tweets, hashtags, and links were engaged within the entire train dataset. So while features from subset \ref{secShortTimePopularityEstimates} represented the user's momentarily interests, these features are a proxy for a more long-term user interest and their engagement history. We also want to estimate the general ``engageability'' of the tweet element present in the target tweet.

In the original challenge, test and val datasets did not contain target labels. Thus counting engagement labels for them would not have been possible for the duration of the challenge. To stay in line with those constraints, for creating engagement history, we created proxies as follows:

\begin{itemize}
    \item For val, test, and val+test datasets (and their subsets), we want to use the entire train dataset  (or the entire corresponding train subset) as the engagement history from which the features for the val, test, or val+test datasets (and their subsets) are to be created and then later used for testing.
    \item For the training dataset (and its subsets), we want to use the first three days of the training dataset (or subsets) as the engagement history to be saved for the last four days  (i.e. Thursday, Friday, Saturday $\rightarrow$ using as history; Sunday, Monday, Tuesday, Wednesday $\rightarrow$ saving engagement history statistics). This was done to avoid contaminating our results based on the training dataset. Unfortunately, this also means that the first three days of the train dataset cannot be used for training and testing purposes in the remainder of the machine learning pipeline.
\end{itemize}

Partially inspired by the work of \cite{[CP6]}, we began by counting the number of positive and negative engagements and authored tweets for all users in the corresponding engagement history (i.e. how many of the tweet instances were engaged with or not in the given manner). Since we count for both the engaged and the engaged-with user and since there are two possible outcomes for five different engagement types, we thus get $2 \cdot 2 \cdot 5 = 20$ count-based features.

\resume{enumerate}
    \item \emph{engaging\_count\_positive\_tweet\_\textbf{engagements}}: a set of features representing counts of \emph{engagements} $\in\ \{$\emph{like, reply, retweet, quote, react}$\}$  by the tweet's viewer in the engagement history dataset, i.e. the number of times the user had engaged with tweets in the given manner.
    \item \emph{engaged\_with\_count\_positive\_tweet\_\textbf{engagements}}: a set of features representing counts of \emph{engagements} $\in\ \{$\emph{like, reply, retweet, quote, react}$\}$  the target tweet's author has gotten in the past, i.e. the number of times users have engaged with the author's tweets in the given manner.
    \item \emph{engaging\_count\_negative\_tweet\_\textbf{engagement}}: a set of features representing counts of negative \emph{engagements} $\in\ \{$\emph{like, reply, retweet, quote, react}$\}$ by the target tweet's viewer in the engagement history dataset, i.e. the times the user had seen a tweet but did \emph{not} interact in the given manner.
    \item \emph{engaged\_with\_count\_negative\_tweet\_\textbf{engagement}}: a set of features representing counts of negative \emph{engagements} $\in\ \{$\emph{like, reply, retweet, quote, react}$\}$ the target tweet's author had in the engagement history dataset, i.e. the number of times other users had seen the author's tweets but did \emph{not} interact in the given manner.
\suspend{enumerate}

Also inspired by a similar idea in \cite{[CP6]} on modelling individual users' preferences for each hashtag or link, we created the count-based features for tweet elements. Again, unlike what was done with features in short-term time features in subsection \ref{secShortTimePopularityEstimates}, we are looking for the entire engagement history, not just a small period before the target tweet from the same dataset. Unlike the ``oracle'' features from subsection \ref{secOracleFeatures}, here we are not looking in the future because the engagement history is always from days prior to the day of the target tweet, and not even from the same dataset as the target tweet. 

Also, it should be noted that we had to count tweet frequencies based on the \textit{engaged\_with\_user\_id} and not \textit{tweet\_id}. We had to do so because there are no tweets from train dataset in test or val dataset. But the same design decision was made by some of the RecSys 2020 winning teams (cf. subsection \ref{secRecSys2020FeatureEngineering}). Moreover, it is reasonable to assume that if someone has used popular hashtags/links in the past, they are likely to keep on doing so in the future. To indicate that we have taken \textit{engaged\_with\_user\_id} as a proxy for not \textit{tweet\_id}, these features contain \textit{\_user\_proxy\_} as part of their names. 

In total, we thus get $3 \cdot 5 \cdot 2 = 30$ new features. We have three different tweet elements (hashtags, links, domains), five engagement types, and two  outcomes (positive and negative). Here, we only counted what the engaging user has seen in their engagement history and we do not count the total views for hashtags, links, or domains of the engaged user. 

\resume{enumerate}
    \item \emph{links\_user\_proxy\_count\_negative\_tweets\_\textbf{engage}}: a set of features representing the number of times the viewing/engaging user had seen tweets from the corresponding engagement history subset that contained at least one of the links from the target tweet and did \textit{not} \emph{engage} $\in\ \{$\emph{like, reply, retweet, quote, react}$\}$ with it.
    \item \emph{hashtags\_user\_proxy\_count\_negative\_tweets\_\textbf{engage}}: defined in the same manner as \emph{links\_user\_proxy\_count\_negative\_tweets\_\textbf{engage}} but with hashtags instead of links.
    \item \emph{domains\_user\_proxy\_count\_negative\_tweets\_\textbf{engage}}: defined in the same manner as \emph{links\_user\_proxy\_count\_negative\_tweets\_\textbf{engage}} but with domains instead of links.
    \item \emph{links\_user\_proxy\_count\_positive\_tweets\_\textbf{engage}}: a set of features representing the number of times the viewing/engaging user had seen tweets from the corresponding engagement history subset that contained at least one of the links from the target tweet and did \emph{engage} $\in\ \{$\emph{like, reply, retweet, quote, react}$\}$ with it.
    \item \emph{hashtags\_user\_proxy\_count\_positive\_tweets\_\textbf{engage}}: defined in the same manner as\emph{links\_user\_proxy\_count\_negative\_tweets\_\textbf{engage}} but with hashtags instead of links.
    \item \emph{domains\_user\_proxy\_count\_positive\_tweets\_\textbf{engage}}: defined in the same manner as \emph{links\_user\_proxy\_count\_negative\_tweets\_\textbf{engage}} but with domains instead of links.
\suspend{enumerate}

We also added features that are not user-specific but tweet-element-specific. These features provide information about hashtag, link, and domain popularity in the corresponding engagement history datasets. Firstly, we added $3$ simple features that count the number of tweets with at least one hashtag from the target domain in the engagement history dataset.

\resume{enumerate}
    \item \emph{hashtags\_count\_all\_tweets}: the number of tweets in the engagement history with at least one hashtag from the hashtags in the target tweet; if the target tweet has no hashtags, this feature is set to zero.
    \item \emph{links\_count\_all\_tweets}: defined in the same manner as \emph{hashtags\_count\_all\_tweets} but for hashtags instead of hashtags.
    \item \emph{domains\_count\_all\_tweets}:  defined inas  \emph{hashtags\_count\_all\_tweets} but for domains instead of hashtags.
\suspend{enumerate}

How the engagement history creation for each of the datasets and subsets was created can be seen in the notebook \texttt{PySpark-FeatureEng-04-EngagementFeatures.ipynb}. The same notebook also contains the engineering of features from this subsection as well as subsections \ref{secLanguageHistory} and \ref{secRatios}. In fact, since features represented in these sections are absolute and relative engagement counts, they are affected by the size of the engagement history subsets, which differ dramatically depending on the target dataset for which the features are built. Therefore, we have not used these features directly for feature selection and classification. Instead, we used ratios, as is discussed later in this section.

\subsection{Language History}
\label{secLanguageHistory}

The language of the tweet was one of the original features provided contained in the dataset. As described in section \ref{secTweetElementsEngagementHistory}, for each dataset and subset, we designated (a part of) the train dataset as the data for engagement history. We counted the number of positive and negative past reactions (how many times the user had engaged with a tweet they had seen and how many times they had not engaged). As is described above, we did this first for the hashtags, links, and domains in the tweet.

This subsection describes how we used the engagement history to infer information about the language preferences of the engaging and the engaged-with users. First of all, we looked at the information in column \textit{language} that was provided in the original dataset. As \cite{davidsThesis} described, the names of languages are encoded with no immediately obvious decodings. For instance, English is represented as \textit{ECED\-8A16\-BE2A\-5E88\-71FD\-55F4\-842F\-16B1}. We thus manually decoded the 10 most frequent languages in the dataset and the code for the unknown language (used when Twitter could not determine the language or the tweet only contained non-language-specific symbols). These are: 1) English; 2) Japanese; 3) Spanish; 4) Portuguese; 5) Unknown; 6) Turkish; 7) Arabic; 8) Korean; 9) Thai; 10) French; and 11) Indonesian. We then added a new Boolean feature that is true only if the language is unknown. This Boolean feature would then be used in the machine learning pipeline.

\resume{enumerate}
    \item \emph{language\_unknown}: set to True if the language is unknown (i.e. the value of the column \emph{language} equals to \textit{B917\-5601\-E871\-01A9\-84A5\-0F8A\-62A1\-C374}) and set as False otherwise.
\suspend{enumerate}

A set of more complex features was also engineered. The leitmotiv in creating these features was to identify the user's language preferences. We did this by collecting all languages the engaging users have seen in a new column \emph{seen\_languages} and all languages the engaged-with user has written in in the column \emph{authored\_languages}. For each user, we then also created a column that contains counts of languages they have seen or authored. Specifically, the column \emph{seen\_languages\_dict} contained a dictionary (a key-value map) summarising the engagement history of the viewer of the target tweet. Its keys are languages that the user has seen in the corresponding engagement history subset, while the values are the numbers of tweets in that language in the engagement history that the user has seen. Similarly, the column \emph{authored\_languages\_dict} provided the counts of tweet languages the author of the target tweet has written in the engagement history subset. These were then used to get $2$ count columns that represent the engaging and engaged-with users' language preferences. 

\resume{enumerate}
    \item \emph{this\_language\_seen\_count}: represents the number of tweets from the corresponding engagement history subset in the same language as the target tweet that the engaging user has seen.
    \item \emph{this\_language\_authored\_count}: represents the number of tweets from the corresponding engagement history subset in the same language as the target tweet that the engaged-with user has authored.
\suspend{enumerate}

The implementation of the engineering of these three language-based features, like the tweet element engagement features from the subsection \ref{secTweetElementsEngagementHistory}, can be found in the notebook \texttt{PySpark-FeatureEng-04-EngagementFeatures.ipynb}.

\subsection{Ratios}
\label{secRatios}

Features from the subsections \ref{secTweetElementsEngagementHistory} and \ref{secLanguageHistory} above consisted mostly of counts and sums for the subsets designated as engagement history. As such, they could not be used directly for classification tasks in the machine learning pipeline (with the exception of the Boolean feature \emph{language\_unknown}), as the counts are affected not only by the individual user preference but also by the size of the corresponding engagement history subsets. Therefore, we needed relative features that would account for these differences in history sizes. We needed ratios. 

To build these ratio features, we needed two further counts of tweets from the engagement history. The first is the number of tweets that the engaging user has seen in the given engagement history, while the second is the number of views the engaged-with user has had in the engagement history.  

\resume{enumerate}
    \item \emph{ratio\_all\_to\_engaging\_count\_positive\_tweets\_\textbf{engagement}}: the number of tweets from the engagement history subset that the viewer of the target tweet had seen.
    \item \emph{engaged\_\-saw\_\-tweets\_\-count\_\-test}: the number of views that tweets written by the author of the target tweet got in the engagement history subset.
\suspend{enumerate}

With these features to serve as denominators in our ratios, we can define $2 \cdot 2 \cdot 5 = 20$ new features -- $10$ for the author and $10$ for the viewer of the target tweet -- that represent the activity of the users by seeing how many of the interactions in the engagement history are related to them.

\resume{enumerate}
    \item \emph{ratio\_all\_to\_engaging\_count\_positive\_tweets\_\textbf{engagement}}: a set of features representing the ratios of past positive engagements for the viewer of the target tweet (\emph{enga\-ging\_count\_\-po\-si\-ti\-ve\_\-tweets\_\-\textbf{en\-gage\-ment}}) to all tweets in the corresponding engagement history (\emph{en\-gag\-ing\_\-count\_\-all\_\-tweets}) for \emph{en\-gage\-ment} $\in\ \{$\emph{like, reply, retweet, quote, react}$\}$.
    \item \emph{ratio\_all\_to\_engaging\_count\_negative\_tweets\_\textbf{en\-gage\-ment}}: defined in analogues manner as \emph{ratio\_\-all\_\-to\_\-en\-gag\-ing\_\-count\_\-po\-si\-ti\-ve\_\-tweets\_\-\textbf{en\-gage\-ment}} but for negative engagements instead of positive.
    \item \emph{ratio\_all\_to\_engaged\_with\_count\_positive\_tweets\_\textbf{en\-gage\-ment}}: defined in analogues manner as \emph{ratio\_\-all\_\-to\_\-en\-gaged\_\-with\_\-count\_\-po\-si\-ti\-ve\_\-tweets\_\-\textbf{en\-gage\-ment}} but for the author of the target tweet and not the viewer. That is, this feature is the ratio of \emph{en\-gaged\_\-with\_\-count\_\-po\-si\-ti\-ve\_\-tweets\_\-\textbf{en\-gage\-ment}} to \emph{en\-gag\-ing\_\-count\_\-all\_\-tweets}.
    \item \emph{ratio\_all\_to\_engaged\_with\_count\_negative\_tweets\_\textbf{en\-gage\-ment}}: defined in analogues manner as \emph{ratio\_\-all\_\-to\_\-en\-gag\-ed\_\-with\_\-count\_\-po\-si\-ti\-ve\_\-tweets\_\textbf{en\-gage\-ment}} but for negative engagements instead of positive.
\suspend{enumerate}

In addition to the ratio of seen and authored tweets to all the tweets in the engagement history, we also calculated the ratios concerning the hashtags, links, and domains. We again did this only for the engaging user, but both for positive and negative interaction, as well as the five engagement types, resulting in $3 \cdot 1 \cdot 2 \cdot 5 = 30$ additional ratio features. As discussed in section \ref{secTweetElementsEngagementHistory}, we used \emph{engaged\_with\_user\_id} as proxies for \emph{tweet\_id} because there are no overlapping tweets between the train dataset (from which the engagement history is taken) and the val and test datasets. This is why the features listed below contain \emph{\_user\_proxy\_} in their names.

\resume{enumerate}
    \item \emph{ratio\_all\_to\_hashtags\_user\_proxy\_count\_positive\_tweets\_\textbf{engagement}}: a set of features representing the ratios of \emph{hashtags\_\-user\_\-proxy\_\-count\_\-po\-si\-ti\-ve\_tweets\_\-\textbf{en\-gage\-ment}} (tweets from the engagement history subset that share at least one hashtag with the target tweet and which the viewer of the target tweet has seen and interacted with in the given manner) to \emph{hashtags\_count\_all\_tweets} (all tweets in the engagement history subset that share at least one  of the hashtags with the target tweet) for \emph{en\-gage\-ment} $\in\ \{$\emph{like, reply, retweet, quote, react}$\}$. 
    \item \emph{ratio\_all\_to\_hashtags\_user\_proxy\_count\_negative\_tweets\_\textbf{engagement}}: as \emph{ra\-tio\_\-all\_\-to\_\-hash\-tags\_\-user\_\-proxy\_\-count\_\-po\-si\-ti\-ve\_\-tweets\_\-\textbf{en\-gage\-ment}} but with the count of tweets from the engagement history subset, that the viewer of the target tweet has also seen but did \textit{not} have an \emph{engagement} $\in\ \{$\emph{like, reply, retweet, quote, react}$\}$ with. That is, the ratio of negative rather than positive tweets counts is in the numerator. This corresponds to the ratio of \emph{hash\-tags\_\-user\_\-proxy\_\-count\_\-ne\-ga\-ti\-ve\_tweets\_\-\textbf{en\-gage\-ment}} to \emph{hash\-tags\_\-count\_\-all\_\-tweets}.
    \item \emph{ratio\_all\_to\_links\_user\_proxy\_count\_positive\_tweets\_\textbf{en\-gage\-ment}}: defined analogously to \emph{ra\-tio\_\-all\_\-to\_\-links\_\-user\_\-proxy\_\-count\_\-po\-si\-ti\-ve\_\-tweets\_\-\textbf{en\-gage\-ment}} but for links instead of hashtags.
    \item \emph{ratio\_all\_to\_links\_user\_proxy\_count\_negative\_tweets\_\textbf{en\-gage\-ment}}: defined analogously to \emph{ra\-tio\_\-all\_\-to\_\-links\_\-user\_\-proxy\_\-count\_\-ne\-ga\-ti\-ve\_\-tweets\_\-\textbf{en\-gage\-ment}} but for links instead of hashtags.
    \item \emph{ratio\_all\_to\_domains\_user\_proxy\_count\_positive\_tweets\_\textbf{en\-gage\-ment}}: defined analogously to \emph{ra\-tio\_\-all\_\-to\_\-domains\_\-user\_\-proxy\_\-count\_\-po\-si\-ti\-ve\_\-tweets\_\-\textbf{en\-gage\-ment}} but for domains instead of hashtags.
    \item \emph{ratio\_all\_to\_domains\_user\_proxy\_count\_negative\_tweets\-\_\textbf{en\-gage\-ment}}: defined analogously to \emph{ra\-tio\_all\_to\_do\-mains\_\-user\_\-proxy\_\-count\_\-ne\-ga\-ti\-ve\_\-tweets\_\-\textbf{en\-gage\-ment}} but for domains instead of hashtags.
\suspend{enumerate}

For the two language features presented in the previous subsection, we defined the $2$ following ratios.

\resume{enumerate}
  \item \emph{ratio\_seen\_tweets\_in\_this\_langauge\_to\_total\_seen\_tweets}: the ratio of the number all of the tweets the engaging user has seen (\emph{en\-gag\-ing\_\-count\_\-all\_\-tweets}) to the number of tweets they saw just in the relevant language (\emph{this\_\-lan\-guage\_\-seen\_\-count}).
  \item \emph{ratio\_seen\_tweets\_in\_this\_langauge\_to\_total\_seen\_tweets}: the ratio of the number all of the tweets the engaged-with user has authored (\emph{this\_\-lan\-guage\_\-author\-ed\_\-count}) to the number of tweets they  authored just in the relevant language (\emph{engag\-ed\_with\_\-count\_\-all\_\-tweets}).
\end{enumerate}

Like the tweet element engagement history and the base language features from subsections \ref{secTweetElementsEngagementHistory} and \ref{secLanguageHistory}, the implementation of ratio features from this subsection can also be found in the notebook \texttt{PySpark-FeatureEng-04-EngagementFeatures.ipynb}. However, unlike the absolute counts from those two subsections, the ratios from this subsection can and were used in the machine learning pipeline for the classification tasks. The exact manner in which this was done is described in the next sections of this chapter.

\section{Feature Selection}
\label{secFeatureSelection}

This section describes how the engineered features were merged and prepared for feature selection. Due to the now-outdated version of PySpark available on the cluster, combined with our desire to contain the scope of the work to pre-provided MLLib methods, we could apply only one feature selector. This selector is based on the $\chi^2$ (chi-squared) metric. Furthermore, the implementation of this selector forced us to scale or bin all continuous features and features with too many values.

\subsection{Merging and Categorising the Extracted Features}
\label{secFinalFeatureEngineeringAndCategorisation}

To make the machine learning pipeline extendable and modular, we implemented feature engineering steps in separate stages, as is elaborated further in section \ref{secStagesOfTheMachineLearningPipeline}. Thus we first needed to merge the outputs of individual feature engineering stages before moving to feature selection. As discussed in subsection \ref{secTweetElementsEngagementHistory}, we dedicated the first three days of the train subsets to be used for engagement history inferences for the remainder of the train subsets. Therefore, we had to remove these three days from further consideration in the machine learning pipeline. Importantly, the three days are Thursday, Friday, and Saturday, so both weekdays and weekends are contained in both the engagement history and the remainder of the train dataset and its subsets. Contrastingly, for the features that represent short-term trends in the dataframe from subsections \ref{secShortTimePopularityEstimates} and \ref{secOracleFeatures}, we added the last 48 hours of the corresponding train subsets to the beginning of the val, test, and val+test datasets. These extra two days were removed before merging the feature engineering module outputs into singular dataframes. This ``trimming'' and merging, alongside with exploratory statistics on all of the engineered features can be seen in the notebook \texttt{PySpark-FeatureEng-05-Final.ipynb}. It is this merged output that was then passed to the next stage of the machine learning pipeline: categorisation and binning.

To prepare the pre-provided and engineered features for feature selection, we first had to make string columns (i.e. text-based columns), such as \emph{tweet\_type} (which indicates if the engaged-with user is the author of the target tweet or if they retweeted someone else's tweet on their timeline) and \emph{language}, categorical. Categorical data is one which takes one value from a limited set of values, as was further explained in subsection \ref{secScales}. PySpark MLLib has a pre-built method of turning textual columns into categorical data and then representing it in a manner that its machine-learning algorithms can accept. To prevent increasing the dimensionality of our data even further (mainly through many individual languages present in the feature \emph{language}), we decided to map individual textual categories to a single numerical representation ordered by frequency. Concretely, this was achieved by utilising the \emph{StringIndexer} method with the ``frequencyDesc'' set as the ordering argument.

Due to the limitations of PySpark MLLib function \cite{pysparkMLLIBdocumentation, chiSqInSpark}, we also had to discretise and bin all continuous features. The technical limitations which necessitated this design choice, as well as the exact features that are affected, are described in subsection \ref{secUsedFrameworksAndLibraries}. These categorised, discretised, and binned features, as well as those already in appropriate forms for feature selection and fitting of classifiers, were then vectorised. This means that they were transformed in a numerical form that PySpark MLLib functions can use. All processing steps described in this paragraph can be seen in notebook \texttt{PySpark-FeatureSelection-00-Categorisation-Local}. The resulting datasets were then used in the following step of the pipeline: chi-squared ($\chi^2$) feature selection. 

\subsection{Chi-Squared ($\chi^2$) Feature Selection}
\label{secChiSquaredFeatureSelection}

Feature selection is finding a subset of the original feature set, so that the dimensionality of the dataset is reduced, enabling better computational performance of the ensuing machine learning prediction tasks as well as reduction of noise \cite{mlbLearningPyspark}. Chi-squared ($\chi^2$) feature selection is the only non-trivial feature selection method in PySpark MLLib version 2.4.1 \cite{pysparkMLLIBdocumentation}. Chi-squared is a very popular statistical feature selection method, most frequently used in supervised machine learning tasks \cite{chiSqInSpark}. The score of the $\chi^2$ statistic measures the deviation of the expected count $E$ from the observed count $O$ \cite{chiSqInSpark, yang1997comparative}. As is further described in \cite{chiSqInSpark}, let $count(f=v_i )$ be the count of occurrences of a category $v_i$ in the feature $f$ across all instances in a dataset. Further, let $count(c)$ be the count of occurrences of instances that belong to the class $c$ in the dataset and let $P(c)$ be the likelihood that an instance chosen at random from the dataset belongs to class $c$. Lastly, let $n$ be the total number of instances in the dataset. The expected sum $E$ of the $\chi^2$ statistic assumes that features $f$ are independent from the class $c$ and is defined as the count of occurrences of category $v_i$ for feature $f$ weighted by the frequency of class $c$ in the dataset. 

$$E(f=v_i , c) = count(f=v_i ) \cdot \frac{count(c)}{n}=count(f = v_i)P(c)$$

On the other hand, the observed sum $O$ is the count of occurrences of category $v_i$ for the features $f$ across all instances in the dataset that belongs to class $c$ \cite{chiSqInSpark}.

$$O(f=v_i , c) = count(f=v_i, c)$$

If the dataset has $k$ different classes and the feature $f$ has $m$ different categories, the overall $\chi^2$ is defined as follows:

\begin{equation*}
\begin{split}
\chi^2 (f) &= \sum^k _{c=1} \sum^m _{i=1} \frac{(O(f=v_i , c) - E(f=v_i ,c))^2}{E(f=v_i ,c)} \\
 &= \sum^k _{c=1} \frac{1}{P(c)} \sum^m _{i=1} \frac{( count(f=v_i, c) - count(f = v_i)P(c) )^2}{count(f=v_i ,c)}
\end{split}
\end{equation*}

This formula of the $\chi^2$ is also used by the MLLib's version of the chi-squared feature selector \cite{chiSqInSpark, pysparkMLLIBdocumentation}. This is also the version that we used. Due to time constraints, we only used the pre-provided chi-square selector, rather than using a version of the selector from \cite{chiSqInSpark} that would allow us to select features without categorising and binning them first.

We used the statistic to calculate the most informative features to be selected. We ran the selector for the train dataset's last four days (cf. subsection \ref{secFinalFeatureEngineeringAndCategorisation}) of all train subsets. The same features were then selected in the corresponding val, test, and val+test datasets. This is a common approach in practice, as otherwise, information about the correct classification would ``leak'' in the validation and test datasets through the selection process. Moreover, in the RecSys challenge, the correct labels for test and val datasets were unavailable until the challenge had been completed. So, by only calculating the $\chi^2$ statistic on the train subsets, we emulate this constraint. We selected top features for two distinct sets of features: excluding and including so-called oracle features that contain information about engagements that happened after the target tweet. The feature sets that exclude the ``oracle'' features (which were described in more detail in subsection \ref{secOracleFeatures}) are more realistic in the sense that online prediction algorithms employed by Twitter would not have access to the information they convey. Yet, since the dataset was originally provided for a challenge asking the participants to maximise their prediction score, we also wanted to test feature sets that include this information as well, to make our predictions that much more comparable with the challenge winners.

The code implementing the chi-square selection that chooses the top $5$, $10$, $25$, and $50$ most informative features can be seen in notebook \texttt{PySpark-Feature\-Se\-lec\-tion-00-Ca\-te\-go\-risation-Local}. In contrast, additional technical design choices are described in section \ref{secStagesOfTheMachineLearningPipeline}. The relevant features, both all of them and only the selected top features, were then used to fit classifier models, as described in the next subsection.

\section{Utilised Algorithms}
\label{secAlgorithms}

The main task of the Challenge was to classify whether the viewing user had liked, replied to, retweeted, or quoted the target tweet. For this, we utilised all classifiers from the PySpark 2.4.1 MLlib library \cite{pysparkMLLIBdocumentation}. These are, in order of implementation, decision trees, logistic regression (LR), support vector machine classifier (SVC), na\"ive Bayes, random forests, gradient boosting trees, and multi-layer perceptron (MLP). The theoretical framework and the learning mechanisms of these classifiers are described in subsections below. Due to an error on the LBD cluster that could not be resolved even with the help of TU Wien's support team, we could not test LR, SVC, and MLP on the cluster, as is further elaborated in subsection \ref{secClassificationImplementation}. Furthermore, MLP was later completely omitted after it proved computationally too expensive for the local machine. The subsections below describe how each employed machine learning algorithm models training data and creates predictions.

\subsection{Na\"ive Bayes}
\label{secBayes}

Na\"ive Bayes (NB) is one of the simplest classification methods that is widely used in practice \cite[chapter 10]{mlbBayesianReasoningAndML2012}. Let $x$ be an input vector (whose individual dimensions or features are $x_1,x_2,\dotsc, x_D$) and $c$ a class label (i.e. the correct value of the target variable for $x$) be $c$. Then we define a joint model of input vectors and corresponding class labels as:

$$p(x,c)=p(c)\prod_{i=1}^{D}p(x_i | c)$$

That is to say, the probability of an instance having input vector $x$ and belonging to class $c$ equals the probability that the instance belongs to the class $c$ multiplied by the probabilities that each of the features in the input vector would have the value it has for the given instance were the instance to belong to the class $c$. As in \cite[chapters 10]{mlbBayesianReasoningAndML2012}, from this step, the Bayes' theorem can be applied, which states that for two events $a$ and $b$ it is the case that $p(a|b)=\frac{p(b|a)\cdot p(a)}{p(b)}$. Thus we get a probability of a new instance (whose classification label we want to predict) with the input vector $y$ belonging to class $c$:

$$p(c|y) = \frac{p(y|c)\cdot p(c)}{p(y)} = \frac{p(y|c)\cdot p(c)}{\sum_c p(y|c) \cdot p(c)}$$

The central assumption of the Na\"ive Bayes classifier is that explanatory variables (also called attributes) $x_i$ are independent given glass $c$, which is also where the algorithm derives its name from \cite[chapter 10]{mlbBayesianReasoningAndML2012}. Moreover, it is assumed that the data is independent and identically distributed\footnote{If data is independent and identically distributed, abbreviated as i.i.d., this means that the drawing of one data point does not affect the drawing of other data points and that all of the data points stem from the same distribution \cite[p. 26]{mlbPatternRecognitionAndML2006}.}. From there, using NB equates to having the Maximum Likelihood estimator learn the optimal parametres of distributions $p(c)$ and $p(x_i|c)$.

The likelihood function expresses how probable the observed dataset is for a set of given parametres, i.e. $p(\text{dataset}|\text{parametres})$ \cite[p. 22]{mlbPatternRecognitionAndML2006}. The maximum likelihood estimator finds the values of parametres so that the likelihood function is maximised \cite[p. 23]{mlbPatternRecognitionAndML2006}. However, as is mathematically derived in \cite[chapter 10]{mlbBayesianReasoningAndML2012}, in the case of NB, the maximum likelihood estimator comes down to counting the appearances of instances for individual classes. That is, for each class $c$ we have that:

$$p(c) = \frac{\text{number of times class c occurs}}{\text{total number of instances in the dataset}},$$

while for the probability of binary attribute $x_i \in \{0, 1\}$ we have:

$$p(x_i=1|c)=\frac{\text{number of times } x_1=1 \text{ for class } c}{\text{number of instances in class }c}.$$

Since the attribute $x_i$ is binary and the sum of all possible probability events is $1$, we have $p(x_i=0|c)=1-p(x_i=1|c)$. Now that we have $p(x_i|c)$ and $p(c)$, we can calculate $p(c|x_1)$ using Bayes' theorem. For multi-state explanatory variables, the maximum likelihood estimate is calculated based on the relative number of times the attribute $x_i$ takes a given value for the class $c$. The NB then assigns the new instance $y$ into that class for which $p(c|y)$ is maximised.

In the technical part of this thesis (cf. sections \ref{secSpark} and \ref{secMachineLearningPipelineDesign}), we implemented NB using PySpark. The two parametres we specified explicitly are \texttt{modelType} and \texttt{smoothing}. The former variable states that multinomial na\"ive Bayes is to be used, allowing the arguments of input vectors to stem from a multinomial distribution, i.e. to assume one value from a set of potentially more than two values \cite{pysparkMLLIBdocumentation}. The \texttt{smoothing} parametres refers to the additive or Laplace smoothing, which controls how aggressively the noise in the data is filtered out \cite{pysparkMLLIBdocumentation}. Specifically, given a set of $n$ input vectors $x$ with individual dimensions or attributes $x_1, x_2, \dotsc , x_D$, the smoothing transforms these attributes into $\hat{x}_1, \hat{x}_2, \dotsc , \hat{x}_D$, where

$$\hat{x}_i = n\cdot\frac{x_i+\alpha}{n+\alpha\cdot d}.$$

The parametre \texttt{smoothing} corresponds to the factor $\alpha$ in this formula \cite{pysparkMLLIBdocumentation}. $\alpha=0$ means that there is no smoothing, and $\alpha$ can take any value in the inclusive interval between $0$ and $1$.

\subsection{Logistic Regression}
\label{secLR}

Despite its name suggesting otherwise, logistic regression (LR) is an example of a generalised linear model for classification, rather than regression \cite[section 4.3.2]{mlbPatternRecognitionAndML2006}. It relies on the logistic sigmoid function $\sigma(a)$, which is defined as:

$$\sigma(a)=\frac{1}{1+\mathrm{e}^a}$$

where $\mathrm{e}$ is the Euler's number \cite[section 4.2]{mlbPatternRecognitionAndML2006}. As Bishop \cite[section 4.2]{mlbPatternRecognitionAndML2006} explains further, the sigmoid function satisfies the symmetry property

$$\sigma(-a) = 1 - \sigma(a)$$

and is known as a squashing function because its domain is all real numbers $(-\infty,\ \infty)$ while its codomain is finite, $(0,\ 1)$ specifically.  Furthermore, the inverse of the sigmoid function is the famous \textit{logit} function, used by many classification algorithms, which is given by the equation 

$$a=\ln{(\frac{\sigma}{1-\sigma})} \text{ for } \sigma \in (0,1)$$

for any real value $a$. For binary classification problems with classes $C_1$ and $C_2$, the logit function thus represents the log-ratio of probabilities that the instance with explanatory variables $\phi$ belongs to one class over the probability that belongs to the other, that is $\ln{\frac{p(C_1|\phi)}{p(C_2|\phi)}}$. This relation is also known as \emph{log odds} \cite[p.197]{mlbPatternRecognitionAndML2006}.

From this relation, logistic regression can be defined, following the \cite[section 4.3.2]{mlbPatternRecognitionAndML2006}. For two-class classifications, like in the case of the RecSys 2020 Challenge, the posterior probability of class $C_1$ can be written as a logistic sigmoid for a vector with explanatory variables $\phi$:

$$p(C_1 | \phi) = y (\phi) = \sigma (w^T\phi) $$

where the $w$ is a vector of weights which are learnt during the logistic regression training and with which the explanatory variables are multiplied \cite{mlbPatternRecognitionAndML2006}. The number of dimensions corresponds to the number of explanatory variables. Since in binary classification problems, the probability for the opposite class is $p(C_2 | \phi) = 1 - p(C_1 | \phi)$. Logistic regression can be extended to problems with multiple classes by using multiple linear functions, one for each class, as shown in \cite[section 4.4]{mlbTheElementsOfStatisticalLearning2009Chapter4linearMethods}. 

As discussed in subsection \ref{secClassification}, when we have many instances and we fit the model as closely as possible to those instances, we run the risk of overfitting to the training data. To mitigate this issue, in the case of logistic regression, we can \emph{regularise} the weights, or make the weights closer to zero \cite[p. 60]{mlbIntroductionToMLWithPython2016}. The two most frequently used regularisation approaches are L2 or Ridge and L1 or Lasso \cite[p. 60]{mlbIntroductionToMLWithPython2016}. The main difference between the two is that Ridge regression tends to shrink the coefficients towards zero while keeping all explanatory variables in the model, whereas Lasso regression may shrink coefficients to zero and thus, essentially, perform feature selection \cite{mlbTheElementsOfStatisticalLearning2009Chapter5expensionsAndRegularisation}.  

The regularisation approach is one of the hyperparametres that can be tuned in the implementation of logistic regression in PySpark \cite{pysparkMLLIBdocumentation}. Specifically, this is done with parametres \texttt{elastic\-Net\-Param} and \texttt{reg\-Param}. Both take values in $[0,\ 1]$. For parametre \texttt{elastic\-Net\-Param} we added both $0,$ corresponding to L2 regularisation penalty, and $1$, corresponding to L1 penalty, in the hyperparametre search grid. In addition to the two, we also tested the regularisation with the value $0.5$ for \texttt{elastic\-Net\-Param}. Similarly, we experimented with values $0$, $0.5$, and $1$ for the \texttt{reg\-Param} as well. The argument \texttt{fit\-Inter\-cept} determines whether the model is also to fit the intercept term, which determines the baseline log odds of the outcome when all predictors are zero. We tried tuning both with true and false. Lastly, like for most other classification models in MLLib \cite{pysparkMLLIBdocumentation}, the argument \texttt{max\-Iter} determines the maximum number of training iterations. For logistic regression hyperparametre-tuning, we used values $10$, $50$, and $100$ for the argument in the hyperparametre search grid.

\subsection{Decision Trees}
\label{secDecisionTrees}

Decision trees own their name to the fact that the model's decision -- which can be either a value for regression or a class assignment for classification tasks -- is based on a set of rules structured in a tree-like hierarchy where the leaves contain the model's prediction \cite{mlbMLWithPyspark2019chapter6forest}. The algorithm used to create these structures of rules is called ID3 and it was developed by J.R. Quinlan \cite{decisionTrees}. A very simple decision tree can be seen in figure \ref{figExampleDecisionTree} below.

\usetikzlibrary{shapes}
\usetikzlibrary{trees}

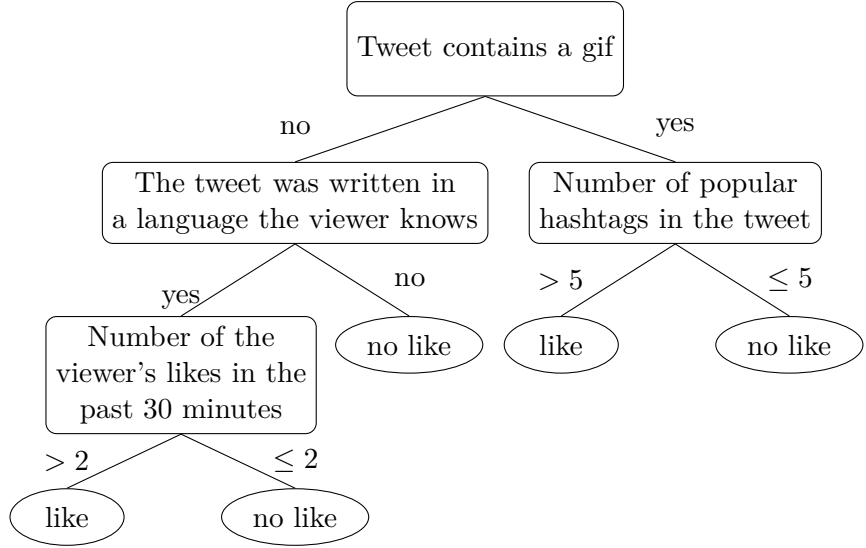
\begin{figure}[ht]
\caption{A simple example of a decision tree indicating whether the viewer would like the target tweet}
\label{figExampleDecisionTree}
\centering
\begin{tikzpicture}[%
level 1/.style={sibling distance=5cm},
level 2/.style={sibling distance=3cm},
every node/.style = {draw, minimum width=1.5cm, minimum height=.75cm, anchor=north},
edge from parent path={(\tikzparentnode.south) -- (\tikzchildnode.north)}]
]
\node[shape=rectangle, minimum height=1.25cm, minimum width=2cm, rounded corners, align=center] {Tweet contains a gif}
child { node (l1) [shape=rectangle, rounded corners, align=center] {The tweet was written in \\a language the viewer knows}
    child { node (ch) [shape=rectangle, rounded corners, align=center] {Number of the \\viewer's likes in the \\past 30 minutes} 
        child { node [shape=ellipse] {like} edge from parent node[left,draw=none] {$> 2$}}
        child { node [shape=ellipse] {no like} edge from parent node[right,draw=none] {$\leq 2$}}
    }
    child { node [shape=ellipse] {no like} edge from parent node[right,draw=none] {no}}
}
child { node (r1) [shape=rectangle, rounded corners, align=center] {Number of popular \\hashtags in the tweet} 
    child { node [shape=ellipse] {like} edge from parent node[left,draw=none] {$> 5$}}
        child { node [shape=ellipse] {no like} edge from parent node[right,draw=none] {$\leq 5$}}
};

\node [draw=none, above of=r1] (no) {yes};
\node [draw=none, above of=l1] (yes) {no};
\node [draw=none, above of=ch] (chyes) {yes};

\end{tikzpicture}
\end{figure}

As is explained conceptually in \cite{mlbMLWithPyspark2019chapter6forest} and in more detail in \cite[section 9.2]{mlbTheElementsOfStatisticalLearning2009Chapter9additiveModelsTrees}, decision trees split data into subsets of data, so that each subset contains only one value for the target variable. To come to these splits, several different approaches exist. One of them is based on entropy which measures the purity or homogeneity of each target class as well as each feature and target class pair. The general definition of entropy in binary classification tasks would be: 

$$Entropy_{C} = - p(C_1) \log{p(C_1)} - p(C_2) \log{p(C_2)}.$$

In the formula above, $p(C_1)$ and $p(C_2)$ indicate the probability that a randomly chosen instance belongs to class $C_1$ and $C_2$ respectively. In the simplest case, these probabilities can by simply dividing the number of instances in the class by the total number of instances. As we show in subsection \ref{secRCE}, the base of entropy can be arbitrarily chosen. Similarly, the entropy of a binary classification when a feature $F$ takes the value $x$ is defined as:

$$Entropy_{C}(F_i =x) = - p(C_1|F_i =x) \log{p(C_1)} - p(C_2|F_i =x) \log{p(C_2|F_i =x)}.$$

In other words, we calculate the entropy only for those instances where the feature $F$ takes the value $x$ and ignore the rest. Again, we get $p(C_1|F=x)$ and $p(C_2|F=x)$ by counting and dividing the instances in $C_1$ and $C_2$ but this time, only considering instances where $F_i=x$. Next, we would calculate the entropy for each feature and target class pair. As further shown in \cite{mlbMLWithPyspark2019chapter6forest}, this is simply calculated as:

$$Entropy_{(F_i, C)} = \sum^{x \in F_i} p(x) \cdot Entropy_{C}(F_i =x),$$

where $p(x)$ corresponds to the posterior probability that the feature $F$ has value $x$ and we can again get it by counting and dividing relative to all instances in the dataset. Following this approach, we would then indeed calculate $Entropy_{(F_i, C)}$ for all features or attributes $F_i$ in the dataset and their respective values. 

In this approach, the next step would then be to calculate the information gain ($IG$), that actually determines the nodes in the decision tree \cite{mlbMLWithPyspark2019chapter6forest}. Information gain is calculated as 

$$IG_{(F_i , C)} = Entropy_{C} - Entropy_{(F_i , C)}.$$

The feature $F_i$ with the biggest information gain is where the next split would happen. Depending on the design choice, there could be one child for each value in feature $F_i$ or a single split maximising the information gain that could be made (as is the case in figure \ref{figExampleDecisionTree}).

If such splitting of the decision tree continues until completion, i.e. until all subsets contain only one the same class\footnote{This, of course, is not possible if there are multiple instances where all features have matching values but different target classes. Nevertheless, this approach of creating decision trees would still be applicable}, it would lead to overfitting \cite{mlbTheElementsOfStatisticalLearning2009Chapter9additiveModelsTrees}. A usual approach is thus to only split the tree up to a predetermined maximal dept. Thereafter, to further increase the generalisation of the model and decrease the level of overfitting, \emph{pruning} is performed. This means that branches of the tree are removed using a complexity cost metric, such as misclassification rate, as explained in \cite{mlbTheElementsOfStatisticalLearning2009Chapter9additiveModelsTrees}.

The implementation of decision trees we used from the MLLib PySpark library \cite{pysparkMLLIBdocumentation} allows us to specify the homogeneity metric via the parametre \texttt{impurity} as well as the maximal tree depth via \texttt{model\-Type}. In our hyperparametre tuning search-grid, we included the Gini index (cf., for example, \cite[p. 15]{mlbTheElementsOfStatisticalLearning2009Chapter9additiveModelsTrees}) as well as the above-described entropy for the former and depts of $5$, $15$, and $30$ levels for the latter parametre.

\subsection{Random Forests}
\label{secRF}

Random forests consist of a large ensemble of decision trees and they are, as stated by Sing \cite{mlbTheElementsOfStatisticalLearning2009Chapter9additiveModelsTrees, randomForests}, one of the most widely-used algorithms in supervised machine learning. Each decision tree in the ensemble is ``grown'' on a random subset of features and training examples. This enables the random forest as a whole to capture diverse relationships between input and output variables. The predictions from these individual trees are then combined to form the collective output of the random forest determined through (weighted) majority voting \cite{mlbTheElementsOfStatisticalLearning2009Chapter9additiveModelsTrees}.

For the hyperparametre tuning in the MLLIb PySpark method that we used \cite{pysparkMLLIBdocumentation}, we again specified Gini index and entropy as the two possible choices for the \texttt{impurity} argument. We also used the argument \texttt{num\-Trees} to limit the maximal number of trees in the forest to $10$, $50$, $100$ in the hyperparametre tuning grid. Lastly, we dynamically limited the number of features available to each of the trees in the forest through the argument \texttt{feature\-Subset\-Strategy}. Specifically, we added three arguments to the search grid: ``log2'' (use $\log_2{|F|}$ features, where $|F|$ is the total number of features), ``sqrt'' (use $\sqrt{|F|}$ features) and ``all'' (use all features).

\subsection{Gradient Boosting Trees}
\label{secGradientBoosting}

The general idea behind all boosting algorithms is to combine multiple ``weak'' classifiers to provide a powerful ensemble \cite{mlbTheElementsOfStatisticalLearning2009Chapter10boostingAndAdditiveTrees}. The aforementioned random forest implements a so-called ``bagging'' ensemble technique, wherein multiple decision trees are trained independently and parallelly on randomised subsets of features and training instances \cite{mlbTheElementsOfStatisticalLearning2009Chapter10boostingAndAdditiveTrees}. Conversely, the boosted trees are constructed sequentially or iteratively, with each subsequent tree attempting to rectify the errors made by preceding trees. The final prediction is obtained by summing the predictions of all individual trees, with weights assigned based on their performance \cite{baggingBoostingAndRandomisationTrees}. Thus gradient boosting trees focus on minimising the residual errors made by prior trees, resulting in a sequential reduction of the training error, as shown in \cite[section 2.2]{comperativeAnalysisOfGradientBoosting}. These residuals are the gradient of the loss function to be minimised \cite{stochasticGradientBoosting}. This iterative process would clearly lead to overfitting if it is not regularised \cite{comperativeAnalysisOfGradientBoosting, stochasticGradientBoosting}. 

In the PySpark MLLib implementation of the gradient boosting trees we used \cite{pysparkMLLIBdocumentation}, the only supported loss function (argument \texttt{lossType}) is ``logistic''. This refers, of course, to logistic regression that uses a criterion similar to binomial log-likelihood \cite{additiveLogisticRegresion}. Superficially, given class probabilities $p_m(x)$, the additive logistic regression approximates the expression $p_m(x)/(1-p_m(x))$  \cite{additiveLogisticRegresion}.

In our implementation, we tuned four hyperparametres. First, for the argument \texttt{min\-Instances\-Per\-Node}, that determines the smallest number of instances each child must have after split \cite{pysparkMLLIBdocumentation}, values $1$, $5$, and $10$ were tried. All of the remaining three arguments put in the search grid for the hyperparameter tuning may take values in the interval $(0,\ 1]$ and we each used the values $0.1$, $0.5$, and $1.0$. These are \texttt{subsampling\-Rate} (specifying the part of the data to be sampled and used in each iteration of mini-batch gradient descent), \texttt{mini\-Info\-Gain} (minimum information gain for a split to be considered at a tree node), and \texttt{step\-Size} (which is the learning rate determining the level of shrinking of the estimator in each step).

\subsection{Support Vector Machines Classifier}
\label{secSVM}

The support vectors machines are a family of algorithms that produce boundaries between, in general, nonseparable classes \cite{mlbMLWithPyspark2019chapter7svm}. It does so by creating a non-linear boundary by transforming the features and then creating a linear boundary for these features. The exact formulae used in this transformation and setting of the hyperplane are rather complex, and they can be seen in \cite{mlbMLWithPyspark2019chapter7svm}. The main high-level idea of the algorithm, as explained by Hastie et al. \cite{mlbMLWithPyspark2019chapter7svm}, is to  find a hyperplane boundary that maximises the distance between the nearest data points from different classes. The wider this margin is, the less prone to overfitting and the more general the model is.

The technical part of this thesis used the implementation of the SVM classifier in \texttt{MLLib}. Thereby, four parameters were tuned before models were fitted. These include the regularisation parameter \texttt{reg\-Param} and the parameter \texttt{threshold} that is applied to the linear model prediction in binary classification. For both of these, values $0$, $0.5$, and $1$ were tested. Moreover, two Boolean parameters that indicate whether to standardise the training features before fitting the model and whether to also fit the intercept, were also added to the search grid. These two parameters are called \texttt{standardization} and \texttt{fit\-Intercept}.






\section{Machine Learning Pipeline Design}
\label{secMachineLearningPipelineDesign}

This subsection describes the manner in which the hyperparametres of the classifiers described in section \ref{secAlgorithms} were tuned, as well as how we then proceeded with the model training, prediction generation, and prediction evaluation.

\subsection{Hyperparametre Tuning}
\label{secHyperparametreTuning}

Which hyperparametres were tuned for specific models was described in the corresponding subsections of section \ref{secAlgorithms}. For all models, however, the general method of hyperparametre tuning remained the same. This subsection describes this method.

The input point for the prediction stage is a dataset whose textual features were categorised and continuously featured and binned into 100 bins. These features were also vectorised in a single column, as is required for MLLib classifiers \cite{mlbMLWithPyspark2019chapter4linearReg, pysparkMLLIBdocumentation}. We performed hyperparametre tuning using only the corresponding train subsets in order not to ``leak'' engagement information into model hyperparametres. We then mapped a grid of parameters to be tuned a performed four-fold cross-validation to find the best set of parameters by searching the hyperparametre space fully using four-fold cross-validation. Cross-validation refers to multiple splitting of the data into multiple folds so that one fold is used for testing and the rest are used for training \cite{mlbTheElementsOfStatisticalLearning2009Chapter7modelAssessmentAndSelection}. The results are then averaged, and the best combination of hyperparametres is saved. 

The cross-validation function in PySpark requires a single goal metric \cite[chapter 6]{mlbLearningPyspark}, so to compare the performance of different sets of hyperparametres, we decided to use only the RCE metric (cf. subsection \ref{secRCE}) through a custom-defined estimator. Further work could explore whether using PRAUC or a combination of the two would result in better end results. Moreover, had we not been constrained with limited computing power both on the cluster and locally, we would have used cross-validation with more folds, as five or ten folds tend to be recommended as best practice \cite[p. 217]{mlbTheElementsOfStatisticalLearning2009Chapter7modelAssessmentAndSelection}.

\subsection{Engagement Prediction}
\label{secEngagementPredection}

After the hyperparametres were tuned, these parameters are used to build prediction models (if they had not already existed and could not be reloaded). To test the effect of training on the train subsets only versus training on the val and test datasets, we implemented fitting the model on both the val, test and val+test datasets and subsets as well as the corresponding train datasets (cf. section \ref{secSampling} for more details). The intent was to create models based on all subsets created and on all extracted and provided features and only on top $5$, $10$, $25$, and $50$ most informative features, as selected by the chi-square selector.

The notebook \texttt{PySpark-Predictions-00-InitialClassification.ipynb} contains the implementation of this approach for all classifiers, and it relies primarily on the function from Python module \texttt{pp\_mllib\_predict\_evaluate.py}. These also have the preliminary evaluation based on the PRAUC and RCE metrics. Most of the evaluations, however, are created in the last stage of the machine learning pipeline -- evaluation. 

\subsection{Evaluation}
\label{secEvaluation}

An intuitive approach for evaluating datasets from the RecSys 2020 Challenge is to use (subsets of) the train dataset for the training of models and (the corresponding subsets of) the val and the test datasets for testing and validating those models. However, since the main motivation of this thesis was not to beat the scores achieved by the Challenge winners but to see the importance of contextual features and the effects of different sampling methods and machine learning algorithms, this simple approach proves lacking. Namely, to perform statistical tests which would determine the significance levels between different combinations of factors (data sampling technique and size, the inclusion of oracle features, feature selection, and the prediction algorithm), we needed more than one evaluation per model and we needed the same evaluation targets for all approaches. 

One possible approach to have multiple evaluations for each combination of the factors is to re-sample from the val and test datasets several times and then evaluate model performance on each of these new subsets. However, one should note that this would result in further sampling after we had already created $84$ samples (cf. appendix \ref{appSubsets}). Moreover, this choice would require further design decisions, such as what sampling technique to perform (as random sampling does not retain the ratio of unique users to unique tweets or total instances) and from which datasets to sample (val, test, val+test). 

Thus, a decision was made to first evaluate all fitted models on themselves (serving only as a primary evaluation, as mentioned in subsection \ref{secEngagementPredection}). But then, the main set of evaluations, which would serve as the basis for statistical significance testing, would be on 1\% and 2\% subsets for all sampling techniques gathered from all sources. As detailed in subsection \ref{secSampling}, we have five sampling techniques (random sampling, sampling that preserves the ratio of engaging users to all instances, sampling that preserves the ratio of engaged-with users to all instances, sampling that preserves the ratios of both the engaging users and the engaged-with users to all instances, and sampling that preserves the ratio of unique tweets to all instances). Moreover, there are four sampling sources (train, val, and test datasets, as well as the merged test+val dataset). Thus, we have $2 \cdot 5 \cdot 4 = 40$ sampling datasets.

This choice of evaluation methodology was partially shaped by technical constraints as well. Particularly, we initially wanted to test on all $84$ subsets to get a complete picture. Yet, the closure of the LBD server made local testing for subsets greater than 2\% in size infeasible within a reasonable timeframe. Moreover, due to computing constraints, some of the combinations remained untested, accommodations for which had to be made in our statistical significance testing, as is elaborated in chapter \ref{chEvaluation}. We thus had two parts of main results -- the first of which was partially done on the cluster and contained bigger subsets as well. The second part was done only locally, but was more complete and tested oracle features too. 

The implementation of the first round of the main evaluations can be seen in the notebook \texttt{First-Round-of-Results/\-PySpark-Predictions-10-Evaluation}. The second round of evaluations with local evaluations only can be seen in the notebook \texttt{PySpark-Predictions-01-OtherClassifications-Local}. In both cases, the main Python module that checks for results still missing and then transforms and evaluates given target models previously created (as described in the subsection above) is \texttt{pp\_mllib\_evaluate\_all.py}. Further implementation details regarding our evaluation method are provided in the following chapter.
\chapter{Implementation}
\label{chImplementation}

This chapter details the implementation of the technical part of this thesis. First, we specify the cluster's architecture and the local machine in section \ref{secArchitecture}. Section \ref{secUsedFrameworksAndLibraries} then focuses on the utilised PySpark and Python packages. Crucially, section \ref{secCode} links and details the git repository containing the source code of our implementation. The instructions to run the code are also provided. The reasons for running a part of the computing tasks locally and further significant implementation decisions are provided in section \ref{secStagesOfTheMachineLearningPipeline}. Finally, comments on the computation performance of individual aspects of the implementation are provided in section \ref{secComputationalPerformance}. 

\section{Computational Architecture}
\label{secArchitecture}

For data preprocessing, creation of features, feature selection, and the initial round of evaluations, we used PySpark on TU Wien's data cluster called ``Little Big Data''. For the second round of evaluations, we had to resort to local machine evaluation on 1\% and 2\% subsets of the data, due to the unavailability of the cluster.

\subsection{Little Big Data Cluster}
\label{secLBD}

The majority of the work was done on the ``Little Big Cluster'' (LBD for short), a Hadoop cluster established in 2017 and maintained by the High-Performance Computing group at TU Wien\footnote{Cluster details are available at \href{https://colab.tuwien.ac.at/pages/viewpage.action?pageId=15209703}{https://colab.tuwien.ac.at/pages/viewpage.action?pageId=15209703}, \href{https://colab.tuwien.ac.at/display/Datalab/TU.it+dataLab}{https://colab.tuwien.ac.at/display/Datalab/TU.it+dataLab}, and \href{https://lbd.zserv.tuwien.ac.at/}{https://lbd.zserv.tuwien.ac.at/} (last access: 2023-04-20)}. The cluster has 20 nodes, each of them having two XeonE5-2650vC CPUs with 24 cores and 256GM of RAM. However, only half of the RAM was available for the purposes of working on the thesis. More memory-intensive aspects of implementation, like the creation of graph-based features, indeed required the using all the available RAM. 

Moreover, 157.2TB of disk space was allocated for the purposes of implementing this thesis, of which 73.4TB or 46\% was utilised. A small minority of this disk space (around 1GB) was used for saving the source code and models, with the rest being used for storing subsets of data after each round of feature selection. Thus, most of the memory was used for saving datasets after each stage of the machine-learning pipeline. As elaborated in section \ref{secStagesOfTheMachineLearningPipeline}, we wanted each stage of the machine learning pipeline to be as independent from others as possible to enable a greater extent of expandability and flexibility. Of course, the downside of this approach is a considerable redundancy of used space. To see the sizes of subsets after joining all columns with feature engineering, see section \ref{appSubsets} in the appendix, which provides the list of all dataset names with the Final\_ prefix. For the implementation, the cluster's default block size of 128MB with the replication factor 3 was not changed.

PySpark version 2.4.1 with Python 3.6 based on Cloudera was the primary tool on the LBD cluster used on the server to implement our proposed method. The filesystem underneath was the Hadoop Distributed File System \cite{hadoop}, which allowed for efficient and robust data storage. The cluster also provided a Jupyter-as-a-service view of the cluster, which offers access to Jupyter Notebooks, in which most of the code was tested and executed. Moreover, this environment also allowed for Kernel management and served as a local file explorer. We also utilised YARN for the resource manager, and its logs were used to change SPARK session parameters where necessary. 

\subsection{Local Machine}
\label{secLocalMachine}

A localised installation of PySpark 2.4.2 with Python 3.6 was installed on the local machine to allow for seamless integration with the environment on the server\footnote{Due to a bug in PySpark 2.4.1 that made using the locally installed version impossible, the subsequent version 2.4.2 had to be used locally instead.}. Python environments on the server and on the local machines were created and an identical set of libraries was installed. 

The specifications of the local computer are as follows: 32GB RAM with 11th Gen Intel(R) Core(TM) i5-11400H, 2.70GHz/2.69GHz processor. The device has a 64-bit-based processor and is run on a 64-bit-based Windows 11 operating system.

The local version was ran only for 1\% and 2\% per cent datasets. Even though this has resulted in much smaller disk memory requirements, especially as the redundant files were deleted whenever possible (unlike on the server) due to much more constrained space, the working directory was around 200GB in size by the end of the implementation.

\section{Used Frameworks and Packages}
\label{secUsedFrameworksAndLibraries}

Since one of the inherent hardships regarding the RecSys2020 challenge that we wanted to tackle, is the great size of the Dataset, Apache Spark for Python was indented to the main framework. Specifically, we used PySpark 2.4.1, since this is the version installed on TU Wien's LBD cluster\footnote{Locally, PySpark 2.4.2 was used instead, due to a known bug that prevented a successful local installation.}. For the same reason, the whole project was implemented using Python 3.6. On the cluster, a new environment was created to implement our approach. The full list of Python packages installed on the cluster and locally can be seen in appendix \ref{appEnvironment}. Notably, we used \texttt{Pandas} and \texttt{Numpy} for simpler data manipulation that does not require using the PySpark (such as for evaluation results). To decode tweet text, we had to make use of \texttt{BertTokenizer} and \texttt{BertModel} from the \texttt{transformers} package. The package \texttt{pyarrow} was used so that the intermediary dataframes can be saved to .parquet files\footnote{Apache Parquet is an open source column-oriented file format, that is frequently used with Apache Pyspark; more information at \url{https://parquet.apache.org/} (last access 2023-05-03).}. We also used \texttt{scikit-learn} to calculate the evaluation metrics that measure the model predictions and to access the functions implementing statistical significance test. We also used \texttt{statsmodels} for \texttt{pingouin} for significance testing.

Whenever possible, we used the native PySpark functions to utilise the framework's computational performance optimisations fully. Notably, we used MLLib 2.4.1\footnote{The documentation used for the implementation, which is also the main source of information for this section, can be found at \href{https://spark.apache.org/docs/2.4.2/ml-guide.html}{https://spark.apache.org/docs/2.4.2/ml-guide.html} (last access 2023-05-01)}, the PySpark machine learning library, for the implementation of all feature selection and classification tasks. MLLib \cite{mllibThePaper} offers not only machine learning algorithms for classification that we used but also functions making feature extraction, transformation, and selection easier. While we did not use the pre-provided feature extraction functions but instead engineered our own features, as described in section \ref{secFeatureEngineering}, we did make use of the feature selection and prediction functions as is further described below. We used the dataframe API of MLLib. The older API option uses RDDs, and it was removed from PySpark version 3.0 onward. Both dataframes and RDDs are datafiles in PySpark, as previously explained in section \ref{secSpark}. The main difference is that RDDs, or Resilient Distributed Datasets, are schema-less data structures. In contrast, dataframes are organised in named columns, similar to the relational databases, which makes data transformations easier to formulate and quicker to compute \cite[chapter 1]{mlbLearningPyspark}.

\section{Source Code Repository}
\label{secCode}

\begin{figure}[htp]
\noindent\makebox[\textwidth]{\includegraphics[width=\textwidth]{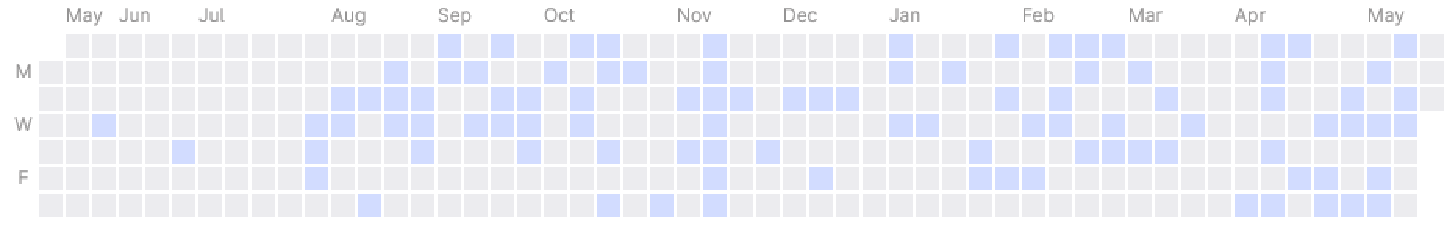}}
\caption{A visualisation of git commits from the primary local machine over the final 12 months}
\label{figGitCommits}
\end{figure}

The source code for the approaches presented in this diploma thesis, alongside code documentation and a guideline for running the code, is available on the author's GitLab repository: \url{https://gitlab.com/Jovan_NS/2020recsystwitter}. In total, over the two years of work on the implementation of the practical part of the thesis, there have been over $270$ commits to the repository, the last year of which is visualised in figure \ref{figGitCommits}. The figure showcases a break in work over the conference period in the Summer followed by steady progress towards the finalisation of practical work. The visualisation in the figure only includes commits made from the primary local machine and excludes those made from the LBD cluster or the secondary local machine used mainly for code refactoring. 

The git repository contains nine folders. Folder \texttt{Data} should contain all the Data files needed for running the code from the local filesystem. If Hadoop distributed file system (HDFS) \cite{hadoop} is used instead, the notebook in \texttt{PS\_Data\-Prep} titled \texttt{PySpark\--Data\-Prep\-00\--Da\-ta\-Sampling\-.ipynb} contains the code that attempts to put the appropriate files from this local directory into HDFS before doing rudimentary data preprocessing and sampling. The scripts which were originally used for getting the data from the challenge website and initiating spark jobs from Python modules can be seen in the directory \texttt{Scripts}. It should be noted that the original data is no longer available from the Challenge website but can be downloaded for reproducibility purposes from the author's Google Drive: \url{https://drive.google.com/drive/folders/19mJ3G8Cdp1Qc7Z4lPBjrTtGeu4tcOXfl?usp=sharing}. 

All directories with the prefix \texttt{PS\_} contain code that requires PySpark to be ran. The un-refactored code used to get the baseline results relying solely on pre-provided features can be seen in the directory \texttt{PS\-\_Baseline}. The refactored code used for running the final data processing, feature engineering, and the ML pipeline -- employing methods described in chapter \ref{chMethodology} -- are split across multiple directories. In order of the expected program execution, these directories are: \texttt{PS\-\_Data\-Prep}, \texttt{PS\-\_Feature\-Eng}, \texttt{PS\-\_Feature\-Selection}, \texttt{PS\-\_Prediction\-Pipeline}, and \texttt{PS\-\_Statistics}. Most of these directories contain two further subdirectories: \texttt{Old} with older versions of notebooks and python modules that were preserved as they might provide an insight into the development process and serve as an inspiration for further development as well as \texttt{Functions} which contains refactored Python modules used to define functions for main programs in the corresponding super-directories. Lastly, directory \texttt{PS\-\_Experimental\-Code} contains Python notebooks that contain simple illustrative examples of some of the functionalities done in other programs (e.g. the use of the Window PySpark function that was employed to extract features on current short-term trends on Twitter).

All outputs of the programs in the \texttt{PS\_} directories are saved in directory \texttt{Results}, which contains further subdirectories. The most important of these subdirectories are \texttt{Pkls}, \texttt{Models}, and \texttt{Evals}. The directory \texttt{Pkls} contains pickle files\footnote{Pickle files are serialised byte streams that preserve python object hierarchy and the contained information, more information at \url{https://docs.python.org/3/library/pickle.html} (last access 2023-05-03).}  with exports of various information relevant for feature extraction and feature selections that allow the programs to continue where the last run left of in case of memory overruns or Spark job kills. Similarly, the directory \texttt{Models} contains all the fitted machine learning models, with each algorithm having its own subdirectory. Binners, string indexers, and chi-square selector models are all saved as well and can be found in the \texttt{Results} directory too. These models are the major reason why the repository is almost 1GB in size, even though the datasets are not contained in the repository. This inclusion of models allowed for an easier integration of development on the local computer and TU Wien's LBD cluster. Namely, the initial versions of the code as well as major refactoring jobs were mostly done locally and then run for all datasets on the cluster. Lastly, the \texttt{Evals} subdirectory contains all of the evaluation results (apart from the baseline evaluations, which are saved in their own subdirectory in \texttt{Results}). This subdirectory contains further directories for all of the algorithm's initial predictions, whereas the final evaluation results are in the subdirectory \texttt{Final} (cf. section \ref{secEvaluation} for additional clarification). 


This concludes the overview of the source code repository; section \ref{secStagesOfTheMachineLearningPipeline} elaborates more on implementation decisions that were made as the source code was written and refactored. 
The most interesting results are reported and discussed in section \ref{secEvaluationResults}. If faced with issues with regard to repository access, running the code, or resulting in reproducibility, you can contact the author at \href{mailto:jovaan.ns@gmail.com}{jovaan.ns@gmail.com}.

\section{Stages of the Feature Engineering and Machine Learning Pipeline}
\label{secStagesOfTheMachineLearningPipeline}

It is significant to note that each phase of feature preprocessing, feature engineering, and feature selection created a unique version of the datasets to allow for a completely modular approach to the work. Table \ref{tabStages} provides the versions of the datasets that were created as well as the prefixes of the corresponding input and output files for each of the modules. To see the file names for datasets to which the prefixes in the table were adjunct, see section \ref{appSubsets} in the appendix. It provides the list of all dataset names with the Final\_ prefix.

\begin{longtable}{|>{\hspace{0pt}}m{0.127\linewidth}|>{\hspace{0pt}}m{0.456\linewidth}|>{\hspace{0pt}}m{0.137\linewidth}|>{\hspace{0pt}}m{0.219\linewidth}|}
\caption{Stages of the machine learning pipeline: preprocessing, feature engineering, feature selection, prediction, and evaluation.}
\label{tabStages}\\ 
\hline
\textbf{File Name}                     & \textbf{Role}                                                                                                                                                                                              & \textbf{Input Files}                                          & \textbf{Output Files}                                                                           \endfirsthead 
\hline
Data Prep: 00 Data Sampling   & Takes the original dataset, assigns the correct schema, names the columns, and creates dataset samples.                                                                                                    & Original TSV files from the challenge                         & All samples files (to which prefixes are then added)                                            \\ 
\hline
Feature Eng: 00 Feature Eng            & Basic data preparation and feature engineering, such as transforming engagement targets in binary flags.                                                                                                   & Output of 00 Data Sampling                                    & Files with the FE\_ prefix                                                                        \\ 
\hline
Feature Eng: 01 Encoding               & Further simple transformations such as counts of media in the tweet and calculating the age of account.                                                                                                    & Files with the FE\_ prefix                    & Files with the Encoding\_ prefix                                                                  \\ 
\hline
Feature Eng: 02 Graph-Based            & Creates graph features for first and second-degree following and engaging connections.                                                                                                                     & Files with the Encoding\_ prefix              & Files with the GraphBased\_ prefix                                              \\ 
\hline
Feature Eng: 03 Time Features          & Creates personalised and general time-based features that measure the recent popularity of hashtags, links, and domains.                                                                                   & Files with the Encoding\_ prefix              & Files with the Time\_ prefix                                                    \\ 
\hline
Feature Eng: 04 Engagement History     & Overall positive and negative engagement counts for both the author and the viewer, as well as positive and negative language-specific engagement counts for the viewer.                                   & Files with the Encoding\_ prefix              & Files with the Engagement\_ prefix                                              \\ 
\hline
Feature Eng: 05 Final                  & Combines all engineering features into a single dataset, i.e. combined Encoding\_, GraphBased\_, Time\_, and Engagement\_ datasets in one. & Files with all feature engineering prefixes.                  & Files with the Final\_ prefix                                                   \\ 
\hline
Feature Selection: 00 Categorisation   & Categorise and bin features, as required for the chi-squared selector.                                                                                                                                     & Files with the Final\_ prefix                 & Files with the Categorised\_ prefix; PKL files indicating transformed features  \\ 
\hline
Feature Selection: 01 ChiSq Selector   & Vectorises explanatory features, applies the chi-square selector and selects top 5, 10, 25, and 50 features (including or excluding oracle features)                                                       & Files with the Categorised\_ prefix           & Files with the ChiSq\_ prefix                                                   \\ 
\hline
Predictions: 00 Initial Classification & Tunes the hyperparameters based on the corresponding train (sub)set and then fits and evaluates a model based on each of the datasets                                                                      & Files with the ChiSq\_ prefix                 & Tuned parameters; models fitted and evaluated on each of the datasets                           \\ 
\hline
Predictions: 01 Other Classifications  & Uses the initially tuned and fitted models and evaluates them on all other datasets.                                                                                                                       & Files with the ChiSq\_ prefix; initial models & Models and evaluations for all other datasets based on the initial models                       \\
\hline
\end{longtable}

As can be seen in the repository linked in subsection \ref{secCode}, the code is in organised Jupyter notebooks, for an easier overview of intermediary steps. A better performance may be achieved if Spark jobs were to be run via the command line instead, as we did for feature extraction tasks. In this case, configurations such as memory allocation and dependencies should be explicitly provided. For instance, to run the Feature Eng file 02 that creates graph features, we ran the command in listing \ref{lstRunGraph}.

\begin{lstlisting}[language=bash,caption={The command used to run the \texttt{PySpark-02-GraphBased.py} file}, label=lstRunGraph]
spark-submit \
  --master yarn \
  --deploy-mode cluster \
   --config "spark.dynamicAllocation.initialExecutors=4" \
   --config "spark.sql.broadcastTimeout=720" \
   --config "spark.sql.session.timeZone=UTC" \
   --config "spark.driver.extraJavaOptions=-Duser.timezone=UTC" \
   --config "spark.executor.extraJavaOptions=-Duser.timezone=UTC" \
   --config "spark.driver.memory=16G" \
   --config "spark.driver.maxResultSize=120G" \
   --config "executor-cores=16" \
   --config "executor-memory=120G" \
  PySpark-FeatureEng-02-GraphBased.py

--py-files Functions.import_dataframes.py, \
  Functions.export_dataframes.py, \
  Functions.show_final_statistic.py, \
  Functions.prepare_for_new_columns.py, \
  Functions.check_if_columns_are_missing.py, \ 
  Functions.pyspark_df_shape.py, \
  Functions.fe02_extract_follows.py, \
  Functions.fe02_extract_follows_2d.py, \
  Functions.fe02_extract_graph_relations.py, \
  Functions.fe02_extract_engagements.py, \
  Functions.fe02_extract_engagements_2d.py, \
  Functions.fe02_extract_graph_relations.py, \
  Functions.fe0X_ratio_of_two_features.py, \
  Functions.fe0X_ratio_of_two_features.py 
\end{lstlisting}

Should some of the Python modules not be distributed explicitly to worker nodes as shown above, this may result in a \texttt{ModuleNotFoundError}.

\subsection{Data Preprocessing and Sampling}
\label{secFeaturePrepImplementation}

Implementation of data preprocessing and sampling was the entry point of the pipeline, where the ``raw'' data from the Challenge first enters the pipeline, as described in section \ref{secSampling}. This part of the pipeline is not particularly memory-intensive or computationally demanding, and 5GB of driver memory and 16GB of executor memory were sufficient. It should be noted that the program expects the initial data to be already in the Hadoop distributed file system. 

This program also shows that the tweets from both the train as well as val and test datasets were sampled from the beginning of the day on a Thursday until the end of the following Wednesday. The dataset also creates subsets corresponding to each day (ultimately, these subsets remained unutilised, as the features regarding the user's engagement history were implemented using a different approach). Notably, to split the dataset in seven days, the time zone had to be set to UTC, which can be achieved by setting \texttt{"Z"} as the value of the argument \texttt{"spark.sql.session.timeZone"} in the active PySpark session. 

Lastly, the manner in which we created \emph{\_EU\_}, \emph{\_EWU\_}, \emph{\_EWU\_}, \emph{\_inter\_EWU+EU\_}, and \emph{\_tweet\_} subsets -- that is, subsets that preserve the ratios of, respectively, the number of unique engaging users, unique engaged-with users, unique engaging and engaged-with users, and unique tweets to the total number of instances -- is by first sampling the desired percentage of the user or tweet IDs and then semi-joining the dataset with those IDs to get the corresponding subset. For the $\_inter\_EWU+EU\_$ subsets, we first sampled the square root of the desired percentage of the engaged-with and engaging user IDs. Thus, for instance, for 10\% $\_inter\_EWU+EU\_$ subsets, we sampled $\sqrt{10\%}=\sqrt{0.1}\approx 0.3162$ of all unique engage-with user IDs and $0.3162$ of engaging user IDs. Then we semi-joined the source dataset with both of these subsets so that only the instances where both the engaged-with and the engaging user ID has been priorly selected remain. As seen in the notebook \texttt{PySpark\--Data\-Prep\--00-Data\-Sampling}, these approaches were verified to lead to approximately the same ratios in the subsets as in the full datasets.

\subsection{Feature Engineering}
\label{secFeatureEngineerngImplementation}

This subsection identifies and explains important or potentially intriguing design choices made in the programs that perform feature engineering\footnote{As in, for instance, \cite{mlbIntroductionToMLWithPython2016} we interchangeably refer to this process of extracting new features as \textit{feature engineering} or \textit{feature extraction}, rather than considering feature engineering a broader term that includes both feature extraction and feature extraction (as was done in \cite{mlbLearningPyspark}).} which can be seen in the directory \texttt{PS\_FeatureEng}.  As can be seen in table \ref{tabStages} and the GitLab repository with the source code that was linked in section \ref{secCode}, there are six programs whose names are enumerated 00--05. 

All feature extraction programs follow the same structure that allows for the computation to continue from where the program last stopped. This was decisive, as memory and time constraints on the cluster frequently caused the programs to shut down before they were finished with all subsets. The structure of each of the programs contains four main parts: user-set program-wide options, data importing, feature extraction, and exporting of results. 

The settings contain general flags as well as import-export settings. The general flags include \texttt{\-CREATE\-\_EVEN\-\_IF\-\_AL\-READY\-\_EXIST}, which specifies whether the program should recreate and rewrite the features, in case the corresponding columns already exist and \texttt{DEV} signals whether all subsets and datasets should be included or just the 1\% datasets. The latter option of \texttt{DEV} was useful for new feature development or local runs of the program. The import-export indicates which import and export prefixes the program should use (again, in line with table \ref{tabStages}) as well as what sampling techniques, percentages, and sources to use (cf. subsection \ref{secSampling}).

The main ideas behind the feature extraction of each program is described in detail in section \ref{secFeatureEngineering}. Specifically, the program with the number 00 in its name corresponds to the preamble in section \ref{secFeatureEngineering}, and its output is used as the input to program 01. The output of program 01, whose approach is described in subsection \ref{secEncoding}, is used as the input for programs 02, 03, and 04. The program in 02 contains graph-based features and its features are described in subsection \ref{secGraphBasedFeatures}. Program 03 utilises the Window PySpark function and creates features that describe short-term trends, as described in subsection \ref{secShortTimePopularityEstimates}, as well as the so-called oracle features, that look into events that took place after the tweet was seen as well and which were described in subsection \ref{secOracleFeatures}. Features created in 04 are extracted from the data declared as \textit{engagement history} data, as is elaborated further in a paragraph below. The premise behind these features is described in subsections \ref{secTweetElementsEngagementHistory}, \ref{secLanguageHistory}, and \ref{secRatios}. Lastly, the outputs of programs 02, 03, and 04 are merged in program 05, as described in subsection \ref{secFinalFeatureEngineeringAndCategorisation}.

The implementation of feature engineering in \texttt{00\--Feature\-Eng} and \texttt{01\--Encoding} is rather straightforward. Of interest for future potential extensions of this work might be the way in which we decoded the text from the tokenised representation. This may be seen in that program. 

Implementing graph-based features from program 02 was one of the major ``roadblocks'' in the realisation of the dissertation. This is because getting the second-degree connections required joining the list of following relations with itself (with the joining key being the followed users on one side and following users on the other) and then adding the new list's followed users as the old list's following users' second-degree connections. We applied an analogues approach to the engagement graph relations as well (cf. subsection \ref{secGraphBasedFeatures}). This was done with the function \texttt{extract\_graph\_relations}
. However, this joining proved extremely memory-intensive and we were unable to perform it for the full datasets, even after applying optimisations such as saving intermediary results and running Spark jobs via Python scripts rather than in Notebooks. We used also used the maximal amount of available memory and experimented with different settings for the number of cores and executors. In the end, we had to create estimates for values of the graph relations for the full train, test, val, and val+test datasets. We created these estimates from graph columns of instances in all subsets. For the Boolean features representing the follows-relation, we know that the full dataset value should be true if there is at least one instance with the corresponding engaging user ID and engaged-with user ID. For the count-based engagement features, we again joined the ID columns and used a weighted sum for the count value. We weighted the counts in 1\% and 2\% subsets with 10, weighted the sums in the 5\% subsets by 6, and the 10\% subset counts were multiplied by 4. This was done for all subsets across the five sampling techniques. As can be seen in the notebook \texttt{Py\-Spark\--Feature\-Eng\--02\--Graph\-Based\--Proxy\-4FullDs}, these weights were chosen to match the observed trend of how the number of instances in the subset corresponds to the average value of engagement counts.

The main idea for features in \texttt{03\--Time\-Features} was to investigate trends in the previous 30 minutes, 60 minutes, and 120 minutes, as well as 12 hours, 24 hours, and 48 hours before the tweet was seen. Since val, and test datasets contain tweets from the week following the week in the train dataset, the first step was to append the last two days from train to val, test, val+test and all their subsets
It should be noted that using the window function was challenging, as its PySpark 2.4 API was still incomplete\footnote{More details about it can be found on StackOverflow: \url{https://stackoverflow.com/a/33226511} (last access 2023-05-04).}. Thus we had to join twice and explode the intermediary results to circumvent the missing functionality. 
In its comments, it also contains links that showcase what the result of each step is. This function might also be helpful for further extension attempts. Of all feature extraction features, \texttt{03-TimeFeatures} was the most time-intensive.  In fact, the speed at which a task is performed decreases with every iteration, likely due to a garbage collection issue. The order of magnitude would rise from minutes per feature to hours per feature until, eventually, the whole Spark job prematurely terminates. Since a lengthy attempt to resolve the issue directly had proved unsuccessful, we eventually worked around the issue by using a bash script that utilises the \texttt{schtasks} command to kill and restarts  the Spark job with this feature creation script every hour. 

As stated previously, the unifying characteristic for features that are engineered in \texttt{Py\-Spark\--Feature\-Eng\--04\--Engage\-ment\-Features} is that they are based on data designated to be the engagement history. For the full test, val, and val+test datasets, the whole train dataset is seen as the engagement history. For the subsets of test, val, and val+test datasets, the corresponding train subsets are the engagement history (e.g. the whole of \texttt{train\-\_EU\-\_sam\-ple\-\_ 5pct} is the engagement history for \texttt{val\-+test\-\_EU\-\_sam\-ple\-\_ 5pct}). For the train dataset and its subsets, the first three days were used as the engagement history for the remaining four days. For this reason, the first three days had to be omitted from the train dataset and all its subsets in the following phases of the machine learning pipeline. The implementation of feature extraction itself in this program is rather straightforward. The only exception may be the ``user-proxy engagement'' for counts of tweet elements in engagement history. As was elaborated in subsection \ref{secTweetElementsEngagementHistory}, there are no tweets which appear in both the train dataset as well as the val and test datasets, which is why we used the \texttt{engaged\-\_with\-\_user\-\_id} instead of the \texttt{tweet\-\_id}. 

Finally, the outputs of previous programs were merged in unified datasets in \texttt{05\--Final}. Before merging the inputs, however, the program had to remove the first days from the train dataset and its subsets, to match the design constraint from the previous paragraph. Moreover, the 48 hours added to val, test, and val+test datasets and their subsets were removed as well before the merger. 

\subsection{Feature Selection}
\label{secFeatureSelectionImplementation}

Since the PySpark version on the TU Wien was 2.4.1, the only non-trivial feature selection function was a chi-square ($\chi ^2$) feature selector. Future work might consider using selectors that were added to PySpark's latter versions or externally developed methods, such as the chi-square selector developed in \cite{chiSqInSpark} that does not require prior feature categorisation.

Feature categorisation, i.e. discretisation of continuous attributes and binning of attributes with too many values, was the first necessary step for chi-square feature selection. This necessary preparation of all merged features was implemented in the notebook \texttt{00\--Ca\-te\-gori\-sa\-tion}. We began by manually recognising and grouping all pre-provided and extracted features that can be used for prediction. We did this to exclude columns which are not directly appropriate for prediction because they are counts dependent on the size of the dataset, are in an inappropriate format (timestamps), or are IDs.  We thus recognise and group all pre-provided and extracted features that can be used for prediction. There are 185 such relevant features and they can be seen in listing \ref{lstRelevantFeatures}. In addition, we singled out 8 ``oracle'' features, which also provide information on events that may have happened after the potential interaction has occurred. These features were described conceptually in subsection \ref{secOracleFeatures} and their names can be seen in listing \ref{lstOracleFeatures}.

Since the PySpark implementation of the chi-squared selector cannot work with attributes with too many distinct values, we had to bin all features with type Double as well as features other numerical features, the counts of which showed over 100 distinct values. Specifically, this affected all ``oracle'' features as well as 131 out of 185 of the remaining relevant features. All these features were binned into up to 100 bins using the \texttt{Quant\-ile\-Discret\-izer} from the PySpark MLLib library \cite{pysparkMLLIBdocumentation}. Lastly, \texttt{00\--Ca\-te\-gori\-sa\-tion} also made textual features \texttt{tweet\_type} and \texttt{language} categorical by using yet another MLLib function, \texttt{StringIndexer}. We decided to use this method of coding textual categories into numerical representation given by categorical frequency rather than to one-hot-encode the categories since we did not want to further increase the number of dimensions of the datasets.

The output of the \texttt{00\--Ca\-te\-gori\-sa\-tion} is the input of \texttt{01\--Chi\-Sq\-Select\-or}, that actually performs chi-square feature selection for top 5, 10, 25, 50 most informative features. The exact mechanism behind this approach was explained in subsection \ref{secChiSquaredFeatureSelection}. Before the selection is done, due to yet another constraint of the PySpark implementation, all empty values had to be padded with zeros. After the selection was done, each subset gets individual; vectorised columns for each of the four selection limits (top 5, 10, 25, and 50 most informative features) as well as for all relevant features, both including and excluding the oracle features, and for each of the five engagement targets (like, reply, retweet, quote, react). 
The following vectorisation of selected (or all) relevant explanatory columns is then done by method \texttt{fs\-\_vector\-ise\-\_select\-ed\-\_columns}
. Vectorised columns are columns with the \texttt{Vector} data type, an array or list-like Spark data type. \texttt{Vector} is the required input format for MLLib classifiers (cf. \cite{mlbMLWithPyspark2019chapter3dataProcessing}). Thus we created vectorised columns for all relevant explanatory features, one excluding and one including the oracle features. We created $(4+1) \cdot 2 \cdot \cot 5 = 50$ vector columns for which classifications were to be done. Moreover, the information on what columns were selected as most informative for which subsets were additionally exported into .pkl files and can be seen in appendix \ref{appFurtherFeatureSelection}. 

Similarly, as had to be done for the feature engineering in \texttt{Py\-Spark\--Feature\-Eng\--04\--Engage\-ment\-Features}, in both \texttt{Py\-Spark\--Feature\-Se\-lection\--00\--Ca\-te\-gori\-sa\-tion} and \texttt{Py\-Spark\--Feature\-Se\-lection\--01\--Chi\-Sq\-Select\-or} exporting intermediary results had to be implemented. This was a way of accommodating the high computational intensity of the transformation. Thus all subsets can be transformed over multiple program runs that continue on from where the last Spark job left off. Moreover, created models for binning, categorisation, and chi-squared feature selection are also saved externally and reloaded with reruns, unless the flag \texttt{RE\-WRITE\-\_EXIST\-ING\-\_MODELS} is set to True.

\subsection{Classification}
\label{secClassificationImplementation}

The implementation of classification itself, i.e. predicting whether the engagements took place or not, consists of two parts. The first part tuned the classifier's hyperparametres on all train subsets and fitted initial models on all subsets. The second part then took these initial models and computed and evaluated predictions based on all other subsets. The reasoning for this approach to evaluating results is elaborated in subsection \ref{secEvaluation}, whereas the results are presented in section \ref{secEvaluationResults}. This subsection describes the implementation itself and clarifies how the corresponding programs can be rerun for reproducibility or extension.

The first part relies on the program in \texttt{Py\-Spark\-Pre\-dictions\--00-Initial\-Classi\-fi\-cation}. This program follows a similar structure to the programs which implement feature extraction and which were discussed earlier in this section. Specifically, the user can change settings through flags such as \texttt{RE\-WRITE\-\_EXIST\-ING\-\_MODELS} that determine whether to reuse the previously saved classification models of transformers and \texttt{RE\-CREATE\-\_MISS\-ING\-\_MODELS} that controls whether to recreate classification models which are missing from the file-system, but whose evaluation already exists (which is overridden by \texttt{RE\-WRITE\-\_EXIST\-ING\-\_MODELS} if it is set to True). The user may also specify what factors, or combinations of features, classifiers, and data subsets, are to be used for fitting the initial classifiers. For instance, \texttt{CLASSIFI\-ER\-\_NAMES} lists all MLLib classification algorithms to be fitted, \texttt{TOP\-\_NS} lists all feature selections approaches to be tested, and \texttt{FEATURES\-\_NOTES} lists all feature sets to be used (specifically, whether to include or exclude the ``oracle'' features). Moreover, \texttt{IMPORT\-\_DATA\-SETS}, \texttt{SAMPLING\-\_TECHNIQUES}, and \texttt{SAMPLING\-\_PERCENTAGES} control which datasets and subsets to import base predictions on, in the same manner as was done in programs described in subsections above. 

After importing the specified datasets and subsets, \texttt{Py\-Spark\-Pre\-dictions\--00\--In\-i\-tial\-Classification} loads a pickle file generated by the feature selection programs. These files also contain a dictionary (i.e. a key-value map) with containing columns to be used as explanatory values in classification tasks. Given this dictionary and the specified combinations in \texttt{TOP\-\_NS} and \texttt{FEATURES\-\_NOTES}, the program checks if the corresponding columns had already been vectorised. If they had not, vectorisation is done on-the-spot using the methods in the module \texttt{pp\-\_vectorise\-\_explanatory\-\_features}, that is very similar to \texttt{fs\-\_vector\-ise\-\_select\-ed\-\_columns}
. The most important difference between the two is that the \texttt{pp\-\_vectorise\-\_explanatory\-\_features} method vectorises only the immediately needed columns rather than all of the feature selection combinations listed in appendix \ref{appAllFeaturesUsedforPredictionTasks}.

Then, the central python module \texttt{pp\-\_ml\-lib\-\_predict\-\_evaluate} is called for each combination of \texttt{TOP\-\_NS}, \texttt{FEATURES\-\_NOTES}, and classifiers. The main method in the module also takes as arguments further options such as which hyperparametres to tune for which classifier, where to save the results, etc. The function then fits models for all dataframes in the corresponding dictionary of dataframes with vectorised explanatory feature columns. An initial set of predictions and evaluations is also created. These initial predictions and evaluations are based on the same dataset on which the model was fitted on. In the next evaluation step, these hyperparametres and models are reevaluated in a more systematic, as we explain in the next paragraph. The fitted models thus exported (by default to \emph{Results/Models/<Classifier name>/csv}, while the corresponding hyperparametres are saved in another dictionary (by default to \emph{Results/Models/<Classifier name>/pkl}). The name of the exported models indicates what it was fitted on, and it follows the format in listing \ref{lstModelNameFormat} below. 
For instance, one of the models fitted locally is named \emph{classifier\-\_model\-\_of\-\_type\--tree\--for\-\_features\--all\--oracle\-\_scaled\--\--for\-\_dataset\--test\-\_random\-\_sample\-\_1pct\--based\-\_on\-\_dataset\--test\-\_random\-\_sample\-\_1pct\--predicting\-\_target\--like\--ht}. This external saving of intermediary evaluations and initial models also enables the classification to restart when the previous Spark jobs die. 

\begin{lstlisting}[language=Python,caption={The naming of individual classificaiton patters}, label=lstModelNameFormat]
hypertune_suffix = "-ht" if (hypertune_params is not None) else ""
full_model_name = "classifier_model_of_type-" + classifier_prefix + \
                                  "-for_features-" + top_n_selected + \
                                  features_note_model_name_addition + \
                                  "-for_dataset-" + sample + \
                                  "-based_on_dataset-" + corresponding_train_key + \
                                  "-predicting_target-" + target + \
                                  hypertune_suffix
\end{lstlisting}

These preliminary results and initial models are reused in a more complete evaluation in the next step that is implemented in \texttt{Py\-Spark\-Pre\-dictions\--01\--Other\-Class\-i\-fi\-cation}. This too starts with the option for the user to specify which evaluations to make and in which order (through the variable \texttt{SORTING\-\_ORDER}). Similarly as was described previously in this subsection, the needed features are then vectorised if that has not been done already. Finally, the function from the module \texttt{pp\-\_ml\-lib\-\_evaluate\-\_all} is called. 
It creates and evaluates further predictions, which are ultimately used for statistical significance testing and inferring conclusion about the relative utility of individual contextual features and factors. As further explained in subsection \ref{secEvaluation}, based on each model, we created and evaluated precision for each 1\% and 2\% subsets for all sampling techniques gathered from all sources. Evaluations on bigger subsets were also created in the first major evaluation round which was done while the LBD cluster was still available. All results are discussed in section \ref{secEvaluationResults}.

\section{Computational Performance}
\label{secComputationalPerformance}

Table \ref{tabComputationalPerformance} contains the approximate computation duration on the cluster and the local machine. The specifications of both are listed in section \ref{secArchitecture} and more details about what each stage in the pipeline does can be found in table \ref{tabStages}.

\begin{longtable}{|>{\hspace{0pt}}m{0.133\linewidth}|>{\hspace{0pt}}m{0.304\linewidth}|>{\hspace{0pt}}m{0.213\linewidth}|>{\hspace{0pt}}m{0.288\linewidth}|}
\caption{Orders Of Magnitute (OOM) for computational performance for each stage of the machine learning pipeline}
\label{tabComputationalPerformance}\\ 
\hline
\textbf{File Name}                          & \textbf{OOM runtime on the LBD cluster}                                                                                                        & \textbf{OOM local runtime}                                                                                  & \textbf{Note}                                                                                                          \endfirsthead 
\hline
Data Preprocessing: 00 Data Sampling        & \textbf{Hours }for all datasets                                                                                                                & Not ran.                                                                                                    & Creating the samples takes the greatest amount of time.                                                                \\ 
\hline
Feature Eng: 00 Feature Eng                 & \textbf{Minutes }for all datasets                                                                                                              & Not ran.                                                                                                    & Simple transformation that do not require shuffling.                                                                   \\ 
\hline
Feature Eng: 01 Encoding                    & \textbf{Minutes }for all datasets                                                                                                              & Not ran.                                                                                                    & Simple transformation that do not require shuffling.                                                                   \\ 
\hline
Feature Eng: 02 Graph-Based                 & \textbf{Days }for all subsets; unfeasible for full datasets                                                                                    & \textbf{Hours }for 1\% random subsets                                                                       & Creating and joining all the graph-like relations failed on the LBD server.                                            \\ 
\hline
Feature Eng: 02 Graph-Based Proxy 4 full DS & \textbf{Hours }for all datasets\textbf{}                                                                                                       & Not applicable                                                                                              & We had to use a combination of subsets to create the graph features for the full dataset.                              \\ 
\hline
Feature Eng: 03 Time Features               & \textbf{Weeks }with periodical restarts to clear the memory                                                                                    & \textbf{Weeks }for  1\% random subsets                                                                      & The most computationally-intensive part of the work due to computationally expensive use of the Window function.       \\ 
\hline
Feature Eng: 04 Engagement History          & \textbf{Hours }for all datasets                                                                                                                & \textbf{Hours }for  1\% random subsets                                                                      & Once the engagement history datasets are created and languages extracted, feature engineering is rather fast.          \\ 
\hline
Feature Eng: 05 Final                       & \textbf{Hours }for all datasets                                                                                                                & \textbf{Hours }for  1\% random subsets                                                                      & Transformations and filtering are rudimental, joining dataset is the only aspect that is more time-intensive.          \\ 
\hline
Feature Selection: 00 Categorisation        & \textbf{Hours }for all datasets                                                                                                                & \textbf{Hours }for all 1\% and 2\% subsets                                                                  & Binning and categorising is not highly computationally intensive, but there is a lot of data to be transformed.        \\ 
\hline
Feature Selection: 01 ChiSq Selector        & \textbf{Hours }for all datasets                                                                                                                & \textbf{Hours }for all 1\% and 2\% subsets                                                                  & The final vectorisation was not done for all combinations, because that step fills up all memory and requires resets.  \\ 
\hline
Predictions: 00 Initial Classification      & \textbf{Weeks} for multiple combinations, unfeasible for full dataset; the LBD cluster was closed after the first round of evluations\textbf{} & \textbf{Weeks} for multiple combinations based on 1\% and 2\% subsets                                       & Very algorithm-specific, look at the repository for records of individual performances.                                \\ 
\hline
Predictions: 01 Other Classifications       & \textbf{Minutes }for individual evaluations, \textbf{weeks }for multiple combinations\textbf{ }                                                & \textbf{Minutes-Hours }for individual evaluations, \textbf{weeks }for multiple combinations\textbf{ }\par{} & Individual requirements again very algorithm-specific, e.g. svc can take hours locally for individual predicitons.     \\
\hline
\end{longtable}

\chapter{Evaluation}
\label{chEvaluation}

In this chapter, we present the most significant evaluation scores of the implemented methods based on area under the precision-recall curve (PRAUC) and relative cross-entropy (RCE) metrics. The chapter first provides a theoretical background for these two evaluation metrics in section \ref{secMetrics}. The baseline results follow in section \ref{secEvalBaseline} as well as the main evaluation results in section \ref{secMainEvaluation}. Due to a high number of evaluated combinations of data sampling methods, feature sets, and developed models, only some results are reported and expounded on in this chapter. The remaining results can be found in appendix \ref{appAdditionalResults}. We discuss which factors significantly influenced the prediction results as well as which features were found to be the most informative in sections \ref{secSignificanceTestingforIndividualFactors} and \ref{secMostSignificantContextualFeatures}, respectively. Finally, our results are contextualised within prior works in sections \ref{secContrastingResultsWithOtherWorks} and \ref{secDiscussion of Results}.

\section{Evaluation Metrics}
\label{secMetrics}

The RecSys 2020 Challenge evaluated the submitted solutions based on two metrics: the area under the precision-recall curve and the relative cross-entropy. The two were calculated for each of the four engagement types separately and the final ranking was then based on the eight measures. In the following subsections, we elaborate on the metrics themselves.

\subsection{Area under the Precision-Recall Curve (PRAUC)}
\label{secPRAUC}

The Area Under the Precision-Recall Curve\footnote{in this thesis abbreviated as PRAUC, while PR-AUC,  AUCPR, and AUC-PR can also commonly be found in literature, e.g. \cite{recsys2020overview, PrecisionRecall3, PrecisionRecall4}.} is the area under the threshold-dependent curve depending on the performance of a model with regards to precision and recall, as described in \cite{PrecisionRecall4}. Following the reasoning from \cite{PrecisionRecall2}, precision and recall for binary classification tasks can be written in the form of a two-by-two table. This table is often referred to as a confusion matrix or contingency table \cite{PrecisionRecall1}. One dimension of the table (e.g. columns) corresponds to the ground truth, i.e. the ``correct'' classes the instances belong to. The other dimension (e.g. rows) corresponds to the class predictions by the model that is to be evaluated. What the two classes represent is task-specific; in the case of tweet engagement prediction, the classes are ``engagement occurred'' and ``engagement did not occur''. When discussing binary classification tasks in general, the two classes are usually labelled as ``positive'' and ``negative'', as in \cite{PrecisionRecall1, PrecisionRecall3, PrecisionRecall4}. The fields of the confusion matrix can then be identified as true positives (TP), false positives (FP), true negatives (TN), and false negatives (FN). True positives and true negatives refer to the counts of instances which were correctly classified by the model, whereas the predictions of false positives and false negatives are the counts of wrong predictions. Since in the case of RecSys 2020 Challenge, the authors of \cite{recsys2020overview} label tweets with engagements as positive instances, a tweet for which the model predicted no engagement that actually was engaged with would be seen as a false negative and would increase the corresponding count in the confusion matrix. The confusion table structure is explained visually in table \ref{tabGenericConfusionMatrix}.

\begin{table}[h]
    \centering
\begin{tabular}{cccc}
                             &                               & \multicolumn{2}{c}{\begin{tabular}[c]{@{}l@{}}Ground truth (correct classes)\end{tabular}}                                                                                                                                                                                       \\
                             & \multicolumn{1}{c|}{}         & \multicolumn{1}{c|}{positive}                                                                                                                      & negative                                                                                                                      \\ \cline{2-4} 
\parbox[t]{2mm}{\multirow{2}{*}{\rotatebox[origin=c]{90}{Predictions}}} & \multicolumn{1}{c|}{positive} & \multicolumn{1}{c|}{\begin{tabular}[c]{@{}c@{}}\textbf{True Positive (TP)}\\ Number of correctly classified \\positive instances\end{tabular}}                & \begin{tabular}[c]{@{}c@{}}\textbf{False Positive (FP)}\\ Number of negative instances\\ incorrectly classified as positive\end{tabular} \\ \cline{2-4} 
                             & \multicolumn{1}{c|}{negative} & \multicolumn{1}{c|}{\begin{tabular}[c]{@{}c@{}}\textbf{False Negative (FN)}\\ Number of positive instances\\ incorrectly classified as negative\end{tabular}} & \begin{tabular}[c]{@{}c@{}}\textbf{True Negative (TN)}\\ Number of correctly classified\\ negative instances\end{tabular}               
\end{tabular}
    \caption{Structure of a confusion matrix}
    \label{tabGenericConfusionMatrix}
\end{table}

The sum of values of all four cells of the confusion matrix is equal to the total number of instances in the dataset.

$$|\textrm{Total instances}| = TP + FP + FN + TN$$

Precision (also called sensitivity, cf. \cite{PrecisionRecall3, PrecisionRecall4, PrecisionRecall5}) is defined as the ratio of true positive instances to all instances marked as positive by the model. Intuitively, this measure evaluates how clean or unpolluted the prediction with regard to the positive instances is.

$$\textrm{Precision} = \frac{TP}{TP+FP}$$

Recall (also called true positive rate, cf. \cite{PrecisionRecall1} or confidence, cf. \cite{PrecisionRecall5}) is defined as the ratio of true positive instances to all positive instances in the dataset. In other words, recall presents the power of the model to find all positive instances.

$$\textrm{Recall} = \textrm{TPR} = \frac{TP}{TP+FN}$$

Conversely, the false positive rate (also called specificity, cf. \cite{PrecisionRecall1, PrecisionRecall3}) is defined as the proportion of false positive instances to all negative instances in the dataset, i.e. it can be seen as the power of the model to find all negative instances.

$$\textrm{FPR} = \frac{FP}{FP+TN}$$

Another common metric is accuracy, which presents the ratio of correctly classified instances to the total number of instances (eg. cf. \cite{PrecisionRecall5}).

$$\textrm{Accuracy} = \frac{TP+TN}{|\textrm{Total instances}|}$$

Using accuracy for skewed datasets (meaning that there are many more instances belonging to one class than the other) can be misleading, as elaborated in \cite{PrecisionRecall1, PrecisionRecall5}. For instance, if only five per cent of instances in a dataset are positive, then a model assigning all instances to be negative would achieve an accuracy score of $95\%$. Potentially more useful models (especially if detecting the positive examples is more important than negatives, for instance by automatic detection of fraudulent behaviour or with for a system that proposes whether a medical screening for a certain disease would be necessary), could consequently achieve lower accuracy scores and those seem less useful. 

For this reason, reporting precision and accuracy is often preferred over-reporting accuracy. Yet, while the perfect model would correctly classify all instances resulting in both recall and precision being equal to one, in reality, there is a trade-off between the two, and this inverse relation is investigated in \cite{PrecisionRecall2, PrecisionRecall1}.  Nevertheless, in some scenarios, having a single measure instead of two can be desired (e.g. for a ranking of various models). One of these approaches is also F-score or F-measure, the most well-known version of which is marked as F1 and is presented in e.g. \cite{PrecisionRecall5} as:

$$F_1 = \frac{TP}{TP+(FN+FP)/2} = 2 \cdot \frac{\textrm{Precision} \cdot \textrm{Recall}}{\textrm{Precision} + \textrm{Recall}}$$

As investigated in \cite{PrecisionRecall1, PrecisionRecall3}, many binary classification models provide not only the label predictions but also a level of confidence (often in the forms of weights, for example on a continuous scale from zero to one, so that the values represent the estimated likelihood of instances belonging to the class in question). The threshold value below which one class is assigned and at or above which the other class is assigned can vary. This enables defining metrics which showcase not only how the model performs for a single threshold but over the whole domain of values. For instance, Receiver Operator Characteristic (ROC) curves showcase the relation between correctly classified positive examples and incorrectly classified negative examples. More formally, the curve is plotted so that each point represents the values of true positive rate (TPR, i.e. recall) and false positive rate (FPR) for some threshold value of the confidence interval. This plot can then be condensed into a single number metric by calculating the corresponding area under the curve (abbreviated as AUC-ROC).

It has been noted, however (cf. \cite{PrecisionRecall1, PrecisionRecall3}), that AUC-ROC too can be misleading and too optimistic for skewed data. A proposed alternative, especially in information retrieval tasks, is to use a curve which plots not recall and false positive rate, but recall and precision. Based on this pair of values for various thresholds of confidence values for classification, we define the precision-recall curve (PRC) and based on the area under it the area under the precision-recall curve (PRAUC). And this perception of PRAUC as a metric that is expected to provide a good measure of both recall and precision of a model is what is cited in \cite{tweetPap} as the explanation as to why it was chosen for the solution evaluation.

\subsection{Relative Cross-Entropy (RCE)}
\label{secRCE}

In his seminal work \cite{shannonEntropyRCE1}, Shannon explored fundamental problems of communication and defined entropy as a measure of the rate at which information is produced. Entropy measures the level of uncertainty of the outcome of a set of possible events with probabilities $p_1, p_2,...,p_n$. Shannon \cite{shannonEntropyRCE1} formally defined entropy as the negative sum of these probabilities and their logarithms.

$$H = -K\sum_{i=1}^n p_i\ log\ p_i$$

The coefficient $K$ in the formula of entropy represents the unit of measurement used for entropy. In information theory, entropy is typically measured in bits, so K would be equal to 1 if we are measuring entropy in bits. However, K can also be used to measure entropy in other units, in which case K would take a different value. K can also be used to calculate entropy in different logarithm bases, as discussed further below in this subsection. Shannon further notes in \cite[p.~11-12]{shannonEntropyRCE1} that entropy is the only measure that fulfils the following three properties:

\begin{enumerate}
    \item H is continuous in the $p_i$.
    \item If all events are equiprobable ($p_i = \frac{1}{n},\ i \in \{1,2,...,n\}$), then H is a monotonically increasing function of n (since the events are equiprobable, there is more uncertainty regarding the outcome as more events are considered).
    \item The value of H for a choice is equal to the weighted sum of individual values of H for sub-choices the original choice can be broken down into.
\end{enumerate}

\begin{figure}[htp]
\noindent\makebox[\textwidth]{\includegraphics[width=\textwidth]{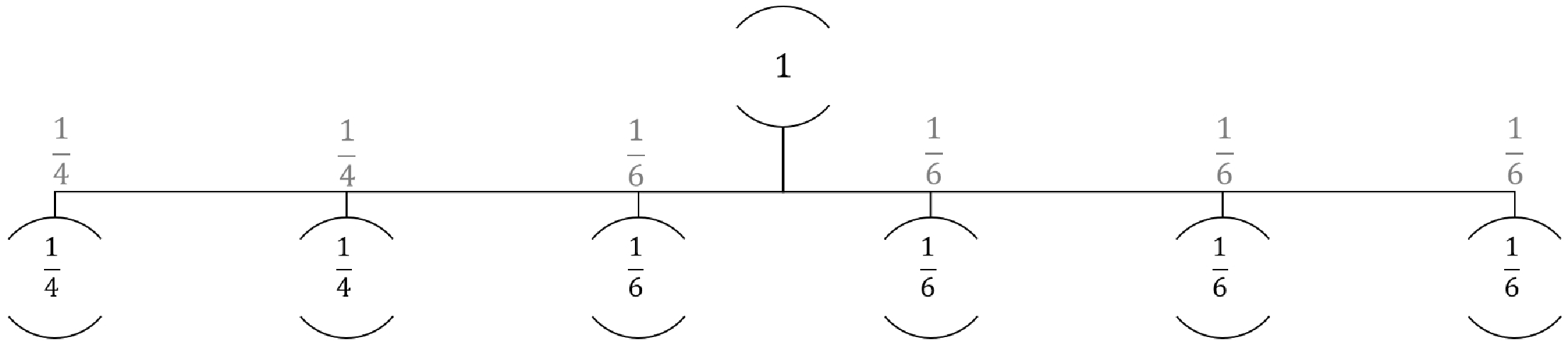}}
\caption{Entropy, simple case with no sub-choices}
\label{figEntropy1}

\noindent\makebox[\textwidth]{\includegraphics[width=0.8\textwidth]{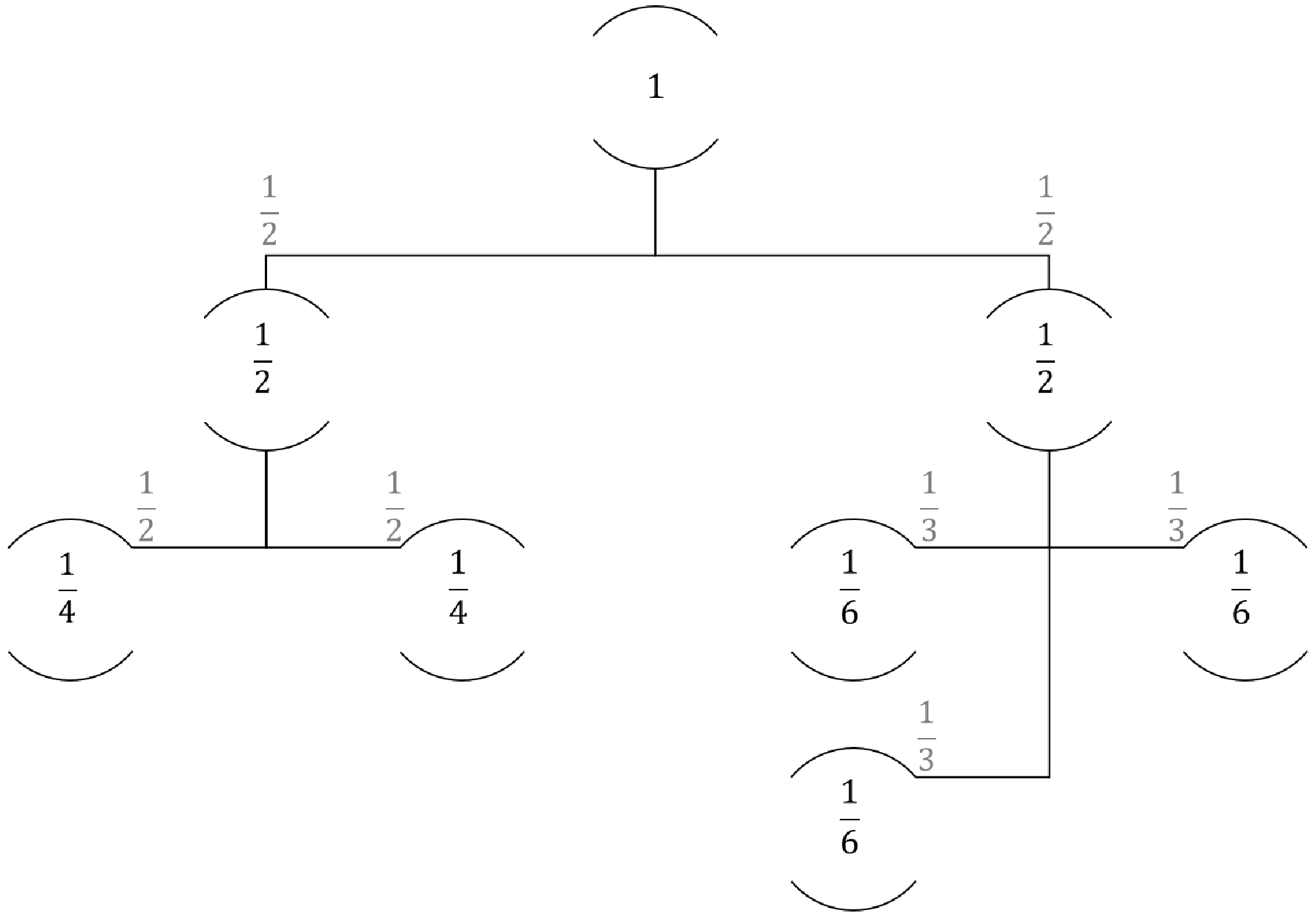}}
\caption{Entropy, second case}
\label{figEntropy2}

\noindent\makebox[\textwidth]{\includegraphics[width=0.95\textwidth]{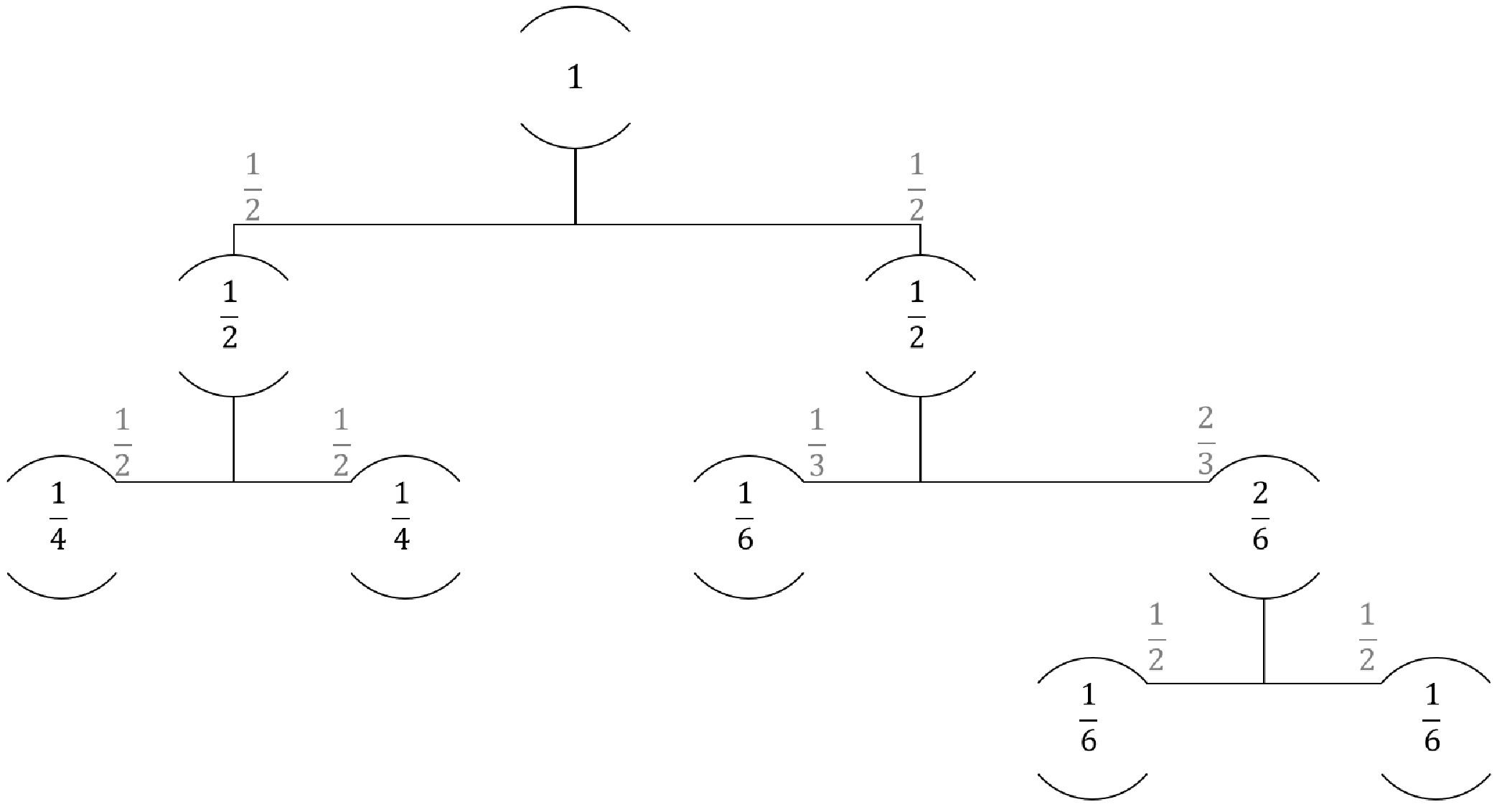}}
\caption{Entropy, third case}
\label{figEntropy3}
\end{figure}

To clarify the last property, let us look at a choice in figure \ref{figEntropy1}, and the choice broken in sub-choices in figures \ref{figEntropy2} and \ref{figEntropy3}.

Now, let us consider the formula for entropy $H$ and calculate further for these three illustrative cases. For the simple case in figure \ref{figEntropy1} when $K = 1$, we have:


$$H(\frac{1}{4},\frac{1}{4},\frac{1}{6},\frac{1}{6},\frac{1}{6},\frac{1}{6}) = -(2 \cdot \frac{1}{4}\log\frac{1}{4} + 3 \cdot \frac{1}{6}) = -(2 \cdot (-0.3465\ldots) + 3 \cdot (-0.2986\ldots)) = 1.5890\ldots$$

For the second case in figure \ref{figEntropy2} we have:

$$H(\frac{1}{2}, \frac{1}{2}) + \frac{1}{2} H(\frac{1}{2}, \frac{1}{2}) + \frac{1}{2}H(\frac{1}{3},\mathbf{\frac{1}{3},\frac{1}{3}})=$$
$$\frac{3}{2}H(\frac{1}{2}, \frac{1}{2}) + \frac{1}{2} H(\frac{1}{3},\frac{1}{3},\frac{1}{3})=$$
$$\frac{3}{2} (-(\cancel{2\cdot\frac{1}{2}}\log\frac{1}{2})) + \frac{1}{2} (-(\cancel{3\cdot\frac{1}{3}}\log\frac{1}{3})) = $$
$$\frac{3}{2}\cdot0.6931\ldots + \frac{1}{2} \cdot 1.0986\ldots = $$
$$1.0397\ldots+0.5493\ldots=$$
$$1.5890\ldots$$

Lastly, if we branch out even further as in the case in figure \ref{figEntropy3}, we get:

$$H(\frac{1}{2}, \frac{1}{2}) + \frac{1}{2} H(\frac{1}{2}, \frac{1}{2}) + \frac{1}{2}H(\frac{1}{3},\mathbf{\frac{2}{3}}) + \mathbf{\frac{1}{2}\cdot \frac{2}{3} H(\frac{1}{2}, \frac{1}{2})} = $$
$$\frac{11}{6}H(\frac{1}{2}, \frac{1}{2}) + \frac{1}{2} H( \frac{1}{3}, \frac{2}{3}) = $$
$$\frac{11}{6} (-(\cancel{2\cdot\frac{1}{2}}\log\frac{1}{2})) + \frac{1}{2} (-(\frac{1}{3}\log\frac{1}{3}+\frac{2}{3}\log\frac{2}{3})) = $$
$$\frac{11}{6}\cdot0.6931\ldots + \frac{1}{2} \cdot 0.6365\ldots = $$
$$1.5890\ldots$$

We see that as long as there is the same number of outcomes with equal probabilities, the number of sub-decisions does not matter. Note that these outcomes cannot be combined, as one can see by looking at the bolded parts of the second and third parts of the example above. If we were to consider, $H(\frac{1}{2}, \frac{1}{2}) + \frac{1}{2} H(\frac{1}{2}, \frac{1}{2}) + \frac{1}{2}H(\frac{1}{3},\frac{2}{3})$, i.e. to sum the last two outcomes in advance, we would get $\mathbf{\frac{3}{2}}H(\frac{1}{2}, \frac{1}{2}) + \frac{1}{2} H( \frac{1}{3}, \frac{2}{3})$, which is clearly less than $\mathbf{\frac{11}{6}}H(\frac{1}{2}, \frac{1}{2}) + \frac{1}{2} H( \frac{1}{3}, \frac{2}{3})$ and thus the total entropy is not the same.

In our calculations, we have used logarithm with Euler's number as the base, since this base was used in the challenge as well. The base properties of entropy hold for all bases. In fact, Shannon in \cite[p.~3-4]{shannonEntropyRCE1} recalls that converting from logarithm in base $b$ to logarithm in base $a$ amounts to multiplying the original value with $\frac{1}{\log_b a}$:

$$\log_a x = \frac{\log_b x}{\log_b a}$$

It is noteworthy that the term relative cross-entropy is somewhat ambiguous. For instance, while entropy is defined as a measure of uncertainty measuring the amount of information necessary on average to describe a single variable, relative entropy in \cite[p.~19-20, 54-55]{bookInfromationTheoryRCE3} is described as a metric for the distance between two probability distributions. The book also notes that the relative entropy, Kullback-–Leibler distance, and cross-entropy are just different names for the metric defined to be $D(p||q) = \sum_{i=1}^n p \log \frac{p}{q}$, as was first described in \cite{kullback1951information}. Another definition can also be found in e.g. \cite{RCE4}, which differentiates between cross-entropy and relative entropy. However, neither of the two definitions presented in these works corresponds to the evaluation approach selected by the challenge organisers.

Namely, the code snippet provided by the challenge organisers is based on Scikit-learn's function \verb=log_loss=. The documentation\footnote{cf.  \url{https://scikit-learn.org/stable/modules/generated/sklearn.metrics.log_loss.html} and \url{https://scikit-learn.org/stable/modules/model_evaluation.html\#log-loss} (access date 2021-05-10)} indicates that the cross-entropy actually refers to the logarithmic entropy loss, as defined in \cite[p.~209]{mlbPatternRecognitionAndML2006}. This book defines cross-entropy error measure as a negative log-likelihood of the probability of the ground truth given model parameters $p(T|w_1,\ldots,w_K)$. In the general case, with multiple target classes, we have:

\begin{equation*}
\begin{split}
CE(w_1,\ldots,w_K) &= -\log Pr(T|w_1,...,w_K) \\
&= -\log \prod_{n=1}^N \prod_{k=1}^K y_{n,k}^{t_{n,k}} \\
&= - \sum_{n=1}^{N}\sum_{k=1}^{K}t_{n,k}\log y_{n,k}
\end{split}
\end{equation*}

In this formula, we assume a 1-of-K coding scheme, which means that each target vector $t_n$ has value one for the dimensions corresponding to the correct class w.r.t the ground truth, with the other dimension being zero. Furthermore, $T$ is the indicator matrix of size  $N \times K$ containing $N$ target vectors $t_{n}$; in other words, $t_{n,k}$ is equal to one (and not zero) if the $n$-th instance belongs to class $k$. Vector $y_{n,k} = y_k(\phi_n)$ contains the probability that instance $n$ belongs to class $k$ given some feature vector $\phi_n$ and a model with parameters $\{w_1,\ldots,w_K\}$.

Given the challenge description, we are more interested in a binary classification case, where the true label can only have values $0$ or $1$, i.e. $l_n \in \{0, 1\}$ for all instances $n$. Let then $p_n=Pr(l_{n}=1)$, i.e. let $p_n$ be the probability that the instance $n$ belongs to the positive class $1$ (i.e. the class of instances where an engagement occurred in the context of this challenge). In that case, $1-p$ is the probability that the instance belongs to the opposing negative class $0$. We then have:

\begin{equation*}
\begin{split}
CE(w_1,\ldots,w_K) &= -\log Pr(l|w_1,\ldots,w_K) \\
&= -\sum_{n=1}^N (l_n\log p_n + (1-l_n) \log (1-p_n))
\end{split}
\end{equation*}

Where $l$ is the vector of length $N$ containing ground truth labels $l_n$. Lastly, it is worth noting that the scikit-learn's corresponding \verb+log_loss+ function also normalises the value for the number of instances by multiplying the above two expressions with $\frac{1}{N}$.

The \emph{relative} part of the relative cross-entropy comes by comparing the cross-entropy value (as it was defined above) of the submitted prediction model with the cross-entropy of a straw man model based on click-through rate (CTR), i.e. the ratio of positive to negative instances. As stated in \cite{tweetPap}, this straw man model ``predicts the average (observed) CTR of the training set''. However, upon observing the code the organisers submitted on the challenge website, we see that the straw man model is actually based on the CTR of the ground truth labels. In other words, the straw man has insight into the labels of the test set, not the train set. As this is, of course, not the case with the proposed model, the straw man is not that easy to beat. From the original challenge description, however, it was not clear whether the actual challenge used CTR of the train set as stated in \cite{tweetPap} or CTR of the test set as indicated by the provided code snippet (cf. subsection \ref{secRanking} for the response we got from the challenge organisers regarding this ambiguity). In either case, the relative cross-entropy is then calculated as:

$$RCE = \frac{(CE_{straw\ man} - CE_{predictions})}{CE_{straw\ man}} \cdot 100$$

where $CE_{predictions}$ is the cross-entropy of the model that is to be ranked, and $CE_{straw\ man}$ is the cross-entropy of the straw man model. The lower $CE_{predictions}$, the higher the total RCE. As stated in \cite{tweetPap} this metric was chosen as it provides a sense of how better the proposed model is than the baseline model.

\subsection{Metric Implementation}
\label{secMetricImplementation}

Below we document two code snippets originally published by the challenge organisers on the challenge website. There are two reasons why the code is recognised as significant enough to be provided in the main body of the work rather than in the appendix. Most importantly, two rather substantial discrepancies between the claimed challenge evaluation protocol and the actual evaluation implementation were identified by analysing the snippets below. Additionally, the evaluation code was removed from the website after the subsequent RecSys 2021 Challenge commenced, so copying and explaining the code here might prove deeply relevant for later research on the RecSys 2020 Challenge. The provided snippets are in Python and use metrics functions\footnote{cf. \url{https://scikit-learn.org/stable/modules/classes.html\#module-sklearn.metrics} (access date 2021-05-11)} from scikit-learn package presented in \cite{scikit-learn}.

\begin{lstlisting}[language=Python, caption=Code calculating the area under the precision-recall curve (PRAUC), label=listPRAUC]

from sklearn.metrics import precision_recall_curve, auc

def compute_prauc(pred, gt):
    prec, recall, thresh = precision_recall_curve(gt, pred)
    prauc = auc(recall, prec)
    return prauc
\end{lstlisting}

The function \verb+precision_recall_curve+ in listing \ref{listPRAUC} computes precision-recall pairs for different probability thresholds in a binary classification task. The implementation of the function corresponds completely with the explanation provided in \cite{tweetPap}. Next, the \verb+auc+ function below calculates the area under the curve using trapezoid approximation.

\begin{lstlisting}[language=Python, caption=Code calculating relative cross entropy (RCE), label=listRCE]
from sklearn.metrics import log_loss

def calculate_ctr(gt):
    positive = len([x for x in gt if x == 1])
    ctr = positive / float(len(gt))
    return ctr


def compute_rce(pred, gt):
    cross_entropy = log_loss(gt, pred)
    data_ctr = calculate_ctr(gt)
    strawman_cross_entropy = log_loss(gt, [data_ctr for _ in range(len(gt))])
    return (1.0 - cross_entropy / strawman_cross_entropy) * 100.0
    
\end{lstlisting}

Here, in listing \ref{listRCE}, we see the first discrepancy between what was stated by challenge organises in \cite{tweetPap, recsys2020overview} and the actual implementation. Namely, the function \verb+calculate_ctr+ calculates the straw man model using the ground truth labels of the test/val dataset, not the train dataset, as elaborated in section \ref{secRCE}. A statement on the official competition forums\footnote{cf. \url{https://groups.google.com/u/1/g/recsys-challenge2020/c/sY9ATUTlCqk} (access date 2021-05-17)} asserting that the code snippets are also used for the evaluation of submitted models, made the likelihood that this is indeed an oversight very likely. This assumption was kindly confirmed by Gabriele Sottocornola, one of the authors of the papers, after an email exchange on the issue.  Casting to float the length of the set before using it as the divisor of the number of positive instances forces actual division instead of whole-number division. The scikit-learn's \verb=log_loss= function calculates the cross-entropy of the model to be scored and the straw man model. Lastly, the two cross-entropy values are used to calculate the relative cross-entropy using the reported formula.

\subsection{Competition Ranking}
\label{secRanking}

The leaderboard with the ten best solutions at the end of the challenge was provided in \cite{tweetPap}, whereas the ranking of the solutions with published papers is provided in section \ref{secRecSys2020Results} of this thesis.

The frequently asked questions section of the challenge website\footnote{cf. \url{https://web.archive.org/web/20200621055111id_/https://recsys-twitter.com/faq/} (access date 2021-05-22)} provided the following answer to the question of how the final ranking is calculated: ``The RCE and PR-AUC values are averaged across the four different engagements. Participants will then be ranked based on those averaged metrics and the sum of these two ranks will be used to obtain their overall ranking score''. However, a simple sum of the two ranks cannot be the way in which the score was obtained, since in the case of the winning team, thus calculated final score would be two, as they had the highest averaged values of both PRAUC and RCE, causing them to be ranked one in both cases. The simple sum should thus be two, not nine, as their actual ranking is. 

A possible alternative explanation could be that the two metrics were computed for each class independently and each model was ranked according to these eight criteria (four engagement types and two metrics). Thus, in the case of the winning team, they achieved the best results for all the metrics except for Reply PRAUC for which they got second place. But this explanation, too, would remain inconsistent because the second team would then have to have a score that is not lower than $7\cdot2+1=15$ (for some metrics, they would actually be ranked worse, so the sum would have to be strictly greater than $15$). Yet, the second team on the leaderboard has a score of just $14$. A plea for clarification was sent to the challenge chairs who stated that they were not tasked with the practical implementation of the challenge. Thus they referred the questions to two organisers from Twitter, who stated that they, too, were confused by the discovery and forwarded the query to a third colleague who had been tasked with the implementation of the leaderboard specifically. 

Finally, in the response we received from the person responsible for the implementation of the scoreboard, they clarified that the simple sum of just two ranks had indeed been the primary mode of computing the final score for the competition ranking. Thus, ideally, the highest ranking position would have a score of $2$, as it would have the highest ranking average score for both PRAUC and RCE. But, as they further stated in the email, ``a lot of teams'' had exploited the fact that the challenge organisers had used an implementation of the PRAUC metric from the scikit-library that ``has a drawback that if you submit constant predictions, you will get a PRAUC of $0.5$''. Thus these teams would still receive a bad RCE score. But $0.5$ was still a very good PRAUC score for the standards of the challenge, especially for engagements other than like. These teams would indeed be ranked high for the PRAUC score but very low overall because of the bad RCE score. Thus, essentially, there were seven submissions that ``exploited'' this particularity PRAUC metric implementation, causing the winning team to get the rank $8$ for the PRAUC. Since the RCE could not be increased in a similar manner, the winning team still got the rank $1$ for RCE. The final score is thus indeed the sum of the two initial ranks (averaged over the four engagements), i.e. $8+1=9$, Yet the seven submissions with higher PRAUC scores likely had much lower RCE scores, which is why they were not first. Following the same logic from this explanation would explain why the second team had a score of, presumably, $12+2=14$, rather than $2+2=4$. 

\section{Evaluation Results}
\label{secEvaluationResults}

Shortly after the engagements for the val dataset became available, we created baseline predictions using only the pre-provided features. These results can be seen in subsection \ref{secEvalBaseline}. We then proceeded with more elaborate feature dataset sampling, feature engineering, and with multiple algorithms. The main results of this approach is in subsection \ref{secEvaluationResults}.

\subsection{First Round of Evaluations: Baseline}
\label{secEvalBaseline}

For the baseline results, we fitted the models based only on the initially provided features and used only two prediction algorithms -- decision trees and logistic regression. Additionally, both of these algorithms' hyperparametres were not tuned. Instead, the default values from MLLib \cite{pysparkMLLIBdocumentation} were taken.

\begin{table}[htb]
    \centering
\begin{tabular}{lrrrrr}
\toprule
{} &   like &  reply &  retweet & quote &  react \\
\midrule
all\_scaled            & \textbf{-1928.887545} &  -643.145029 &    -960.266779 &                 -478.385215 &        -2039.375523 \\
all\_binned            & \textbf{-1891.491721} &  -643.145029 &    -960.266779 &                 -478.385215 &        -1996.574991 \\
engaged\_with          & \textbf{-1953.090152} &  -643.145029 &    -960.266779 &                 -478.385215 &        -2075.582714 \\
engaging              & \textbf{-1903.638460} &  -643.145029 &    -960.266779 &                 -478.385215 &        -2147.797617 \\
engaging+engaged & \textbf{-1908.501838} &  -643.145029 &    -960.266779 &                 -478.385215 &        -2089.597918 \\
timestamps            & \textbf{-1961.683423} &  -643.145029 &    -960.266779 &                 -478.385215 &        -2157.793471 \\
media                 & \textbf{-1955.720727} &  -643.145029 &    -960.266779 &                 -478.385215 &        \-2093.000447 \\
dummy                 & -2028.924259 &  -643.145029 &    -960.266779 &                 -478.385215 &        -2487.537127 \\
\bottomrule
\end{tabular}
    \caption{RCE for full train and val datasets using baseline decision trees}
    \label{tabBaselineDTfullRCE}
\end{table}

\begin{table}[ht]
    \centering
\begin{tabular}{lrrrrr}
\toprule
{} &   like &  reply &  retweet & quote &  react \\
\midrule
all\_scaled            & \textbf{-1851.173521} &  -643.347603 &    -960.266779 &                 -478.572703 &        -1916.529666 \\
all\_binned            & -\textbf{1839.933656} &  -643.423569 &    -960.266779 &                 -478.531945 &        -1916.004898 \\
engaged\_with          & \textbf{-1960.959541} &  -643.145029 &    -960.266779 &                 -478.385215 &        -2064.192775 \\
engaging              & -2028.955908 &  -643.192859 &    -960.449628 &                 -478.466732 &        -1978.915417 \\
engaging+engaged & \textbf{-1901.839232} &  -643.167537 &    -960.423656 &                 -478.442277 &        -2016.205784 \\
timestamps            & \textbf{-1893.985138} &  -643.145029 &    -960.266779 &                 -478.385215 &        -2011.660096 \\
media                 & \textbf{-1996.038670} &  -643.145029 &    -960.292752 &                 -478.385215 &        -2137.726157 \\
dummy                 & -2028.924259 &  -643.145029 &    -960.266779 &                 -478.385215 &        -2487.537127 \\
\bottomrule
\end{tabular}
    \caption{RCE for full train and val datasets using baseline logistic regression}
    \label{tabBaselineLRfullRCE}
\end{table}

Table \ref{tabBaselineDTfullRCE} shows RCE results for the full train and validation datasets using a decision tree model, whereas table \ref{tabBaselineLRfullRCE} is for the logistic regression. We see that in both cases, only for like there was a significant improvement over the dummy model that simply predicted the majority class (i.e. the likelihood of zero for all engagements). For all engagement types, with the exception of like, the decision tree actually classified all instances as negative. This is not surprising, given the dataset imbalance. In the case of logistic regression, just taking into account the engaging users' features on their own was not enough to beat the model based solely on the intercept value.

\begin{table}[ht!]
    \centering
\begin{tabular}{lrrrrr}
\toprule
{} &   like &  reply &  retweet & quote &  react \\
\midrule
all\_scaled            &    0.578481 &     0.512858 &       0.549682 &                    0.503454 &            0.672206 \\
all\_binned            &    0.598095 &     0.512858 &       0.549682 &                    0.503454 &            0.690480 \\
engaged\_with          &    0.545419 &     0.512858 &       0.549682 &                    0.503454 &            0.672184 \\
engaging              &    0.565760 &     0.512858 &       0.549682 &                    0.503454 &            0.649040 \\
engaging+engaged      &    0.574432 &     0.512858 &       0.549682 &                    0.503454 &            0.663664 \\
timestamps            &    0.532203 &     0.512858 &       0.549682 &                    0.503454 &            0.589639 \\
media                 &    0.547546 &     0.512858 &       0.549682 &                    0.503454 &            0.719188 \\
dummy                 &    0.709577 &     0.512858 &       0.549682 &                    0.503454 &            0.740627 \\
\bottomrule
\end{tabular}
    \caption{PRAUC for full train and val datasets using baseline decision trees}
    \label{tabBaselineDTfullAUC}
\end{table}

\begin{table}[ht]
    \centering
\begin{tabular}{lrrrrr}
\toprule
{} &   like &  reply &  retweet & quote &  react \\
\midrule
all\_scaled            &    0.584699 &     0.043355 &       0.549682 &                    0.023461 &            0.681477 \\
all\_binned            &    0.595620 &     0.039896 &       0.549682 &                    0.028461 &            0.692099 \\
engaged\_with          &    0.541828 &     0.512858 &       0.549682 &                    0.503454 &            0.651735 \\
engaging              &    0.351472 &     0.039176 &       0.116363 &                    0.003454 &            0.673105 \\
engaging+engaged &    0.561851 &     0.062860 &       0.115547 &                    0.003454 &            0.657345 \\
timestamps            &    0.565128 &     0.512858 &       0.549682 &                    0.503454 &            0.655918 \\
media                 &    0.509485 &     0.512858 &       0.121113 &                    0.503454 &            0.730702 \\
dummy                 &    0.709577 &     0.512858 &       0.549682 &                    0.503454 &            0.740627 \\
\bottomrule
\end{tabular}
    \caption{PRAUC for full train and val datasets using baseline logistic regression}
    \label{tabBaselineLRfullAUC}
\end{table}

Tables \ref{tabBaselineDTfullAUC} and \ref{tabBaselineLRfullAUC} show the same setting but with PRAUC as the evaluation metric. We see that for like engagement in the case of decision trees, the better RCE resulted in a lower PRAUC. This was true for logistic regression across all engagements.  We also note that we got a PRAUC of around 0.5, which is a very high score for non-like engagements (cf. table \ref{tabLeaderBoard2} and \ref{tabLeaderBoard3}). This is the result of the phenomenon discussed in subsection \ref{secRanking} that constant predictions result in a PRAUC of around $0.5$. We thus know that these results indicate that the underlying models actually produce constant predictions (always predicting negative) rather than actually being more informative than the challenge models.

We also ran the baseline models for just subsets of pre-provided features and for combinations of 1\% and 10\% random samples of the train dataset as well as the 1\% and 5\% of random samples of the val dataset. These results can be seen in appendix \ref{appEvalSampleBaselineEcals}. It can be concluded that the patterns remain the same according to all sampling combinations. This is not surprising given that the pre-provided features are very simple and that the decision trees and logistic regression are too regimental to capture complex user-based or temporal features from these features alone. The performance remained poor even on the full dataset.

\subsection{Main Evaluation Results}
\label{secMainEvaluation}

The main results were based on predictions that were created and evaluated based on engineered contextual features. Several prediction algorithms were employed. Specifically, from mid-November until mid-March 2023 we ran na\"ive Bayes, decision tree, random forest, and gradient boosting algorithms on the cluster. In that period, we also ran the four algorithms as well as the logistic regression, multilayered perceptron, and support vector machines classifier locally\footnote{Running the latter four algorithms resulted in the following cryptic error message: \texttt{Py4JError: An error occurred while calling o8618.fit}. After a significant amount of time unsuccessfully trying to resolve the issue, it was decided to run these classifiers only locally.}. All experiments ran in this period excluded the so-called ``oracle'' features, which contain information on events that happened after the viewing of the tweet. The classification was done based both on the whole dataset as well as the five, ten, 25, and 50 most informative features according to the $\chi^2$ feature selector. After that period and the closing of TU Wien's LBD cluster, we ran additional experiments locally, this time focusing on features that include the oracle, but excluding prior feature selection.

\subsubsection{Distribution of the Results Over Factors}
\label{secDistributionOfTheResultsOverFactors}

In total, the main evaluation resulted in roughly hundred and twenty thousand results ($59690$ for both PRAUC and RCE). This number excludes the intermediary results used for hyperparametre-tuning as well as preliminary results which were based on models whose parameters were not tuned. While this amount of evaluations is high, it should be noted that this is still just a fraction of all possible combinations. Since there are 84 datasets (cf. appendix \ref{appSubsets}, each could theoretically be seen as both a base for training and testing), two feature sets (with and without oracle features), seven classification algorithms, five feature selection approaches (all features plus prior $\chi^2$-selected features with four limits), five target variables (like, reply, retweet, quote, and react), and 2 evaluation metrics, the total search space was $84 \cdot 84 \cdot \ 2 \cdot 7 \cdot 5 \cdot 5 \cdot 5 \cdot 2 = 24\ 696\ 000$. Thus we had to limit our exploration and to extrapolate the results. Unfortunately, partially due to technical constraints, this is not a balanced exploration. Let us now explore this into more detail.

\begin{figure}[htp]
\noindent\makebox[\textwidth]{\includegraphics[width=0.6\textwidth]{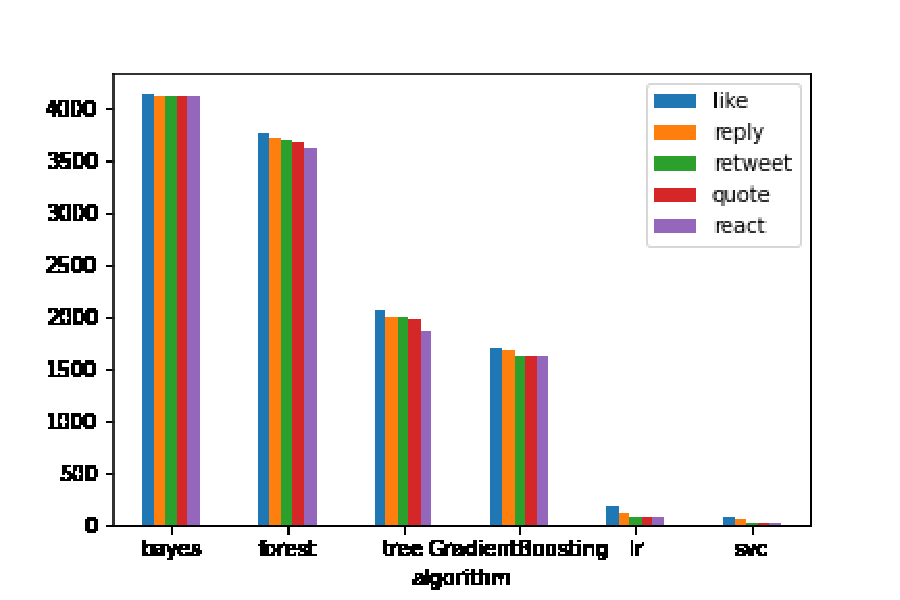}}
\caption{Number of evaluations of predictions for each algorithm per engagement type}
\label{figEvalValueConts-algorithm}
\end{figure}

Figure \ref{figEvalValueConts-algorithm} shows that most results are available for the na\"ive Bayes and random forest algorithms. While this is partially a result of the fact that these algorithms (and na\"ive Bayes in particular) are faster to train than, e.g. SVC, this is not the main cause of the discrepancy. Instead, logistic regression (LR) and SVC were only done locally, due to an opaque error preventing its use on the TU Wien's LBD server. And the reason why there are more results for random forests than for decision trees (even though the latter is conceptually a simpler algorithm), is that local prediction and evaluation were done sequentially. This contrasts the approach performed on the cluster, where all four classifiers were run in parallel. Due to an oversight in an earlier phase of this experimentation, more initial models for forest and Bayes were created than for the other algorithms. This was later changed to balance the output in terms of algorithms, but this initial advantage remained.

\begin{figure}[htp]
\noindent\makebox[\textwidth]{\includegraphics[width=0.6\textwidth]{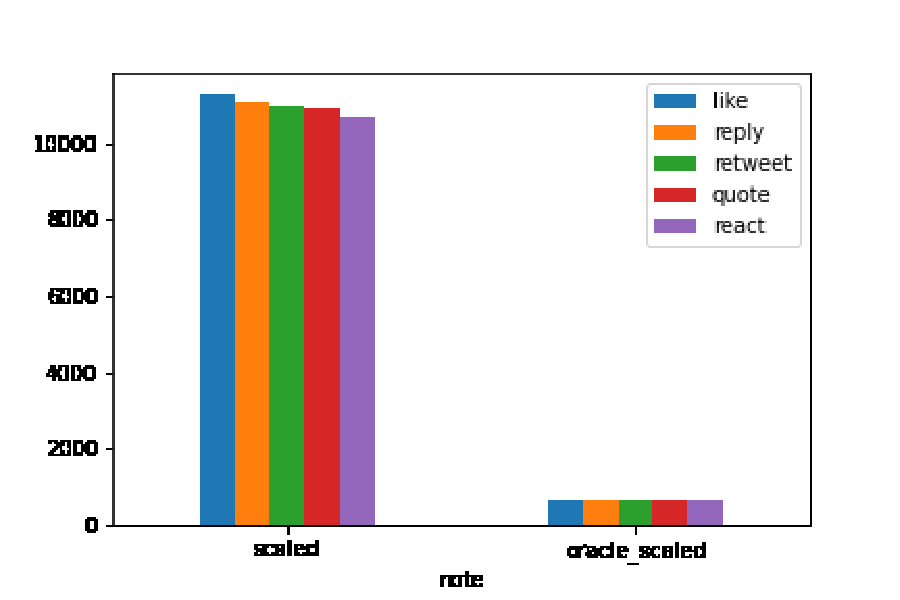}}
\caption{Number of evaluations of predictions between two feature sets: scaled features excluding the ``oracle'' features on the left-hand side and including the ``oracle'' features on the right-hand side; the ``oracle'' features contain information about the event after the tweet was seen, as discussed in subsection \ref{secOracleFeatures}}
\label{figEvalValueConts-feature note}
\end{figure}

The ratio of scaled non-oracle features to scaled oracle features can be seen in figure \ref{figEvalValueConts-feature note}. We did not test algorithms on features which were not scaled due to time constraints, even though these feature sets were created in the feature selection stage of the pipeline (cf. subsection \ref{secStagesOfTheMachineLearningPipeline}). From this figure, we can also read the ratio of initial predictions created and evaluated before mid-March on the cluster and locally (oracle excluded) to the predictions done locally after mid-march (oracle included). This disbalance is thus the result of the difference in runtime and the fast difference in computational capabilities (cf. \ref{secArchitecture}) and also showcases how well Spark can be scaled horizontally. 

\begin{figure}[htp]
\noindent\makebox[\textwidth]{\includegraphics[width=0.6\textwidth]{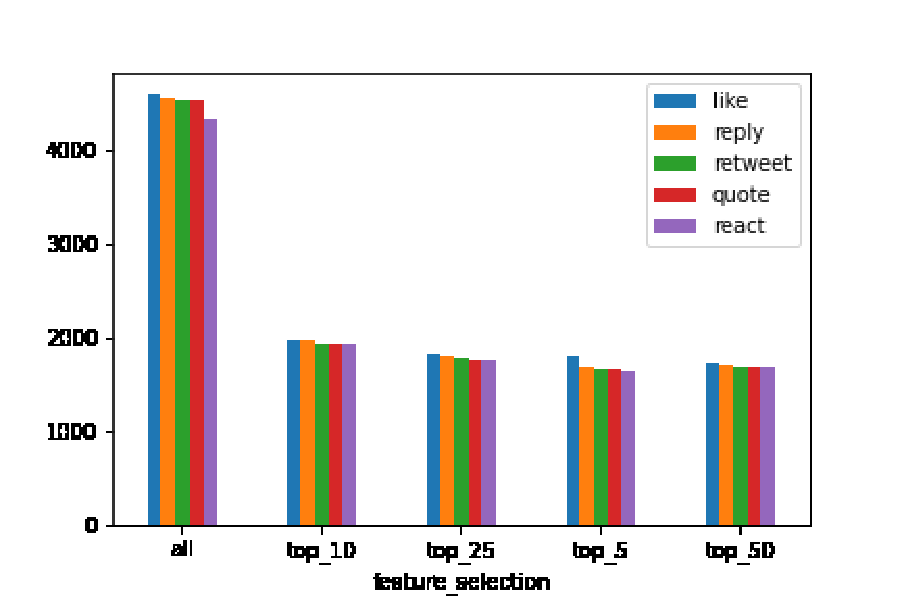}}
\caption{Number of evaluations of predictions for each feature selection approach per engagement type}
\label{figEvalValueConts-feature selection approach}
\end{figure}

As elaborated in subsection \ref{secFeatureSelectionImplementation}, we used $\chi^2$ feature selector to choose the top 5, 10, 25, and 50 features for each engagement type and train subset. We also tried using classifiers on all engineered features without feature selection. After the initial few months of training initial classifications on all five feature selection factors, we switched to training classifiers on all features only, as preliminary results have shown that feature selection is not contributing to a better result. Thus we wanted to use the remaining time and computational resources to get more instances of other factors, which were believed to be more interesting.

\begin{figure}[htp]
\centering
\includegraphics[width=.32\textwidth]{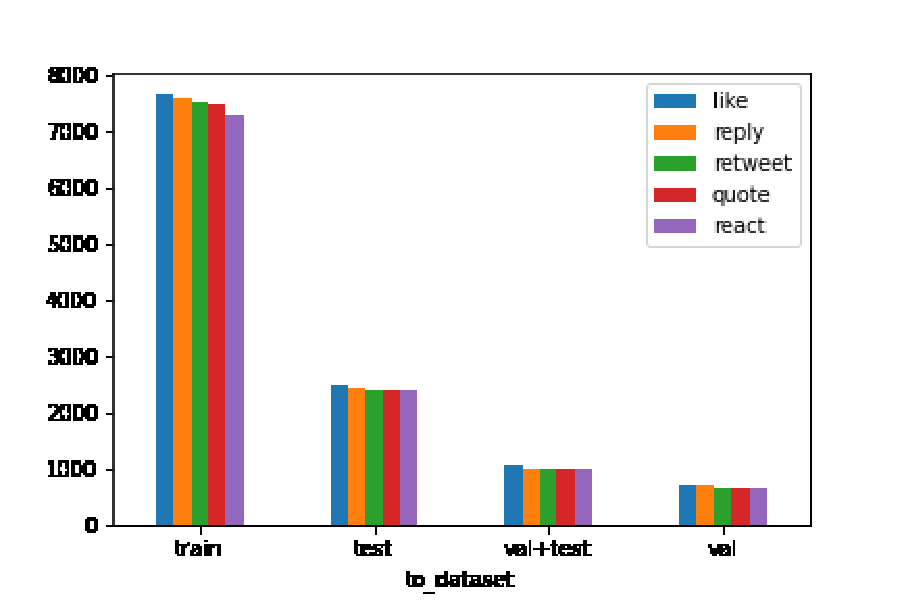} 
\includegraphics[width=.32\textwidth]{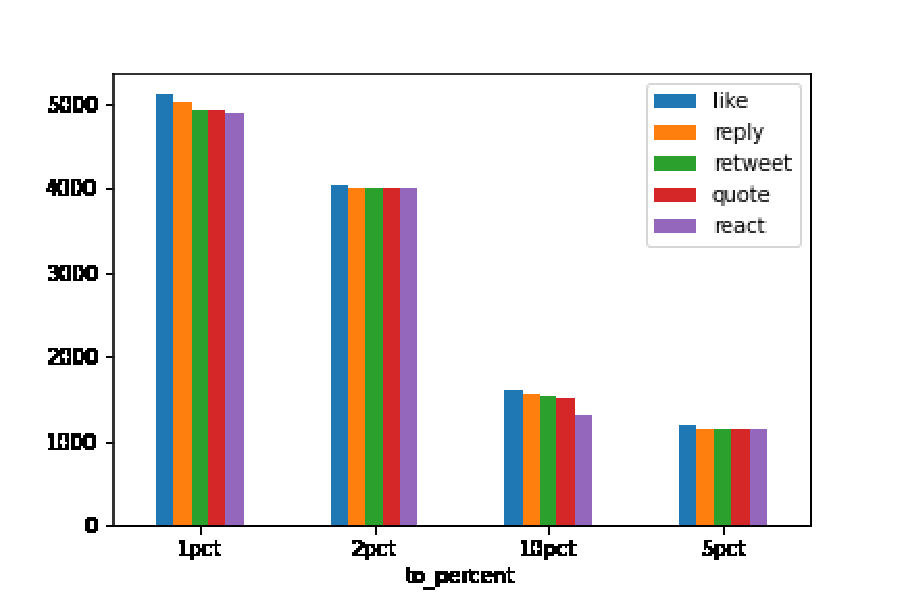} 
\includegraphics[width=.32\textwidth]{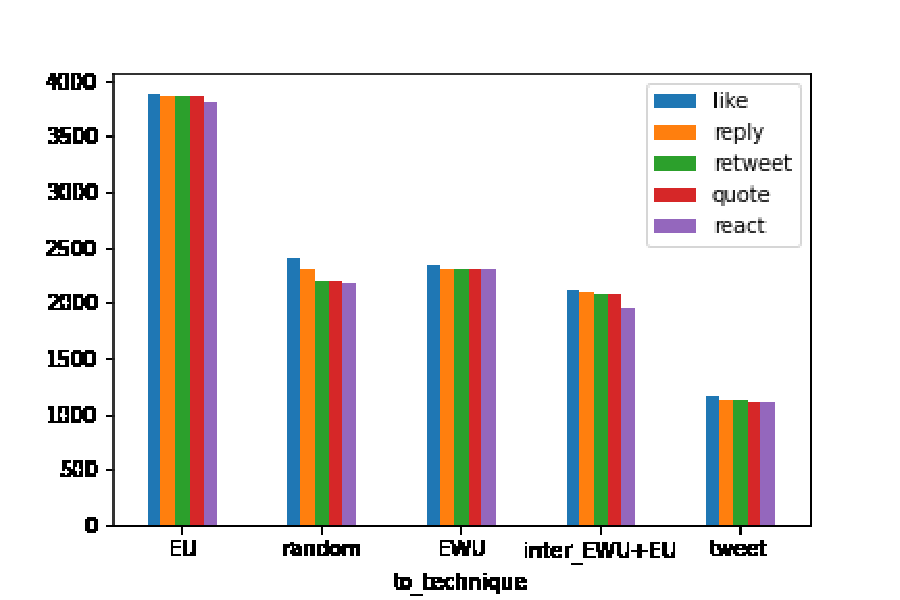}
\caption{Number of evaluations of predictions by the subset used for the training (from left to right: the source dataset, the per cent of values sampled, and the technique used for sampling)}
\label{figEvalValueConts-to}
\end{figure}

Bar plots in figure \ref{figEvalValueConts-to} showcase what subsets were used for training the models. The vast majority of them were based on the train subsets, with only some on test, val+test, and val subset which were added later for an ad-hoc comparison. Unfortunately, fitting classifiers on the full dataset failed due to memory issues on the TU Wien cluster. While we tried to use the full dataset at least for the most informative features (as selected by the $\chi^2$ selector), we, in the end, decided to use the remaining time available before the announced shutdown of the LBD cluster (which was expected in December 2022 already) to get results based on 5\% and 10\% subsets instead. The reason why the distribution of the sampling techniques is unequal (with tweet-ratio-preserving subsets being underrepresented and engagement-user-ratio-preserving sampling being overrepresented) is again due to the sequential nature of local classifier training that was initially unnoticed but was later remedied. 

\begin{figure}[htp]
\centering
\includegraphics[width=.32\textwidth]{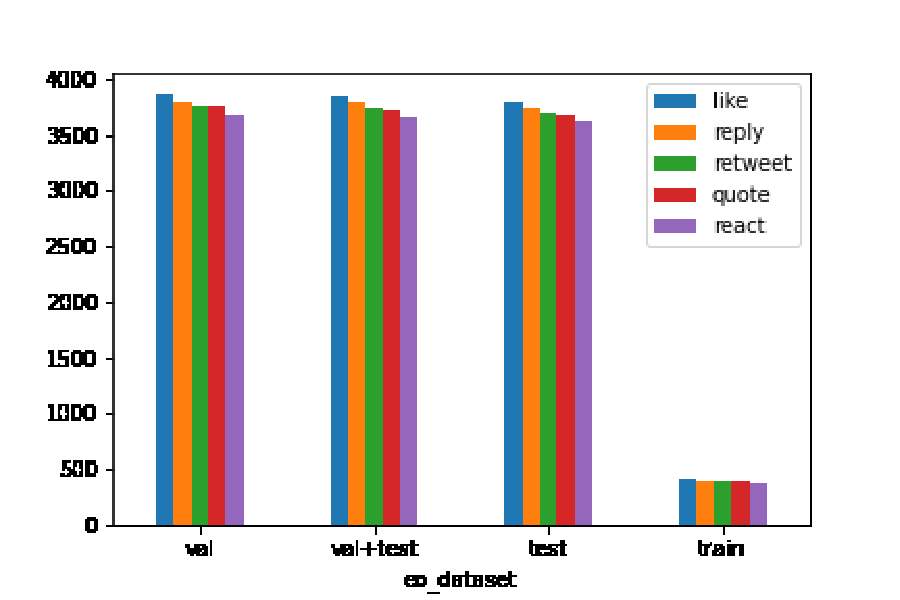} 
\includegraphics[width=.32\textwidth]{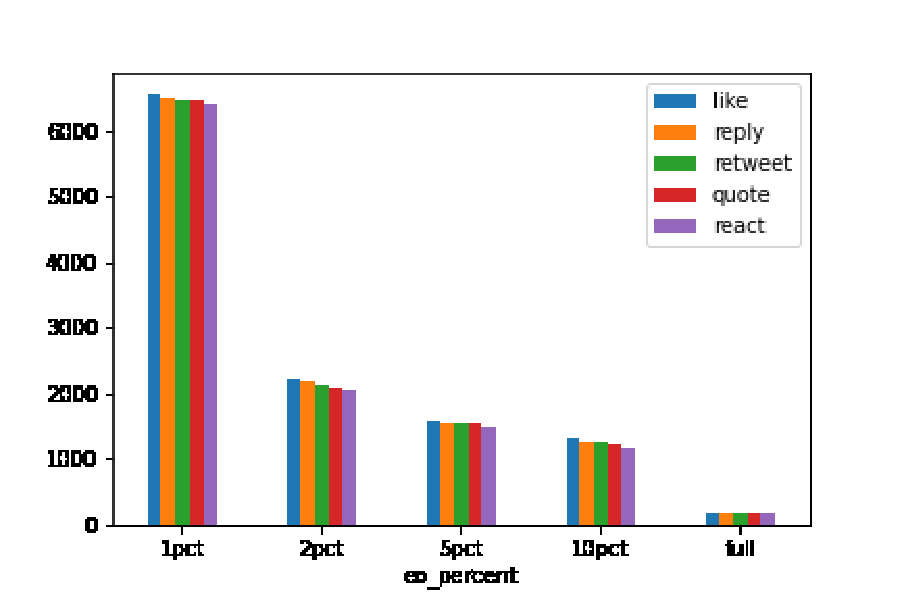} 
\includegraphics[width=.32\textwidth]{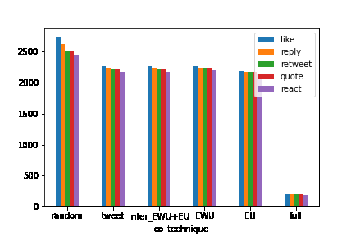}
\caption{Number of evaluations of predictions by the subset used for the evaluation (from left to right: the source dataset, the per cent of values sampled, the technique used for sampling)}
\label{figEvalValueConts-eo}
\end{figure}

Lastly, the plots in figure \ref{figEvalValueConts-eo} showcase the subsets on which the classifiers were evaluated. We see that the random sampling technique and 1\% datasets are overrepresented. This is because, during the period from mid-November 2022 to mid-March 2023, the local running of experiments only used those subsets. From mid-March to mid-May, both 1\% and 2\% datasets across all sampling techniques were used. It is important to note that on the cluster full datasets were evaluated as well, which is important because it allows us to compare our context-based results and content-based results in \cite{davidsThesis} and the performance achieved by challenge winners. 

Due to the fact that experiments had to be rerun after kernels had been killed or had run out of memory, a mechanism had to be implemented that would check if a classifier based on a given combination of factors had already been created. An early implementation of the engagement stage of the machine learning pipeline, had a bug which would skip a combination of factors if at least one engagement prediction (e.g. that for \emph{like}) existed, rather than checking if models for each engagement type for the given combination of factors existed. This issue was discovered and resolved, yet this early inequality has partially remained.

It should also be noted that the final results were checked and duplicates were deleted. These duplicates were created in an early phase of the result generation, while the experiments were simultaneously run locally and on the TU Wien's cluster. These duplicates were created intentionally to verify that there are no unaccounted elements affecting results (i.e. to ensure reproducibility). We also removed results for which only RCE or only PRAUC existed, but not both. This discrepancy was not intentional and is likely a result of the unplanned killings of kernels before the other metric could be calculated. The numbers in the plots below thus represent the numbers of prediction models for which both evaluation metrics were computed and the number of RCE evaluations is equal to the displayed number of PRAUC results in the plots above.

Due to incompleteness and inequality of distribution of results over factors, we needed to compare factors only on equal intersections. This was indeed done, as will be discussed later in this chapter. It should be noted that the counts in the plots above represent counts of unique PRAUC evaluations. 

\subsubsection{Best Achieved Results}
\label{secBestAchievedResults}

Let us now look at the best results which were achieved in terms of both metrics. First, we would look at the best result for any combination of factors.

\begingroup 
\setlength\tabcolsep{2pt}
\footnotesize

\setlength\LTcapwidth{\textwidth} 

\setlength\LTleft{0pt}            
\setlength\LTright{0pt}           

  \begin{longtable}[ht]{@{\extracolsep{\fill}}llllllr}
\caption{Best PRAUC overall} \label{tabBestResultsPRAUC} \\
\toprule
target & algorithm & note & fs & trained & evaluated & PRAUC \\
\midrule
\endfirsthead
\caption[]{Best PRAUC overall} \\
\toprule
target & algorithm & note & fs & trained & evaluated & PRAUC \\
\midrule
\endhead
\midrule
\multicolumn{7}{r}{Continued on next page} \\
\midrule
\endfoot
\bottomrule
\endlastfoot
like & tree & scaled & top\_50 & train\_tweet\_sample\_10pct & train\_tweet\_sample\_10pct & 0.7969 \\
reply & GradientBoosting & scaled & top\_10 & train\_EWU\_sample\_1pct & train\_EWU\_sample\_1pct & 0.5142 \\
retweet & bayes & scaled & all & train\_EWU\_sample\_1pct & train\_EWU\_sample\_1pct & 0.5562 \\
quote & tree & oracle\_scaled & all & test\_EWU\_sample\_1pct & test\_EWU\_sample\_1pct & 0.5082 \\
react & tree & oracle\_scaled & all & train\_EWU\_sample\_1pct & train\_EWU\_sample\_1pct & 0.8090 \\
\end{longtable}
\endgroup

As we see in table \ref{tabBestResultsPRAUC}, the area below the precision-recall curve is rather modest, even for the best performing predictions. In general, the best performance was achieved when the model was trained and evaluated on the same dataset, often train. We cannot see a clear best performer in terms of classifier algorithms or regarding the inclusion of the ``oracle'' features. Feature selection does not seem to have helped, since almost all best-performing models were fit either on all features or 50 best features (which is a high proportion of all features). The sole exception is for reply, where gradient boosting trees performed the best, which achieved the best result when only ten features were selected. Thus, the interaction between algorithms and feature selection might deserve a closer inspection (cf. subsection \ref{secSignificanceTestingforIndividualFactors}). Now, let us also look into the best results and the corresponding parameters for the relative cross-entropy.

\begingroup 
\setlength\tabcolsep{2pt}
\footnotesize

\setlength\LTcapwidth{\textwidth} 

\setlength\LTleft{0pt}            
\setlength\LTright{0pt}           

  \begin{longtable}[ht]{@{\extracolsep{\fill}}llllllr}
\caption{Best RCE overall} \label{tabBestResultsRCE} \\
\toprule
target & algorithm & note & fs & trained & evaluated & RCE \\
\midrule
\endfirsthead
\caption[]{Best RCE overall} \\
\toprule
target & algorithm & note & fs & trained & evaluated & RCE \\
\midrule
\endhead
\midrule
\multicolumn{7}{r}{Continued on next page} \\
\midrule
\endfoot
\bottomrule
\endlastfoot
like & tree & scaled & top\_50 & train\_tweet\_sample\_10pct & train\_tweet\_sample\_10pct & -1022.9 \\
reply & tree & scaled & top\_5 & val\_EU\_sample\_1pct & val\_EU\_sample\_1pct & -634.7 \\
retweet & tree & scaled & all & train\_EU\_sample\_10pct & train\_EU\_sample\_10pct & -794.8 \\
quote & tree & scaled & all & test\_EWU\_sample\_1pct & test\_EWU\_sample\_1pct & -465.0 \\
react & tree & oracle\_scaled & all & val\_EWU\_sample\_1pct & val\_EWU\_sample\_1pct & -1133.4 \\
\end{longtable}
\endgroup

As we can see in table \ref{tabBestResultsRCE}, for RCE, the best-performing algorithm was always the tree algorithm. Like in the case of PRAUC, we again cannot see clearly the effect of the inclusion of the ``oracle'' features. Feature selection seems to again be of little use, apart from -- rather curiously -- in the case of reply interaction again. It might be possible that the utility of feature selection is thus related to the target more than the algorithm.

However, in table \ref{tabBestResultsPRAUC} and \ref{tabBestResultsRCE}, we looked for the highest evaluation among any computed combinations of parameters, which is not very realistic. Instead, tables \ref{tabBestResultstrained on a train subset and evaluated on full val+testPRAUC} and \ref{tabBestResultstrained on a train subset and evaluated on full val+testRCE} are more realistic and in line with the competition. They represent the best-achieved results for the full val+test dataset, i.e. the dataset used for the final evaluation by RecSys2020 participants as well as Gradinariu \cite{davidsThesis}. We also limited the subsets on which the models were fitted to the train dataset, again in line with the conditions of the challenge. In appendix \ref{appBestAchievedResults}, additional tables can be seen. These correspond to the best results on val+train subsets (rather than full datasets) both with and without exclusion of models which were not trained on train subsets. 

\begingroup 
\setlength\tabcolsep{2pt}
\footnotesize

\setlength\LTcapwidth{\textwidth} 

\setlength\LTleft{0pt}            
\setlength\LTright{0pt}           

\begin{longtable}[ht]{@{\extracolsep{\fill}}llllllr}
\caption{Best PRAUC achieved when trained on a train subset and evaluated on full val+test dataset} \label{tabBestResultstrained on a train subset and evaluated on full val+testPRAUC} \\
\toprule
target & algorithm & note & fs & trained & evaluated & PRAUC \\
\midrule
\endfirsthead
\caption[]{Best PRAUC achieved when trained on a train subset and evaluated on full val+test dataset} \\
\toprule
target & algorithm & note & fs & trained & evaluated & PRAUC \\
\midrule
\endhead
\midrule
\multicolumn{7}{r}{Continued on next page} \\
\midrule
\endfoot
\bottomrule
\endlastfoot
like & GradientBoosting & scaled & all & train\_EU\_sample\_5pct & val+test & 0.7273 \\
reply & forest & scaled & all & train\_EU\_sample\_10pct & val+test & 0.5129 \\
retweet & bayes & scaled & all & train\_EU\_sample\_1pct & val+test & 0.5497 \\
quote & bayes & scaled & top\_10 & train\_EU\_sample\_10pct & val+test & 0.5035 \\
react & GradientBoosting & scaled & all & train\_EU\_sample\_5pct & val+test & 0.7623 \\
\end{longtable}
\endgroup

As we see in tables \ref{tabBestResultstrained on a train subset and evaluated on full val+testPRAUC} and \ref{tabBestResultstrained on a train subset and evaluated on full val+testRCE}, with the exception of reply for RCE and retweet for PRAUC, all best-performing models were trained on either 10\% or 5\% subsets (i.e. the largest) subsets and have been sampled so that they preserve the ratio of the engaging user (or both the engaging user and the engaged-with user in case of react for RCE). There are no such clear patterns for the best classifier algorithm, with even na\"ive Bayes appearing in some cases. However, as noted before, while these results are certainly better than the baseline, they remain significantly worse than the competition-winning models that relied on neural networks. 

\begingroup 
\setlength\tabcolsep{2pt}
\footnotesize

\setlength\LTcapwidth{\textwidth} 

\setlength\LTleft{0pt}            
\setlength\LTright{0pt}           

\begin{longtable}[ht]{@{\extracolsep{\fill}}llllllr}
\caption{Best RCE achieved when trained on a train subset and evaluated on full val+test dataset} \label{tabBestResultstrained on a train subset and evaluated on full val+testRCE} \\
\toprule
target & algorithm & note & fs & trained & evaluated & RCE \\
\midrule
\endfirsthead
\caption[]{Best RCE achieved when trained on a train subset and evaluated on full val+test dataset} \\
\toprule
target & algorithm & note & fs & trained & evaluated & RCE \\
\midrule
\endhead
\midrule
\multicolumn{7}{r}{Continued on next page} \\
\midrule
\endfoot
\bottomrule
\endlastfoot
like & GradientBoosting & scaled & all & train\_EU\_sample\_5pct & val+test & -1418.2 \\
reply & GradientBoosting & scaled & all & train\_EU\_sample\_2pct & val+test & -643.5 \\
retweet & GradientBoosting & scaled & all & train\_EU\_sample\_5pct & val+test & -865.2 \\
quote & bayes & scaled & top\_10 & train\_EU\_sample\_10pct & val+test & -478.4 \\
react & forest & scaled & all & train\_EWU\_sample\_5pct & val+test & -1450.4 \\
\end{longtable}
\endgroup

\begin{figure}[htp]
\centering
\includegraphics[width=.45\textwidth]{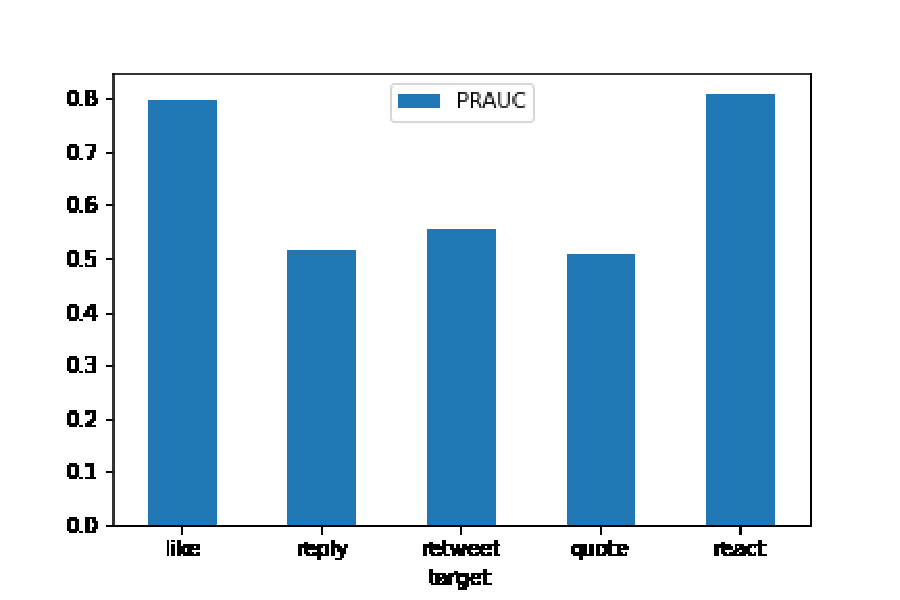} 
\caption{Bar plot indicating how the best overall PRAUC scores from table \ref{tabBestResultsPRAUC} differ from best scores achieved by models trained on train subsets and evaluated on full val+train datasets from table \ref{tabBestResultstrained on a train subset and evaluated on full val+testPRAUC}}
\label{figBestResultsPRAUC}
\end{figure}

It should also be noted that these results, which were obtained by training classifiers exclusively on train subsets and then evaluating them on full val+test datasets are indeed worse than the best overall results achieved. This can be seen more clearly for the area under the precision-recall curve in figure \ref{figBestResultsPRAUC}.

\begin{figure}[htp]
\centering
\includegraphics[width=.45\textwidth]{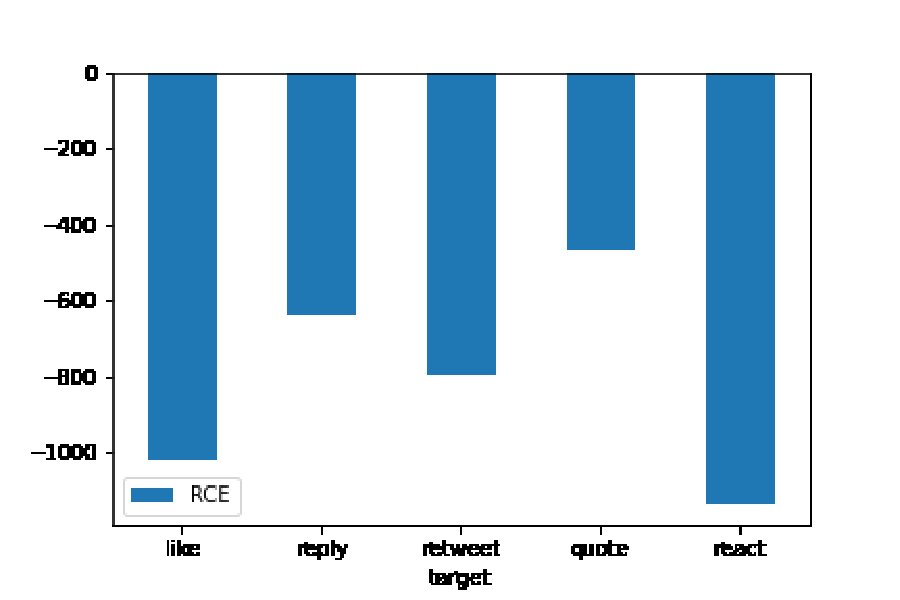} 
\caption{Bar plot indicating how the best overall RCE scores from table \ref{tabBestResultsRCE} differ from best scores achieved by models trained on train subsets and evaluated on full val+train datasets from table \ref{tabBestResultstrained on a train subset and evaluated on full val+testRCE}}
\label{figBestResultsRCE}
\end{figure}

Similarly, figure \ref{figBestResultsRCE} compares the best-achieved results for each engagement type overall to the best performance among the models trained on train subsets and evaluated on the full val+train dataset exclusively. Unfortunately, we see that neither of the results defeats the straw man model. This fact is further discussed in subsection \ref{secContrastingResultsWithOtherWorks}. 

Next, let us see how results differ between factors.

\subsubsection{Individual Factors' Statistics}
\label{secIndividualFactorsStatistics}

As mentioned previously, we due to incompleteness and inequality of distribution of results over factors, we needed to compare factors only on equal intersections. Specifically, we did this by grouping the results by the individual factors in question and then maintaining only those combinations of remaining factors that exist in all groups. For instance, when considering performance for different algorithms, our grouping factor would be the algorithms. We have seven different algorithms which whose hyperparametres were tuned and that are included in these final results: gradient boosting trees, na\"ive Bayes, random forests, logistic regression, support vector classifiers, and decision trees. To create the descriptive statistics below, we only looked at the intersection of remaining factors (in this case: whether the ``oracle'' features were included or not, what subset the data was trained on, what subset the data was evaluated on and what feature selection approach was taken). Thus we get a much smaller number of results than the total of sixty thousand PRAUC results reported above. Yet, comparing the results without finding this intersection would be biased by other factors. For instance, if na\"ive Bayes was trained on all subset sizes and svc only on 1\% datasets, we would expect the former to perform better even if it is not the better fit for the data at hand. Now, let us comment on the statistics for some of the factors. 

\begin{longtable}[ht]{llrrrrrr}
\caption{Statistics for PRAUC of common results for algorithm; there are 612 such PRAUC results in total, i.e. 102 instances per group.} \label{tabDescStatisticalgorithmPRAUC} \\
\toprule
 & algorithm & GradientBoosting & bayes & forest & lr & svc & tree \\
target & statistic &  &  &  &  &  &  \\
\midrule
\endfirsthead
\caption[]{Statistics for PRAUC of common results for algorithm; there are 612 such PRAUC results in total, i.e. 102 instances per group.} \\
\toprule
 & algorithm & GradientBoosting & bayes & forest & lr & svc & tree \\
target & statistic &  &  &  &  &  &  \\
\midrule
\endhead
\midrule
\multicolumn{8}{r}{Continued on next page} \\
\midrule
\endfoot
\bottomrule
\endlastfoot
\multirow[c]{4}{*}{like} & max & 0.7109 & 0.7100 & 0.7100 & 0.7110 & 0.7138 & \textbf{0.7151} \\
 & mean & 0.6194 & 0.6130 & \textbf{0.6613} & 0.6183 & 0.6090 & 0.6274 \\
 & min & 0.3798 & 0.2095 & \textbf{0.4060} & 0.3345 & 0.2964 & 0.2095 \\
 & std & 0.1144 & 0.1004 & 0.0789 & 0.0821 & 0.1021 & \textbf{0.1275} \\
\multirow[c]{4}{*}{quote} & max & 0.5033 & 0.5033 & 0.5033 & 0.5033 & 0.5033 & \textbf{0.5044} \\
 & mean & \textbf{0.5033} & 0.2307 & \textbf{0.5033} & \textbf{0.5033} & \textbf{0.5033} & 0.3407 \\
 & min & \textbf{0.5033} & 0.0884 & \textbf{0.5033} & \textbf{0.5033} & \textbf{0.5033} & 0.0140 \\
 & std & 0.0000 & 0.2361 & 0.0000 & 0.0000 & 0.0000 & \textbf{0.2829} \\
\multirow[c]{4}{*}{react} & max & \textbf{0.7544} & 0.7019 & 0.7324 & 0.7270 & 0.7252 & 0.7348 \\
 & mean & \textbf{0.7489} & 0.6958 & 0.7315 & 0.7202 & 0.7204 & 0.7306 \\
 & min & \textbf{0.7455} & 0.6858 & 0.7311 & 0.7147 & 0.7171 & 0.7284 \\
 & std & 0.0048 & \textbf{0.0088} & 0.0007 & 0.0063 & 0.0042 & 0.0036 \\
\multirow[c]{4}{*}{reply} & max & \textbf{0.5138} & 0.5131 & 0.5137 & 0.5137 & 0.5137 & 0.5137 \\
 & mean & 0.4228 & 0.2107 & \textbf{0.5131} & 0.4359 & \textbf{0.5131} & 0.4137 \\
 & min & 0.0129 & 0.0306 & \textbf{0.5129} & 0.0129 & \textbf{0.5129} & 0.0129 \\
 & std & 0.1848 & \textbf{0.2177} & 0.0003 & 0.1808 & 0.0003 & 0.2030 \\
\multirow[c]{4}{*}{retweet} & max & 0.4169 & 0.5494 & 0.4529 & \textbf{0.5553} & \textbf{0.5553} & 0.3774 \\
 & mean & 0.4017 & 0.2905 & 0.4009 & 0.3813 & \textbf{0.5518} & 0.3556 \\
 & min & 0.3717 & 0.1325 & 0.3595 & 0.2231 & \textbf{0.5494} & 0.3273 \\
 & std & 0.0184 & \textbf{0.1554} & 0.0389 & 0.1178 & 0.0032 & 0.0190 \\
\end{longtable}

From table \ref{tabDescStatisticalgorithmPRAUC}, we see that no algorithm is best across all engagement types. Moreover, there is no overlap between the best-performing algorithm for the best-achieved result (max), the worst-achieved result (min), and the mean result. In fact, while results generally beat the baseline, they remain rather unsatisfactory, especially relative to the results achieved at the competition. We will discuss this fact in section \ref{secContrastingResultsWithOtherWorks}. We see that results are particularly bad in the case of the quote interaction, which is also the sparsest of all engagement types, indicating what the likely reason might be.

\begin{longtable}[ht]{llrrrrrr}
\caption{Statistics for RCE of common results for algorithm; there are 612 such RCE results in total, i.e. 102 instances per group.} \label{tabDescStatisticalgorithmRCE} \\
\toprule
 & algorithm & GradientBoosting & bayes & forest & lr & svc & tree \\
target & statistic &  &  &  &  &  &  \\
\midrule
\endfirsthead
\caption[]{Statistics for RCE of common results for algorithm; there are 612 such RCE results in total, i.e. 102 instances per group.} \\
\toprule
 & algorithm & GradientBoosting & bayes & forest & lr & svc & tree \\
target & statistic &  &  &  &  &  &  \\
\midrule
\endhead
\midrule
\multicolumn{8}{r}{Continued on next page} \\
\midrule
\endfoot
\bottomrule
\endlastfoot
\multirow[c]{4}{*}{like} & max & -1483.9 & -1905.3 & -1545.0 & -1592.7 & -1650.9 & \textbf{-1427.3} \\
 & mean & -1733.2 & -2088.3 & \textbf{-1716.9} & -1927.3 & -1966.6 & -1769.9 \\
 & min & -2346.9 & -2345.1 & \textbf{-2043.9} & -2763.9 & -2197.8 & -2623.0 \\
 & std & 257.9 & 99.9 & 186.3 & \textbf{316.6} & 125.2 & 274.4 \\
\multirow[c]{4}{*}{quote} & max & -474.5 & -474.5 & -474.5 & -474.5 & -474.5 & \textbf{-473.2} \\
 & mean & \textbf{-474.5} & -5327.5 & \textbf{-474.5} & \textbf{-474.5} & \textbf{-474.5} & -512.4 \\
 & min & \textbf{-474.5} & -8752.5 & \textbf{-474.5} & \textbf{-474.5} & \textbf{-474.5} & -589.9 \\
 & std & 0.0 & \textbf{4319.8} & 0.0 & 0.0 & 0.0 & 67.1 \\
\multirow[c]{4}{*}{react} & max & \textbf{-1449.7} & -1979.7 & -1576.1 & -1610.2 & -1625.1 & -1560.6 \\
 & mean & \textbf{-1483.9} & -1991.4 & -1581.4 & -1646.9 & -1664.1 & -1580.4 \\
 & min & \textbf{-1506.0} & -2009.2 & -1588.7 & -1681.0 & -1704.1 & -1591.3 \\
 & std & 30.0 & 15.7 & 6.5 & 35.5 & \textbf{39.5} & 17.1 \\
\multirow[c]{4}{*}{reply} & max & \textbf{-643.7} & \textbf{-643.7} & \textbf{-643.7} & \textbf{-643.7} & \textbf{-643.7} & \textbf{-643.7} \\
 & mean & -849.7 & -1159.7 & \textbf{-646.5} & -646.7 & \textbf{-646.5} & -648.1 \\
 & min & -2491.6 & -5915.0 & \textbf{-653.8} & \textbf{-653.8} & \textbf{-653.8} & -664.2 \\
 & std & 493.9 & \textbf{1257.6} & 3.4 & 3.4 & 3.4 & 5.8 \\
\multirow[c]{4}{*}{retweet} & max & \textbf{-924.6} & -958.5 & -936.3 & -943.4 & -958.5 & -943.3 \\
 & mean & \textbf{-950.6} & -1809.0 & -961.0 & -982.0 & -974.5 & -964.3 \\
 & min & \textbf{-980.1} & -2522.3 & -985.6 & -1072.9 & -998.5 & -990.9 \\
 & std & 22.1 & \textbf{674.1} & 22.6 & 55.6 & 21.9 & 21.6 \\
\end{longtable}

In table \ref{tabDescStatisticalgorithmRCE} with results for RCE we see a similar pattern to the results in PRAUC from table \ref{tabDescStatisticalgorithmPRAUC} with very few exceptions. For this reason, we will continue with reporting only tables for RCE here and we would include the remaining PRAUC tables in appendix \ref{appIndividualFactorsStatistics}.

\begin{longtable}[ht]{llrrrrr}
\caption{Statistics for PRAUC of common results for feature\_selection; there are 37,845 such PRAUC results in total, i.e. 7,569 instances per group.} \label{tabDescStatisticfeatureselectionPRAUC} \\
\toprule
 & feature\_selection & all & top\_10 & top\_25 & top\_5 & top\_50 \\
target & statistic &  &  &  &  &  \\
\midrule
\endfirsthead
\caption[]{Statistics for PRAUC of common results for feature\_selection; there are 37,845 such PRAUC results in total, i.e. 7,569 instances per group.} \\
\toprule
 & feature\_selection & all & top\_10 & top\_25 & top\_5 & top\_50 \\
target & statistic &  &  &  &  &  \\
\midrule
\endhead
\midrule
\multicolumn{7}{r}{Continued on next page} \\
\midrule
\endfoot
\bottomrule
\endlastfoot
\multirow[c]{4}{*}{like} & max & 0.7681 & 0.7216 & 0.7334 & 0.7138 & \textbf{0.7969} \\
 & mean & \textbf{0.6541} & 0.5940 & 0.6283 & 0.5915 & 0.6168 \\
 & min & 0.3732 & 0.3190 & 0.2095 & \textbf{0.3798} & 0.2079 \\
 & std & 0.0646 & 0.0924 & 0.0963 & 0.0848 & \textbf{0.1073} \\
\multirow[c]{4}{*}{quote} & max & 0.5056 & 0.5042 & \textbf{0.5059} & 0.5041 & 0.5056 \\
 & mean & \textbf{0.4874} & 0.4604 & 0.4447 & 0.4766 & 0.4513 \\
 & min & 0.0032 & 0.0032 & 0.0032 & 0.0032 & \textbf{0.0034} \\
 & std & 0.0806 & 0.1343 & \textbf{0.1434} & 0.1108 & 0.1308 \\
\multirow[c]{4}{*}{react} & max & \textbf{0.8086} & 0.7761 & 0.7849 & 0.7440 & 0.7897 \\
 & mean & \textbf{0.7089} & 0.6755 & 0.6973 & 0.6873 & 0.6992 \\
 & min & \textbf{0.4721} & 0.3402 & 0.2402 & 0.4331 & 0.2383 \\
 & std & 0.0515 & \textbf{0.0823} & 0.0730 & 0.0649 & 0.0731 \\
\multirow[c]{4}{*}{reply} & max & \textbf{0.5142} & \textbf{0.5142} & \textbf{0.5142} & \textbf{0.5142} & \textbf{0.5142} \\
 & mean & \textbf{0.4571} & 0.3589 & 0.4185 & 0.3797 & 0.4273 \\
 & min & \textbf{0.0123} & \textbf{0.0123} & \textbf{0.0123} & \textbf{0.0123} & \textbf{0.0123} \\
 & std & 0.1445 & \textbf{0.2157} & 0.1750 & 0.2137 & 0.1670 \\
\multirow[c]{4}{*}{retweet} & max & \textbf{0.5562} & 0.5553 & 0.5525 & 0.5560 & 0.5525 \\
 & mean & \textbf{0.4110} & 0.2702 & 0.3616 & 0.2360 & 0.3271 \\
 & min & \textbf{0.0495} & 0.0490 & 0.0489 & 0.0485 & 0.0485 \\
 & std & 0.1464 & 0.1846 & \textbf{0.1889} & 0.1702 & 0.1717 \\
\end{longtable}

Regarding the results for feature selection shown in table \ref{tabDescStatisticfeatureselectionPRAUC}, we see that using no feature selection resulted in the highest mean performance for all engagement types. The min and max results are not as straightforward to generalise. There are two things to note here, however. On the one hand, the utility of feature selection is not exclusively to improve the classifier's prediction accuracy (which it could achieve by removing columns with low information gain or high noise). Feature selection is also employed to lower the computational intensity. This effect of feature selection was not investigated in detail in this thesis, and is one of the potential future works, as elaborated in section \ref{secFutureWork}. On the other hand, feature selection is more useful for some types of classification algorithms than others. For instance, as we have discussed in section \ref{secClassification}, decision trees by design select the most informative features and thus have a feature selection integrated into their mode of operations. So the interaction of the algorithm and feature selection should be investigated further.

\begin{longtable}[ht]{llrr}
\caption{Statistics for PRAUC of common results for note; there are 5182 such PRAUC results in total, i.e. 2591 instances per group.} \label{tabDescStatisticnotePRAUC} \\
\toprule
 & note & oracle\_scaled & scaled \\
target & statistic &  &  \\
\midrule
\endfirsthead
\caption[]{Statistics for PRAUC of common results for note; there are 5,182 such PRAUC results in total, i.e. 2,591 instances per group.} \\
\toprule
 & note & oracle\_scaled & scaled \\
target & statistic &  &  \\
\midrule
\endhead
\midrule
\multicolumn{4}{r}{Continued on next page} \\
\midrule
\endfoot
\bottomrule
\endlastfoot
\multirow[c]{4}{*}{like} & max & \textbf{0.7684} & 0.7681 \\
 & mean & \textbf{0.6657} & 0.6556 \\
 & min & 0.3234 & \textbf{0.4592} \\
 & std & \textbf{0.0766} & 0.0690 \\
\multirow[c]{4}{*}{quote} & max & \textbf{0.5082} & 0.5051 \\
 & mean & \textbf{0.4992} & 0.4980 \\
 & min & \textbf{0.0035} & 0.0033 \\
 & std & 0.0360 & \textbf{0.0481} \\
\multirow[c]{4}{*}{react} & max & \textbf{0.8090} & 0.8086 \\
 & mean & \textbf{0.7209} & 0.7177 \\
 & min & \textbf{0.5560} & 0.5491 \\
 & std & 0.0395 & \textbf{0.0439} \\
\multirow[c]{4}{*}{reply} & max & \textbf{0.5138} & \textbf{0.5138} \\
 & mean & 0.4917 & \textbf{0.4995} \\
 & min & 0.0123 & \textbf{0.0125} \\
 & std & \textbf{0.0942} & 0.0743 \\
\multirow[c]{4}{*}{retweet} & max & \textbf{0.5560} & \textbf{0.5560} \\
 & mean & 0.4584 & \textbf{0.4935} \\
 & min & 0.0485 & \textbf{0.0860} \\
 & std & \textbf{0.1270} & 0.1046 \\
\end{longtable}

Table \ref{tabDescStatisticnotePRAUC} compares the performance achieved by classifiers trained on features that include ``oracle'' information on events happening after the tweet was seen (cf. subsection \ref{secOracleFeatures}) named \emph{oracle\_scaled} as well as feature sets that do not include this information, named \emph{scaled}. This naming indicates that features in both cases were scaled before prediction. Investigating the role this scaling has on predictions was, unfortunately, not investigated due to time constraints. Between the two options for the factor (dubbed ``note'' in the implemented programs), it is quite clear that ``oracle'' features have contributed to better prediction results in virtually all cases. However, it must be noted the achieved results remain modest, as the improvements are sometimes limited to the third digit behind the decimal point. This necessitates that we calculate test statistics to see if these differences between factors are significantly different. 

\section{Statistical Significance Testing}
\label{secStat}

We wanted to check whether the means of individual factors from the previous subsection indeed differ significantly. Moreover, we wanted to investigate not only the difference \emph{within} individual factor groups but also the interactions \emph{between} groups of factors. The usual manner to achieve this, as described in e.g. \cite{learningStatisticsWithR}, would be to use factorial ANOVA with Turkey's HSD posthoc test if the conditions of spherasticity, normality, and homoscedasticity were satisfied and Kruskal-Wallis test followed by Wilcoxon pairwise tests otherwise.

However, as was argued in subsection \ref{secEvaluation}, we believed that evaluating on already sampled subsets would be a better approach than re-sampling from already existing subsets. Thus, we could use the evaluated-on subsets as the \emph{subjects}. Each factor would then be considered a different treatment, and the PRAUC and RCE evaluation a repeated measurement. In designing this paired approach, we relied on the pingouin package, which provides user-friendly alternatives or wrappers for statistical tests implemented in scipy and statsmodels \cite{pingouinDocu, learningStatisticsWithPython}. Since preliminary examination of data showed that conditions for using repeated-measurement ANOVA are not satisfied, we had to instead use the Friedman nonparametric test, followed by Wilcoxon pairwise tests. The results of these significance tests can be seen in subsection \ref{secSignificanceTestingforIndividualFactors}. 

It should be noted that running multiple significance tests can result in positive results due to chance alone \cite[pp. 443--446]{learningStatisticsWithR}. Therefore, whenever running multiple tests, as we do here, the p-value has to be adjusted to account for the false positive rate (Type I-error).  Following the suggested steps from \cite[part V.15]{learningStatisticsWithPython}, we used the Hohn correction \cite{holm1979correction} procedure. This procedure can be seen as sequential Bonferroni correction from \cite{dunn1961Bonferronicorrection}, with the main difference being that not all p-values are multiplied by the total number of performed tests \cite[pp. 444--445]{learningStatisticsWithR}. Instead, the tests are observed as if they are done sequentially, and the corrected p-value for the j-th largest uncorrected p-value ($p_j^{uncorr}$ is defined as

$$ p_j^{corr} = \max (j \cdot p_j^{uncorr},\ p_{j-1}^{corr}) $$

where $p_{j-1}^{corr}$ is the correct p-value for the previous smallest uncorrected p-value \cite[pp. 444--445]{learningStatisticsWithR}. As shown in \cite[pp. 445--446]{learningStatisticsWithR} in reference to \cite{holm1979correction}, the Holm correction has higher power (i.e. lower Type II error rate) than the Bonferroni correction, but the same Type I error rate, which is why we employed this correction. In doing so, we mostly relied on the method \texttt{multicomp} for p-adjustment for multiple comparisons and posthoc tests from the Pingouin package with argument \texttt{method} set to {"holm"} \cite{pingouinDocu}.

A downside of this approach was that the effects of interactions between groups of factors could not be investigated this way, necessitating the need for further tests \cite{pingouinDocu}, which is one of the potential directions for future work, as elaborated in subsection \ref{secFutureWork}.


\subsection{Finding Common Factor Combinations}
\label{secFindingCommonFactorCombinations}

As was indicated in subsection \ref{secDistributionOfTheResultsOverFactors}, there are 84 datasets (cf. appendix \ref{appSubsets}, each of which could theoretically be seen as both a base for training and testing), two feature sets (with and without oracle features), seven classification algorithms, five feature selection approaches (all features plus prior $\chi^2$-selected features with four limits), five target variables (like, reply, retweet, quote, and react), and 2 evaluation metrics. Thus, if were to exhaustively evaluate all of these combinations of factors, we would have to fit and evaluate  $84 \cdot 84 \cdot \ 2 \cdot 7 \cdot 5 \cdot 5 \cdot 5 \cdot 2 = 24\ 696\ 000$ models. Thus the first issue we had to resolve before running statistical tests significantly is this inherent incompleteness of results. The second issue -- partially as a consequence of the fact that the LBD cluster closed before the evaluation was complete and partially a result of the time constraints as well as the implementation of the ordering of evaluations in the machine learning pipeline -- is that the distribution of results across factors is also unbalanced. 

The approach we used to resolve these issues was to find common combinations of other factors for each value for the target factor factors. For instance, a dataframe has three columns with factors $A,\ B,$ and $C$ a single column with evaluations $PRAUC$. Let the values be $A=[tree, tree, tree, lr, lr],\ B=[1,2,3,1,2],\ C=[a,b,c,a,d],$ and $D=[0.1, 0.2, 0.3, 0.01, 0.05]$. The function we implemented
would then find three sets of intersections since there are three columns with factors. Specifically for factor $A$ the intersection would be $A=[tree, lr],\ B=[1,1],\ C=[a,a], D=[0.1, 0.05]$ because just for those two rows the both values in $A$ ($tree$ and $lr$) have corresponding matches in the within columns $B$ and $C$ ($1$ and $a$, respectively). In an analogous manner, common combinations of $A$ and $C$ would be found for values in $B$ as well as for $A$ and $B$ for factor $C$. In this case, both for $B$ and $C$, the resulting dataframe subset would be empty, as the intersection for the corresponding two cases is an empty set. The notebook \texttt{Statistics\--Experiment\-al\-Code\--Balancing\-Functions} contains a bit more complex illustrative example showcasing how the method transforms the data and finds the common combinations.

Naturally, excluding non-common combinations of remaining factors reduces  the size of evaluations to be used for statistical testing was decreased significantly. In fact, we omitted results for logistic regression and support vector classifier, because they reduced the size of common factors from several thousand per engagement type to several dozen per engagement type. Moreover, for statistical testing, we limited results to those based on models that were trained on train subsets and evaluated on subsets based on val, train, or val+train datasets. Through this exclusion, we were left with a total of $35790$ PRAUC and as many RCE results. We then found common combinations for each of the factors.

The five factors we systematically investigated for statistical significance are:

\begin{itemize}
    \item \texttt{algorithm} with values \texttt{[GradientBoosting, bayes, forest, tree]} corresponding to the four of the classification algorithm used for the generation of predictions;
    \item \texttt{note} with values \texttt{[oracle\_scaled, scaled]} corresponding to feature sets with and without ``oracle'' features, as discussed in subsection \ref{secMachineLearningPipelineDesign};
    \item \texttt{feature\_selection} with values \texttt{[all, top\_10, top\_25, top\_5, top\_50]}, corresponding to the four feature set created by a $\chi^2$ selector, as described in subsection \ref{secFeatureSelectionImplementation};
    \item \texttt{to\_technique} and \texttt{to\_percent}, where the prefix \texttt{to} stand for ``trained on'' and which corresponds to the sampling technique and size, as explained in subsection \ref{secSampling}.
\end{itemize}

\begin{figure}[htp]
\noindent\makebox[\textwidth]{\includegraphics[width=0.6\textwidth]{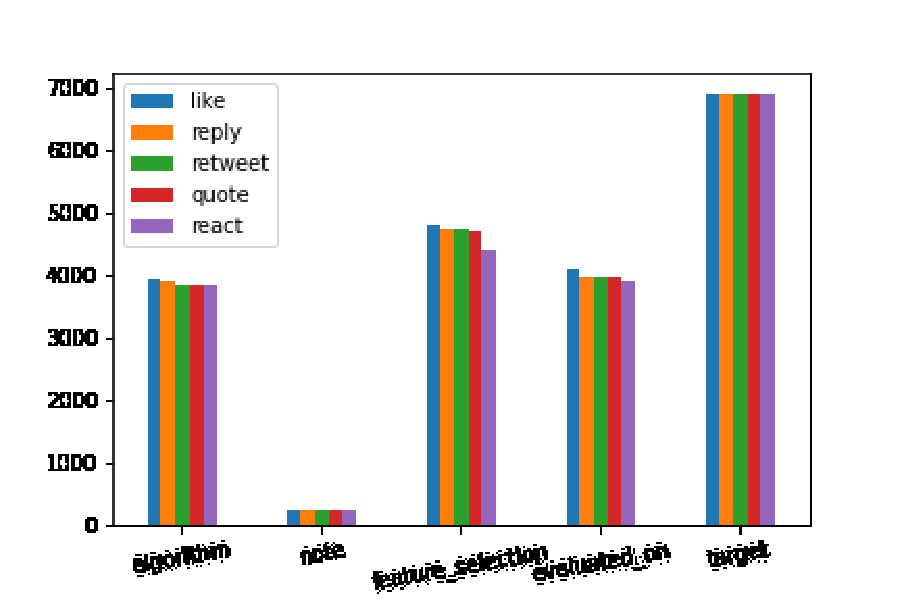}}
\caption{Number of combinations of other factors for each element of the target factor (horizontal axis) and the engagement type (colour)}
\label{figCommongFacotrs}
\end{figure}

Due to the aforementioned incompleteness and imbalance of results distributions, the combinations of factors in common differ between target factors and individual engagements. Figure \ref{figCommongFacotrs} showcases how many combinations were found. As we can see, the number of common factors corresponding to the factor \texttt{note} is smaller than the rest, which is not surprising given that the featuresets that include ``oracle'' features were evaluated locally (cf. subsection \ref{secDistributionOfTheResultsOverFactors}).  

Moreover, as we can read in figure \ref{figCommongFacotrs}, there are no common combinations for all values in \texttt{to\_technique} and \texttt{to\_percent}. Thus, for those two factors, we actually took the common combinations for \texttt{evaluated\_on}. Since this is the \emph{subject} column, performing Friedman tests is still possible, since even though the total number of each \texttt{to\_technique} and \texttt{to\_percent} is unequal, it is balanced across. This difference and its effect on the test statistic was examined in notebook \texttt{Sta\-tistics\--03\--Preliminary\-Significance\-Tests}.

\subsection{Significance Testing for Individual Factors}
\label{secSignificanceTestingforIndividualFactors}

Since we have five factors (as in the bullet list in subsection \ref{secFindingCommonFactorCombinations}) with five target engagement types, and two evaluation metrics (PRAUC and RCE), we ran a total of $5 \cdot 5 \cdot 2 = 50$ Friedman tests. Let us first look at the test results which indicate that the null hypothesis (that the means are equal within all values or groups in the factor) can be rejected at the $\alpha = 0.05$ significance level. These results can be seen in table \ref{tabFriedmanPos} below.

\begin{longtable}[ht]{llllrrrr}
\caption{Results of the Friedman tests for individual facotrs where \texttt{p-corr}$<0.05$} \label{tabFriedmanPos} \\
\toprule
metric & target & within & balanced on & W & ddof1 & Q & p-corr \\
\midrule
\endfirsthead
\caption[]{Results of the Friedman tests for individual facotrs where \texttt{p-corr}$<0.05$} \\
\toprule
metric & target & within & balanced on & W & ddof1 & Q & p-corr \\
\midrule
\endhead
\midrule
\multicolumn{8}{r}{Continued on next page} \\
\midrule
\endfoot
\bottomrule
\endlastfoot
PRAUC & like & algorithm & algorithm & 0.5509 & 3 & 104.1238 & 0.0000 \\
RCE & like & algorithm & algorithm & 0.5509 & 3 & 104.1238 & 0.0000 \\
PRAUC & reply & algorithm & algorithm & 0.7978 & 3 & 150.7905 & 0.0000 \\
RCE & reply & algorithm & algorithm & 0.7978 & 3 & 150.7905 & 0.0000 \\
PRAUC & retweet & algorithm & algorithm & 0.4556 & 3 & 86.1048 & 0.0000 \\
RCE & retweet & algorithm & algorithm & 0.4556 & 3 & 86.1048 & 0.0000 \\
PRAUC & quote & algorithm & algorithm & 0.5275 & 3 & 99.6881 & 0.0000 \\
RCE & quote & algorithm & algorithm & 0.5275 & 3 & 99.6881 & 0.0000 \\
PRAUC & react & algorithm & algorithm & 0.5281 & 3 & 99.8190 & 0.0000 \\
RCE & react & algorithm & algorithm & 0.5281 & 3 & 99.8190 & 0.0000 \\
PRAUC & like & note & note & 1.0000 & 1 & 15.0000 & 0.0013 \\
RCE & like & note & note & 1.0000 & 1 & 15.0000 & 0.0013 \\
PRAUC & like & feature\_selection & feature\_selection & 0.6309 & 4 & 158.9841 & 0.0000 \\
RCE & like & feature\_selection & feature\_selection & 0.6309 & 4 & 158.9841 & 0.0000 \\
PRAUC & reply & feature\_selection & feature\_selection & 0.6043 & 4 & 152.2921 & 0.0000 \\
RCE & reply & feature\_selection & feature\_selection & 0.6043 & 4 & 152.2921 & 0.0000 \\
PRAUC & retweet & feature\_selection & feature\_selection & 0.7177 & 4 & 180.8635 & 0.0000 \\
RCE & retweet & feature\_selection & feature\_selection & 0.7177 & 4 & 180.8635 & 0.0000 \\
PRAUC & quote & feature\_selection & feature\_selection & 0.1591 & 4 & 40.0879 & 0.0000 \\
RCE & quote & feature\_selection & feature\_selection & 0.1591 & 4 & 40.0879 & 0.0000 \\
PRAUC & react & feature\_selection & feature\_selection & 0.4398 & 4 & 110.8317 & 0.0000 \\
RCE & react & feature\_selection & feature\_selection & 0.4398 & 4 & 110.8317 & 0.0000 \\
PRAUC & like & to\_technique & evaluated\_on & 0.6533 & 2 & 82.3175 & 0.0000 \\
RCE & like & to\_technique & evaluated\_on & 0.6533 & 2 & 82.3175 & 0.0000 \\
PRAUC & retweet & to\_technique & evaluated\_on & 0.6621 & 2 & 83.4286 & 0.0000 \\
RCE & retweet & to\_technique & evaluated\_on & 0.6621 & 2 & 83.4286 & 0.0000 \\
PRAUC & quote & to\_technique & evaluated\_on & 0.2475 & 2 & 31.1878 & 0.0000 \\
RCE & quote & to\_technique & evaluated\_on & 0.2475 & 2 & 31.1878 & 0.0000 \\
PRAUC & react & to\_technique & evaluated\_on & 0.5359 & 2 & 67.5238 & 0.0000 \\
RCE & react & to\_technique & evaluated\_on & 0.5359 & 2 & 67.5238 & 0.0000 \\
PRAUC & like & to\_percent & evaluated\_on & 0.5166 & 3 & 97.6286 & 0.0000 \\
RCE & like & to\_percent & evaluated\_on & 0.5166 & 3 & 97.6286 & 0.0000 \\
PRAUC & reply & to\_percent & evaluated\_on & 0.6512 & 3 & 123.0762 & 0.0000 \\
RCE & reply & to\_percent & evaluated\_on & 0.6512 & 3 & 123.0762 & 0.0000 \\
PRAUC & retweet & to\_percent & evaluated\_on & 0.7044 & 3 & 133.1333 & 0.0000 \\
RCE & retweet & to\_percent & evaluated\_on & 0.7044 & 3 & 133.1333 & 0.0000 \\
PRAUC & quote & to\_percent & evaluated\_on & 0.5506 & 3 & 104.0684 & 0.0000 \\
RCE & quote & to\_percent & evaluated\_on & 0.5506 & 3 & 104.0684 & 0.0000 \\
PRAUC & react & to\_percent & evaluated\_on & 0.5102 & 3 & 96.4286 & 0.0000 \\
RCE & react & to\_percent & evaluated\_on & 0.5102 & 3 & 96.4286 & 0.0000 \\
\end{longtable}

We see in table \ref{tabFriedmanPos} that 40 of the 50 tests have multi-test adjusted p-value at less than the significance threshold of $0.05$. In fact, with the exception of the factor \texttt{note} which indicates whether the ``oracle'' features are included for the engagement type like, all p-values are smaller than $10^{-5}$. In contrast, let us look at factors where the null hypothesis could not be rejected in table ref \ref{tabFriedmanNeg}.

\begin{longtable}[ht]{llllrrrr}
\caption{Results of the Friedman tests for individual facotrs where \texttt{p-corr}$>0.05$} \label{tabFriedmanNeg} \\
\toprule
metric & target & within & balanced on & W & ddof1 & Q & p-corr \\
\midrule
\endfirsthead
\caption[]{Results of the Friedman tests for individual facotrs where \texttt{p-corr}$>0.05$} \\
\toprule
metric & target & within & balanced on & W & ddof1 & Q & p-corr \\
\midrule
\endhead
\midrule
\multicolumn{8}{r}{Continued on next page} \\
\midrule
\endfoot
\bottomrule
\endlastfoot
PRAUC & reply & note & note & 0.0051 & 1 & 0.0769 & 1.0000 \\
RCE & reply & note & note & 0.0051 & 1 & 0.0769 & 1.0000 \\
PRAUC & retweet & note & note & 0.3600 & 1 & 5.4000 & 0.2014 \\
RCE & retweet & note & note & 0.3600 & 1 & 5.4000 & 0.2014 \\
PRAUC & quote & note & note & 0.0667 & 1 & 1.0000 & 1.0000 \\
RCE & quote & note & note & 0.0667 & 1 & 1.0000 & 1.0000 \\
PRAUC & react & note & note & 0.3600 & 1 & 5.4000 & 0.2014 \\
RCE & react & note & note & 0.3600 & 1 & 5.4000 & 0.2014 \\
PRAUC & reply & to\_technique & evaluated\_on & 0.0199 & 2 & 2.5079 & 1.0000 \\
RCE & reply & to\_technique & evaluated\_on & 0.0199 & 2 & 2.5079 & 1.0000 \\
\end{longtable}

From table \ref{tabFriedmanNeg}, we can read that only the factors \texttt{note} and \texttt{to\_technique} (which indicates the sampling technique based on the training subset) are rejected. The fact that for some target engagements, the zero hypotheses could be rejected with a high degree of confidence and in other targets of the same factors the adjusted p-value is close to $1$ indicates that further exploration of the effects is necessary. This is because it might be the case that the very high number of instances led to a Type I-error. Moreover, it might also be that the utilised algorithms are simply too simple to create meaningful predictions for engagement types different from like and react. We are going to elaborate on this hypothesis further in subsection \ref{secContrastingResultsWithOtherWorks}. 

Thereafter, we proceeded with the posthoc test for all factors where the null hypothesis in Friedman could be rejected for the significance threshold of $0.05$. We used the non-parametric pairwise Wilcoxon signed-rank test with the Holm p-value adjustment from the pingouin package \cite{pingouinDocu}. In all cases, we used the two-sided alternative for simplicity. The results can be seen in appendix \ref{appPostHocTestStatistics}. As we can see, most pairs of values were found to be significant, likely again at least in part due to the great number of instances. Pairs for which the null hypothesis could not be rejected include some trained-on dataset sampling techniques (cf., for example, tables \ref{tabPostHoc-Friedman-RCE-quote-to_technique-evaluated_on-evaluated_on} and \ref{tabPostHoc-Friedman-PRAUC-react-to_technique-evaluated_on-evaluated_on}) as well as some of the combinations of feature selections (e.g. tables \ref{tabPostHoc-Friedman-RCE-retweet-feature_selection-evaluated_on-feature_selection} and \ref{tabPostHoc-Friedman-PRAUC-react-feature_selection-evaluated_on-feature_selection}). 

\section{Most Informative Contextual Features}
\label{secMostSignificantContextualFeatures}

As elaborated in subsection \ref{secChiSquaredFeatureSelection}, $\chi^2$ selector was used to choose the most informative features for each of the target engagement types and for each train subset as well as the full dataset. In table \ref{tabOverview of the 5 most informative feature as selected by ChiSq-selector on the train dataset and its 20 subsets} we can see which columns were selected for non-oracle values.

\begin{longtable}[ht]{llr}
\caption{Overview of the 5 most informative feature as selected by ChiSq-selector on the train dataset and its 20 subsets} \label{tabOverview of the 5 most informative feature as selected by ChiSq-selector on the train dataset and its 20 subsets} \\
\toprule
 &  & count \\
target & selected feature &  \\
\midrule
\endfirsthead
\caption[]{Overview of the 5 most informative feature as selected by ChiSq-selector on the train dataset and its 20 subsets} \\
\toprule
 &  & count \\
target & selected feature &  \\
\midrule
\endhead
\midrule
\multicolumn{3}{r}{Continued on next page} \\
\midrule
\endfoot
\bottomrule
\endlastfoot
\hline \multirow[c]{6}{*}{like} & language\_indexed & 21 \\
 & tweet\_type\_indexed & 21 \\
 & user\_domains\_frequency\_12h\_binned & 21 \\
 & ratio\_all\_to\_engaged\_with\_count\_negative\_tweets\_reply\_binned & 21 \\
 & ratio\_all\_to\_links\_user\_proxy\_count\_negative\_tweets\_quote\_binned & 17 \\
 & user\_hashtags\_frequency\_1h\_binned & 4 \\
\hline \multirow[c]{6}{*}{reply} & language\_indexed & 21 \\
 & tweet\_type\_indexed & 21 \\
 & user\_domains\_frequency\_12h\_binned & 21 \\
 & ratio\_all\_to\_engaged\_with\_count\_negative\_tweets\_reply\_binned & 21 \\
 & ratio\_all\_to\_links\_user\_proxy\_count\_negative\_tweets\_quote\_binned & 14 \\
 & user\_hashtags\_frequency\_1h\_binned & 7 \\
\hline \multirow[c]{6}{*}{retweet} & language\_indexed & 21 \\
 & tweet\_type\_indexed & 21 \\
 & user\_domains\_frequency\_12h\_binned & 21 \\
 & ratio\_all\_to\_engaged\_with\_count\_negative\_tweets\_reply\_binned & 21 \\
 & ratio\_all\_to\_links\_user\_proxy\_count\_negative\_tweets\_quote\_binned & 17 \\
 & user\_hashtags\_frequency\_1h\_binned & 4 \\
\hline \multirow[c]{18}{*}{quote} & language\_indexed & 21 \\
 & tweet\_type\_indexed & 18 \\
 & ratio\_all\_to\_links\_user\_proxy\_count\_negative\_tweets\_quote\_binned & 15 \\
 & ratio\_all\_to\_engaged\_with\_count\_negative\_tweets\_reply\_binned & 15 \\
 & user\_domains\_frequency\_12h\_binned & 10 \\
 & user\_hashtags\_frequency\_1h\_binned & 5 \\
 & engaging\_user\_follower\_count\_binned & 4 \\
 & ratio\_all\_to\_hashtags\_user\_proxy\_count\_positive\_tweets\_react\_binned & 3 \\
 & ratio\_all\_to\_hashtags\_user\_proxy\_count\_positive\_tweets\_quote\_binned & 3 \\
 & ratio\_all\_to\_engaging\_count\_positive\_tweets\_retweet\_binned & 2 \\
 & ratio\_all\_to\_engaged\_with\_count\_positive\_tweets\_retweet\_binned & 2 \\
 & ratio\_all\_to\_engaged\_with\_count\_positive\_tweets\_react\_binned & 1 \\
 & domains\_frequency\_24h\_binned & 1 \\
 & ratio\_all\_to\_engaging\_count\_positive\_tweets\_reply\_binned & 1 \\
 & ratio\_all\_to\_domains\_user\_proxy\_count\_negative\_tweets\_retweet\_binned & 1 \\
 & engageds\_tweets\_views\_count\_12h\_binned & 1 \\
 & domains\_frequency\_48h\_binned & 1 \\
 & ratio\_all\_to\_engaging\_count\_negative\_tweets\_quote\_binned & 1 \\
\hline \multirow[c]{6}{*}{react} & language\_indexed & 21 \\
 & tweet\_type\_indexed & 21 \\
 & user\_domains\_frequency\_12h\_binned & 21 \\
 & ratio\_all\_to\_engaged\_with\_count\_negative\_tweets\_reply\_binned & 21 \\
 & ratio\_all\_to\_links\_user\_proxy\_count\_negative\_tweets\_quote\_binned & 17 \\
 & user\_hashtags\_frequency\_1h\_binned & 4 \\
\end{longtable}

We see in table \ref{tabOverview of the 5 most informative feature as selected by ChiSq-selector on the train dataset and its 20 subsets} that the two features selected most often are actually pre-provided and indicate the target tweet's language and type (whether the target tweet is ``top-level'' tweet, a response, a quote, or a retweeted tweet). Other than that, various time-window-based features can be seen, and in for both their personalised (with the prefix \emph{user\_}) and non-personalised variations (cf. subsection \ref{secShortTimePopularityEstimates}). We also see quite a few features beginning with the prefix \emph{ratio\_all\_to}, which indicate the relative popularity of a tweet element or activity or popularity of a user, as elaborated in subsection \ref{secTweetElementsEngagementHistory}. It is quite intriguing to see that the most informative features for predicting like including past negative instances with reply and quote can be found for each engagement type, not just the two. This might be indicative of the fact that the engaging users for whom these values are not zero (of which there are few since those are the scarcest of the engagement types) are most likely to engage. This may also be indicative of a deeper connection that may exist between engagement types, which was examined by some teams in the RecSys challenge. Specifically, authors of \cite{[CP4], [CP6]} created a two-stage model which enables preliminary predictions for all four engagement types to be used for the final prediction for each engagement type. Let us now look at selected features for \texttt{top\_10} threshold.

\begin{longtable}[ht]{llr}
\caption{Overview of the 10 most informative feature as selected by ChiSq-selector on the train dataset and its 20 subsets} \label{tabOverview of the 10 most informative feature as selected by ChiSq-selector on the train dataset and its 20 subsets} \\
\toprule
 &  & count \\
target & selected feature &  \\
\midrule
\endfirsthead
\caption[]{Overview of the 10 most informative feature as selected by ChiSq-selector on the train dataset and its 20 subsets} \\
\toprule
 &  & count \\
target & selected feature &  \\
\midrule
\endhead
\midrule
\multicolumn{3}{r}{Continued on next page} \\
\midrule
\endfoot
\bottomrule
\endlastfoot
\hline \multirow[c]{11}{*}{like} & language\_indexed & 21 \\
 & tweet\_type\_indexed & 21 \\
 & user\_domains\_frequency\_12h\_binned & 21 \\
 & ratio\_all\_to\_engaged\_with\_count\_negative\_tweets\_reply\_binned & 21 \\
 & user\_hashtags\_frequency\_1h\_binned & 21 \\
 & hashtags\_frequency\_1h\_binned & 21 \\
 & domains\_frequency\_48h\_binned & 21 \\
 & hashtags\_frequency\_12h\_binned & 21 \\
 & ratio\_all\_to\_hashtags\_user\_proxy\_count\_positive\_tweets\_react\_binned & 21 \\
 & ratio\_all\_to\_links\_user\_proxy\_count\_negative\_tweets\_quote\_binned & 17 \\
 & user\_links\_frequency\_2h\_binned & 4 \\
\hline \multirow[c]{11}{*}{reply} & language\_indexed & 21 \\
 & tweet\_type\_indexed & 21 \\
 & user\_domains\_frequency\_12h\_binned & 21 \\
 & ratio\_all\_to\_engaged\_with\_count\_negative\_tweets\_reply\_binned & 21 \\
 & user\_hashtags\_frequency\_1h\_binned & 21 \\
 & hashtags\_frequency\_1h\_binned & 21 \\
 & domains\_frequency\_48h\_binned & 21 \\
 & hashtags\_frequency\_12h\_binned & 21 \\
 & ratio\_all\_to\_hashtags\_user\_proxy\_count\_positive\_tweets\_react\_binned & 21 \\
 & ratio\_all\_to\_links\_user\_proxy\_count\_negative\_tweets\_quote\_binned & 14 \\
 & user\_links\_frequency\_2h\_binned & 7 \\
\hline \multirow[c]{11}{*}{retweet} & language\_indexed & 21 \\
 & tweet\_type\_indexed & 21 \\
 & user\_domains\_frequency\_12h\_binned & 21 \\
 & ratio\_all\_to\_engaged\_with\_count\_negative\_tweets\_reply\_binned & 21 \\
 & user\_hashtags\_frequency\_1h\_binned & 21 \\
 & hashtags\_frequency\_1h\_binned & 21 \\
 & domains\_frequency\_48h\_binned & 21 \\
 & hashtags\_frequency\_12h\_binned & 21 \\
 & ratio\_all\_to\_hashtags\_user\_proxy\_count\_positive\_tweets\_react\_binned & 21 \\
 & ratio\_all\_to\_links\_user\_proxy\_count\_negative\_tweets\_quote\_binned & 17 \\
 & user\_links\_frequency\_2h\_binned & 4 \\
\hline \multirow[c]{36}{*}{quote} & language\_indexed & 21 \\
 & tweet\_type\_indexed & 18 \\
 & ratio\_all\_to\_hashtags\_user\_proxy\_count\_positive\_tweets\_react\_binned & 16 \\
 & ratio\_all\_to\_links\_user\_proxy\_count\_negative\_tweets\_quote\_binned & 15 \\
 & ratio\_all\_to\_engaged\_with\_count\_negative\_tweets\_reply\_binned & 15 \\
 & user\_hashtags\_frequency\_1h\_binned & 14 \\
 & hashtags\_frequency\_1h\_binned & 13 \\
 & hashtags\_frequency\_12h\_binned & 13 \\
 & domains\_frequency\_48h\_binned & 12 \\
 & user\_domains\_frequency\_12h\_binned & 10 \\
 & engaging\_user\_follower\_count\_binned & 7 \\
 & ratio\_all\_to\_hashtags\_user\_proxy\_count\_positive\_tweets\_quote\_binned & 6 \\
 & ratio\_all\_to\_engaged\_with\_count\_positive\_tweets\_react\_binned & 5 \\
 & ratio\_all\_to\_engaging\_count\_positive\_tweets\_retweet\_binned & 5 \\
 & ratio\_all\_to\_domains\_user\_proxy\_count\_negative\_tweets\_retweet\_binned & 4 \\
 & links\_frequency\_48h\_binned & 4 \\
 & ratio\_all\_to\_domains\_user\_proxy\_count\_negative\_tweets\_quote\_binned & 3 \\
 & ratio\_all\_to\_engaged\_with\_count\_positive\_tweets\_retweet\_binned & 3 \\
 & ratio\_all\_to\_engaging\_count\_negative\_tweets\_quote\_binned & 3 \\
 & ratio\_all\_to\_links\_user\_proxy\_count\_positive\_tweets\_react\_binned & 2 \\
 & ratio\_all\_to\_domains\_user\_proxy\_count\_positive\_tweets\_retweet\_binned & 2 \\
 & ratio\_all\_to\_engaging\_count\_positive\_tweets\_reply\_binned & 2 \\
 & ratio\_all\_to\_engaged\_with\_count\_negative\_tweets\_retweet\_binned & 2 \\
 & engaging\_user\_following\_count\_binned & 2 \\
 & ratio\_engaged\_to\_engaging\_follower\_counts\_binned & 2 \\
 & domains\_frequency\_24h\_binned & 1 \\
 & ratio\_all\_to\_engaging\_count\_negative\_tweets\_retweet\_binned & 1 \\
 & ratio\_all\_to\_domains\_user\_proxy\_count\_positive\_tweets\_quote\_binned & 1 \\
 & engageds\_tweets\_views\_count\_12h\_binned & 1 \\
 & engageds\_tweets\_views\_count\_24h\_binned & 1 \\
 & engageds\_tweets\_views\_count\_05h\_binned & 1 \\
 & engageds\_tweets\_views\_count\_48h\_binned & 1 \\
 & ratio\_all\_to\_hashtags\_user\_proxy\_count\_negative\_tweets\_quote\_binned & 1 \\
 & user\_links\_frequency\_2h\_binned & 1 \\
 & engageds\_tweets\_views\_count\_2h\_binned & 1 \\
 & engaged\_with\_user\_follower\_count\_binned & 1 \\
\hline \multirow[c]{11}{*}{react} & language\_indexed & 21 \\
 & tweet\_type\_indexed & 21 \\
 & user\_domains\_frequency\_12h\_binned & 21 \\
 & ratio\_all\_to\_engaged\_with\_count\_negative\_tweets\_reply\_binned & 21 \\
 & user\_hashtags\_frequency\_1h\_binned & 21 \\
 & hashtags\_frequency\_1h\_binned & 21 \\
 & domains\_frequency\_48h\_binned & 21 \\
 & hashtags\_frequency\_12h\_binned & 21 \\
 & ratio\_all\_to\_hashtags\_user\_proxy\_count\_positive\_tweets\_react\_binned & 21 \\
 & ratio\_all\_to\_links\_user\_proxy\_count\_negative\_tweets\_quote\_binned & 17 \\
 & user\_links\_frequency\_2h\_binned & 4 \\
\end{longtable}

Table \ref{tabOverview of the 10 most informative feature as selected by ChiSq-selector on the train dataset and its 20 subsets} showcases more clearly another interesting finding: the selected features for all subsets mostly overlap for like, reply, rewteet, react, but not for quote. This is likely due to the scarcity of that particular engagement type. Therefore, future work might investigate appropriate oversampling techniques or generation of simulated positive instances based on the remaining engagement types.  Furthermore, we see that the second-degree following reaction (the engineering of which proved rather difficult, cf. subsections \ref{secGraphBasedFeatures} and \ref{secFeatureEngineerngImplementation}) was never selected as the most informative by the $\chi^2$ selector. A penitential reason for this is that the following relation in general is not very indicative of the engagement likelihood. Evidence for this view could be that any other following relations (e.g. follower count) were not frequently selected either. Another feature which was not frequently selected is the number of recent tweet views or any popularity estimate for a period of less than 1 hour. While the former might indicate that the number of views alone cannot determine the likelihood of a particular engagement type. The latter might entail that 1h or 12h  windows simply better reflect what trends are of interest at the moment. To definitely verify or refute these hypotheses, however, furthermore, targeted investigations are necessary. The 25 and 50 most informative features can be seen in the appendix, in tables \ref{tabOverview of the 25 most informative feature as selected by ChiSq-selector on the train dataset and its 20 subsets} and \ref{tabOverview of the 50 most informative feature as selected by ChiSq-selector on the train dataset and its 20 subsets} respectively. 

Finally, let us contrast the achieved performance for our context-based models to previously published works.

\section{Contrasting Results with Other Works}
\label{secContrastingResultsWithOtherWorks}

\begin{figure}[htp]
\noindent\makebox[\textwidth]{\includegraphics[width=\textwidth]{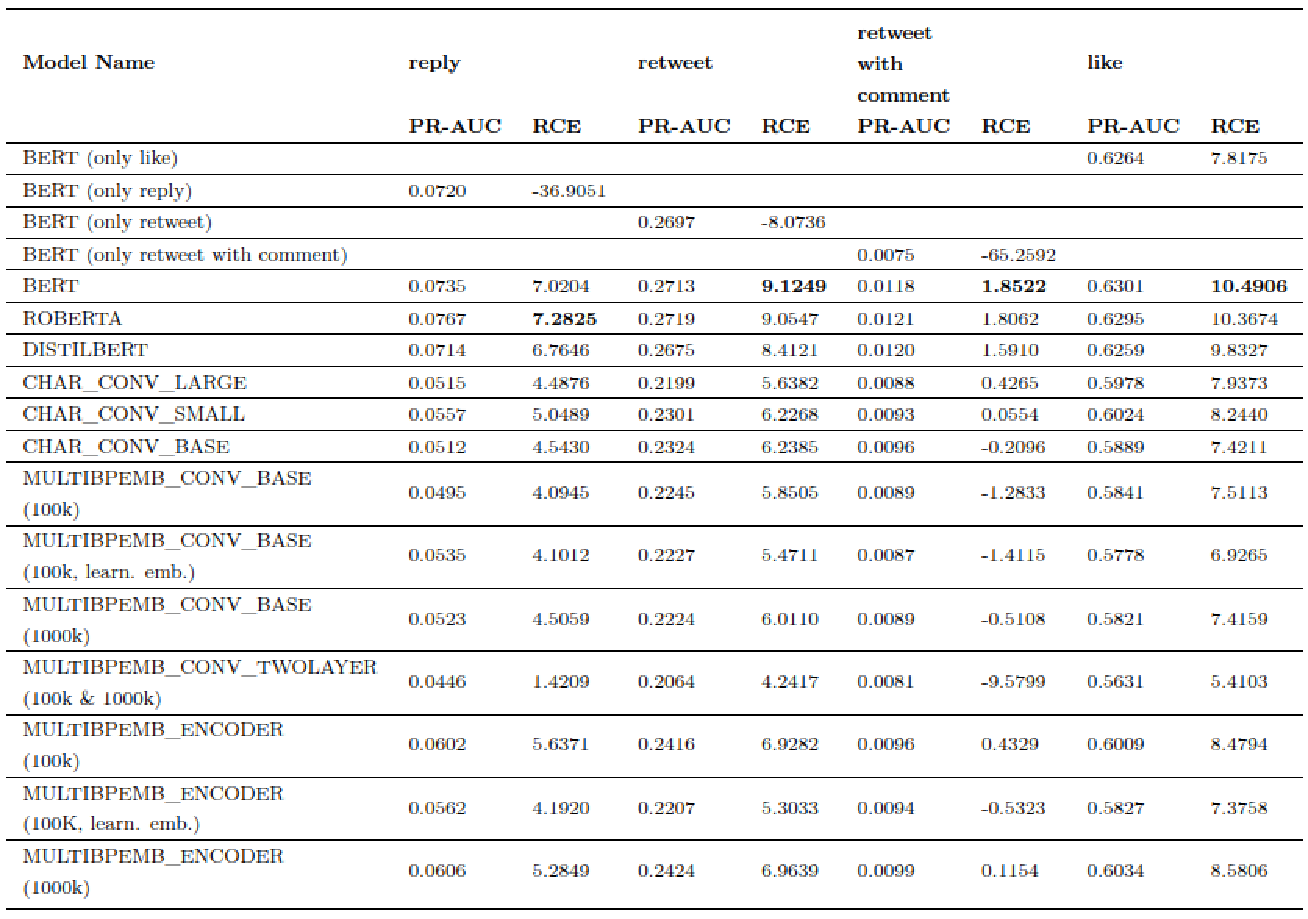}}
\caption{Reproduction of table 7.2 from \cite{davidsThesis} (included with the author's permission), indicating the content-only results obtained from 1\% training subset using transformers BERT and RoBERTa as well as reduced subword embedding models \texttt{MUTI\-BPEMB\-\_CONV\-\_BASE}, two-layer network architecture models \texttt{MUTI\-BPEMB\-\_CONV\-\_TWO\-LAYER} and encoder architecture \texttt{MUTI\-BPEMB\-\_ENCODER}}
\label{figDavidTable72}
\end{figure}

At first glance, especially for the RCE values that remained consistently negative, our achieved results may seem unsatisfactory. For this reason, this section situates the achieved results with prior reported results for the RecSys 2020 Challenge. It does so by contrasting the best achieved context-based results -- in tables \ref{tabBestResultstrained on a train subset and evaluated on full val+testPRAUC} and \ref{tabBestResultstrained on a train subset and evaluated on full val+testRCE} -- to the reported results in content-based results \cite{davidsThesis} in figure \ref{figDavidTable72} and amongst the all-including RecSys 2020 winners in tables \ref{tabLeaderBoard2} and \ref{tabLeaderBoard3}.

Figure \ref{figDavidTable72} reproduces (with the author's permission), the main results from Gradinariu's thesis \cite{davidsThesis}, which focused on content-based features for tweet engagement prediction. In addition to these results, Gradinariu also reports results for the subword embedding models \texttt{MUTI\-BPEMB\-\_CONV\-\_BASE}, two-layer network architecture models \texttt{MUTI\-BPEMB\-\_CONV\-\_TWO\-LAYER} and encoder architecture \texttt{MUTI\-BPEMB\-\_ENCOD\-\_ER} based on 10\% per cent of the training dataset based on ten per cent of the training dataset, which are slighlty better than the corresponding results reported here, but are still worse than results relying on the transformers BERT \cite{bert} and RoBERTa \cite{liu2019roberta} which could only be run on the 1\% dataset. 

It should be noted that, despite both Gradinariu's and our focus on the big-data aspect of the challenge, and using TU Wien's Little Big Data Spark cluster and a powerful GPU architecture respectively, neither of us succeeded in training the best performing models on the full dataset. This indicates that there is a gap in the computational capabilities available (at the point in time) to many postgraduate students at universities and researchers that won the challenge, including the NVIDIA RAPIDS.AI team \cite{[CP7]} that could make us of the full training data made available.

Regarding the achieved results themselves, we see a significant divergence between the two metrics. For the like engagement, with our context-only approach utilising gradient boosting trees we got a PRAUC score of $0.7273$, which is approximately as high as learner\_recsys. The content-based results lag behind with the best like PRAUC being around 10 times smaller, at $0.0721$. However, for the RCE engagement and the like engagement, our context-based models achieved at most $-1418.2$. Yet, Gradinariu's content-based like predicting models scored as high as $6.8$, almost achieving the score of the team saaay71. From here, it can be concluded that further work for both context and content-based models ought to aspire to improve performance for the weaker-performing metric, which is the RCE and PRAUC, respectively. In our case, the lower RCE results might in part be a consequence of the hyperparametre-tuning of the models that was limited to four-fold cross-validation based only on train subsets. Thus a potential improvement might be coming up with a more thorough tuning of hyperparametres. Further improvements in context-based results might have also been achieved through further fine-tuning of best-performing models, creating an ensemble meta-classifier, or utilising deep learning approaches. This is discussed in the next section.

\section{Discussion of Results}
\label{secDiscussion of Results}

This section is primarily inspired by \cite{areWeReallyMakingProgress}, an awarded paper by Ferrari Dacrema et al. that presented an extensive reproducibility study of publications from four top-level conferences. The findings reported in the paper signalled to the recommender systems research community, that a majority of publications introducing neural approaches for top-n recommendations are not reproducible with a reasonable amount of effort. The most frequent reasons include unobtainable data, the lack of train-test splits (or the generating code), as well as unpublished code for preprocessing and evaluation. Crucially, from the papers for which the results could be reproduced, it was found that well-tuned, conceptually simple heuristic baseline models outperform all but one of the proposed neural architectures. The authors found that simple baselines would achieve higher scores with regard to most of the evaluation measures proposed in these papers, with more complex non-neural approaches attaining leading results in many of the remaining metrics. Their approach consisted of downloading the available resources, refactoring the code to apply identical implementation of evaluation measures through the study, tuning the hyperparameters of the baseline models and (if applicable) the proposed architecture, and then comparing the results. When appropriate, they then further analysed the datasets and the train-test splits to identify the causes of conspicuously diverging results. For instance, they calculated the Gini index and Shannon entropy and discovered that the popularity distribution in a test subset differs significantly from the training subset, indicating improper random sampling. Moreover, several common methodological deficiencies were identified and criticised. Crucially for the above-mentioned findings, many papers did not fine-tune the parameters of baseline methods, while some compared results exclusively with previous neural approaches. Given that these neural approaches seem to also have methodological issues, outperforming them does not necessarily entail actual progress within the field.  Furthermore, the authors note that the choice of datasets, accuracy measures, as well as measures’ cut-off values often seem arbitrary and based solely on what has been used previously rather than on application context. These findings might be interpreted to indicate that papers on deep learning in recommender systems commonly focus on “publishability” rather than on reproducibility and soundness of results. As an alternative to such a trajectory, Ferrari Dacrema et al. also outlined steps to restrain further “phantom progress”. 

With thoee messages in mind, we must begin by acknowledging that the models we fitted using algorithms in PySpark's MLLib \cite{pysparkMLLIBdocumentation} have indeed significantly underperformed relative to the state-of-the-art, mostly neural, models used by the Challenge winners. Yet, as Ferrari Dacrema et al.'s work in \cite{areWeReallyMakingProgress} had shown, we should be wary of proclaiming that this discrepancy is due to inherently lower complex-pattern-discovering capabilities in the algorithms used. Instead, we propose that further, more-closely-monitored, model tuning and further feature extraction along the path we set might yet lead to significant prediction improvements. Some of the concrete ways in which this could be done are discussed in chapter \ref{chCriticalReflection}. Similarly, we also cannot definitely conclude that the worse performance achieved is definitive proof that context-only approaches are inherently less successful in terms of maximising the relative cross-entropy than content-only approaches. Instead, these results indicate that there might be an upper bound in predictive capabilities in purely manual feature engineering for complex classification tasks such as tweet engagement prediction. Further work is necessary to verify or refute this indication as well as to judge its generality. To that end, we invested a great amount of time and effort into refactoring and documenting the code to make potential reproducibility and extendability endavours easier, as elaborated in subsection \ref{secCode}. 
\chapter{Critical Reflection}
\label{chCriticalReflection}

This chapter reflects on some of the contributions and limitations of the work. In addition, we also discuss possible directions for future work. This thesis then concludes with final thoughts on the role of context in tweet engagement prediction and recommender systems in general. 

\section{Limitations}
\label{secLimitations}

The limitations of the work can be grouped into seven conceptual themes, each corresponding to one of the paragraphs below.

First of all, there are limitations inherent to the data used in the Challenge. Due to the desire to preserve user privacy \cite{tweetPap,recsys2020overview}, instances that had no interactions were simulated based on tweets that the user might not have seen at all, as was mentioned in subsection \ref{secRecSys}. Thus the conclusions made by this work -- and all other experiments that rely on the data from the Challenge -- are unavoidably affected by these potential false signals stemming from the data. Moreover, as shown in subsection \ref{secDataSet}, the data is unbalanced, especially for quote and reply engagements. Unfortunately, we did not look into the effects this has on algorithm performance. More generally, focusing on any one dataset or any one task is a limiting factor; to make more general evaluations about the role of context in engagement predictions, more diverse scenarios ought to be considered.

Second, our feature engineering process was limited by the available hardware and software on TU Wien's Little Big Data Cluster. Concretely, this has led us to estimate some of the features (such as second-degree graph-based features) for the full dataset based on 10\% subsets. Moreover, due to technical difficulties partially caused by opaque cluster-specific limitations, some of the initially envisioned context-related features and MLLib algorithms could not be computed. We discuss some of these features in subsection \ref{secFutureWork} below.

Due to the now-outdated version of PySpark available on the cluster combined with our desire to contain the scope of the work to pre-provided MLLib methods, we applied only one feature selector, based on $\chi^2$ selection. Furthermore, this selector necessitated us to scale or bin all continuous features or features with too many values, which could have been avoided using non-native implementations such as \cite{chiSqInSpark}.  

On a related note, the prediction algorithms we used were also exclusively MLLib-based. The section on future work elaborates further on possible better approaches which were limited by this design decision. Moreover, we performed hyperparametre tuning using RCE as the only metric based on cross-validation with just four folds. There was also no further fine-tuning of the best-performing models. As already indicated in sections \ref{secContrastingResultsWithOtherWorks} and \ref{secDiscussion of Results}, this is suboptimal and might have resulted in a decreased overall performance.

Regarding the evaluation metrics in general, the work has not implemented methods that would try to balance the two separate evaluation targets and that would try to account for the rather underwhelming RCE score. Therefore, it is difficult to generalise these results to real-world scenarios.

The last step, i.e. our implementation of test statistics also remains limited by the fact that chosen non-parametric test statistics cannot investigate interactions between factors, but only differences within each factor. Furthermore, due to the great number of instances, it is possible that some positive outcomes might, in fact, be Type I-errors. 

Lastly, our work was rather algorithm and process-centric. We did not consider user opinions or social factors. Moreover, we did not consider whether displaying tweets which prompt the user to engage with social media more is ultimately in the user's interest. We also did not consider the broader implications of recommender systems or social media on society as a whole. These limitations were unfortunate given their importance, yet it was seen as a necessary decision to keep the scope of the work from inflating even further.

\section{Contributions of the Work}
\label{secContributions}

Even with the limitations disclosed in the subsection above, there are multiple contributions to this work. These contributions might be split into several general groups.

The first contribution is the implementation per se. The code and the entire pipeline for data exploration, data sampling, feature engineering, feature selection, model training, prediction generation, evaluation of results, and statistical testing of the evaluations are openly available, as explained in section \ref{secCode}. Notably, all results and trained models are also available in the repository. Moreover, great effort was put into refactoring and documenting the code to make it more understandable, reproducible, and expendable. We hope this work might serve as inspiration or a practical foundation for future work. We believe that the contribution of this code is made more valuable because examples of complex PySpark code are less widely available than, for instance, for less computationally intensive Scikit implementations. More locally, during the implementation of the work and through many communications with the members of the Centre for High-Performance Computing at TU Wien, several issues with the cluster were recognised and flagged for corrections. 

Second, this thesis provided a contrasting overview of all methods used by teams published at the RecSys 2020 conference (cf. subsection \ref{secRecSys2020Results}) and situated our approaches within related work on tweet engagement prediction. Moreover, the results we achieved with our models are compared to the content-based models in \cite{davidsThesis} and the results achieved by the Challenge participants who published their results. Thus, this thesis provides an insight into the utility of context within the use-case, as well as specific suggestions (in subsections \ref{secContrastingResultsWithOtherWorks}, \ref{secDiscussion of Results}, and \ref{secFutureWork}) on how these results might be verified, interpreted, and expanded upon.

Moreover, the results we reported offer an insight into the relative utility of extracted and provided contextual features, as well as an insight into the appropriateness of data sampling techniques. As elaborated in chapter \ref{chEvaluation}, we saw that preserving the ratio of users in sampling subsets can increase the performance of the classifiers. Contrastingly, our results indicate that the users' following relations are not highly indicative of the engagement likelihood (cf. subsection \ref{secMostSignificantContextualFeatures}).

Lastly, by having around $120\ 000$ evaluation results distributed unevenly across two metrics for five targets and five factors, we have created a meta-dataset that can be further explored. This might be an interesting test suite for exploring the powers and errors of different test statistics, for instance. 

\section{Future Work}
\label{secFutureWork}

There are several potential further directions our work can be expended into, and these correspond roughly to the recognised limitations in section \ref{secLimitations} as well as to the individual stages in the machine learning pipeline from section \ref{secStagesOfTheMachineLearningPipeline}. For instance, additional sampling methods that combine unique tweet and user ratios could be implemented.

One of the main aspects of this work was feature engineering. Nevertheless, there are many further possible context-based features that can be extracted and added into the machine learning pipeline as well. For instance, the graph-based features from subsection \ref{secGraphBasedFeatures} could be enhanced through the use of page-rank \cite{page1999pagerank}, similarly as was done by done in \cite{[CP4]}. Furthermore, user similarity is a contextual feature that Twitter's current algorithm utilises \ref{secTwittersCurrentRS} and that \cite{[CP4]} included as well. Regarding user interactions, future work might consider collaborative filtering approaches  for user-tweet and viewer-author pairs to model interactions, potentially by looking at the approaches in \cite{[CP9]} for inspiration. How (and if) these features should then be selected before they are passed to the classifier is yet another area of potential future work that might improve predicting tweet engagement likelihood.

While section \ref{secMostSignificantContextualFeatures} lists and discusses the most informative features as indicated by repeated $\chi^2$ feature selection, this is not the only approach to assessing feature utility. Future work might, for instance, examine transparent or explainable models, such as decision trees, to see what feature sets led to particular predictions.

Of course, applying state-of-the-art classifiers, such as various deep neural networks and more powerful boosting algorithms could help unearth more complex context-based engagement patterns. Contrastingly, a meta-classifier that would take the outputs of multiple simpler first-round algorithms (like the ones implemented in this work) as its input and then learn to weight them and use them to create better predictions would be another (potentially non-neural) approach to increasing performance of the classifiers. 

Regarding the evaluation of results, statistical tests that evaluate the interactions between factors would be welcome as well. The most obvious next step in this direction would be to investigate the interactions of factors that are likely to influence the performance of the model together (like the feature selector and the classification algorithm combination). 

Regarding the broader health and social consequences of social networks, the influence of recommender systems on mental well-being and worldview could be investigated. For example, it is conceivable that context-only recommendations might lead to more diverse recommendations that would decrease the so-called ``filter bubbles''. This denomination is used for the phenomenon, where users only see content that is similar to one another and thus may narrows the user's ability to interpret events from different angles.

Lastly, the context-based recommendation can be generalised to context-based user modelling. The work on this thesis is one of the influencing factors that has led the author to co-found and co-organise the first edition of the Workshop on Context Representation in User Modelling (CRUM 2023), a half-day workshop that is set to take place in June 2023 adjunct to the thirty-first edition of the ACM Conference on User Modeling and Personalization (UMAP 2023). The workshop, in turn, aims to catalyse further utilisation of context in all adaptive applications, including in recommender systems \cite{crum2023}.

\section{Conclusion}
\label{secConclusion}

The work presented in this thesis has investigated the role of context in tweet engagement prediction. Specifically, we used Apache PySpark on TU Wien's computing cluster to perform a set of experiments that investigated the role of several factors. These factors include sampling sizes and sampling technique of the dataset used for model training; the feature sets used as explanatory variables; the role of feature selection; and utilised the classifier algorithm. The end results were then compared with previously published work that used all available data content of the tweet exclusively. In relation to most of those, our predictions underperformed in terms of relative cross-entropy, one of the two target metrics. This might be interpreted as an indicator of an upper bound in the predictive utility of context-based approaches or manually engineering features for complex classification tasks. Alternatively (as indicated by Ferrari Dacrema et al. \cite{areWeReallyMakingProgress}), it might be the case that further hyperparametre tuning or accounting for different data distributions could lead to better results even when relying solely on manual feature engineering and simple, explainable, machine learning models. Furthermore, using an ensemble meta-model on top of these simpler models could improve the results further. We thus hope that this thesis would inspire future students or researchers to explore the role of context and feature engineering in recommender systems, rather than relying solely on opaque deep neural models.

\begin{appendices}

\chapter{Environment}
\label{appEnvironment}

This part of the appendix provides more information about the virtual environment used for the implementation of the practical part of the project.

\section{LBD Environment}
\label{appLBDEnvironment}

A new environment was created on the TU Wien's Little Big Data cluster (cf. section \ref{secLBD}). It was set up as can be seen in listing \ref{lstEnvironmentLBD} below.

\begin{lstlisting}[language=bash,showstringspaces=false,caption={The environment used on LBD}, label=lstEnvironmentLBD]
{
    "argv": [
        "/home/adbs20/e01528091/.conda/envs/envJovan1/bin/python3.6",
        "-m",
        "ipykernel_launcher",
        "-f",
        "{connection_file}"
        ],
        "display_name": "jovan2",
        "language": "python",
        "env": {
            "SPARK_HOME": "/opt/cloudera/parcels/CDH/lib/spark",
            "PYSPARK_PYTHON": "/home/adbs20/e01528091/.conda/envs/envJovan1/bin/python3.6",
            "PYSPARK_DRIVER_PYTHON":"/home/adbs20/e01528091/.conda/envs/envJovan1/bin/python3.6" }
}
\end{lstlisting}

\section{Installed Packages}
\label{appInstalledPackages}

Table \ref{tabPackages} lists all installed packages with their versions. The most important packages and their use are described in section \ref{secUsedFrameworksAndLibraries}. Environments on the local machine and on the cluster contained mostly the same packages.

\begin{longtable}{llll}
\caption{The complete list of installed packages}
\label{tabPackages}\\ 
\hline
\textbf{Name} &                   \textbf{Version} &                  \textbf{Build} & \textbf{Channel} \\
\hline
alabaster                &  0.7.10 &          py36hcd07829\_0 &  \\
anaconda                 &  5.2.0 &                   py36\_3 &  \\
anaconda-client          &  1.6.14 &                  py36\_0 &  \\
anaconda-project         &  0.8.2 &           py36hfad2e28\_0 &  \\
asn1crypto               &  0.24.0 &                  py36\_0 &  \\
astroid                  &  1.6.3 &                   py36\_0 &  \\
astropy                  &  3.0.2 &           py36h452e1ab\_1 &  \\
attrs                    & 18.1.0 &                  py36\_0 &  \\
babel                    &  2.5.3 &                   py36\_0 &  \\
backcall                 &  0.1.0 &                   py36\_0 &  \\
backports                &  1.0 &             py36h81696a8\_1 &  \\
backports.shutil\_get\_terminal\_size &  1.0.0 &           py36h79ab834\_2 &  \\
beautifulsoup4           &  4.6.0 &           py36hd4cc5e8\_1 &  \\
bitarray                 &  0.8.1 &           py36hfa6e2cd\_1 &  \\
bkcharts                 &  0.2 &             py36h7e685f7\_0 &  \\
blas                     &  1.0 &                        mkl &  \\
blaze                    &  0.11.3 &          py36h8a29ca5\_0 &  \\
bleach                   &  2.1.3 &                   py36\_0 &  \\
blosc                    &  1.14.3 &              he51fdeb\_0 &  \\
bokeh                    &  0.12.16 &                 py36\_0 &  \\
boto                     &  2.48.0 &          py36h1a776d2\_1 &  \\
bottleneck               &  1.2.1 &           py36hd119dfa\_0 &  \\
bzip2                    &  1.0.6 &               hfa6e2cd\_5 &  \\
ca-certificates          & 2018.03.07 &                   0  &  \\
cachetools               &  4.2.4 &                   pypi\_0 &   pypi  \\
certifi                  & 2018.4.16 &               py36\_0 &  \\
cffi                     &  1.11.5 &          py36h945400d\_0 &  \\
chardet                  &  3.0.4 &           py36h420ce6e\_1 &  \\
click                    &  6.7 &             py36hec8c647\_0 &  \\
cloudpickle              &  0.5.3 &                   py36\_0 &  \\
clyent                   &  1.2.2 &           py36hb10d595\_1 &  \\
colorama                 &  0.3.9 &           py36h029ae33\_0 &  \\
comtypes                 &  1.1.4 &                   py36\_0 &  \\
console\_shortcut         &  0.1.1 &               h6bb2dd7\_3 &  \\
contextlib2              &  0.5.5 &           py36he5d52c0\_0 &  \\
cryptography             &  2.2.2 &           py36hfa6e2cd\_0 &  \\
curl                     &  7.60.0 &              h7602738\_0 &  \\
cycler                   &  0.10.0 &          py36h009560c\_0 &  \\
cython                   &  0.28.2 &          py36hfa6e2cd\_0 &  \\
cytoolz                  &  0.9.0.1 &         py36hfa6e2cd\_0 &  \\
dask                     &  0.17.5 &                  py36\_0 &  \\
dask-core                &  0.17.5 &                  py36\_0 &  \\
datashape                &  0.5.4 &           py36h5770b85\_0 &  \\
decorator                &  4.3.0 &                   py36\_0 &  \\
distributed              &  1.21.8 &                  py36\_0 &  \\
docutils                 &  0.14 &            py36h6012d8f\_0 &  \\
entrypoints              &  0.2.3 &           py36hfd66bb0\_2 &  \\
et\_xmlfile               &  1.0.1 &           py36h3d2d736\_0 &  \\
fastcache                &  1.0.2 &           py36hfa6e2cd\_2 &  \\
filelock                 &  3.0.4 &                   py36\_0 &  \\
flask                    &  1.0.2 &                   py36\_1 &  \\
flask-cors               &  3.0.4 &                   py36\_0 &  \\
flask-login              &  0.5.0 &                   pypi\_0 &   pypi  \\
freetype                 &  2.8 &                 h51f8f2c\_1 &  \\
get\_terminal\_size        &  1.0.0 &               h38e98db\_0 &  \\
gevent                   &  1.3.0 &           py36hfa6e2cd\_0 &  \\
glob2                    &  0.6 &             py36hdf76b57\_0 &  \\
google-api-core          &  2.8.2 &                   pypi\_0 &   pypi  \\
google-api-python-client &  2.52.0 &                  pypi\_0 &   pypi  \\
google-auth              &  2.11.1 &                  pypi\_0 &   pypi  \\
google-auth-httplib2     &  0.1.0 &                   pypi\_0 &   pypi  \\
google-auth-oauthlib     &  0.5.3 &                   pypi\_0 &   pypi  \\
googleapis-common-protos &  1.56.3 &                  pypi\_0 &   pypi  \\
greenlet                 &  0.4.13 &          py36hfa6e2cd\_0 &  \\
h5py                     &  2.7.1 &           py36h3bdd7fb\_2 &  \\
hdf5                     &  1.10.2 &              hac2f561\_1 &  \\
heapdict                 &  1.0.0 &                   py36\_2 &  \\
html5lib                 &  1.0.1 &           py36h047fa9f\_0 &  \\
httplib2                 &  0.20.4 &                  pypi\_0 &   pypi  \\
icc\_rt                   & 2017.0.4 &            h97af966\_0 &  \\
icu                      & 58.2 &                ha66f8fd\_1 &  \\
idna                     &  2.6 &             py36h148d497\_1 &  \\
imageio                  &  2.3.0 &                   py36\_0 &  \\
imagesize                &  1.0.0 &                   py36\_0 &  \\
intel-openmp             & 2018.0.0 &                     8 &  \\
ipykernel                &  4.8.2 &                   py36\_0 &  \\
ipython                  &  6.4.0 &                   py36\_0 &  \\
ipython\_genutils         &  0.2.0 &           py36h3c5d0ee\_0 &  \\
ipywidgets               &  7.2.1 &                   py36\_0 &  \\
isort                    &  4.3.4 &                   py36\_0 &  \\
itsdangerous             &  0.24 &            py36hb6c5a24\_1 &  \\
jdcal                    &  1.4 &                     py36\_0 &  \\
jedi                     &  0.12.0 &                  py36\_1 &  \\
jinja2                   &  2.10 &            py36h292fed1\_0 &  \\
jpeg                     & 9b &                  hb83a4c4\_2 &  \\
jsonschema               &  2.6.0 &           py36h7636477\_0 &  \\
jupyter                  &  1.0.0 &                   py36\_4 &  \\
jupyter\_client           &  5.2.3 &                   py36\_0 &  \\
jupyter\_console          &  5.2.0 &           py36h6d89b47\_1 &  \\
jupyter\_core             &  4.4.0 &           py36h56e9d50\_0 &  \\
jupyterlab               &  0.32.1 &                  py36\_0 &  \\
jupyterlab\_launcher      &  0.10.5 &                  py36\_0 &  \\
jupyterthemes            &  0.20.0 &                    py\_0 &   conda-forge  \\
kiwisolver               &  1.0.1 &           py36h12c3424\_0 &  \\
lazy-object-proxy        &  1.3.1 &           py36hd1c21d2\_0 &  \\
lesscpy                  &  0.15.0 &            pyhd8ed1ab\_0 &   conda-forge  \\
libcurl                  &  7.60.0 &              hc4dcbb0\_0 &  \\
libiconv                 &  1.15 &                h1df5818\_7 &  \\
libpng                   &  1.6.34 &              h79bbb47\_0 &  \\
libsodium                &  1.0.16 &              h9d3ae62\_0 &  \\
libssh2                  &  1.8.0 &               hd619d38\_4 &  \\
libtiff                  &  4.0.9 &               hb8ad9f9\_1 &  \\
libxml2                  &  2.9.8 &               hadb2253\_1 &  \\
libxslt                  &  1.1.32 &              hf6f1972\_0 &  \\
littleutils              &  0.2.2 &                   pypi\_0 &   pypi  \\
llvmlite                 &  0.23.1 &          py36hcacf6c6\_0 &  \\
locket                   &  0.2.0 &           py36hfed976d\_1 &  \\
lxml                     &  4.2.1 &           py36heafd4d3\_0 &  \\
lzo                      &  2.10 &                h6df0209\_2 &  \\
m2w64-gcc-libgfortran    &  5.3.0 &                        6 &  \\
m2w64-gcc-libs           &  5.3.0 &                        7 &  \\
m2w64-gcc-libs-core      &  5.3.0 &                        7 &  \\
m2w64-gmp                &  6.1.0 &                        2 &  \\
m2w64-libwinpthread-git  &  5.0.0.4634.697f757               2 &  \\
markupsafe               &  1.0 &             py36h0e26971\_1 &  \\
matplotlib               &  3.3.4 &                   pypi\_0 &   pypi  \\
mccabe                   &  0.6.1 &           py36hb41005a\_1 &  \\
menuinst                 &  1.4.14           py36hfa6e2cd\_0 &  \\
mistune                  &  0.8.3 &           py36hfa6e2cd\_1 &  \\
mkl                      & 2018.0.2 &                     1 &  \\
mkl-service              &  1.1.2 &           py36h57e144c\_4 &  \\
mkl\_fft                  &  1.0.1 &           py36h452e1ab\_0 &  \\
mkl\_random               &  1.0.1 &           py36h9258bd6\_0 &  \\
more-itertools           &  4.1.0 &                   py36\_0 &  \\
mpmath                   &  1.0.0 &           py36hacc8adf\_2 &  \\
msgpack-python           &  0.5.6 &           py36he980bc4\_0 &  \\
msys2-conda-epoch        & 20160418                      1 &  \\
multipledispatch         &  0.5.0 &                   py36\_0 &  \\
nbconvert                &  5.3.1 &           py36h8dc0fde\_0 &  \\
nbformat                 &  4.4.0 &           py36h3a5bc1b\_0 &  \\
networkx                 &  2.1 &                     py36\_0 &  \\
nltk                     &  3.3.0 &                   py36\_0 &  \\
nose                     &  1.3.7 &           py36h1c3779e\_2 &  \\
notebook                 &  5.5.0 &                   py36\_0 &  \\
numba                    &  0.38.0 &          py36h830ac7b\_0 &  \\
numexpr                  &  2.6.5 &           py36hcd2f87e\_0 &  \\
numpy                    &  1.19.5 &                  pypi\_0 &   pypi  \\
numpydoc                 &  0.8.0 &                   py36\_0 &  \\
oauthlib                 &  3.2.1 &                   pypi\_0 &   pypi  \\
odo                      &  0.5.1 &           py36h7560279\_0 &  \\
olefile                  &  0.45.1 &                  py36\_0 &  \\
openpyxl                 &  2.5.3 &                   py36\_0 &  \\
openssl                  &  1.0.2o &               h8ea7d77\_0 &  \\
outdated                 &  0.2.2 &                   pypi\_0 &   pypi  \\
packaging                & 17.1 &                    py36\_0 &  \\
pandas                   &  1.1.5 &                   pypi\_0 &   pypi  \\
pandas-flavor            &  0.2.0 &                   pypi\_0 &   pypi  \\
pandoc                   &  1.19.2.1 &            hb2460c7\_1 &  \\
pandocfilters            &  1.4.2 &           py36h3ef6317\_1 &  \\
parso                    &  0.2.0 &                   py36\_0 &  \\
partd                    &  0.3.8 &           py36hc8e763b\_0 &  \\
path.py                  & 11.0.1 &                  py36\_0 &  \\
pathlib2                 &  2.3.2 &                   py36\_0 &  \\
patsy                    &  0.5.0 &                   py36\_0 &  \\
pep8                     &  1.7.1 &                   py36\_0 &  \\
pickleshare              &  0.7.4 &           py36h9de030f\_0 &  \\
pillow                   &  8.4.0 &                   pypi\_0 &   pypi  \\
pingouin                 &  0.3.12 &                   pypi\_0 &   pypi  \\
pip                      & 21.3.1 &                  pypi\_0 &   pypi  \\
pkginfo                  &  1.4.2 &                   py36\_1 &  \\
pluggy                   &  0.6.0 &           py36hc7daf1e\_0 &  \\
ply                      &  3.11 &                    py36\_0 &  \\
prompt\_toolkit           &  1.0.15 &          py36h60b8f86\_0 &  \\
protobuf                 &  3.19.5 &                  pypi\_0 &   pypi  \\
psutil                   &  5.4.5 &           py36hfa6e2cd\_0 &  \\
py                       &  1.5.3 &                   py36\_0 &  \\
py4j                     &  0.10.7 &                  pypi\_0 &   pypi  \\
pyarrow                  &  6.0.1 &                   pypi\_0 &   pypi  \\
pyasn1                   &  0.4.8 &                   pypi\_0 &   pypi  \\
pyasn1-modules           &  0.2.8 &                   pypi\_0 &   pypi  \\
pycodestyle              &  2.4.0 &                   py36\_0 &  \\
pycosat                  &  0.6.3 &           py36h413d8a4\_0 &  \\
pycparser                &  2.18 &            py36hd053e01\_1 &  \\
pycrypto                 &  2.6.1 &           py36hfa6e2cd\_8 &  \\
pycurl                   &  7.43.0.1 &        py36h74b6da3\_0 &  \\
pyflakes                 &  1.6.0 &           py36h0b975d6\_0 &  \\
pygments                 &  2.2.0 &           py36hb010967\_0 &  \\
pylint                   &  1.8.4 &                   py36\_0 &  \\
pyodbc                   &  4.0.23 &          py36h6538335\_0 &  \\
pyopenssl                & 18.0.0 &                  py36\_0 &  \\
pyparsing                &  3.0.7 &                   pypi\_0 &   pypi  \\
pyqt                     &  5.9.2 &           py36h1aa27d4\_0 &  \\
pysocks                  &  1.6.8 &                   py36\_0 &  \\
pyspark                  &  2.4.4 &                   pypi\_0 &   pypi  \\
pytables                 &  3.4.3 &           py36he6f6034\_1 &  \\
pytest                   &  3.5.1 &                   py36\_0 &  \\
pytest-arraydiff         &  0.2 &                     py36\_0 &  \\
pytest-astropy           &  0.3.0 &                   py36\_0 &  \\
pytest-doctestplus       &  0.1.3 &                   py36\_0 &  \\
pytest-openfiles         &  0.3.0 &                   py36\_0 &  \\
pytest-remotedata        &  0.2.1 &                   py36\_0 &  \\
python                   &  3.6.5 &               h0c2934d\_0 &  \\
python-dateutil          &  2.8.2 &                   pypi\_0 &   pypi  \\
pytz                     & 2022.7.1 &                pypi\_0 &   pypi  \\
pywavelets               &  0.5.2 &           py36hc649158\_0 &  \\
pywin32                  & 223 &             py36hfa6e2cd\_1 &  \\
pywinpty                 &  0.5.1 &                   py36\_0 &  \\
pyyaml                   &  3.12 &            py36h1d1928f\_1 &  \\
pyzmq                    & 17.0.0 &          py36hfa6e2cd\_1 &  \\
qt                       &  5.9.5 &           vc14he4a7d60\_0 &  \\
qtawesome                &  0.4.4 &           py36h5aa48f6\_0 &  \\
qtconsole                &  4.3.1 &           py36h99a29a9\_0 &  \\
qtpy                     &  1.4.1 &                   py36\_0 &  \\
requests                 &  2.18.4 &          py36h4371aae\_1 &  \\
requests-oauthlib        &  1.3.1 &                   pypi\_0 &   pypi  \\
rope                     &  0.10.7 &          py36had63a69\_0 &  \\
rsa                      &  4.9 &                     pypi\_0 &   pypi  \\
ruamel\_yaml              &  0.15.35          py36hfa6e2cd\_1 & \\
scikit-image             &  0.13.1 &          py36hfa6e2cd\_1 & \\
scikit-learn             &  0.19.1 &          py36h53aea1b\_0 & \\
scipy                    &  1.5.4 &                   pypi\_0 &   pypi  \\
seaborn                  &  0.11.2 &                  pypi\_0 &   pypi  \\
send2trash               &  1.5.0 &                   py36\_0 &  \\
setuptools                59.6.0 &                  pypi\_0 &   pypi  \\
simplegeneric            &  0.8.1 &                   py36\_2 &  \\
singledispatch           &  3.4.0.3 &         py36h17d0c80\_0 &  \\
sip                      &  4.19.8 &          py36h6538335\_0 &  \\
six                      &  1.16.0 &                  pypi\_0 &   pypi  \\
snappy                   &  1.1.7 &               h777316e\_3 &  \\
snowballstemmer          &  1.2.1 &           py36h763602f\_0 &  \\
sortedcollections        &  0.6.1 &                   py36\_0 &  \\
sortedcontainers         &  1.5.10 &                  py36\_0 &  \\
sphinx                   &  1.7.4 &                   py36\_0 &  \\
sphinxcontrib            &  1.0 &             py36hbbac3d2\_1 &  \\
sphinxcontrib-websupport &  1.0.1 &           py36hb5e5916\_1 &  \\
spyder                   &  3.2.8 &                   py36\_0 &  \\
sqlalchemy               &  1.2.7 &           py36ha85dd04\_0 &  \\
sqlite                   &  3.23.1 &              h35aae40\_0 &  \\
statsmodels              &  0.12.2 &                  pypi\_0 &   pypi  \\
style                    &  1.1.0 &                   pypi\_0 &   pypi  \\
sympy                    &  1.1.1 &           py36h96708e0\_0 &  \\
tabulate                 &  0.8.10 &                   pypi\_0 &   pypi  \\
tbb                      & 2018.0.5 &            he980bc4\_0 &  \\
tbb4py                   & 2018.0.5 &        py36he980bc4\_0 &  \\
tblib                    &  1.3.2 &           py36h30f5020\_0 &  \\
terminado                &  0.8.1 &                   py36\_1 &  \\
testpath                 &  0.3.1 &           py36h2698cfe\_0 &  \\
tk                       &  8.6.7 &               hcb92d03\_3 &  \\
toolz                    &  0.9.0 &                   py36\_0 &  \\
tornado                  &  5.0.2 &                   py36\_0 &  \\
traitlets                &  4.3.2 &           py36h096827d\_0 &  \\
typing                   &  3.6.4 &                   py36\_0 &  \\ 
unicodecsv               &  0.14.1 &          py36h6450c06\_0 &  \\
update                   &  0.0.1 &                   pypi\_0 &   pypi  \\
uritemplate              &  4.1.1 &                   pypi\_0 &   pypi  \\
urllib3                  &  1.22 &            py36h276f60a\_0 &  \\
vc                       & 14 &                  h0510ff6\_3 &  \\
vs2015\_runtime           & 14.0.25123 &                   3 &  \\ 
wcwidth                  &  0.1.7 &           py36h3d5aa90\_0 &  \\
webencodings             &  0.5.1 &           py36h67c50ae\_1 &  \\
werkzeug                 &  0.14.1 &                  py36\_0 &  \\
wheel                    &  0.31.1 &                  py36\_0 &  \\
widgetsnbextension       &  3.2.1 &                   py36\_0 &  \\
win\_inet\_pton            &  1.0.1 &           py36he67d7fd\_1 &  \\
win\_unicode\_console      &  0.5 &             py36hcdbd4b5\_0 &  \\
wincertstore             &  0.2 &             py36h7fe50ca\_0 &  \\
winpty                   &  0.4.3 &                        4 &  \\
wrapt                    &  1.10.11 &         py36he5f5981\_0 &  \\
xarray                   &  0.16.2 &                  pypi\_0 &   pypi  \\
xlrd                     &  1.1.0 &           py36h1cb58dc\_1 &  \\
xlsxwriter               &  1.0.4 &                   py36\_0 &  \\
xlwings                  &  0.11.8 &                  py36\_0 &  \\
xlwt                     &  1.3.0 &           py36h1a4751e\_0 &  \\
yaml                     &  0.1.7 &               hc54c509\_2 &  \\
zeromq                   &  4.2.5 &               hc6251cf\_0 &  \\
zict                     &  0.1.3 &           py36h2d8e73e\_0 &  \\
zlib                     &  1.2.11 &              h8395fce\_2 &  \\
\hline
\end{longtable}

\chapter{Datasets}
\label{appDatasets}

This chapter provides additional information about the datasets and the created subsets.

\section{Subsets}
\label{appSubsets}

The list of 84 datasets and sampled subsets with their sizes when exported to .parquet files, as elaborated in section \ref{secSampling}, is given below:

\begin{enumerate}
\item 747M    Final\_test\_EU\_sample\_10pct.parquet
\item 76M     Final\_test\_EU\_sample\_1pct.parquet
\item 146M    Final\_test\_EU\_sample\_2pct.parquet
\item 369M    Final\_test\_EU\_sample\_5pct.parquet
\item 423M    Final\_test\_EWU\_sample\_10pct.parquet
\item 55M     Final\_test\_EWU\_sample\_1pct.parquet
\item 105M    Final\_test\_EWU\_sample\_2pct.parquet
\item 261M    Final\_test\_EWU\_sample\_5pct.parquet
\item 616M    Final\_test\_inter\_EWU+EU\_sample\_10pct.parquet
\item 64M     Final\_test\_inter\_EWU+EU\_sample\_1pct.parquet
\item 118M    Final\_test\_inter\_EWU+EU\_sample\_2pct.parquet
\item 290M    Final\_test\_inter\_EWU+EU\_sample\_5pct.parquet
\item 7.8G    Final\_test.parquet
\item 750M    Final\_test\_random\_sample\_10pct.parquet
\item 76M     Final\_test\_random\_sample\_1pct.parquet
\item 145M    Final\_test\_random\_sample\_2pct.parquet
\item 368M    Final\_test\_random\_sample\_5pct.parquet
\item 706M    Final\_test\_tweet\_sample\_10pct.parquet
\item 69M     Final\_test\_tweet\_sample\_1pct.parquet
\item 131M    Final\_test\_tweet\_sample\_2pct.parquet
\item 340M    Final\_test\_tweet\_sample\_5pct.parquet
\item 4.8G    Final\_train\_EU\_sample\_10pct.parquet
\item 464M    Final\_train\_EU\_sample\_1pct.parquet
\item 939M    Final\_train\_EU\_sample\_2pct.parquet
\item 2.4G    Final\_train\_EU\_sample\_5pct.parquet
\item 4.0G    Final\_train\_EWU\_sample\_10pct.parquet
\item 480M    Final\_train\_EWU\_sample\_1pct.parquet
\item 922M    Final\_train\_EWU\_sample\_2pct.parquet
\item 2.3G    Final\_train\_EWU\_sample\_5pct.parquet
\item 4.7G    Final\_train\_inter\_EWU+EU\_sample\_10pct.parquet
\item 480M    Final\_train\_inter\_EWU+EU\_sample\_1pct.parquet
\item 973M    Final\_train\_inter\_EWU+EU\_sample\_2pct.parquet
\item 2.4G    Final\_train\_inter\_EWU+EU\_sample\_5pct.parquet
\item 53G     Final\_train.parquet
\item 4.7G    Final\_train\_random\_sample\_10pct.parquet
\item 447M    Final\_train\_random\_sample\_1pct.parquet
\item 906M    Final\_train\_random\_sample\_2pct.parquet
\item 2.3G    Final\_train\_random\_sample\_5pct.parquet
\item 4.3G    Final\_train\_tweet\_sample\_10pct.parquet
\item 384M    Final\_train\_tweet\_sample\_1pct.parquet
\item 794M    Final\_train\_tweet\_sample\_2pct.parquet
\item 242M    Final\_train\_tweet\_sample\_5pct.parquet
\item 743M    Final\_val\_EU\_sample\_10pct.parquet
\item 75M     Final\_val\_EU\_sample\_1pct.parquet
\item 145M    Final\_val\_EU\_sample\_2pct.parquet
\item 365M    Final\_val\_EU\_sample\_5pct.parquet
\item 540M    Final\_val\_EWU\_sample\_10pct.parquet
\item 58M     Final\_val\_EWU\_sample\_1pct.parquet
\item 109M    Final\_val\_EWU\_sample\_2pct.parquet
\item 262M    Final\_val\_EWU\_sample\_5pct.parquet
\item 608M    Final\_val\_inter\_EWU+EU\_sample\_10pct.parquet
\item 64M     Final\_val\_inter\_EWU+EU\_sample\_1pct.parquet
\item 117M    Final\_val\_inter\_EWU+EU\_sample\_2pct.parquet
\item 286M    Final\_val\_inter\_EWU+EU\_sample\_5pct.parquet
\item 7.7G    Final\_val.parquet
\item 745M    Final\_val\_random\_sample\_10pct.parquet
\item 75M     Final\_val\_random\_sample\_1pct.parquet
\item 145M    Final\_val\_random\_sample\_2pct.parquet
\item 366M    Final\_val\_random\_sample\_5pct.parquet
\item 1.4G    Final\_val+test\_EU\_sample\_10pct.parquet
\item 137M    Final\_val+test\_EU\_sample\_1pct.parquet
\item 277M    Final\_val+test\_EU\_sample\_2pct.parquet
\item 704M    Final\_val+test\_EU\_sample\_5pct.parquet
\item 899M    Final\_val+test\_EWU\_sample\_10pct.parquet
\item 101M    Final\_val+test\_EWU\_sample\_1pct.parquet
\item 201M    Final\_val+test\_EWU\_sample\_2pct.parquet
\item 509M    Final\_val+test\_EWU\_sample\_5pct.parquet
\item 1.2G    Final\_val+test\_inter\_EWU+EU\_sample\_10pct.parquet
\item 114M    Final\_val+test\_inter\_EWU+EU\_sample\_1pct.parquet
\item 142M    Final\_val+test\_inter\_EWU+EU\_sample\_2pct.parquet
\item 652M    Final\_val+test\_inter\_EWU+EU\_sample\_5pct.parquet
\item 16G     Final\_val+test.parquet
\item 1.5G    Final\_val+test\_random\_sample\_10pct.parquet
\item 137M    Final\_val+test\_random\_sample\_1pct.parquet
\item 277M    Final\_val+test\_random\_sample\_2pct.parquet
\item 705M    Final\_val+test\_random\_sample\_5pct.parquet
\item 1.4G    Final\_val+test\_tweet\_sample\_10pct.parquet
\item 123M    Final\_val+test\_tweet\_sample\_1pct.parquet
\item 248M    Final\_val+test\_tweet\_sample\_2pct.parquet
\item 654M    Final\_val+test\_tweet\_sample\_5pct.parquet
\item 701M    Final\_val\_tweet\_sample\_10pct.parquet
\item 69M     Final\_val\_tweet\_sample\_1pct.parquet
\item 130M    Final\_val\_tweet\_sample\_2pct.parquet
\item 339M    Final\_val\_tweet\_sample\_5pct.parquet
\end{enumerate}

The local work was done on 1\% and 2\% subsections. For the \textit{Final\_} versions of the dataset only, these parquet files are around 11GB in size. Of course, there are other versions of the datasets as well, corresponding to each stage of feature engineering and feature selection, as elaborated in chapters \ref{chMethodology} and \ref{chImplementation}.

\section{All Features Used for Prediction Tasks}
\label{appAllFeaturesUsedforPredictionTasks}

Not all of the pre-provided and extracted features can be directly used for predictions. This may be because they are counts dependent on the size of the dataset, are in an inappropriate format (timestamps), or are IDs. We thus recognised and grouped all pre-provided and extracted features that can be used for prediction. There are 185 relevant features and they can be seen in listing \ref{lstRelevantFeatures}.

\begin{lstlisting}[language=Python,basicstyle=\small,caption={How the 185 features relevant for predictions were manually grouped}, label=lstRelevantFeatures]

relevant_features["raw"] = ['tweet_type', 'language', 'engaged_with_user_follower_count', 'engaged_with_user_following_count', 'engaged_with_user_is_verified', 'engaging_user_follower_count', 'engaging_user_following_count', 'engaging_user_is_verified', 'engagee_follows_engager', ]
relevant_features["graph-based"] = ['graph_engagee_follows_engager_2d', 'graph_engager_follows_engagee_2d', 'graph_engaging_flag_like_from_engaged_1d', 'graph_engaging_flag_reply_from_engaged_1d', 'graph_engaging_flag_retweet_from_engaged_1d', 'graph_engaging_flag_quote_from_engaged_1d', 'graph_engaging_flag_react_from_engaged_1d', 'graph_engaging_count_like_from_engaged_1d', 'graph_engaging_count_reply_from_engaged_1d', 'graph_engaging_count_retweet_from_engaged_1d', 'graph_engaging_count_quote_from_engaged_1d', 'graph_engaging_count_react_from_engaged_1d', 'graph_engaged_flag_like_from_engaging_1d', 'graph_engaged_flag_reply_from_engaging_1d', 'graph_engaged_flag_retweet_from_engaging_1d', 'graph_engaged_flag_quote_from_engaging_1d', 'graph_engaged_flag_react_from_engaging_1d', 'graph_engaged_count_like_from_engaging_1d', 'graph_engaged_count_reply_from_engaging_1d', 'graph_engaged_count_retweet_from_engaging_1d', 'graph_engaged_count_quote_from_engaging_1d', 'graph_engaged_count_react_from_engaging_1d', 'graph_engaging_flag_like_from_engaged_2d', 'graph_engaging_flag_reply_from_engaged_2d', 'graph_engaging_flag_retweet_from_engaged_2d', 'graph_engaging_flag_quote_from_engaged_2d', 'graph_engaging_flag_react_from_engaged_2d', 'graph_engaging_count_like_from_engaged_2d', 'graph_engaging_count_reply_from_engaged_2d', 'graph_engaging_count_retweet_from_engaged_2d', 'graph_engaging_count_quote_from_engaged_2d', 'graph_engaging_count_react_from_engaged_2d', 'graph_engaged_flag_like_from_engaging_2d', 'graph_engaged_flag_reply_from_engaging_2d', 'graph_engaged_flag_retweet_from_engaging_2d', 'graph_engaged_flag_quote_from_engaging_2d', 'graph_engaged_flag_react_from_engaging_2d', 'graph_engaged_count_like_from_engaging_2d', 'graph_engaged_count_reply_from_engaging_2d', 'graph_engaged_count_retweet_from_engaging_2d', 'graph_engaged_count_quote_from_engaging_2d', 'graph_engaged_count_react_from_engaging_2d',  ]
relevant_features["graph-based ratios"] = ['ratio_engaged_to_engaging_follower_counts', 'ratio_engaged_to_engaging_following_counts', ]
relevant_features["time"] = ['hashtags_frequency_05h', 'links_frequency_05h', 'domains_frequency_05h', 'hashtags_frequency_1h', 'links_frequency_1h', 'domains_frequency_1h', 'hashtags_frequency_2h', 'links_frequency_2h', 'domains_frequency_2h', 'hashtags_frequency_12h', 'links_frequency_12h', 'domains_frequency_12h', 'hashtags_frequency_24h', 'links_frequency_24h', 'domains_frequency_24h', 'hashtags_frequency_48h', 'links_frequency_48h', 'domains_frequency_48h', 'user_hashtags_frequency_05h', 'user_links_frequency_05h', 'user_domains_frequency_05h', 'user_hashtags_frequency_1h', 'user_links_frequency_1h', 'user_domains_frequency_1h', 'user_hashtags_frequency_2h', 'user_links_frequency_2h', 'user_domains_frequency_2h', 'user_hashtags_frequency_12h', 'user_links_frequency_12h', 'user_domains_frequency_12h', 'user_hashtags_frequency_24h', 'user_links_frequency_24h', 'user_domains_frequency_24h', 'user_hashtags_frequency_48h', 'user_links_frequency_48h', 'user_domains_frequency_48h', 'engaging_saw_tweets_count_05h', 'engaging_saw_tweets_count_1h', 'engaging_saw_tweets_count_2h', 'engaging_saw_tweets_count_12h', 'engaging_saw_tweets_count_24h', 'engaging_saw_tweets_count_48h',  'engageds_tweets_views_count_05h', 'engageds_tweets_views_count_1h', 'engageds_tweets_views_count_2h', 'engageds_tweets_views_count_12h', 'engageds_tweets_views_count_24h', 'engageds_tweets_views_count_48h', ]
relevant_features["engagement ratios"] = ['ratio_all_to_engaging_count_positive_tweets_like', 'ratio_all_to_engaging_count_positive_tweets_reply', 'ratio_all_to_engaging_count_positive_tweets_retweet', 'ratio_all_to_engaging_count_positive_tweets_quote', 'ratio_all_to_engaging_count_positive_tweets_react', 'ratio_all_to_engaging_count_negative_tweets_like', 'ratio_all_to_engaging_count_negative_tweets_reply', 'ratio_all_to_engaging_count_negative_tweets_retweet', 'ratio_all_to_engaging_count_negative_tweets_quote', 'ratio_all_to_engaging_count_negative_tweets_react', 'ratio_all_to_engaged_with_count_positive_tweets_like', 'ratio_all_to_engaged_with_count_positive_tweets_reply', 'ratio_all_to_engaged_with_count_positive_tweets_retweet', 'ratio_all_to_engaged_with_count_positive_tweets_quote', 'ratio_all_to_engaged_with_count_positive_tweets_react', 'ratio_all_to_engaged_with_count_negative_tweets_like', 'ratio_all_to_engaged_with_count_negative_tweets_reply', 'ratio_all_to_engaged_with_count_negative_tweets_retweet', 'ratio_all_to_engaged_with_count_negative_tweets_quote', 'ratio_all_to_engaged_with_count_negative_tweets_react', 'ratio_all_to_hashtags_count_positive_tweets_like', 'ratio_all_to_hashtags_count_positive_tweets_reply', 'ratio_all_to_hashtags_count_positive_tweets_retweet', 'ratio_all_to_hashtags_count_positive_tweets_quote', 'ratio_all_to_hashtags_count_positive_tweets_react', 'ratio_all_to_hashtags_count_negative_tweets_like', 'ratio_all_to_hashtags_count_negative_tweets_reply', 'ratio_all_to_hashtags_count_negative_tweets_retweet', 'ratio_all_to_hashtags_count_negative_tweets_quote', 'ratio_all_to_hashtags_count_negative_tweets_react', 'ratio_all_to_hashtags_user_proxy_count_positive_tweets_like', 'ratio_all_to_hashtags_user_proxy_count_positive_tweets_reply', 'ratio_all_to_hashtags_user_proxy_count_positive_tweets_retweet', 'ratio_all_to_hashtags_user_proxy_count_positive_tweets_quote', 'ratio_all_to_hashtags_user_proxy_count_positive_tweets_react', 'ratio_all_to_hashtags_user_proxy_count_negative_tweets_like', 'ratio_all_to_hashtags_user_proxy_count_negative_tweets_reply', 'ratio_all_to_hashtags_user_proxy_count_negative_tweets_retweet', 'ratio_all_to_hashtags_user_proxy_count_negative_tweets_quote', 'ratio_all_to_hashtags_user_proxy_count_negative_tweets_react', 
'ratio_all_to_links_count_positive_tweets_like', 'ratio_all_to_links_count_positive_tweets_reply', 'ratio_all_to_links_count_positive_tweets_retweet', 'ratio_all_to_links_count_positive_tweets_quote', 'ratio_all_to_links_count_positive_tweets_react', 'ratio_all_to_links_count_negative_tweets_like', 'ratio_all_to_links_count_negative_tweets_reply', 'ratio_all_to_links_count_negative_tweets_retweet', 'ratio_all_to_links_count_negative_tweets_quote', 'ratio_all_to_links_count_negative_tweets_react', 'ratio_all_to_links_user_proxy_count_positive_tweets_like', 'ratio_all_to_links_user_proxy_count_positive_tweets_reply', 'ratio_all_to_links_user_proxy_count_positive_tweets_retweet', 'ratio_all_to_links_user_proxy_count_positive_tweets_quote', 'ratio_all_to_links_user_proxy_count_positive_tweets_react', 'ratio_all_to_links_user_proxy_count_negative_tweets_like', 'ratio_all_to_links_user_proxy_count_negative_tweets_reply', 'ratio_all_to_links_user_proxy_count_negative_tweets_retweet', 'ratio_all_to_links_user_proxy_count_negative_tweets_quote', 'ratio_all_to_links_user_proxy_count_negative_tweets_react', 'ratio_all_to_domains_count_positive_tweets_like', 'ratio_all_to_domains_count_positive_tweets_reply', 'ratio_all_to_domains_count_positive_tweets_retweet', 'ratio_all_to_domains_count_positive_tweets_quote', 'ratio_all_to_domains_count_positive_tweets_react', 'ratio_all_to_domains_count_negative_tweets_like', 'ratio_all_to_domains_count_negative_tweets_reply', 'ratio_all_to_domains_count_negative_tweets_retweet', 'ratio_all_to_domains_count_negative_tweets_quote', 'ratio_all_to_domains_count_negative_tweets_react', 'ratio_all_to_domains_user_proxy_count_positive_tweets_like', 'ratio_all_to_domains_user_proxy_count_positive_tweets_reply', 'ratio_all_to_domains_user_proxy_count_positive_tweets_retweet', 'ratio_all_to_domains_user_proxy_count_positive_tweets_quote', 'ratio_all_to_domains_user_proxy_count_positive_tweets_react', 'ratio_all_to_domains_user_proxy_count_negative_tweets_like', 'ratio_all_to_domains_user_proxy_count_negative_tweets_reply', 'ratio_all_to_domains_user_proxy_count_negative_tweets_retweet', 'ratio_all_to_domains_user_proxy_count_negative_tweets_quote', 'ratio_all_to_domains_user_proxy_count_negative_tweets_react', ]
relevant_features["languages"] = ['this_language_seen_count', 'this_language_authored_count', ]
relevant_features["language ratios"] = ['ratio_seen_tweets_in_this_langauge_to_total_seen_tweets', 'ratio_authored_tweets_in_this_langauge_to_total_authored_tweets', ]
\end{lstlisting}

In addition, we singled out 8 ``oracle'' features, that may provide also provide information concerning events that happened after the potential interaction had already occurred. These features were described conceptually in subsection \ref{secOracleFeatures} and were led to a set of additional experiments. These features can be seen in listing \ref{lstOracleFeatures}.

\begin{lstlisting}[language=Python,caption={The list of 8 so-called ``Otacle'' fetures that provide insight in events after the tweet was seen as well}, label=lstOracleFeatures]
oracle_frequencies = ['hashtags_frequency', 'links_frequency', 'domains_frequency', 'user_hashtags_frequency', 'user_links_frequency', 'user_domains_frequency', 'engaging_saw_tweets_count', 'engageds_tweets_views_count', ]
\end{lstlisting}

The vectorised columns are named using the following scheme: \texttt{"ev\_\_"} \texttt{+} \texttt{top\-\_n\-\_select\-ed} \texttt{+} \texttt{"\_\_"} \texttt{+} \texttt{note} \texttt{+} \texttt{"\_\_"} \texttt{+} \texttt{target} \texttt{engagement}  \texttt{+} \texttt{"sdotd"}. Specifically, we created vectorised columns for all relevant explanatory features, one excluding and one including oracle features. We thus formed $4 \cdot 2 \cdot \cot 5 + 2 = 42$ vector columns for which classifications were to be done. The actual names of these columns can be seen in listing \ref{lstVectorised}. The abbreviation \texttt{sdotd} stands for ``Selection Done On Train Dataset'' and it signifies that feature selection was not done on train and val datasets, to prevent ``leaking'' of information about the engagements in those datasets.

\begin{lstlisting}[language=Python, caption={The complete list of vectorised columns to be used as input for classifiers}, label=lstVectorised]
['ev__top_5__scaled__like__sdotd',
 'ev__top_10__scaled__like__sdotd',
 'ev__top_25__scaled__like__sdotd',
 'ev__top_50__scaled__like__sdotd',
 'ev__all__scaled__like__sdotd',
 'ev__top_5__scaled__reply__sdotd',
 'ev__top_10__scaled__reply__sdotd',
 'ev__top_25__scaled__reply__sdotd',
 'ev__top_50__scaled__reply__sdotd',
 'ev__all__scaled__reply__sdotd',
 'ev__top_5__scaled__retweet__sdotd',
 'ev__top_10__scaled__retweet__sdotd',
 'ev__top_25__scaled__retweet__sdotd',
 'ev__top_50__scaled__retweet__sdotd',
 'ev__all__scaled__retweet__sdotd',
 'ev__top_5__scaled__quote__sdotd',
 'ev__top_10__scaled__quote__sdotd',
 'ev__top_25__scaled__quote__sdotd',
 'ev__top_50__scaled__quote__sdotd',
 'ev__all__scaled__quote__sdotd',
 'ev__top_5__scaled__react__sdotd',
 'ev__top_10__scaled__react__sdotd',
 'ev__top_25__scaled__react__sdotd',
 'ev__top_50__scaled__react__sdotd',
 'ev__all__scaled__react__sdotd',
 'ev__top_5__oracle_scaled__like__sdotd',
 'ev__top_10__oracle_scaled__like__sdotd',
 'ev__top_25__oracle_scaled__like__sdotd',
 'ev__top_50__oracle_scaled__like__sdotd',
 'ev__all__oracle_scaled__like__sdotd',
 'ev__top_5__oracle_scaled__reply__sdotd',
 'ev__top_10__oracle_scaled__reply__sdotd',
 'ev__top_25__oracle_scaled__reply__sdotd',
 'ev__top_50__oracle_scaled__reply__sdotd',
 'ev__all__oracle_scaled__reply__sdotd',
 'ev__top_5__oracle_scaled__retweet__sdotd',
 'ev__top_10__oracle_scaled__retweet__sdotd',
 'ev__top_25__oracle_scaled__retweet__sdotd',
 'ev__top_50__oracle_scaled__retweet__sdotd',
 'ev__all__oracle_scaled__retweet__sdotd',
 'ev__top_5__oracle_scaled__quote__sdotd',
 'ev__top_10__oracle_scaled__quote__sdotd',
 'ev__top_25__oracle_scaled__quote__sdotd',
 'ev__top_50__oracle_scaled__quote__sdotd',
 'ev__all__oracle_scaled__quote__sdotd',
 'ev__top_5__oracle_scaled__react__sdotd',
 'ev__top_10__oracle_scaled__react__sdotd',
 'ev__top_25__oracle_scaled__react__sdotd',
 'ev__top_50__oracle_scaled__react__sdotd',
 'ev__all__oracle_scaled__react__sdotd']
\end{lstlisting}

\chapter{Additional Results}
\label{appAdditionalResults}

This appendix contains most of the evaluation scores obtained during the realisation of this diploma thesis. Most informative scores for a number of significant combinations of data sampling methods, sets of features, and developed models are provided in chapter \ref{chEvaluation}.

\section{Baseline}
\label{appEvalSampleBaselineEcals}

The tables below contain the evaluation of the baseline decision tree and logistic regression models for samples of train and validation datasets. For an interpretation of these results, refer to section \ref{secEvalBaseline}.

The results for the full train and validation datasets using decision trees can be seen in tables \ref{tabBaselineDTfullAUC} and \ref{tabBaselineDTfullRCE} on page \pageref{tabBaselineDTfullRCE}.

\begin{table}[ht!]
    \centering
\begin{tabular}{lrrrrr}
\toprule
{} &   like &  reply &  retweet & quote &  react \\
\midrule
all\_scaled            &    0.572097 &     0.512918 &       0.549428 &                    0.503316 &            0.672774 \\
all\_binned            &    0.597840 &     0.512918 &       0.549428 &                    0.503316 &            0.691049 \\
engaged\_with          &    0.546069 &     0.512918 &       0.549428 &                    0.503316 &            0.671823 \\
engaging              &    0.582698 &     0.512918 &       0.549428 &                    0.503316 &            0.670157 \\
engaging+engaged      &    0.571157 &     0.512918 &       0.549428 &                    0.503316 &            0.671579 \\
timestamps            &    0.532687 &     0.512918 &       0.549428 &                    0.503316 &            0.591358 \\
media                 &    0.546630 &     0.512918 &       0.549428 &                    0.503316 &            0.718741 \\
dummy                 &    0.709752 &     0.512918 &       0.549428 &                    0.503316 &            0.740188 \\
\bottomrule
\end{tabular}
    \caption{PRAUC for full train and 1\% val datasets using baseline decision trees}
    \label{AUC_train_val1pc_DT}
\end{table}
\begin{table}[ht!]
    \centering
\begin{tabular}{lrrrrr}
\toprule
{} &   like &  reply &  retweet & quote &  react \\
\midrule
all\_scaled            &    0.579288 &     0.512973 &       0.549429 &                    0.503343 &            0.672614 \\
all\_binned            &    0.598403 &     0.512973 &       0.549429 &                    0.503343 &            0.690706 \\
engaged\_with          &    0.546692 &     0.512973 &       0.549429 &                    0.503343 &            0.672713 \\
engaging              &    0.566826 &     0.512973 &       0.549429 &                    0.503343 &            0.649200 \\
engaging+engaged      &    0.575057 &     0.512973 &       0.549429 &                    0.503343 &            0.664114 \\
timestamps            &    0.534134 &     0.512973 &       0.549429 &                    0.503343 &            0.590192 \\
media                 &    0.546840 &     0.512973 &       0.549429 &                    0.503343 &            0.719336 \\
dummy                 &    0.709970 &     0.512973 &       0.549429 &                    0.503343 &            0.740703 \\
\bottomrule
\end{tabular}
    \caption{PRAUC for full train and 5\% val datasets using baseline decision trees}
    \label{AUC_train_val5pc_DT}
\end{table}
\begin{table}[ht!]
    \centering
\begin{tabular}{lrrrrr}
\toprule
{} &   like &  reply &  retweet & quote &  react \\
\midrule
all\_scaled            &    0.578481 &     0.512858 &       0.549682 &                    0.503454 &            0.672206 \\
all\_binned            &    0.598095 &     0.512858 &       0.549682 &                    0.503454 &            0.690480 \\
engaged\_with          &    0.545419 &     0.512858 &       0.549682 &                    0.503454 &            0.672184 \\
engaging              &    0.565760 &     0.512858 &       0.549682 &                    0.503454 &            0.649040 \\
engaging+engaged      &    0.574432 &     0.512858 &       0.549682 &                    0.503454 &            0.663664 \\
timestamps            &    0.532203 &     0.512858 &       0.549682 &                    0.503454 &            0.589639 \\
media                 &    0.547546 &     0.512858 &       0.549682 &                    0.503454 &            0.719188 \\
dummy                 &    0.709577 &     0.512858 &       0.549682 &                    0.503454 &            0.740627 \\
\bottomrule
\end{tabular}
    \caption{PRAUC for 1\% train and full val datasets using baseline decision trees}
    \label{AUC_train1pc_val_DT}
\end{table}
\begin{table}[ht!]
    \centering
\begin{tabular}{lrrrrr}
\toprule
{} &   like &  reply &  retweet & quote &  react \\
\midrule
all\_scaled                &    0.561205 &     0.512918 &       0.549428 &                    0.503316 &            0.688314 \\
all\_binned                &    0.592029 &     0.512918 &       0.549428 &                    0.503316 &            0.674961 \\
engaged\_with              &    0.546662 &     0.512918 &       0.549428 &                    0.503316 &            0.670856 \\
engaging                  &    0.539514 &     0.512918 &       0.549428 &                    0.503316 &            0.620355 \\
engaging+engaged      &    0.546864 &     0.012918 &       0.549428 &                    0.503316 &            0.648033 \\
timestamps                &    0.481814 &     0.512918 &       0.549428 &                    0.503316 &            0.536718 \\
media                     &    0.545589 &     0.512918 &       0.549428 &                    0.503316 &            0.718740 \\
dummy                     &    0.709752 &     0.512918 &       0.549428 &                    0.503316 &            0.740188 \\
\bottomrule
\end{tabular}
    \caption{PRAUC for 1\% train and 1\% val datasets using baseline decision trees}
    \label{AUC_train1pc_val1pc_DT}
\end{table}

\begin{table}[ht!]
    \centering
\begin{tabular}{lrrrrr}
\toprule
{} &   like &  reply &  retweet & quote &  react \\
\midrule
all\_scaled                &    0.574171 &     0.512973 &       0.549429 &                    0.503343 &            0.691866 \\
all\_binned                &    0.592352 &     0.512973 &       0.549429 &                    0.503343 &            0.674141 \\
engaged\_with              &    0.547364 &     0.512973 &       0.549429 &                    0.503343 &            0.671434 \\
engaging                  &    0.549104 &     0.512973 &       0.549429 &                    0.503343 &            0.628080 \\
engaging+engaged          &    0.566216 &     0.012973 &       0.549429 &                    0.503343 &            0.654467 \\
timestamps                &    0.478768 &     0.512973 &       0.549429 &                    0.503343 &            0.535993 \\
media                     &    0.545813 &     0.512973 &       0.549429 &                    0.503343 &            0.719333 \\
dummy                     &    0.709970 &     0.512973 &       0.549429 &                    0.503343 &            0.740703 \\
\bottomrule
\end{tabular}
    \caption{PRAUC for 1\% train and 5\% val datasets using baseline decision trees}
    \label{AUC_train1pc_val5pc_DT}
\end{table}

\begin{table}[ht!]
    \centering
\begin{tabular}{lrrrrr}
\toprule
{} &   like &  reply &  retweet & quote &  react \\
\midrule
all\_scaled            &    0.578481 &     0.512858 &       0.549682 &                    0.503454 &            0.672206 \\
all\_binned            &    0.598095 &     0.512858 &       0.549682 &                    0.503454 &            0.690480 \\
engaged\_with          &    0.545419 &     0.512858 &       0.549682 &                    0.503454 &            0.672184 \\
engaging              &    0.565760 &     0.512858 &       0.549682 &                    0.503454 &            0.649040 \\
engaging+engaged      &    0.574432 &     0.512858 &       0.549682 &                    0.503454 &            0.663664 \\
timestamps            &    0.532203 &     0.512858 &       0.549682 &                    0.503454 &            0.589639 \\
media                 &    0.547546 &     0.512858 &       0.549682 &                    0.503454 &            0.719188 \\
dummy                 &    0.709577 &     0.512858 &       0.549682 &                    0.503454 &            0.740627 \\
\bottomrule
\end{tabular}
    \caption{PRAUC for 10\% train and full val datasets using baseline decision trees}
    \label{AUC_train10pc_val_DT}
\end{table}

\begin{table}[ht!]
    \centering
\begin{tabular}{lrrrrr}
\toprule
{} &   like &  reply &  retweet & quote &  react \\
\midrule
all\_scaled            &    0.572097 &     0.512918 &       0.549428 &                    0.503316 &            0.672774 \\
all\_binned            &    0.597840 &     0.512918 &       0.549428 &                    0.503316 &            0.691049 \\
engaged\_with          &    0.546069 &     0.512918 &       0.549428 &                    0.503316 &            0.671823 \\
engaging              &    0.582698 &     0.512918 &       0.549428 &                    0.503316 &            0.670157 \\
engaging+engaged      &    0.571157 &     0.512918 &       0.549428 &                    0.503316 &            0.671579 \\
timestamps            &    0.532687 &     0.512918 &       0.549428 &                    0.503316 &            0.591358 \\
media                 &    0.546630 &     0.512918 &       0.549428 &                    0.503316 &            0.718741 \\
dummy                 &    0.709752 &     0.512918 &       0.549428 &                    0.503316 &            0.740188 \\
\bottomrule
\end{tabular}
    \caption{PRAUC for 10\% train and 1\% val datasets using baseline decision trees}
    \label{AUC_train10pc_val1pc_DT}
\end{table}

\begin{table}[ht!]
    \centering
\begin{tabular}{lrrrrr}
\toprule
{} &   like &  reply &  retweet & quote &  react \\
\midrule
all\_scaled            &    0.579288 &     0.512973 &       0.549429 &                    0.503343 &            0.672614 \\
all\_binned            &    0.598403 &     0.512973 &       0.549429 &                    0.503343 &            0.690706 \\
engaged\_with          &    0.546692 &     0.512973 &       0.549429 &                    0.503343 &            0.672713 \\
engaging              &    0.566826 &     0.512973 &       0.549429 &                    0.503343 &            0.649200 \\
engaging+engaged      &    0.575057 &     0.512973 &       0.549429 &                    0.503343 &            0.664114 \\
timestamps            &    0.534134 &     0.512973 &       0.549429 &                    0.503343 &            0.590192 \\
media                 &    0.546840 &     0.512973 &       0.549429 &                    0.503343 &            0.719336 \\
dummy                 &    0.709970 &     0.512973 &       0.549429 &                    0.503343 &            0.740703 \\
\bottomrule
\end{tabular}
    \caption{PRAUC for 10\% train and 5\% val datasets using baseline decision trees}
    \label{AUC_train10pc_val5pc_DT}
\end{table}

\begin{table}[ht]
    \centering
\begin{tabular}{lrrrrr}
\toprule
{} &   like &  reply &  retweet & quote &  react \\
\midrule
all\_scaled            & -1909.807897 &   -643.89454 &    -958.513087 &                 -474.459107 &        -2031.358901 \\
all\_binned            & -1894.040826 &   -643.89454 &    -958.513087 &                 -474.459107 &        -1990.747616 \\
engaged\_with          & -1952.145345 &   -643.89454 &    -958.513087 &                 -474.459107 &        -2076.422364 \\
engaging              & -1885.569428 &   -643.89454 &    -958.513087 &                 -474.459107 &        -2357.017366 \\
engaging+engaged      & -1881.195452 &   -643.89454 &    -958.513087 &                 -474.459107 &        -2274.613484 \\
timestamps            & -1962.688519 &   -643.89454 &    -958.513087 &                 -474.459107 &        -2146.390652 \\
media                 & -1959.907845 &   -643.89454 &    -958.513087 &                 -474.459107 &        -2094.664070 \\
dummy                 & -2030.349025 &   -643.89454 &    -958.513087 &                 -474.459107 &        -2492.172275 \\
\bottomrule
\end{tabular}
    \caption{RCE for full train and 1\% val datasets using baseline decision trees}
    \label{RCE_train_val1pc_DT}
\end{table}
\begin{table}[ht]
    \centering
\begin{tabular}{lrrrrr}
\toprule
{} &   like &  reply &  retweet & quote &  react \\
\midrule
all\_scaled            & -1929.191264 &  -644.591509 &    -958.522579 &                 -475.223883 &        -2038.071814 \\
all\_binned            & -1893.724385 &  -644.591509 &    -958.522579 &                 -475.223883 &        -1996.291861 \\
engaged\_with          & -1952.983010 &  -644.591509 &    -958.522579 &                 -475.223883 &        -2073.355556 \\
engaging              & -1903.736659 &  -644.591509 &    -958.522579 &                 -475.223883 &        -2146.200750 \\
engaging+engaged      & -1909.308174 &  -644.591509 &    -958.522579 &                 -475.223883 &        -2087.569892 \\
timestamps            & -1962.335181 &  -644.591509 &    -958.522579 &                 -475.223883 &        -2156.464907 \\
media                 & -1961.381281 &  -644.591509 &    -958.522579 &                 -475.223883 &        -2092.568009 \\
dummy                 & -2032.120962 &  -644.591509 &    -958.522579 &                 -475.223883 &        -2486.742244 \\
\bottomrule
\end{tabular}
    \caption{RCE for full train and 5\% val datasets using baseline decision trees}
    \label{RCE_train_val5pc_DT}
\end{table}
\begin{table}[ht]
    \centering
\begin{tabular}{lrrrrr}
\toprule
{} &   like &  reply &  retweet & quote &  react \\
\midrule
all\_scaled            & -1928.887545 &  -643.145029 &    -960.266779 &                 -478.385215 &        -2039.375523 \\
all\_binned            & -1891.491721 &  -643.145029 &    -960.266779 &                 -478.385215 &        -1996.574991 \\
engaged\_with          & -1953.090152 &  -643.145029 &    -960.266779 &                 -478.385215 &        -2075.582714 \\
engaging              & -1903.638460 &  -643.145029 &    -960.266779 &                 -478.385215 &        -2147.797617 \\
engaging+engaged      & -1908.501838 &  -643.145029 &    -960.266779 &                 -478.385215 &        -2089.597918 \\
timestamps            & -1961.683423 &  -643.145029 &    -960.266779 &                 -478.385215 &        -2157.793471 \\
media                 & -1955.720727 &  -643.145029 &    -960.266779 &                 -478.385215 &        -2093.000447 \\
dummy                 & -2028.924259 &  -643.145029 &    -960.266779 &                 -478.385215 &        -2487.537127 \\
\bottomrule
\end{tabular}
    \caption{RCE for 1\% train and full val datasets using baseline decision trees}
    \label{RCE_train1pc_val_DT}
\end{table}
\begin{table}[ht]
    \centering
\begin{tabular}{lrrrrr}
\toprule
{} &   like &  reply &  retweet & quote &  react \\
\midrule
all\_scaled                & -1934.584566 &  -643.894540 &    -958.513087 &                 -474.459107 &        -2046.221096 \\
all\_binned                & -2020.221309 &  -643.894540 &    -958.513087 &                 -474.459107 &        -2044.653750 \\
engaged\_with              & -1952.344546 &  -643.894540 &    -958.513087 &                 -474.459107 &        -2073.245187 \\
engaging                  & -1952.389286 &  -643.894540 &    -958.513087 &                 -474.459107 &        -2078.069706 \\
engaging+engaged      & -1940.352382 &  -644.176645 &    -958.513087 &                 -474.459107 &        -2091.812337 \\
timestamps                & -2024.082382 &  -643.894540 &    -958.513087 &                 -474.459107 &        -2275.023279 \\
media                     & -1958.763144 &  -643.894540 &    -958.513087 &                 -474.459107 &        -2094.468567 \\
dummy                     & -2030.349025 &  -643.894540 &    -958.513087 &                 -474.459107 &        -2492.172275 \\
\bottomrule
\end{tabular}
    \caption{RCE for 1\% train and 1\% val datasets using baseline decision trees}
    \label{RCE_train1pc_val1pc_DT}
\end{table}

\begin{table}[ht]
    \centering
\begin{tabular}{lrrrrr}
\toprule
{} &   like &  reply &  retweet & quote &  react \\
\midrule
all\_scaled                & -1905.450174 &  -644.591509 &    -958.522579 &                 -475.223883 &        -2036.803703 \\
all\_binned                & -2019.067134 &  -644.591509 &    -958.522579 &                 -475.223883 &        -2053.134114 \\
engaged\_with              & -1952.577626 &  -644.591509 &    -958.522579 &                 -475.223883 &        -2072.247460 \\
engaging                  & -1933.355616 &  -644.591509 &    -958.522579 &                 -475.223883 &        -2088.803928 \\
engaging+engaged      & -1914.241810 &  -644.815132 &    -958.522579 &                 -475.223883 &        -2070.766058 \\
timestamps                & -2028.315283 &  -644.591509 &    -958.522579 &                 -475.223883 &        -2282.072061 \\
media                     & -1959.926922 &  -644.591509 &    -958.522579 &                 -475.223883 &        -2092.499988 \\
dummy                     & -2032.120962 &  -644.591509 &    -958.522579 &                 -475.223883 &        -2486.742244 \\
\bottomrule
\end{tabular}
    \caption{RCE for 1\% train and 5\% val datasets using baseline decision trees}
    \label{RCE_train1pc_val5pc_DT}
\end{table}

\begin{table}[ht]
    \centering
\begin{tabular}{lrrrrr}
\toprule
{} &   like &  reply &  retweet & quote &  react \\
\midrule
all\_scaled            & -1928.887545 &  -643.145029 &    -960.266779 &                 -478.385215 &        -2039.375523 \\
all\_binned            & -1891.491721 &  -643.145029 &    -960.266779 &                 -478.385215 &        -1996.574991 \\
engaged\_with          & -1953.090152 &  -643.145029 &    -960.266779 &                 -478.385215 &        -2075.582714 \\
engaging              & -1903.638460 &  -643.145029 &    -960.266779 &                 -478.385215 &        -2147.797617 \\
engaging+engaged      & -1908.501838 &  -643.145029 &    -960.266779 &                 -478.385215 &        -2089.597918 \\
timestamps            & -1961.683423 &  -643.145029 &    -960.266779 &                 -478.385215 &        -2157.793471 \\
media                 & -1955.720727 &  -643.145029 &    -960.266779 &                 -478.385215 &        -2093.000447 \\
dummy                 & -2028.924259 &  -643.145029 &    -960.266779 &                 -478.385215 &        -2487.537127 \\
\bottomrule
\end{tabular}
    \caption{RCE for 10\% train and full val datasets using baseline decision trees}
    \label{RCE_train10pc_val_DT}
\end{table}

\begin{table}[ht]
    \centering
\begin{tabular}{lrrrrr}
\toprule
{} &   like &  reply &  retweet & quote &  react \\
\midrule
all\_scaled            & -1909.807897 &   -643.89454 &    -958.513087 &                 -474.459107 &        -2031.358901 \\
all\_binned            & -1894.040826 &   -643.89454 &    -958.513087 &                 -474.459107 &        -1990.747616 \\
engaged\_with          & -1952.145345 &   -643.89454 &    -958.513087 &                 -474.459107 &        -2076.422364 \\
engaging              & -1885.569428 &   -643.89454 &    -958.513087 &                 -474.459107 &        -2357.017366 \\
engaging+engaged      & -1881.195452 &   -643.89454 &    -958.513087 &                 -474.459107 &        -2274.613484 \\
timestamps            & -1962.688519 &   -643.89454 &    -958.513087 &                 -474.459107 &        -2146.390652 \\
media                 & -1959.907845 &   -643.89454 &    -958.513087 &                 -474.459107 &        -2094.664070 \\
dummy                 & -2030.349025 &   -643.89454 &    -958.513087 &                 -474.459107 &        -2492.172275 \\
\bottomrule
\end{tabular}
    \caption{RCE for 10\% train and 1\% val datasets using baseline decision trees}
    \label{RCE_train10pc_val1pc_DT}
\end{table}

\begin{table}[ht]
    \centering
\begin{tabular}{lrrrrr}
\toprule
{} &   like &  reply &  retweet & quote &  react \\
\midrule
all\_scaled            & -1929.191264 &  -644.591509 &    -958.522579 &                 -475.223883 &        -2038.071814 \\
all\_binned            & -1893.724385 &  -644.591509 &    -958.522579 &                 -475.223883 &        -1996.291861 \\
engaged\_with          & -1952.983010 &  -644.591509 &    -958.522579 &                 -475.223883 &        -2073.355556 \\
engaging              & -1903.736659 &  -644.591509 &    -958.522579 &                 -475.223883 &        -2146.200750 \\
engaging+engaged      & -1909.308174 &  -644.591509 &    -958.522579 &                 -475.223883 &        -2087.569892 \\
timestamps            & -1962.335181 &  -644.591509 &    -958.522579 &                 -475.223883 &        -2156.464907 \\
media                 & -1961.381281 &  -644.591509 &    -958.522579 &                 -475.223883 &        -2092.568009 \\
dummy                 & -2032.120962 &  -644.591509 &    -958.522579 &                 -475.223883 &        -2486.742244 \\
\bottomrule
\end{tabular}
    \caption{RCE for 10\% train and 5\% val datasets using baseline decision trees}
    \label{RCE_train10pc_val5pc_DT}
\end{table}

The results for the full train and validation datasets using logistic regression can be seen in tables \ref{tabBaselineLRfullAUC} and \ref{tabBaselineLRfullRCE} on page \pageref{tabBaselineLRfullRCE}.

\begin{table}[htb]
    \centering
\begin{tabular}{lrrrrr}
\toprule
{} &   like &  reply &  retweet & quote &  react \\
\midrule
all\_scaled            &    0.587319 &     0.012918 &       0.549428 &                    0.503316 &            0.669374 \\
all\_binned            &    0.597034 &     0.012918 &       0.549428 &                    0.503316 &            0.690880 \\
engaged\_with          &    0.544407 &     0.512918 &       0.549428 &                    0.503316 &            0.652195 \\
engaging              &    0.709752 &     0.012918 &       0.216139 &                    0.503316 &            0.627378 \\
engaging+engaged      &    0.574112 &     0.012918 &       0.216139 &                    0.003316 &            0.653193 \\
timestamps            &    0.569988 &     0.512918 &       0.549428 &                    0.503316 &            0.655719 \\
media                 &    0.507106 &     0.512918 &       0.549428 &                    0.503316 &            0.729487 \\
dummy                 &    0.709752 &     0.512918 &       0.549428 &                    0.503316 &            0.740188 \\
\bottomrule
\end{tabular}
    \caption{PRAUC for full train and 1\% val datasets using baseline logistic regression}
    \label{AUC_train_val1pc_LR}
\end{table}
\begin{table}[htb]
    \centering
\begin{tabular}{lrrrrr}
\toprule
{} &   like &  reply &  retweet & quote &  react \\
\midrule
all\_scaled            &    0.584921 &     0.012973 &       0.549429 &                    0.003343 &            0.681205 \\
all\_binned            &    0.596048 &     0.084439 &       0.549429 &                    0.003343 &            0.692024 \\
engaged\_with          &    0.543499 &     0.512973 &       0.549429 &                    0.503343 &            0.652519 \\
engaging              &    0.397474 &     0.012973 &       0.140356 &                    0.003343 &            0.672783 \\
engaging+engaged      &    0.563845 &     0.012973 &       0.140356 &                    0.003343 &            0.657369 \\
timestamps            &    0.566717 &     0.512973 &       0.549429 &                    0.503343 &            0.656070 \\
media                 &    0.509447 &     0.512973 &       0.299438 &                    0.503343 &            0.730669 \\
dummy                 &    0.709970 &     0.512973 &       0.549429 &                    0.503343 &            0.740703 \\
\bottomrule
\end{tabular}
    \caption{PRAUC for full train and 5\% val datasets using baseline logistic regression}
    \label{AUC_train_val5pc_LR}
\end{table}
\begin{table}[htb]
    \centering
\begin{tabular}{lrrrrr}
\toprule
{} &   like &  reply &  retweet & quote &  react \\
\midrule
all\_scaled            &    0.584699 &     0.043355 &       0.549682 &                    0.023461 &            0.681477 \\
all\_binned            &    0.595620 &     0.039896 &       0.549682 &                    0.028461 &            0.692099 \\
engaged\_with          &    0.541828 &     0.512858 &       0.549682 &                    0.503454 &            0.651735 \\
engaging              &    0.351472 &     0.039176 &       0.116363 &                    0.003454 &            0.673105 \\
engaging+engaged      &    0.561851 &     0.062860 &       0.115547 &                    0.003454 &            0.657345 \\
timestamps            &    0.565128 &     0.512858 &       0.549682 &                    0.503454 &            0.655918 \\
media                 &    0.509485 &     0.512858 &       0.121113 &                    0.503454 &            0.730702 \\
dummy                 &    0.709577 &     0.512858 &       0.549682 &                    0.503454 &            0.740627 \\
\bottomrule
\end{tabular}
    \caption{PRAUC for 1\% train and full val datasets using baseline logistic regression}
    \label{AUC_train1pc_val_LR}
\end{table}
\begin{table}[htb]
    \centering
\begin{tabular}{lrrrrr}
\toprule
{} &   like &  reply &  retweet & quote &  react \\
\midrule
all\_scaled            &    0.587319 &     0.012918 &       0.549428 &                    0.503316 &            0.669374 \\
all\_binned            &    0.597034 &     0.012918 &       0.549428 &                    0.503316 &            0.690880 \\
engaged\_with          &    0.544407 &     0.512918 &       0.549428 &                    0.503316 &            0.652195 \\
engaging              &    0.709752 &     0.012918 &       0.216139 &                    0.503316 &            0.627378 \\
engaging+engaged      &    0.574112 &     0.012918 &       0.216139 &                    0.003316 &            0.653193 \\
timestamps            &    0.569988 &     0.512918 &       0.549428 &                    0.503316 &            0.655719 \\
media                 &    0.507106 &     0.512918 &       0.549428 &                    0.503316 &            0.729487 \\
dummy                 &    0.709752 &     0.512918 &       0.549428 &                    0.503316 &            0.740188 \\
\bottomrule
\end{tabular}
    \caption{PRAUC for 1\% train and 1\% val datasets using baseline logistic regression}
    \label{AUC_train1pc_val1pc_LR}
\end{table}

\begin{table}[htb]
    \centering
\begin{tabular}{lrrrrr}
\toprule
{} &   like &  reply &  retweet & quote &  react \\
\midrule
all\_scaled            &    0.584921 &     0.012973 &       0.549429 &                    0.003343 &            0.681205 \\
all\_binned            &    0.596048 &     0.084439 &       0.549429 &                    0.003343 &            0.692024 \\
engaged\_with          &    0.543499 &     0.512973 &       0.549429 &                    0.503343 &            0.652519 \\
engaging              &    0.397474 &     0.012973 &       0.140356 &                    0.003343 &            0.672783 \\
engaging+engaged      &    0.563845 &     0.012973 &       0.140356 &                    0.003343 &            0.657369 \\
timestamps            &    0.566717 &     0.512973 &       0.549429 &                    0.503343 &            0.656070 \\
media                 &    0.509447 &     0.512973 &       0.299438 &                    0.503343 &            0.730669 \\
dummy                 &    0.709970 &     0.512973 &       0.549429 &                    0.503343 &            0.740703 \\
\bottomrule
\end{tabular}
    \caption{PRAUC for 1\% train and 5\% val datasets using baseline logistic regression}
    \label{AUC_train1pc_val5pc_LR}
\end{table}

\begin{table}[htb]
    \centering
\begin{tabular}{lrrrrr}
\toprule
{} &   like &  reply &  retweet & quote &  react \\
\midrule
all\_scaled            &    0.584699 &     0.043355 &       0.549682 &                    0.023461 &            0.681477 \\
all\_binned            &    0.595620 &     0.039896 &       0.549682 &                    0.028461 &            0.692099 \\
engaged\_with          &    0.541828 &     0.512858 &       0.549682 &                    0.503454 &            0.651735 \\
engaging              &    0.351472 &     0.039176 &       0.116363 &                    0.003454 &            0.673105 \\
engaging+engaged      &    0.561851 &     0.062860 &       0.115547 &                    0.003454 &            0.657345 \\
timestamps            &    0.565128 &     0.512858 &       0.549682 &                    0.503454 &            0.655918 \\
media                 &    0.509485 &     0.512858 &       0.121113 &                    0.503454 &            0.730702 \\
dummy                 &    0.709577 &     0.512858 &       0.549682 &                    0.503454 &            0.740627 \\
\bottomrule
\end{tabular}
    \caption{PRAUC for 10\% train and full val datasets using baseline logistic regression}
    \label{AUC_train10pc_val_LR}
\end{table}

\begin{table}[htb]
    \centering
\begin{tabular}{lrrrrr}
\toprule
{} &   like &  reply &  retweet & quote &  react \\
\midrule
all\_scaled            &    0.587319 &     0.012918 &       0.549428 &                    0.503316 &            0.669374 \\
all\_binned            &    0.597034 &     0.012918 &       0.549428 &                    0.503316 &            0.690880 \\
engaged\_with          &    0.544407 &     0.512918 &       0.549428 &                    0.503316 &            0.652195 \\
engaging              &    0.709752 &     0.012918 &       0.216139 &                    0.503316 &            0.627378 \\
engaging+engaged      &    0.574112 &     0.012918 &       0.216139 &                    0.003316 &            0.653193 \\
timestamps            &    0.569988 &     0.512918 &       0.549428 &                    0.503316 &            0.655719 \\
media                 &    0.507106 &     0.512918 &       0.549428 &                    0.503316 &            0.729487 \\
dummy                 &    0.709752 &     0.512918 &       0.549428 &                    0.503316 &            0.740188 \\
\bottomrule
\end{tabular}
    \caption{PRAUC for 10\% train and 1\% val datasets using baseline logistic regression}
    \label{AUC_train10pc_val1pc_LR}
\end{table}

\begin{table}[htb]
    \centering
\begin{tabular}{lrrrrr}
\toprule
{} &   like &  reply &  retweet & quote &  react \\
\midrule
all\_scaled            &    0.584921 &     0.012973 &       0.549429 &                    0.003343 &            0.681205 \\
all\_binned            &    0.596048 &     0.084439 &       0.549429 &                    0.003343 &            0.692024 \\
engaged\_with          &    0.543499 &     0.512973 &       0.549429 &                    0.503343 &            0.652519 \\
engaging              &    0.397474 &     0.012973 &       0.140356 &                    0.003343 &            0.672783 \\
engaging+engaged      &    0.563845 &     0.012973 &       0.140356 &                    0.003343 &            0.657369 \\
timestamps            &    0.566717 &     0.512973 &       0.549429 &                    0.503343 &            0.656070 \\
media                 &    0.509447 &     0.512973 &       0.299438 &                    0.503343 &            0.730669 \\
dummy                 &    0.709970 &     0.512973 &       0.549429 &                    0.503343 &            0.740703 \\
\bottomrule
\end{tabular}
    \caption{PRAUC for 10\% train and 5\% val datasets using baseline logistic regression}
    \label{AUC_train10pc_val5pc_LR}
\end{table}

\begin{table}[htb]
    \centering
\begin{tabular}{lrrrrr}
\toprule
{} &   like &  reply &  retweet & quote &  react \\
\midrule
all\_scaled            & -1849.101815 &  -644.458750 &    -958.513087 &                 -474.459107 &        -1865.322461 \\
all\_binned            & -1835.625680 &  -644.458750 &    -958.513087 &                 -474.459107 &        -1921.685641 \\
engaged\_with          & -1954.533321 &  -643.894540 &    -958.513087 &                 -474.459107 &        -2059.457300 \\
engaging              & -2030.349025 &  -644.176645 &    -958.617999 &                 -474.459107 &        -2198.236166 \\
engaging+engaged      & -1882.137873 &  -644.176645 &    -958.617999 &                 -475.307663 &        -1984.871686 \\
timestamps            & -1883.135274 &  -643.894540 &    -958.513087 &                 -474.459107 &        -2007.453648 \\
media                 & -2000.350857 &  -643.894540 &    -958.513087 &                 -474.459107 &        -2146.621173 \\
dummy                 & -2030.349025 &  -643.894540 &    -958.513087 &                 -474.459107 &        -2492.172275 \\
\bottomrule
\end{tabular}
    \caption{RCE for full train and 1\% val datasets using baseline logistic regression}
    \label{RCE_train_val1pc_LR}
\end{table}
\begin{table}[htb]
    \centering
\begin{tabular}{lrrrrr}
\toprule
{} &   like &  reply &  retweet & quote &  react \\
\midrule
all\_scaled            & -1854.025266 &  -644.871038 &    -958.522579 &                 -475.559103 &        -1917.801828 \\
all\_binned            & -1841.347719 &  -644.871039 &    -958.522579 &                 -475.559103 &        -1916.570917 \\
engaged\_with          & -1959.975772 &  -644.591509 &    -958.522579 &                 -475.223883 &        -2060.941221 \\
engaging              & -2032.140745 &  -644.647415 &    -958.668598 &                 -475.391493 &        -1979.330203 \\
engaging+engaged      & -1900.529812 &  -644.647415 &    -958.668598 &                 -475.391493 &        -2016.426955 \\
timestamps            & -1893.645839 &  -644.591509 &    -958.522579 &                 -475.223883 &        -2011.188845 \\
media                 & -1999.662252 &  -644.591509 &    -958.522579 &                 -475.223883 &        -2137.299100 \\
dummy                 & -2032.120962 &  -644.591509 &    -958.522579 &                 -475.223883 &        -2486.742244 \\
\bottomrule
\end{tabular}
    \caption{RCE for full train and 5\% val datasets using baseline logistic regression}
    \label{RCE_train_val5pc_LR}
\end{table}
\begin{table}[htb]
    \centering
\begin{tabular}{lrrrrr}
\toprule
{} &   like &  reply &  retweet & quote &  react \\
\midrule
all\_scaled            & -1851.173521 &  -643.347603 &    -960.266779 &                 -478.572703 &        -1916.529666 \\
all\_binned            & -1839.933656 &  -643.423569 &    -960.266779 &                 -478.531945 &        -1916.004898 \\
engaged\_with          & -1960.959541 &  -643.145029 &    -960.266779 &                 -478.385215 &        -2064.192775 \\
engaging              & -2028.955908 &  -643.192859 &    -960.449628 &                 -478.466732 &        -1978.915417 \\
engaging+engaged      & -1901.839232 &  -643.167537 &    -960.423656 &                 -478.442277 &        -2016.205784 \\
timestamps            & -1893.985138 &  -643.145029 &    -960.266779 &                 -478.385215 &        -2011.660096 \\
media                 & -1996.038670 &  -643.145029 &    -960.292752 &                 -478.385215 &        -2137.726157 \\
dummy                 & -2028.924259 &  -643.145029 &    -960.266779 &                 -478.385215 &        -2487.537127 \\
\bottomrule
\end{tabular}
    \caption{RCE for 1\% train and full val datasets using baseline logistic regression}
    \label{RCE_train1pc_val_LR}
\end{table}
\begin{table}[htb]
    \centering
\begin{tabular}{lrrrrr}
\toprule
{} &   like &  reply &  retweet & quote &  react \\
\midrule
all\_scaled            & -1849.101815 &  -644.458750 &    -958.513087 &                 -474.459107 &        -1865.322461 \\
all\_binned            & -1835.625680 &  -644.458750 &    -958.513087 &                 -474.459107 &        -1921.685641 \\
engaged\_with          & -1954.533321 &  -643.894540 &    -958.513087 &                 -474.459107 &        -2059.457300 \\
engaging              & -2030.349025 &  -644.176645 &    -958.617999 &                 -474.459107 &        -2198.236166 \\
engaging+engaged      & -1882.137873 &  -644.176645 &    -958.617999 &                 -475.307663 &        -1984.871686 \\
timestamps            & -1883.135274 &  -643.894540 &    -958.513087 &                 -474.459107 &        -2007.453648 \\
media                 & -2000.350857 &  -643.894540 &    -958.513087 &                 -474.459107 &        -2146.621173 \\
dummy                 & -2030.349025 &  -643.894540 &    -958.513087 &                 -474.459107 &        -2492.172275 \\
\bottomrule
\end{tabular}
    \caption{RCE for 1\% train and 1\% val datasets using baseline logistic regression}
    \label{RCE_train1pc_val1pc_LR}
\end{table}

\begin{table}[htb]
    \centering
\begin{tabular}{lrrrrr}
\toprule
{} &   like &  reply &  retweet & quote &  react \\
\midrule
all\_scaled            & -1854.025266 &  -644.871038 &    -958.522579 &                 -475.559103 &        -1917.801828 \\
all\_binned            & -1841.347719 &  -644.871039 &    -958.522579 &                 -475.559103 &        -1916.570917 \\
engaged\_with          & -1959.975772 &  -644.591509 &    -958.522579 &                 -475.223883 &        -2060.941221 \\
engaging              & -2032.140745 &  -644.647415 &    -958.668598 &                 -475.391493 &        -1979.330203 \\
engaging+engaged      & -1900.529812 &  -644.647415 &    -958.668598 &                 -475.391493 &        -2016.426955 \\
timestamps            & -1893.645839 &  -644.591509 &    -958.522579 &                 -475.223883 &        -2011.188845 \\
media                 & -1999.662252 &  -644.591509 &    -958.522579 &                 -475.223883 &        -2137.299100 \\
dummy                 & -2032.120962 &  -644.591509 &    -958.522579 &                 -475.223883 &        -2486.742244 \\
\bottomrule
\end{tabular}
    \caption{RCE for 1\% train and 5\% val datasets using baseline logistic regression}
    \label{RCE_train1pc_val5pc_LR}
\end{table}

\begin{table}[htb]
    \centering
\begin{tabular}{lrrrrr}
\toprule
{} &   like &  reply &  retweet & quote &  react \\
\midrule
all\_scaled            & -1851.173521 &  -643.347603 &    -960.266779 &                 -478.572703 &        -1916.529666 \\
all\_binned            & -1839.933656 &  -643.423569 &    -960.266779 &                 -478.531945 &        -1916.004898 \\
engaged\_with          & -1960.959541 &  -643.145029 &    -960.266779 &                 -478.385215 &        -2064.192775 \\
engaging              & -2028.955908 &  -643.192859 &    -960.449628 &                 -478.466732 &        -1978.915417 \\
engaging+engaged      & -1901.839232 &  -643.167537 &    -960.423656 &                 -478.442277 &        -2016.205784 \\
timestamps            & -1893.985138 &  -643.145029 &    -960.266779 &                 -478.385215 &        -2011.660096 \\
media                 & -1996.038670 &  -643.145029 &    -960.292752 &                 -478.385215 &        -2137.726157 \\
dummy                 & -2028.924259 &  -643.145029 &    -960.266779 &                 -478.385215 &        -2487.537127 \\
\bottomrule
\end{tabular}
    \caption{RCE for 10\% train and full val datasets using baseline logistic regression}
    \label{RCE_train10pc_val_LR}
\end{table}

\begin{table}[htb]
    \centering
\begin{tabular}{lrrrrr}
\toprule
{} &   like &  reply &  retweet & quote &  react \\
\midrule
all\_scaled            & -1849.101815 &  -644.458750 &    -958.513087 &                 -474.459107 &        -1865.322461 \\
all\_binned            & -1835.625680 &  -644.458750 &    -958.513087 &                 -474.459107 &        -1921.685641 \\
engaged\_with          & -1954.533321 &  -643.894540 &    -958.513087 &                 -474.459107 &        -2059.457300 \\
engaging              & -2030.349025 &  -644.176645 &    -958.617999 &                 -474.459107 &        -2198.236166 \\
engaging+engaged      & -1882.137873 &  -644.176645 &    -958.617999 &                 -475.307663 &        -1984.871686 \\
timestamps            & -1883.135274 &  -643.894540 &    -958.513087 &                 -474.459107 &        -2007.453648 \\
media                 & -2000.350857 &  -643.894540 &    -958.513087 &                 -474.459107 &        -2146.621173 \\
dummy                 & -2030.349025 &  -643.894540 &    -958.513087 &                 -474.459107 &        -2492.172275 \\
\bottomrule
\end{tabular}
    \caption{RCE for 10\% train and 1\% val datasets using baseline logistic regression}
    \label{RCE_train10pc_val1pc_LR}
\end{table}

\begin{table}[ht]
    \centering
\begin{tabular}{lrrrrr}
\toprule
{} &   like &  reply &  retweet & quote &  react \\
\midrule
all\_scaled            & -1854.025266 &  -644.871038 &    -958.522579 &                 -475.559103 &        -1917.801828 \\
all\_binned            & -1841.347719 &  -644.871039 &    -958.522579 &                 -475.559103 &        -1916.570917 \\
engaged\_with          & -1959.975772 &  -644.591509 &    -958.522579 &                 -475.223883 &        -2060.941221 \\
engaging              & -2032.140745 &  -644.647415 &    -958.668598 &                 -475.391493 &        -1979.330203 \\
engaging+engaged      & -1900.529812 &  -644.647415 &    -958.668598 &                 -475.391493 &        -2016.426955 \\
timestamps            & -1893.645839 &  -644.591509 &    -958.522579 &                 -475.223883 &        -2011.188845 \\
media                 & -1999.662252 &  -644.591509 &    -958.522579 &                 -475.223883 &        -2137.299100 \\
dummy                 & -2032.120962 &  -644.591509 &    -958.522579 &                 -475.223883 &        -2486.742244 \\
\bottomrule
\end{tabular}
    \caption{RCE for 10\% train and 5\% val datasets using baseline logistic regression}
    \label{RCE_train10pc_val5pc_LR}
\end{table}

\section{Main Results}
\label{appMainResults}

The tables below contain additional results from the main round of evaluation.

\subsection{Individual Factors' Statistics}
\label{appIndividualFactorsStatistics}

This appendix includes additional tables with results on different factors. For more details, consider section \ref{secIndividualFactorsStatistics}.

\begin{longtable}[ht]{llrrrrr}
\caption{Statistics for RCE of common results for feature\_selection; there are 37845 such RCE results in total, i.e. 7569 instances per group.} \label{tabDescStatisticfeatureselectionRCE} \\
\toprule
 & feature\_selection & all & top\_10 & top\_25 & top\_5 & top\_50 \\
target & statistic &  &  &  &  &  \\
\midrule
\endfirsthead
\caption[]{Statistics for RCE of common results for feature\_selection; there are 37845 such RCE results in total, i.e. 7569 instances per group.} \\
\toprule
 & feature\_selection & all & top\_10 & top\_25 & top\_5 & top\_50 \\
target & statistic &  &  &  &  &  \\
\midrule
\endhead
\midrule
\multicolumn{7}{r}{Continued on next page} \\
\midrule
\endfoot
\bottomrule
\endlastfoot
\multirow[c]{4}{*}{like} & max & -1164.2 & -1419.9 & -1341.5 & -1615.3 & \textbf{-1022.9} \\
 & mean & -2050.1 & -1994.3 & -2133.4 & \textbf{-1985.2} & -2100.3 \\
 & min & -2974.9 & -2856.6 & -2927.8 & \textbf{-2818.2} & -2906.4 \\
 & std & \textbf{410.0} & 244.8 & 374.9 & 198.9 & 353.6 \\
\multirow[c]{4}{*}{quote} & max & \textbf{-465.0} & -471.6 & -471.6 & -471.6 & -470.6 \\
 & mean & \textbf{-661.0} & -1778.9 & -9391.1 & -1871.7 & -9630.1 \\
 & min & \textbf{-37876.0} & -85935.9 & -88091.3 & -85826.5 & -88062.9 \\
 & std & 2511.3 & 8256.5 & \textbf{23587.4} & 10362.3 & 23459.9 \\
\end{longtable}

\begin{longtable}[ht]{llrr}
\caption{Statistics for RCE of common results for note; there are 5182 such RCE results in total, i.e. 2591 instances per group.} \label{tabDescStatisticnoteRCE} \\
\toprule
 & note & oracle\_scaled & scaled \\
target & statistic &  &  \\
\midrule
\endfirsthead
\caption[]{Statistics for RCE of common results for note; there are 5182 such RCE results in total, i.e. 2591 instances per group.} \\
\toprule
 & note & oracle\_scaled & scaled \\
target & statistic &  &  \\
\midrule
\endhead
\midrule
\multicolumn{4}{r}{Continued on next page} \\
\midrule
\endfoot
\bottomrule
\endlastfoot
\multirow[c]{4}{*}{like} & max & \textbf{-1162.4} & -1164.2 \\
 & mean & \textbf{-2094.0} & -2214.5 \\
 & min & -2895.9 & \textbf{-2883.3} \\
 & std & 410.5 & \textbf{459.0} \\
\multirow[c]{4}{*}{quote} & max & -465.9 & \textbf{-465.0} \\
 & mean & \textbf{-476.8} & -477.5 \\
 & min & \textbf{-494.3} & -606.5 \\
 & std & 4.0 & \textbf{10.3} \\
\multirow[c]{4}{*}{react} & max & \textbf{-1133.4} & -1153.2 \\
 & mean & \textbf{-2078.7} & -2111.6 \\
 & min & -2518.3 & \textbf{-2473.5} \\
 & std & \textbf{311.8} & 277.5 \\
\multirow[c]{4}{*}{reply} & max & \textbf{-635.8} & \textbf{-635.8} \\
 & mean & -644.2 & \textbf{-643.8} \\
 & min & \textbf{-703.9} & -780.3 \\
 & std & 6.9 & \textbf{7.4} \\
\multirow[c]{4}{*}{retweet} & max & \textbf{-861.3} & -862.3 \\
 & mean & -1992.8 & \textbf{-1955.4} \\
 & min & \textbf{-9521.8} & -9695.3 \\
 & std & 2451.9 & \textbf{2645.2} \\
\end{longtable}

\subsection{Best Achieved Results}
\label{appBestAchievedResults}

This appendix includes additional tables with best-achieved results over two more sets of constraints. For the main tables, cf. section \ref{secSignificanceTestingforIndividualFactors}.

\begingroup 
\setlength\tabcolsep{2pt}
\footnotesize

\setlength\LTcapwidth{\textwidth} 

\setlength\LTleft{0pt}            
\setlength\LTright{0pt}           

\begin{longtable}[ht]{@{\extracolsep{\fill}}llllllr}
\caption{Best PRAUC achieved when evaluated on full val+test dataset} \label{tabBestResultsevaluated on full val+testPRAUC} \\
\toprule
target & algorithm & note & fs & trained & evaluated & PRAUC \\
\midrule
\endfirsthead
\caption[]{Best PRAUC achieved when evaluated on full val+test dataset} \\
\caption[]{Best PRAUC achieved when evaluated on full val+test dataset} \\
\toprule
target & algorithm & note & fs & trained & evaluated & PRAUC \\
\midrule
\endhead
\midrule
\multicolumn{7}{r}{Continued on next page} \\
\midrule
\endfoot
\bottomrule
\endlastfoot
like & GradientBoosting & scaled & all & train\_EU\_sample\_5pct & val+test & 0.7273 \\
reply & forest & scaled & all & train\_EU\_sample\_10pct & val+test & 0.5129 \\
retweet & bayes & scaled & all & train\_EU\_sample\_1pct & val+test & 0.5497 \\
quote & bayes & scaled & top\_10 & train\_EU\_sample\_10pct & val+test & 0.5035 \\
react & GradientBoosting & scaled & all & train\_EU\_sample\_5pct & val+test & 0.7623 \\
\end{longtable}
\endgroup

\begingroup 
\setlength\tabcolsep{2pt}
\footnotesize

\setlength\LTcapwidth{\textwidth} 

\setlength\LTleft{0pt}            
\setlength\LTright{0pt}           

\begin{longtable}[ht]{@{\extracolsep{\fill}}llllllr}
\caption{Best RCE achieved when evaluated on full val+test dataset} \label{tabBestResultsevaluated on full val+testRCE} \\
\toprule
target & algorithm & note & fs & trained & evaluated & RCE \\
\midrule
\endfirsthead
\caption[]{Best RCE achieved when evaluated on full val+test dataset} \\
\toprule
target & algorithm & note & fs & trained & evaluated & RCE \\
\midrule
\endhead
\midrule
\multicolumn{7}{r}{Continued on next page} \\
\midrule
\endfoot
\bottomrule
\endlastfoot
like & GradientBoosting & scaled & all & train\_EU\_sample\_5pct & val+test & -1418.2 \\
reply & GradientBoosting & scaled & all & train\_EU\_sample\_2pct & val+test & -643.5 \\
retweet & GradientBoosting & scaled & all & train\_EU\_sample\_5pct & val+test & -865.2 \\
quote & bayes & scaled & top\_10 & train\_EU\_sample\_10pct & val+test & -478.4 \\
react & forest & scaled & all & train\_EWU\_sample\_5pct & val+test & -1450.4 \\
\end{longtable}
\endgroup

\begingroup 
\setlength\tabcolsep{2pt}
\footnotesize

\setlength\LTcapwidth{\textwidth} 

\setlength\LTleft{0pt}            
\setlength\LTright{0pt}           

  \begin{longtable}[ht]{@{\extracolsep{\fill}}llllllr}
\caption{Best PRAUC achieved when trained on a train subset and evaluated on a val+test subset} \label{tabBestResultstrained on a train subset and evaluated on a val+test subsetPRAUC} \\
\toprule
target & algorithm & note & fs & trained & evaluated & PRAUC \\
\midrule
\endfirsthead
\caption[]{Best PRAUC achieved when trained on a train subset and evaluated on a val+test subset} \\
\toprule
target & algorithm & note & fs & trained & evaluated & PRAUC \\
\midrule
\endhead
\midrule
\multicolumn{7}{r}{Continued on next page} \\
\midrule
\endfoot
\bottomrule
\endlastfoot
like & GradientBoosting & scaled & all & train\_EU\_sample\_5pct & val+test & 0.7273 \\
reply & GradientBoosting & scaled & all & train\_EWU\_sample\_2pct & val+test\_random\_sample\_1pct & 0.5132 \\
retweet & GradientBoosting & scaled & top\_25 & train\_inter\_EWU+EU\_sample\_1pct & val+test\_EWU\_sample\_1pct & 0.5505 \\
quote & bayes & scaled & top\_10 & train\_EU\_sample\_10pct & val+test\_EU\_sample\_10pct & 0.5035 \\
react & GradientBoosting & scaled & all & train\_EU\_sample\_5pct & val+test & 0.7623 \\
\end{longtable}
\endgroup

\begingroup 
\setlength\tabcolsep{2pt}
\footnotesize

\setlength\LTcapwidth{\textwidth} 

\setlength\LTleft{0pt}            
\setlength\LTright{0pt}           

  \begin{longtable}[ht]{@{\extracolsep{\fill}}llllllr}
\caption{Best RCE achieved when trained on a train subset and evaluated on a val+test subset} \label{tabBestResultstrained on a train subset and evaluated on a val+test subsetRCE} \\
\toprule
target & algorithm & note & fs & trained & evaluated & RCE \\
\midrule
\endfirsthead
\caption[]{Best RCE achieved when trained on a train subset and evaluated on a val+test subset} \\
\toprule
target & algorithm & note & fs & trained & evaluated & RCE \\
\midrule
\endhead
\midrule
\multicolumn{7}{r}{Continued on next page} \\
\midrule
\endfoot
\bottomrule
\endlastfoot
like & GradientBoosting & scaled & all & train\_EU\_sample\_5pct & val+test & -1418.2 \\
reply & GradientBoosting & scaled & top\_25 & train\_EU\_sample\_2pct & val+test\_EU\_sample\_1pct & -639.0 \\
retweet & GradientBoosting & scaled & all & train\_EU\_sample\_5pct & val+test & -865.2 \\
quote & GradientBoosting & scaled & top\_10 & train\_EU\_sample\_2pct & val+test\_random\_sample\_1pct & -474.9 \\
react & forest & scaled & all & train\_EWU\_sample\_5pct & val+test & -1450.4 \\
\end{longtable}
\endgroup

\section{Statistics}
\label{appStatistics}

This appendix provides tables with additional statistics and summative overviews of the achieved results.

\subsection{Further Feature Selection}
\label{appFurtherFeatureSelection}

This subsection, in tables \ref{tabOverview of the 25 most informative feature as selected by ChiSq-selector on the train dataset and its 20 subsets} and \ref{tabOverview of the 50 most informative feature as selected by ChiSq-selector on the train dataset and its 20 subsets} respectively, provides the 25 and the 50 most informative features from the train dataset and its 20 created subsets. For an analysis of the 5 and 10 most informative features selected, consider subsection \ref{secMostSignificantContextualFeatures}.

\begin{longtable}[ht]{llr}
\caption{Overview of the 25 most informative feature as selected by ChiSq-selector on the train dataset and its 20 subsets} \label{tabOverview of the 25 most informative feature as selected by ChiSq-selector on the train dataset and its 20 subsets} \\
\toprule
 &  & count \\
target & selected feature &  \\
\midrule
\endfirsthead
\caption[]{Overview of the 25 most informative feature as selected by ChiSq-selector on the train dataset and its 20 subsets} \\
\toprule
 &  & count \\
target & selected feature &  \\
\midrule
\endhead
\midrule
\multicolumn{3}{r}{Continued on next page} \\
\midrule
\endfoot
\bottomrule
\endlastfoot
\multirow[c]{27}{*}{like} & language\_indexed & 21 \\
 & tweet\_type\_indexed & 21 \\
 & user\_domains\_frequency\_12h\_binned & 21 \\
 & ratio\_all\_to\_engaged\_with\_count\_negative\_tweets\_reply\_binned & 21 \\
 & user\_hashtags\_frequency\_1h\_binned & 21 \\
 & hashtags\_frequency\_1h\_binned & 21 \\
 & domains\_frequency\_48h\_binned & 21 \\
 & hashtags\_frequency\_12h\_binned & 21 \\
\end{longtable}

\begin{longtable}[ht]{llr}
\caption{Overview of the 50 most informative feature as selected by ChiSq-selector on the train dataset and its 20 subsets} \label{tabOverview of the 50 most informative feature as selected by ChiSq-selector on the train dataset and its 20 subsets} \\
\toprule
 &  & count \\
target & selected feature &  \\
\midrule
\endfirsthead
\caption[]{Overview of the 50 most informative feature as selected by ChiSq-selector on the train dataset and its 20 subsets} \\
\toprule
 &  & count \\
target & selected feature &  \\
\midrule
\endhead
\midrule
\multicolumn{3}{r}{Continued on next page} \\
\midrule
\endfoot
\bottomrule
\endlastfoot
\hline \multirow[c]{52}{*}{like} & language\_indexed & 21 \\
 & tweet\_type\_indexed & 21 \\
 & user\_domains\_frequency\_12h\_binned & 21 \\
 & ratio\_all\_to\_engaged\_with\_count\_negative\_tweets\_reply\_binned & 21 \\
 & user\_hashtags\_frequency\_1h\_binned & 21 \\
 & hashtags\_frequency\_1h\_binned & 21 \\
 & domains\_frequency\_48h\_binned & 21 \\
 & hashtags\_frequency\_12h\_binned & 21 \\
\end{longtable}

\subsection{Post-Hoc Test Statistics}
\label{appPostHocTestStatistics}

The tables below contain results of posthoc Wilcoxon signed-rank test with the Holm p-value adjustment. These tests were run for all cases where the Friedman test indicated that the null hypothesis for the factor can be rejected, as described in subsection \ref{secIndividualFactorsStatistics}.

\begin{longtable}[ht]{llrrrrlr}
\caption{Results of the post hoc Wilcoxon for algorithm and target like} \label{tabPostHoc-Friedman-PRAUC-like-algorithm-evaluated_on-algorithm} \\
A & B & mean(A) & mean(B) & dof & p-corr & BF10 & hedges \\
\endfirsthead
\caption[]{Results of the post hoc Wilcoxon for algorithm and target like} \\
A & B & mean(A) & mean(B) & dof & p-corr & BF10 & hedges \\
\endhead
\multicolumn{8}{r}{Continued on next page} \\
\endfoot
\endlastfoot
GradientBoosting & bayes & 0.6235 & 0.5829 & 62.0000 & 0.0000 & 1.129e+09 & 1.3437 \\
GradientBoosting & forest & 0.6235 & 0.6361 & 62.0000 & 0.0000 & 2.047e+05 & -0.6849 \\
GradientBoosting & tree & 0.6235 & 0.6171 & 62.0000 & 0.0056 & 5.698 & 0.3346 \\
bayes & forest & 0.5829 & 0.6361 & 62.0000 & 0.0000 & 1.441e+14 & -1.6524 \\
bayes & tree & 0.5829 & 0.6171 & 62.0000 & 0.0000 & 2.366e+13 & -1.0480 \\
forest & tree & 0.6361 & 0.6171 & 62.0000 & 0.0000 & 1.262e+07 & 0.8585 \\
\end{longtable}

\begin{longtable}[ht]{llrrrrlr}
\caption{Results of the post hoc Wilcoxon for algorithm and target like} \label{tabPostHoc-Friedman-RCE-like-algorithm-evaluated_on-algorithm} \\
A & B & mean(A) & mean(B) & dof & p-corr & BF10 & hedges \\
\endfirsthead
\caption[]{Results of the post hoc Wilcoxon for algorithm and target like} \\
A & B & mean(A) & mean(B) & dof & p-corr & BF10 & hedges \\
\endhead
\multicolumn{8}{r}{Continued on next page} \\
\endfoot
\endlastfoot
GradientBoosting & bayes & 0.6235 & 0.5829 & 62.0000 & 0.0000 & 1.129e+09 & 1.3437 \\
GradientBoosting & forest & 0.6235 & 0.6361 & 62.0000 & 0.0000 & 2.047e+05 & -0.6849 \\
GradientBoosting & tree & 0.6235 & 0.6171 & 62.0000 & 0.0056 & 5.698 & 0.3346 \\
bayes & forest & 0.5829 & 0.6361 & 62.0000 & 0.0000 & 1.441e+14 & -1.6524 \\
bayes & tree & 0.5829 & 0.6171 & 62.0000 & 0.0000 & 2.366e+13 & -1.0480 \\
forest & tree & 0.6361 & 0.6171 & 62.0000 & 0.0000 & 1.262e+07 & 0.8585 \\
\end{longtable}

\begin{longtable}[ht]{llrrrrlr}
\caption{Results of the post hoc Wilcoxon for algorithm and target reply} \label{tabPostHoc-Friedman-PRAUC-reply-algorithm-evaluated_on-algorithm} \\
A & B & mean(A) & mean(B) & dof & p-corr & BF10 & hedges \\
\endfirsthead
\caption[]{Results of the post hoc Wilcoxon for algorithm and target reply} \\
A & B & mean(A) & mean(B) & dof & p-corr & BF10 & hedges \\
\endhead
\multicolumn{8}{r}{Continued on next page} \\
\endfoot
\endlastfoot
GradientBoosting & bayes & 0.3007 & 0.3525 & 62.0000 & 0.0001 & 254.997 & -0.9496 \\
GradientBoosting & forest & 0.3007 & 0.5043 & 62.0000 & 0.0000 & 8.392e+34 & -4.3333 \\
GradientBoosting & tree & 0.3007 & 0.4327 & 62.0000 & 0.0000 & 6.063e+35 & -2.2616 \\
bayes & forest & 0.3525 & 0.5043 & 62.0000 & 0.0000 & 2.406e+29 & -4.7327 \\
bayes & tree & 0.3525 & 0.4327 & 62.0000 & 0.0000 & 1.997e+07 & -1.6985 \\
forest & tree & 0.5043 & 0.4327 & 62.0000 & 0.0000 & 2.948e+17 & 1.8790 \\
\end{longtable}%

\end{appendices}


\backmatter

\sloppy
\listoffigures 

\cleardoublepage 
\listoftables 
\fussy 


\printindex

\printglossaries

\bibliographystyle{alpha}
\bibliography{thesis}

\end{document}